UNIVERSITÀ DEGLI STUDI DI MILANO

Scuola di Dottorato in Fisica, Astrofisica e Fisica Applicata
Dipartimento di Fisica

Corso di Dottorato in Fisica, Astrofisica e Fisica Applicata
Ciclo XXVI

# A microscopic particle-vibration coupling approach for atomic nuclei

## Giant resonance properties and the renormalization of the effective interaction

Settore Scientifico Disciplinare FIS/04

Supervisore: Professor Gianluca COLÒ

Coordinatore: Professor Marco BERSANELLI

Tesi di Dottorato di:
Marco BRENNA

**Commission of the final examination:**

External Referee:
Professor Elias KHAN
Institut de Physique Nucléaire, Université Paris-Sud, IN2P3-CNRS, F-91406 Orsay Cedex, France

External Member:
Professor Santo LUNARDI
Dipartimento di Fisica e Astronomia, Università di Padova and INFN Sezione di Padova, I-35131 Padova, Italy

Internal Member:
Professor Gianluca COLÒ
Dipartimento di Fisica, Università degli Studi di Milano and INFN Sezione di Milano, I-20133 Milano, Italy

**Final examination:**

Date 07-03-2014

Università degli Studi di Milano, Dipartimento di Fisica, Milano, Italy

# Contents

# Introduction

The atomic nucleus is a finite many-body system governed by the rules of quantum mechanics. To a first approximation it can be described as composed by independent particles, moving in an average potential generated by all the nucleons. This is the fundamental idea of the self-consistent mean-field (SCMF) methods [BHR03]. This approach can produce good results for bulk nuclear properties (e.g. masses, radii and deformations), covering almost the whole chart of nuclides, including the super-heavy and super-deformed regions. The SCMF models are in many respects analogs of the density functional theory (DFT), which gives rather successful results in condensed matter physics. Nuclear SCMF models employ effective interactions which are adjusted by extensive fits to nuclear structure data. These effective interactions can be derived either in a non-relativistic or in a covariant framework, and they have become increasingly sophisticated in the last decade. Throughout this thesis, we adopt the non-relativistic zero-range Skyrme interaction.

The time dependent extension of stationary mean-field models, or time dependent DFT, is formally straightforward. The linearization of the time-dependent mean-field equations bring to the Random Phase Approximation (RPA) equations. Those describe the dynamics of the nucleus as a whole, as a coherent superposition of one-particle –one-hole ($1p-1h$) excitations. In particular, the RPA is one of the most successful theories for the description of the main observables related to the nuclear excitations in the energy range of giant resonances (GR), i.e. between 10 and 30 MeV. The main feature of these collective states is that they can absorb a considerable part of the maximum energy that can be transferred to the system. Accordingly, the macroscopic picture of a giant resonance is often thought to be that of a coherent motion of all the nucleons. These states have finite lifetime and consequently they carry a width of few MeV. The particle-hole states that form the resonance can decay directly into the continuum, producing a free particle and the $A-1$ nucleus in the corresponding hole state. The resonances in a heavy nucleus typically lie in a region of very high level density, thus the simple $1p-1h$ states of the resonance are mixed into the more complex $np-nh$ states which exist at the same excitation energy, resulting in a redistribution of the energy and the formation of a compound nucleus. This damping mechanism, associated to the so-called spreading width, is the most important. Another possible decay process is the $\gamma$ decay, which in heavy nuclei occurs only about once in e3 decays, and because of that it has not been studied extensively in the past [Bee+89; Bee+90]. Recently, the study of this decay branch has been performed in different nuclei with the demonstrator of the segmented germanium array AGATA [Nic+11; BC12] at the Laboratori Nazionali di Legnaro (LNL) of INFN. The interest in the $\gamma$ decay lies in the fact that it is extremely sensitive to the resonance multipolarity and that the electromagnetic interaction is well-known, so the direct electromagnetic deexcitation of resonances offers the possibility of a determination of the resonance strength complementary to that provided by the analysis of inelastic scattering data and free from the ambiguities associated with them. These latter allow extracting the properties of the collective states only through an analysis involving poorly known quantities (the reaction mechanism, the choice of the optical potentials, etc.).

The study of the GRs has been important so far also to give an estimate of some important parameters of the equation of state of nuclear matter. In particular, the determination of the equation of state of isospin-asymmetric nuclear matter, i.e., nuclear matter in which the number of neutrons exceeds the number of protons, is one of the most outstanding problems nowadays. Specifically, it is of paramount importance to constrain the properties of the symmetry energy, which is an estimate of the energy cost to convert a proton into a neutron in symmetric nuclear matter. The knowledge of the nuclear symmetry energy, and in particular of its density dependence, is essential for solving many problems in nuclear physics, such as the dynamics of heavy-ion collisions and the structure of exotic nuclei (their masses, neutron and proton density distributions, mean radii, collective excitations). Besides, the symmetry energy deeply influences a number of important issues in astrophysics, such as the mass-radius relation of neutron stars, the nucleosynthesis during pre-supernova evolution of massive stars, the cooling of proto-neutron stars and the fractional moment of inertia of the neutron star crust.

In the first part of this thesis, we study two different nuclear excitations and their correlation with the symmetry energy: the pygmy dipole resonance (PDR) and the isovector giant quadrupole resonance (IVGQR). The former is a low-energy peak which is found in the dipole response of several neutron-rich nuclei, and recently it has been claimed that it may be correlated to the density dependence of the symmetry energy. In particular, we analyze the isoscalar and isovector RPA spectrum in a stable ($^{208}$Pb) and two unstable ($^{68}$Ni and $^{132}$Sn) systems with three Skyrme interactions characterized by different isovector properties. The spurious state is carefully subtracted from the particle-hole contributions of the RPA states. The nature of the pygmy states is studied by means of the transition densities. Eventually, to assess the collectivity of the pygmy, the number and the coherence of the particle-hole excitations in the RPA states are considered.

The second excitation taken into account, i.e. the isovector giant quadrupole resonance, has remained elusive for quite a long time because of the lack of selective experimental probes that can excite this resonance: only few years ago, a new experimental technique has been implemented at the HI$\vec{\gamma}$S facility at Duke University [Hen+11]. In our work, the predictions for the strength and the transition densities of this mode in $^{208}$Pb from three Skyrme interactions and two relativistic Lagrangians are exposed. Moreover, starting from a macroscopic model, a combination of the energies of the isoscalar and isovector giant quadrupole resonances are found to be correlated with the symmetry energy at the density of $0.1\,\mathrm{fm}^{-1}$. This correlation is analyzed in detail using two families (one non-relativistic and one relativistic) of systematically varied functionals. This allows us to extract a prediction for the slope of the symmetry energy and for the neutron skin thickness in $^{208}$Pb.

However, the SCMF approaches present well-known limitations. For instance, they do not reproduce, as a rule, the level density around the Fermi energy. Moreover, the fragmentation of single-particle states and the GR spreading widths, as well as the width related to $\gamma$ decay to excited states, are outside the framework of these models. To improve further on the quality of the description of the nuclear structure, one of the route that can be undertaken consists in introducing correlations beyond the mean-field, allowing the interweaving between the single-particle degrees of freedom and the dynamics of the mean-field, i.e. the phonons.

In the second part of this thesis, we adopt the nuclear field theory (NFT), which is based on the particle-vibration coupling (PVC) idea. NFT provides us with a consistent and perturbative framework in which phonons and single-particle degrees of freedom are considered as the relevant independent building blocks of the low-lying spectrum of nuclei. In particular, we apply a recently-developed consistent model for the particle-vibration coupling, using the complete Skyrme interaction at the particle-vibration vertex. In this model, the single-particle states are obtained from a Hartree-Fock procedure, while the collective phonons are computed in fully self-consistent RPA. First we implement this model to compute an inclusive observable related to GRs, viz. the spreading width of the isoscalar and isovector giant quadrupole resonances in $^{208}$Pb. We include in the calculation the lowest order contributions beyond the mean-field, namely two-particles –two-holes contributions. Then, we examine a more exclusive observable, specifically the direct $\gamma$ decay of the isoscalar giant quadrupole resonance in $^{208}$Pb and $^{90}$Zr to the ground state and to the first collective octupole state, taking into account properly the polarization of the nuclear medium.

The main limitation of our beyond mean-field model consists in the use of effective interactions fitted at mean-field level to some experimental data. These interactions already comprise in an effective and uncontrolled way some beyond mean-field correlations. This fact may introduce an overcounting of the correlations. Therefore, one should aim at refitting the interaction at the desired level of approximation. Moreover, if zero-range interactions, like the Skyrme force, are used, divergences arise, which should be handled in refitting the interaction. This procedure has been carried out successfully in nuclear matter, with different degrees of local neutron-proton asymmetry, including a cutoff of the transferred momentum between particles. It has been shown that it is possible to obtain a new set of Skyrme parameters, fitted to the total energy of the system up to second order beyond the mean-field, for each chosen momentum cutoff.

In the third part of this thesis, we want to investigate whether it is possible to use these interactions in finite systems to compute the total energy at second order level. In this part we use a simplified Skyrme interaction, with only the velocity independent terms. Since the lack of translational invariance prevents us from defining a transferred momentum in a nucleus, it is necessary to introduce the cutoff on the relative momenta of the interacting particles in the initial and final channels. Therefore, we first perform the calculation of the total energy up to second order in nuclear matter using this new representation in order to determine the relation between the cutoff on the transferred momentum and the one on the relative momentum. The inclusion of the cutoff on relative momenta is performed in the following way: we Fourier transform the interaction written in the center of mass and relative motion coordinate system, we multiply it by two step functions in the relative momentum (one for the initial and one for the final state), and then we Fourier anti-transform the resulting expression. In this way the interaction acquire a finite range. Also the Hartree-Fock wave functions, used in the computation of the matrix elements of the interaction, have to be written in the center of mass and relative motion coordinate system by means of the Brody-Moshinsky transformations. This procedure is then applied to the computation of the second order total energy in $^{16}$O.

The thesis is organized as follows. In chapter 1, the standard formulation of the Skyrme interaction is presented.

Part I, comprising chapters 2, 3, 4, 5, and 6, is devoted to the mean-field approach. In particular, the main feature of the Hartree-Fock and Random Phase Approximation are sketched in chapters 2 and 3, respectively. Chapter 4 is an introduction to the equation of state of nuclear matter, focusing the attention in particular on the symmetry energy.

Chapter 5 gives an overview on the experimental properties of giant resonances, explaining more in detail the ones that are useful for this thesis and their relation with the parameters of the equation of state of nuclear matter. Finally, in chapter 6 our theoretical predictions for the pygmy dipole resonance and for the isovector giant quadrupole resonance are presented; moreover, a study on the relation between these excited states and the symmetry energy is included.

In part II, including chapters 7, 8 and 9, the extension to beyond mean-field models for nuclear structure is discussed. In chapter 7 the tool of the Green's functions is presented and the fundamental concepts of the nuclear field theory are exposed. In chapter 8 the nuclear field theory is used to compute the spreading width of GRs and the width related to the direct $\gamma$ decay of GRs to the ground state of the nucleus or to a low-lying state. Eventually, chapter 9 collects our results for the spreading width of the isoscalar and isovector giant quadrupole resonances in $^{208}$Pb and the $\gamma$-decay width of the isoscalar giant quadrupole resonances in $^{208}$Pb and $^{90}$Zr to the ground state and to the first collective octupole states.

In part III, we detail the problems occurring in beyond mean-field theories when zero-range interactions are used. We briefly present the work done in nuclear matter and the machinery used to treat the issue of divergences in nuclei (chapter 10). Then, in chapter 11 we report on the present first applications of refitted interactions in $^{16}$O.

Chapter 1

# The Skyrme interaction

The study of the atomic nucleus is complicated by the poor knowledge of the nuclear force, due to its complexity. Two main directions have been followed so far. In the former, a bare nucleon-nucleon (NN) interaction, with the addition of a three-nucleon interaction, is used to build the nucleus in an *ab initio* procedure. In the latter, the fact that the nucleons within a nucleus, due to the presence of the many other nucleons, do not feel the bare NN interaction is exploited. Thus, it is possible to introduce an effective NN force, in which all the difficulties and subtleties of the nuclear force is included in few parameters that must be adjusted to a selected set of experimental data. Among them, the most commonly used are the Gogny [DG80] and the Skyrme [Sky56; Sky59a; Sky59b] interactions.

Since in the following we will use always the Skyrme interaction, here we summarize its properties. The generic form of the effective interaction is

$$V = \sum_{i<j} V_{12}(\boldsymbol{r}_i, \boldsymbol{r}_j) + \sum_{i<j<k} V_{123}(\boldsymbol{r}_i, \boldsymbol{r}_j, \boldsymbol{r}_k),$$

where $V_{12}$ represent a two-body interaction and $V_{123}$ is a three-body interaction. The idea which lies at the basis of the Skyrme interaction is that *"it is generally believed that the most important part of the two-body interaction can be represented by a contact potential"* [Sky59a]. Thus,

$$V_{12}(\boldsymbol{r}_i, \boldsymbol{r}_j) = V_{12}(\boldsymbol{k}, \boldsymbol{k}')\, \delta_3\,(\boldsymbol{r}_i - \boldsymbol{r}_j) \qquad V_{123}(\boldsymbol{r}_i, \boldsymbol{r}_j, \boldsymbol{r}_k) = t_3\, \delta_3\,(\boldsymbol{r}_i - \boldsymbol{r}_j)\, \delta_3\,(\boldsymbol{r}_j - \boldsymbol{r}_k),$$

where $t_3$ is a constant, $\boldsymbol{k} = \frac{1}{2i}(\nabla_1 - \nabla_2)$ is the relative momentum of particle 1 and 2 placed on the right of the delta function, while $\boldsymbol{k}'$ denotes the same operator placed on the left. The form of $V_{12}$ has been chosen to be a polynomial expansion in powers of $\boldsymbol{k}$ and $\boldsymbol{k}'$.

It is important, for the last part of this work, to recall that *"this form is unrealistic for large momentum transfers, so that it is not suitable for the discussion of second-order effects, unless some momentum cut-off is introduced"* [Sky59a].

The Skyrme interaction has been improved and developed in the last years. At now, its most widely used form is,

$$\begin{aligned} V_{\text{eff}}(\boldsymbol{r}_1, \boldsymbol{r}_2) = {} & t_0(1 + x_0 P_\sigma)\delta(\boldsymbol{r}) + \frac{1}{2} t_1(1 + x_1 P_\sigma)[\boldsymbol{k}'^2 \delta(\boldsymbol{r}) + \delta(\boldsymbol{r})\boldsymbol{k}^2] \\ & + t_2(1 + x_2 P_\sigma)\boldsymbol{k}' \cdot \delta(\boldsymbol{r})\boldsymbol{k} + \frac{1}{6} t_3(1 + x_3 P_\sigma)\rho^\alpha(\boldsymbol{R})\delta(\boldsymbol{r}) \\ & + iW_0(\boldsymbol{\sigma}_1 + \boldsymbol{\sigma}_2) \cdot [\boldsymbol{k}' \times \delta(\boldsymbol{r})\boldsymbol{k}] \end{aligned} \tag{1.1}$$

**Table 1.1**
Nuclear matter properties as predicted by the Skyrme interactions used in this work, with the bibliographical information.

| Interaction | Ref. | $\rho_\infty$[fm$^{-3}$] | $k_F$[fm$^{-1}$] | $m^*/m$ | $J$ [MeV] | $L$ [MeV] |
|---|---|---|---|---|---|---|
| SIII | [Bei+75] | 0.145 | 1.291 | 0.763 | 28.16 | 9.95 |
| SGII | [VS81] | 0.158 | 1.328 | 0.786 | 26.83 | 37.63 |
| SKM* | [Bar+82] | 0.160 | 1.334 | 0.789 | 30.03 | 45.78 |
| SkP | [DFT84] | 0.162 | 1.340 | 1.000 | 30.00 | 19.68 |
| SkI3 | [RF95] | 0.158 | 1.327 | 0.577 | 34.83 | 100.52 |
| SLy5 | [Cha+98] | 0.160 | 1.333 | 0.697 | 32.02 | 48.27 |
| BSk1 | [Sam+02] | 0.157 | 1.325 | 1.050 | 27.81 | 7.19 |
| SK272 | [ASK03] | 0.155 | 1.320 | 0.773 | 37.39 | 91.67 |
| SK255 | [ASK03] | 0.157 | 1.325 | 0.797 | 37.40 | 95.05 |
| KDE | [ASA05] | 0.164 | 1.345 | 0.756 | 31.97 | 41.42 |
| LNS | [Cao+06] | 0.175 | 1.372 | 0.826 | 33.43 | 61.45 |
| SAMi | [RCS12] | 0.159 | 1.329 | 0.675 | 28.12 | 43.56 |

where $\boldsymbol{r}_i$ and $\boldsymbol{\sigma}_i$ are the space and spin variables of the two nucleons, $\boldsymbol{r} = \boldsymbol{r}_1 - \boldsymbol{r}_2$, $\boldsymbol{R} = \frac{1}{2}(\boldsymbol{r}_1+\boldsymbol{r}_2)$ and $P_\sigma = \frac{1}{2}(1+\boldsymbol{\sigma}_1\cdot\boldsymbol{\sigma}_2)$ is the spin-exchange operator [Cha+97]. Here, $\rho = \rho_n+\rho_p$ is the total nucleon density, and we will use the notation $\rho_q$ to distinguish the neutron ($q = 0$) and proton ($q = 1$) densities. In Eq. (1.1), $t_0$ and $t_3$ are velocity-independent terms, whereas $t_1$ and $t_2$ are velocity-dependent terms; $W_0$ term is a two-body spin-orbit force. In particular, the $t_0$ term is attractive, $t_1$ and $t_2$ terms mimic a finite range and $t_3$ term is repulsive. In the Hartree-Fock approximation, the contribution of the three-body contact term is the same as that given by the $t_3$ term of Eq. (1.1) when $x_3 = 1$ and $\alpha = 1$. Such term provides a simple phenomenological representation of many-body effects, and describes the way in which the interaction between two nucleons is influenced by the presence of others.

The parameters $t_0$, $t_1$, $t_2$, $t_3$, $x_0$, $x_1$, $x_2$, $x_3$ and $\alpha$ are free parameters which can be obtained by fitting both experimental data in a restricted number of nuclei, like binding energies and r.m.s. radii, and theoretical properties of nuclear matter, such as the saturation density $\rho_0$, the incompressibility $K_\infty$ and the symmetry energy at saturation density, $E_{\rm sym}(\rho_0)$.

For the sake of completeness, in Table 1.1 we list the nuclear matter properties interesting for this work as predicted by the Skyrme parameterizations which we will use in the following, together with the references where these parameterizations can be found.

# Part I

# The mean-field approach

Chapter 2

# The static mean-field

The simplest method which can be used to treat a many-body system consists in the reduction of its dynamics to that of independent particles moving in a self-consistent potential. Schematically, we obtain a solvable many-particle problem by considering the following decomposition of the original Hamiltonian

$$\mathcal{H} = T + V = (T + U) + (V - U) = \mathcal{H}_0 + \mathcal{H}_1, \tag{2.1}$$

with $U$ a suitably chosen one-body operator. The mean-field approximation consists in considering $\mathcal{H}_0$, treating $\mathcal{H}_1$ as a perturbation.

The nuclear shell model lays on this basic idea. The first formalization of this model in its simplest non-interacting version was introduced in 1949 by Goeppert-Mayer [Goe49; Goe50], Haxel [HJS49], and since then much work has been done to refine it.

Qualitatively, the reason why the shell model is so reliable is the fact that the nucleus is not a very dense system, thanks to the action of the Pauli principle. Since the nucleus is a self-interacting system, the shape of the one-body potential in which the nucleons move should be determined in a self-consistent way from a microscopic nucleon-nucleon interaction. Nevertheless, for several years a simpler approach has been used, introducing an *ad hoc* potential. This potential must satisfy some characteristics:

- A nucleon in the center of the nucleus should not feel any net force
- The force should become stronger approaching the surface
- The force is short range

Good results can be obtained assuming an analytical ansatz for the potential, known as Woods-Saxon

$$V_{W-S}(r) = -\frac{V_0}{1 + e^{\frac{r-R_0}{a}}}, \tag{2.2}$$

where $R_0 = r_0 A^{1/3} = 1.2\,A^{1/3}$ fm, $a$ is the diffusivity and $V_0$ is the depth of the potential. Commonly used values are $V_0 = -50\,\text{MeV}$ and $a = 0.5\,\text{fm}$.

The experimental findings show that: (1) the single-particle states in which the spin of the particle is parallel to its orbital angular momentum lie at lower energy with respect to the corresponding state in which the spin is anti-parallel to the angular momentum, and (2) this splitting can be rather large. For this reason, to obtain the correct level ordering, a strong attractive spin-orbit interaction should be included in the Hamiltonian of the system. From the diagonalization of such an Hamiltonian, one could get the energy levels, unidentified by the principal quantum numbers $n$ (related to the number of nodes of the wave function), the orbital angular momentum $l$ and the total angular momentum $\boldsymbol{j} = \boldsymbol{l} + \boldsymbol{s}$,where $\boldsymbol{s}$ is the spin of the nucleon.

## 2.1 The self-consistent mean-field

On a microscopic basis, the mean-field approximation is realized in a self-consistent way solving the Hartree-Fock (HF) equations. This approach will be referred to as self-consistent mean-field (SCMF). In this section we sketch this method, addressing the interested reader to Refs. [BR86; BHR03; RS04; Suh07]. Mean-field methods predict ground state properties of nuclei with increasing accuracy as the system becomes larger. Thus, they are particularly useful for describing masses, radii, shapes and deformations of medium-heavy nuclei. In the following we will concentrate on spherical, closed-shell nuclei, in order to avoid complications coming from deformations and pairing correlations, since are beyond the scope of this thesis.

### 2.1.1 Variational derivation

The HF equations can be obtained using variational principles. Let $E[\psi]$ be the (normalized) expectation value in a state $|\psi\rangle$ of the Hamiltonian $\mathcal{H}$. The Ritz variational principle states that any state making the functional $E[\psi]$ stationary is an eigenstate of $\mathcal{H}$ with eigenvalue $E$. If $E_0$ and $|\psi_0\rangle$ are the ground state energy and wave function of the Hamiltonian $\mathcal{H}$, then $E[\psi] \leq E[\psi_0]$. The equality holds only if $|\psi\rangle \equiv |\psi_0\rangle$. We then build a trial wave function in a given class of wave functions. The trial wave function can be determined by making the energy expectation value stationary with respect to infinitesimal variations of $|\psi\rangle$ in the class of wave functions to which $|\psi\rangle$ belongs. The HF approximation consists in assuming that the trial wave function is a Slater determinant

$$|\psi\rangle = \left( \prod_{i=1}^{n} c^\dagger_{\alpha_i} \right) |0\rangle. \tag{2.3}$$

Actually, the use of the single-particle matrix associated to the state $|\psi\rangle$, defined as

$$\rho_{ij} = \langle i|\rho|j\rangle = \langle\psi|c^\dagger_j c_i|\psi\rangle, \tag{2.4}$$

is more convenient. This operator is Hermitian, positive definite and $\mathrm{Tr}\,\rho = A$.

It can be shown [RS04] that $|\psi\rangle$ is a Slater determinant if and only if the corresponding density matrix $\rho$ satisfy the equation

$$\rho^2 = \rho. \tag{2.5}$$

This means that the density operator is a projector onto the space spanned by the occupied orbitals. Let $\mathcal{H}$ be the second quantization form of the Hamiltonian of the system

$$\mathcal{H} = \sum_{\alpha\beta} \langle\alpha|T|\beta\rangle c^\dagger_\alpha c_\beta + \frac{1}{4} \sum_{\alpha\beta\gamma\delta} \langle\alpha\beta|\bar{V}|\gamma\delta\rangle c^\dagger_\alpha c^\dagger_\beta c_\delta c_\gamma. \tag{2.6}$$

Using the Wick theorem, it is possible to write the energy as a functional of $\rho$

$$E[\rho] = \langle\psi|\mathcal{H}|\psi\rangle = \sum_{\alpha\beta} \langle\alpha|T|\beta\rangle \rho_{\beta\alpha} + \frac{1}{2} \sum_{\alpha\beta\gamma\delta} \langle\alpha\beta|\bar{V}|\gamma\delta\rangle \rho_{\gamma\alpha}\rho_{\delta\beta}. \tag{2.7}$$

We now want to minimize the energy functional with respect to the variations of $\rho$

$$\delta \left[ E[\rho] - \mathrm{Tr}\,\Lambda(\rho^2 - \rho) \right] = 0, \tag{2.8}$$

where the matrix $\Lambda$ is a matrix of Lagrange multipliers introduced because of the constraint on $\rho$ (2.5). Defining the HF Hamiltonian as

$$h_{\alpha\beta} = \langle\alpha|h|\beta\rangle = \frac{\partial E[\rho]}{\partial\rho_{\beta\alpha}} = \langle\alpha|T|\beta\rangle + \sum_{\gamma\delta}\langle\alpha\gamma|\bar{V}|\beta\delta\rangle\rho_{\delta\gamma}, \tag{2.9}$$

the variational equation (2.8) leads to the so-called HF equations

$$[h,\rho] = 0 \tag{2.10}$$

$$h|\varphi_\nu\rangle = \varepsilon_\nu|\varphi_\nu\rangle \tag{2.11}$$

Thus, the HF equations can be solved by first diagonalizing $h$ and then constructing $\rho$ from the eigenstates of $h$. The main issue of this problem is the fact that the Hamiltonian depends on the density matrix, so the HF equations are non-linear. The total energy of the system, therefore, becomes

$$E^{\mathrm{HF}} = \sum_{i=1}^{A} t_{ii} + \frac{1}{2}\sum_{ij=1}^{A}\bar{v}_{ijij} = \sum_{i=1}^{A}\varepsilon_i - \frac{1}{2}\sum_{ij=1}^{A}\bar{v}_{ijij} \tag{2.12}$$

The HF equations can be written in the coordinate space

$$-\frac{\hbar^2}{2m}\nabla^2\varphi_\alpha(\boldsymbol{x}) + v_H(\boldsymbol{x})\varphi_\alpha(\boldsymbol{r}) - \int \mathrm{d}_3 x' v_F(\boldsymbol{x}',\boldsymbol{x})\varphi_\alpha(\boldsymbol{x}') = \varepsilon_\alpha\varphi_\alpha(\boldsymbol{x}) \tag{2.13}$$

$$v_H(\boldsymbol{x}) = \sum_{\substack{\beta \\ \varepsilon_\beta<\varepsilon_F}} \int \mathrm{d}_3\boldsymbol{r}'\varphi^\dagger_\beta(\boldsymbol{x}')v(\boldsymbol{x}',\boldsymbol{x})\varphi_\beta(\boldsymbol{x}')$$

$$v_F(\boldsymbol{x}) = \sum_{\substack{\beta \\ \varepsilon_\beta<\varepsilon_F}} \varphi^\dagger_\beta(\boldsymbol{x}')v(\boldsymbol{x}',\boldsymbol{x})\varphi_\beta(\boldsymbol{x})$$

The first term with the local potential $v_H(\boldsymbol{x})$ is called the Hartree or direct term, and the second term with the non-local potential $v_F(\boldsymbol{x}',\boldsymbol{x})$ is called the Fock or exchange term.

### Density dependent interactions

The effective interaction $V(\rho)$ used in the mean-field may depend on the local density of particles, as it is the case in nuclear physics. The energy expectation value then becomes:

$$E[\rho] = \sum_{\alpha\beta}\langle\alpha|T|\beta\rangle\rho_{\beta\alpha} + \frac{1}{2}\sum_{\alpha\beta\gamma\delta}\langle\alpha\beta|\bar{V}(\rho)|\gamma\delta\rangle\rho_{\gamma\alpha}\rho_{\delta\beta}. \tag{2.14}$$

The HF Hamiltonian calculated from Eq. (2.9) is

$$h_{\alpha\beta} = \langle\alpha|T|\beta\rangle + \sum_{\gamma\delta}\langle\alpha\gamma|\bar{V}(\rho)|\beta\delta\rangle\rho_{\delta\gamma} + \frac{1}{2}\sum_{\gamma\delta\eta\zeta}\langle\eta\zeta|\frac{\partial\bar{V}(\rho)}{\partial\rho_{\beta\alpha}}|\gamma\delta\rangle\rho_{\gamma\eta}\rho_{\delta\zeta}, \tag{2.15}$$

where the extra term is called rearrangement term $U_R$.

### 2.1.2 Derivation with the Wick's theorem

The Hartree-Fock equation can also be derived by applying Wick's theorem in the particle-hole representation. This means that we use normal ordering and contraction with respect to the particle-hole vacuum $|\mathrm{HF}\rangle$. However, instead of using the particle-hole representation of the Hamiltonian, we use its particle representation. The Hamiltonian (2.6) can be reduced to

$$\mathcal{H} = \sum_{\alpha\beta}\Bigg(t_{\alpha\beta} + \sum_{\substack{\gamma \\ \varepsilon_\gamma \leq \varepsilon_F}} \bar{v}_{\gamma\alpha\gamma\beta}\Bigg) c^\dagger_\alpha c_\beta - \frac{1}{2}\sum_{\substack{\alpha\beta \\ \varepsilon_{\alpha,\beta} \leq \varepsilon_F}} \bar{v}_{\alpha\beta\alpha\beta} + \frac{1}{4}\sum_{\alpha\beta\gamma\delta} \bar{v}_{\alpha\beta\gamma\delta} : c^\dagger_\alpha c^\dagger_\beta c_\delta c_\gamma :, \tag{2.16}$$

where $: \ldots :$ indicates the normal product. The first term is clearly a one-body operator. To get the HF equations we have to diagonalize this term

$$t_{\alpha\beta} + \sum_{\substack{\gamma \\ \varepsilon_\gamma \leq \varepsilon_F}} \bar{v}_{\gamma\alpha\gamma\beta} = \varepsilon_\alpha \delta_{\alpha\beta}. \tag{2.17}$$

This equation is exactly equal to Eq. (2.11). This method is particularly interesting because it produces automatically the expression for the residual interaction

$$V_{res} = \frac{1}{4}\sum_{\alpha\beta\gamma\delta} \bar{v}_{\alpha\beta\gamma\delta} : c^\dagger_\alpha c^\dagger_\beta c_\delta c_\gamma : \tag{2.18}$$

Also the calculation of the ground state energy is straightforward

$$E^{\mathrm{HF}} = \langle \mathrm{HF}|\mathcal{H}|\mathrm{HF}\rangle = \sum_{\substack{\alpha \\ \varepsilon_\alpha \leq \varepsilon_F}} \epsilon_\alpha - \frac{1}{2}\sum_{\substack{\alpha\beta \\ \varepsilon_{\alpha,\beta} \leq \varepsilon_F}} \bar{v}_{\alpha\beta\alpha\beta}.$$

### 2.1.3 The Hartree-Fock eigenvalue problem

Due to the non-locality of the Fock term, solving the HF equations in the coordinate space can be quite difficult and time consuming. For this reason, it is useful to convert them to an eigenvalue problem of a Hermitian matrix. To do so, the wave functions $\varphi_\alpha(\boldsymbol{r})$ are expanded in terms of some basis states, usually harmonic oscillator wave functions. Thus we seek solutions of the HF equation in the form

$$|\alpha\rangle = \sum_j c_{\alpha,j} |j\rangle, \tag{2.19}$$

where the basis $\{|j\rangle\}$ is orthonormal and complete. The coefficients $c_{\alpha,j}$ are determined by performing a variation of the Hartree-Fock ground-state energy $E^{\mathrm{HF}}$. In this procedure we seek the minimum of $E^{\mathrm{HF}}$ using the $c_{\alpha,j}$ as variational parameters. The variational condition yields

$$\frac{\partial}{\partial c^*_{\alpha,j}}\left[E^{\mathrm{HF}} - \sum_{\alpha' j'} \varepsilon_{\alpha'} c^*_{\alpha',j'} c_{\alpha',j'}\right] = 0, \tag{2.20}$$

where $\varepsilon_{\alpha'}$ are the Lagrange multipliers. Therefore, in the general case of density dependent interaction, we obtain

$$\sum_{j'} h^{(\alpha)}_{jj'} c_{\alpha,j'} = \varepsilon_\alpha c_{\alpha,j},$$

$$h^{(\alpha)}_{jj'} \equiv t_{jj'} + \sum_{\substack{\beta \\ \varepsilon_\beta \le \varepsilon_F}} \sum_{j_1 j_2} c^*_{\beta,j_1} \langle j j_1 | \bar{V} | j' j_2 \rangle c_{\beta,j_2} + \frac{1}{2} \sum_{\substack{\beta\gamma \\ \varepsilon_{\beta,\gamma} \le \varepsilon_F}} \sum_{\substack{j_1 j_2 \\ j_3 j_4}} c^*_{\beta,j_1} c^*_{\gamma,j_2} \langle j_1 j_2 | \frac{\partial \bar{V}}{\partial c^*_{\alpha,j}} | j_3 j_4 \rangle c_{\beta,j_3} c_{\gamma,j_4}, \tag{2.21}$$

In this way we are able to convert the HF equation into an eigenvalue problem for the matrix $h$.

The non-linearity of the HF equation is present irrespective of the method of solution. In particular, the matrix equation must be solved iteratively since the solutions $c_{\beta,j}$ themselves are building blocks of the matrix $h$ to be diagonalized.

### 2.1.4 Derivation for the Skyrme interaction

#### Skyrme Energy Density Functional

For the Skyrme interaction, it is possible to write the energy of the system as the integral of an energy density that is local in the space coordinate

$$E[\rho] = \langle \psi | T + V_{\text{Skyrme}} | \psi \rangle = \int \mathrm{d}_3 r \mathcal{E}(\boldsymbol{r}). \tag{2.22}$$

Assuming that the subspace of occupied single-particle states is invariant under time reversal, i.e. that the nucleus is even-even, the energy density functional $\mathcal{E}(\boldsymbol{r})$ is an explicit function of the nucleon densities $\rho_q$, the kinetic energy $\tau_q$ and spin densities $\boldsymbol{J}_q$, where the label $q$ can indicate neutrons ($n$) or protons ($p$). These quantities depend in turn on the single-particle states $\varphi_i$,

$$\rho_q(\boldsymbol{r}) = \sum_{\alpha,\sigma} |\varphi_\alpha(r,\sigma,q)|^2 \tag{2.23}$$

$$\tau_q(\boldsymbol{r}) = \sum_{\alpha,\sigma} |\nabla\varphi_\alpha(r,\sigma,q)|^2 \tag{2.24}$$

$$\boldsymbol{J}_q(\boldsymbol{r}) = -i \sum_{\alpha,\sigma,\sigma'} \varphi^*_\alpha(r,\sigma,q) \left[\nabla\varphi_\alpha(r,\sigma',q) \times \langle \sigma | \boldsymbol{\sigma} | \sigma' \rangle\right], \tag{2.25}$$

and the sums are taken over all occupied single-particle states; besides $\tau = \tau_n + \tau_p$ and $\boldsymbol{J} = \boldsymbol{J}_n + \boldsymbol{J}_p$. In particular (we drop the dependence on r for simplicity),

$$\mathcal{E} = K + \mathcal{E}_0 + \mathcal{E}_3 + \mathcal{E}_{\text{eff}} + \mathcal{E}_{\text{fin}} + \mathcal{E}_{\text{so}} + \mathcal{E}_{\text{sg}} + \mathcal{E}_{\text{coul}} \tag{2.26}$$

where $K$ is the kinetic energy term, $\mathcal{E}_0$ is a zero-range term, $\mathcal{E}_3$ is the density-dependent term, $\mathcal{E}_{\text{eff}}$ is an effective mass term, $\mathcal{E}_{\text{fin}}$ is a finite range term, $\mathcal{E}_{\text{so}}$ is a spin-orbit term, $\mathcal{E}_{\text{sg}}$ is a term due to the tensor coupling with spin and gradient and $\mathcal{E}_{\text{coul}}$ is a Coulomb term.

Their expressions are [Cha+97]

$$
\begin{aligned}
K =& \frac{\hbar^2}{2m_p}\tau_p + \frac{\hbar^2}{2m_n}\tau_n \\
\mathscr{E}_0 =& \frac{1}{4}t_0\left[(2+x_0)\rho^2 - (2x_0+1)(\rho_p^2+\rho_n^2)\right] \\
\mathscr{E}_3 =& \frac{1}{24}t_3\rho^\alpha\left[(2+x_3)\rho^2 - (2x_3+1)(\rho_p^2+\rho_n^2)\right] \\
\mathscr{E}_{\text{eff}} =& \frac{1}{8}\left[t_1(2+x_1)+t_2(2+x_2)\right]\tau\rho \\
&+ \frac{1}{8}\left[t_2(2x_2+1)-t_1(2x_1+1)\right](\tau_p\rho_p+\tau_n\rho_n) \\
\mathscr{E}_{\text{fin}} =& \frac{1}{32}\left[3t_1(2+x_1)-t_2(2+x_2)\right](\nabla\rho)^2 \\
&- \frac{1}{32}\left[3t_1(2x_1+1)+t_2(2x_2+1)\right]\left[(\nabla\rho_p)^2+(\nabla\rho_n)^2\right] \\
\mathscr{E}_{\text{so}} =& \frac{1}{2}W_0\left[\boldsymbol{J}\cdot\nabla\rho + \boldsymbol{J}_p\cdot\nabla\rho_p + \boldsymbol{J}_n\cdot\nabla\rho_n\right] \\
\mathscr{E}_{\text{sg}} =& -\frac{1}{16}(t_1x_1+t_2x_2)J^2 + \frac{1}{16}(t_1-t_2)[J_p^2+J_n^2].
\end{aligned}
\tag{2.27}
$$

The Coulomb potential requires an approximation for the exchange term contributions in order to keep it local, since the Coulomb force has non-zero range; a local density approximation called the Slater approximation [Sla51] is then used to obtain

$$
\mathscr{E}_{\text{coul}}(\boldsymbol{r}) = \frac{e^2\rho_p(r)}{2}\int d_3r'\frac{\rho_p(\boldsymbol{r}')}{|\boldsymbol{r}-\boldsymbol{r}'|} - \frac{3e^2}{4}\left(\frac{3}{\pi}\right)^{1/3}\rho_p(\boldsymbol{r})^{4/3} \equiv \rho_p(\boldsymbol{r})V_c(\boldsymbol{r}) \tag{2.28}
$$

### Skyrme-Hartree-Fock Equations

Using the Skyrme interaction, it is difficult to express the energy of the system as a functional of the standard density matrix, since there is a dependence on $\tau$ and $\boldsymbol{J}$ (see Eq. (2.26) and (2.27)). In this case, it is necessary to perform the variation of the energy with respect to the single-particle wave functions $\varphi_k$, with the subsidiary condition that the $\varphi_i$ are normalized [VB72]

$$
\delta\left[E - \sum_i \varepsilon_i \int d_3r|\varphi_i(r)|^2\right] = 0. \tag{2.29}
$$

From Eq. (2.26),

$$
\delta E = \sum_q \int d_3r\left[\frac{\hbar^2}{2m_q^*(\boldsymbol{r})}\delta\tau_q(\boldsymbol{r}) + U_q(\boldsymbol{r})\delta\rho_q(\boldsymbol{r}) + \boldsymbol{W}_q(\boldsymbol{r})\cdot\delta\boldsymbol{J}_q(\boldsymbol{r})\right] \tag{2.30}
$$

where the coefficients of the variation are,

$$
\frac{\hbar^2}{2m_q^*} = \frac{\delta E}{\delta\tau_q} = \frac{\hbar^2}{2m_q} + \frac{1}{8}[t_1(2+x_1)+t_2(2+x_2)]\rho + \frac{1}{8}[t_2(2x_2+1)-t_1(2x_1+1)]\rho_q \tag{2.31}
$$

$$
\begin{aligned}
U_q = \frac{\delta E}{\delta \rho_q} =& \frac{1}{2} t_0[(2+x_0)\rho - (2x_0+1)\rho_q^2] \\
&+ \frac{1}{24} t_3 \alpha \rho^{\alpha-1}[(2+x_3)\rho^2 - (2x_3+1)(\rho_p^2+\rho_n^2)] \\
&+ \frac{1}{12} t_3 \rho^{\alpha}[(2+x_3)\rho - (2x_3+1)\rho_q] \\
&+ \frac{1}{8}[t_1(2+x_1) + t_2(2+x_2)]\tau \\
&+ \frac{1}{8}[t_2(2x_2+1) - t_1(2x_1+1)]\tau_q \\
&- \frac{1}{16}[3t_1(2+x_1) - t_2(2+x_2)]\nabla^2\rho + \frac{1}{16}[3t_1(2x_1+1) + t_2(2x_2+1)]\nabla^2\rho_q \\
&- \frac{1}{2} W_0[\nabla\cdot \boldsymbol{J} + \nabla\cdot \boldsymbol{J}_q] + \delta_{1q} V_c
\end{aligned}
\tag{2.32}
$$

$$
\boldsymbol{W}_q = \frac{\delta E}{\delta \boldsymbol{J}_q} = \frac{1}{2} W_0[\nabla\rho + \nabla\rho_q] - \frac{1}{8}(t_1x_1 + t_2x_2)\boldsymbol{J} + \frac{1}{8}(t_1 - t_2)\boldsymbol{J}_q . \tag{2.33}
$$

We then obtain that the single-particle wave functions $\varphi_i$ have to satisfy the following set of equations

$$
\left[-\nabla\cdot\frac{\hbar^2}{2m_q^*(\boldsymbol{r})}\nabla + U_q(\boldsymbol{r}) + \boldsymbol{W}_q(\boldsymbol{r})\cdot(-i)(\nabla\times\sigma)\right]\varphi_i(r,\sigma,q) = \varepsilon_i\varphi_i(r,\sigma,q) \tag{2.34}
$$

These equations, which are known as the Skyrme-Hartree-Fock (SHF) equations, although non-linear, involve only local potentials, and therefore can be solved in coordinate space. This is the major difference with the HF equations corresponding to finite range interactions which give rise to non-local potentials. We can identify $m_q^*(\boldsymbol{r})$ as an effective mass, $U_q(\boldsymbol{r})$ as a central potential and $\boldsymbol{W}_q(\boldsymbol{r})$ as a spin-orbit potential.

In the case of spherical symmetry, the Skyrme-Hartree-Fock equations can be simplified greatly into a set of one-dimensional differential equations in the radial coordinate $r$. In fact, in this case, the single-particle wave functions can be written as

$$
\varphi_{\alpha m}(r,\sigma,\tau) = i^l \frac{u_\alpha(r)}{r}\left[Y_l(\hat{r})\otimes\chi_{\frac{1}{2}}(\sigma)\right]_{jm}\chi_q(\tau), \tag{2.35}
$$

where $\alpha$ stands for the sets of quantum numbers $\alpha = q,n,l,j$. The phase $i^l$ has been introduced following the convention II of the appendix A of Ref. [Row70]. One can thus easily deduce the SHF radial equations, which read

$$
\begin{aligned}
&\frac{\hbar^2}{2m_q^*(r)}\left[-u_\alpha''(r) + \frac{l(l+1)}{r^2}u_\alpha(r)\right] - \frac{\mathrm{d}}{\mathrm{d}r}\left(\frac{\hbar^2}{2m_q^*(r)}\right)u_\alpha'(r) \\
&+\left\{U_q(r) + \frac{1}{r}\frac{\mathrm{d}}{\mathrm{d}r}\left(\frac{\hbar^2}{2m_q^*(r)}\right) + \left[j(j+1) - l(l+1) - \frac{3}{4}\right]\frac{W_q(r)}{r}\right\}u_\alpha(r) = \varepsilon_\alpha u_\alpha(r)
\end{aligned}
\tag{2.36}
$$

where $W_q(r)$ is defined reducing the spin orbit term $\boldsymbol{W}_q(r)\cdot(-i)(\nabla\times\sigma)$ to the usual form $\frac{1}{r}W_q(r)\boldsymbol{l}\cdot\boldsymbol{\sigma}$.

## 2.2 Symmetries and the Hartree-Fock field

Let $|\psi\rangle$ be an independent-particle state and let $\rho$ be the associated single-particle density matrix. Let us consider a unitary transformation in which the state $|\psi\rangle$ becomes

$$|\bar{\psi}\rangle = U|\psi\rangle.$$

We call $\bar{\rho}$ the density associated with the state $|\bar{\psi}\rangle$. It can be shown that $\bar{\rho}$ transforms as

$$\bar{\rho} = U\rho U^{\dagger}.$$

If the transformation produced by the operator $U$ leaves the Hamiltonian invariant, then $U^{\dagger}\mathcal{H}U = \mathcal{H}$ and one has

$$E[\bar{\rho}] = \langle\bar{\psi}|\mathcal{H}|\bar{\psi}\rangle = \langle\psi|U^{\dagger}\mathcal{H}U|\psi\rangle = E[\rho]. \tag{2.37}$$

Consider now an infinitesimal variation $\delta\rho$ of the density matrix, to which corresponds a variation $\delta\bar{\rho} = U\delta\rho U^{\dagger}$. It follows that $E[\bar{\rho} + \delta\bar{\rho}] = E[\rho + \delta\rho]$. Expanding both sides in powers of $\delta\rho$ and equating the first-order terms yields

$$\sum_{ij} \frac{\delta E}{\delta\rho_{ij}} \delta\rho_{ij} = \sum_{ij} \left.\frac{\delta E}{\delta\rho_{ij}}\right|_{\rho=\bar{\rho}} \delta\bar{\rho}_{ij} \tag{2.38}$$

which gives

$$\mathrm{Tr}\,(h[\rho]\delta\rho) = \mathrm{Tr}\,(h[\bar{\rho}]\delta\bar{\rho}) = \mathrm{Tr}\left(U^{\dagger}h[\bar{\rho}]U\delta\rho\right).$$

This equality holds for any $\delta\rho$, hence

$$h[\bar{\rho}] = Uh[\rho]U^{\dagger}. \tag{2.39}$$

Two cases may occur:

(i) The density matrix is invariant under the transformation $U$. In this case $\bar{\rho} = \rho$ and $h[\bar{\rho}] = h[\rho]$ commutes with $U$. We say that $U$ represents a self-consistent symmetry of the Hartree-Fock Hamiltonian.

(ii) The density matrix $\rho$ is not invariant under the transformation $U$, In this case, it may happen that the Hartree-Fock Hamiltonian does not commute with $U$. We then say that $U$ represents a broken symmetry. For example, translational symmetry is always broken by the Hartree-Fock potential of a bound finite system. A broken symmetry always leads to a degeneracy of the variational solutions. It should be emphasized that, in finite systems, broken symmetries arise only as a result of approximations. Typically, broken symmetries occur when the variational principle is applied with a too-restricted trial wave function.

Spherical symmetry is an important example of a self-consistent symmetry which is realized in a closed shell nucleus. In such a nucleus, each of the orbitals $|nljm\tau\rangle$ is occupied. A Slater determinant built from such a set of orbitals is an eigenstate of the total angular momentum with eigenvalue $J = 0$. It follows that the Hartree-Fock potential is spherically symmetric.

Chapter 3

# The time-dependent mean-field

The static SCMF approach discussed in chapter 2 is used to describe systems in equilibrium, that is, in their ground state or, more generally, in stationary states. Besides, the analysis of the spectra of nuclear excitations reveals a series of nuclear excited states which can be very adequately explained as single-particle independent excitations. But there are also many excited states with features that cannot be understood in terms of shell model excitations: these modes can only be explained if we suppose that a coherent participation by many nucleons takes place in the nucleus, resulting in a collective excitation of the system as a whole. Giant resonances (GRs) are a major example of such collective states (their main features will be discussed in chapter 5).

Such excitations can be thought as small-amplitude vibrations of the SCMF about an equilibrium state, after the action of an external perturbation. For this reason, in this chapter we study the time evolution of the SCMF of systems off equilibrium. As in the static case, mean-field approximations are obtained by assuming independent-particle states. In the following we will concentrate on spherical, closed-shell nuclei, in order to avoid complications coming from deformations and pairing correlations, since are beyond the scope of this thesis. For a wider explanation of these topics we refer to Refs. [Row70; BR86; RS04].

## 3.1 Random Phase Approximation

The Random Phase Approximation (RPA) is the simplest theory in which the ground state of the system is not purely HF but may contains correlations. For example, if the excited states are described as vibrational excitations, the ground state correlations may be associated with the vibrational zero-point motions.

Even though RPA plays an important role in the theory of nuclear collective motions (as can be clearly seen in its alternative formulation in terms of time-dependent Hartree-Fock theory, TDHF), this approach was developed for the first time by Bohm and Pines for the plasma oscillations of the electron gas [BP53]. In this theory, the parameters of the electromagnetic field, representing the interaction between the electrons, were quantized and treated as the collective coordinates of the plasma oscillations; the term Random Phase Approximation, as a matter of fact, refers to the neglect of the coupling between plasma vibrations of different momenta.

The Hamiltonian of the system is given in Eq. (2.16).

$$\mathcal{H} = \mathcal{H}_0 + V_{res} = \sum_{\alpha} \varepsilon_\alpha c^\dagger_\alpha c_\alpha - \frac{1}{2} \sum_{\substack{\alpha\beta \\ \varepsilon_{\alpha,\beta} \le \varepsilon_F}} \bar{v}_{\alpha\beta\alpha\beta} + \frac{1}{4} \sum_{\alpha\beta\gamma\delta} \bar{v}_{\alpha\beta\gamma\delta} : c^\dagger_\alpha c^\dagger_\beta c_\delta c_\gamma : . \tag{3.1}$$

The corresponding Schrödinger equation is given by

$$\mathcal{H} |\nu\rangle = E_\nu |\nu\rangle \tag{3.2}$$

where $\{|\nu\rangle\}$ is a set of exact eigenstates of the Hamiltonian $H$ and $\{E_\nu\}$ are the associated eigenvalues. It is possible to define operators $\Gamma_\nu^\dagger$ and $\Gamma_\nu$ in such a way that

$$|\nu\rangle = \Gamma_\nu^\dagger |0\rangle \qquad \text{and} \qquad \Gamma_\nu |0\rangle = 0.$$

From Eq. (3.2), we get the equation of motion

$$\left[H, \Gamma_\nu^\dagger\right] |0\rangle = (E_\nu - E_0)\, \Gamma_\nu^\dagger |0\rangle$$

and multiplying from the left with an arbitrary state of the form $\langle 0| \delta\Gamma$

$$\langle 0| \left[\delta\Gamma, \left[H, \Gamma_\nu^\dagger\right]\right] |0\rangle = (E_\nu - E_0) \langle 0| \left[\delta\Gamma, \Gamma_\nu^\dagger\right] \Gamma|0\rangle. \tag{3.3}$$

Since in RPA the true ground state is not simply the $ph$-vacuum, both $ph$-destruction and $ph$-creation operators are included in the space. Thus, $\Gamma_\nu^\dagger$ is expanded

$$\Gamma_\nu^\dagger = \sum_{ph} X_{ph}^\nu c_p^\dagger c_h - Y_{ph}^\nu c_h^\dagger c_p. \tag{3.4}$$

The RPA ground state is defined by

$$\Gamma_\nu |\text{RPA}\rangle = 0.$$

We have two different kinds of variations $\delta\Gamma|0\rangle$, i.e. $c_p^\dagger c_h|0\rangle$ and $c_h^\dagger c_p|0\rangle$. Thus, from Eq. (3.3), two sets of equations arise:

$$\langle \text{RPA}| \left[c_h^\dagger c_p, \left[H, \Gamma_\nu^\dagger\right]\right] |\text{RPA}\rangle = \hbar\omega_\nu \langle \text{RPA}| \left[c_h^\dagger c_p, \Gamma_\nu^\dagger\right] |\text{RPA}\rangle \tag{3.5a}$$

$$\langle \text{RPA}| \left[c_p^\dagger c_h, \left[H, \Gamma_\nu^\dagger\right]\right] |\text{RPA}\rangle = \hbar\omega_\nu \langle \text{RPA}| \left[c_p^\dagger c_h, \Gamma_\nu^\dagger\right] |\text{RPA}\rangle, \tag{3.5b}$$

where $\hbar\omega_\nu$ is the excitation energy $\hbar\omega_\nu = E_\nu - E_0$ of the state $|\nu\rangle$. Since $|\text{RPA}\rangle$ is not known yet, the evaluation of these equations is very complicated. A possible approximation consists in assuming that the correlated $|\text{RPA}\rangle$ ground state does not differ very much from the HF ground state, so we can calculate all expectation values in the HF approximations. This approach is known as quasi-boson approximation and gives the quasi-boson commutator

$$\left[c_h^\dagger c_p, c_{p'}^\dagger c_{h'}\right] \simeq \delta_{pp'}\delta_{hh'}. \tag{3.6}$$

Within this framework the amplitudes $X_{ph}^\nu$ and $Y_{ph}^\nu$ have a direct meaning: their absolute square give the probability of finding the states $c_p^\dagger c_h|0\rangle$ and $c_h^\dagger c_p|0\rangle$ in the excited state $|\nu\rangle$:

$$\langle 0|c_h^\dagger c_p|\nu\rangle \simeq \langle \text{HF}| \left[c_h^\dagger c_p, \Gamma_\nu^\dagger\right] |\text{HF}\rangle = X_{ph}^\nu \tag{3.7a}$$

$$\langle 0|c_p^\dagger c_h|\nu\rangle \simeq \langle \text{HF}| \left[c_p^\dagger c_h, \Gamma_\nu^\dagger\right] |\text{HF}\rangle = Y_{ph}^\nu. \tag{3.7b}$$

The two amplitudes $X$ and $Y$ are called forward and backward amplitudes, respectively. Hence, Eq. (3.5) can be written in the compact form

$$\begin{pmatrix} A & B \\ -B^* & -A^* \end{pmatrix} \begin{pmatrix} X \\ Y \end{pmatrix} = \hbar\omega \begin{pmatrix} X \\ Y \end{pmatrix} \tag{3.8}$$

The sub-matrices $A$ and $B$ are defined

$$A_{php'h'} = \langle \mathrm{HF} | \left[ c_h^\dagger c_p, \left[ H, c_{p'}^\dagger c_{h'} \right] \right] | \mathrm{HF} \rangle = \left( \epsilon_p - \epsilon_h \right) \delta_{pp'} \delta_{hh'} + \bar{v}_{ph'p'h} \tag{3.9a}$$

$$B_{php'h'} = -\langle \mathrm{HF} | \left[ c_h^\dagger c_p, \left[ H, c_{h'}^\dagger c_{p'} \right] \right] | \mathrm{HF} \rangle = \bar{v}_{pp'h'h}. \tag{3.9b}$$

When there is a good quantum number, such as angular momentum, it is appropriate to exploit it to reduce the dimensions of the matrix to be diagonalized. Thus the definition

$$\Gamma_\nu^\dagger (JM) = \sum_{ph} X_{ph}^{\nu J} A_{ph}^\dagger (JM) - Y_{ph}^{\nu J} A_{ph}(\widetilde{JM}) \tag{3.10}$$

should be used, where $A_{ph}^\dagger (JM)$ and $A_{ph}(\widetilde{JM})$ are the coupled ph creation and destruction operators:

$$A_{ph}^\dagger (JM) = \sum_{m_p m_h} (-)^{j_h - m_h} \langle j_p m_p j_h - m_h | JM \rangle c_{j_p m_p}^\dagger c_{j_h m_h} \tag{3.11a}$$

$$A_{ph}(\widetilde{JM}) = \sum_{m_p m_h} (-)^{J+M+j_h-m_h} \langle j_p m_p j_h - m_h | J - M \rangle c_{j_h m_h}^\dagger c_{j_p m_p} \tag{3.11b}$$

The RPA equation can also be obtained from a more general derivation, in which the response of the system to an external time-dependent field is treated in the time-dependent HF (TDHF) approximation and the resulting equations are linearized. We are not going to present in detail this procedure, addressing the reader to Refs. [BR86; RS04]. We only want to point out that following this route the link between the residual interaction and the HF Hamiltonian and, in turn, the energy density functional is clear:

$$\bar{v}_{\alpha\beta\gamma\delta} = \frac{\partial h_{\alpha\gamma}}{\partial \rho_{\delta\beta}} = \frac{\partial^2 E}{\partial \rho_{\gamma\alpha} \partial \rho_{\delta\beta}}. \tag{3.12}$$

**Properties of the RPA matrix**

For the sake of completeness, we summarize here some properties of the RPA matrices. The proofs of these properties can be found in Ref. [BR86]. Let us define the three matrices $M$, $\gamma$ and $\eta$

$$M = \begin{pmatrix} A & B \\ B^* & A^* \end{pmatrix}, \qquad \gamma = \begin{pmatrix} 0 & 1 \\ 1 & 0 \end{pmatrix}, \qquad \eta = \begin{pmatrix} 1 & 0 \\ 0 & -1 \end{pmatrix}.$$

such that the RPA equations (3.8) can be written as the eigenvalue problem

$$\eta M V^n = \hbar\omega_n V^n,$$

where $V^n$ is the right eigenvector of the non-Hermitian matrix $\eta M$ belonging to the eigenvalue $\hbar\omega_n$,

$$V^n = \begin{pmatrix} X^n \\ Y^n \end{pmatrix}$$

- For each right eigenvector $V^n$ associated with the eigenvalue $\hbar\omega_n$, $\gamma V^n$ is a right eigenvector belonging to the eigenvalue $-\hbar\omega_n^*$ and $\bar{V}^n = V^{n\dagger}\eta$ is a left eigenvector belonging to the eigenvalue $\hbar\omega_n^*$.
- Given a right eigenvector $V^n$ with eigenvalue $\hbar\omega_n$ and a left eigenvector $\bar{V}^m$ with eigenvalue $\omega_m^*$, if $\omega_n \neq \omega_m^*$, then $\bar{V}^m$ and $V^n$ are orthogonal. As a consequence, eigenvectors belonging to complex eigenvalues have zero norm with the metric $\eta$.
- If $P$ is an eigenvector associated with a zero eigenvalue, there exist another vector $Q$ with zero eigenvalue such that $\eta M Q = -\frac{i}{\mu}P$, with $\mu > 0$.
- If the matrix $M$ is semi-positive definite, the eigenvalues of $\eta M$ are all real and the eigenvectors belonging to positive, zero and negative eigenvalues have respectively a positive, zero and negative norm, with the metric $\eta$. As a consequence, the non-zero eigenvalues may be grouped into pairs $\pm\omega_n$.

## 3.2 Broken transformation symmetry and the spurious mode

Consider a unitary transformation of the Hartree-Fock state $|\psi_0\rangle$:

$$|\psi\rangle = e^{i\lambda S}|\psi_0\rangle, \tag{3.13}$$

where $S$ is a Hermitian one-body operator. It can be shown that $S$ satisfies the equations

$$\begin{pmatrix} A & B \\ B^* & A^* \end{pmatrix} \begin{pmatrix} S \\ -S^* \end{pmatrix} = 0, \tag{3.14}$$

where $A$ and $B$ are the RPA matrices. Two cases may occur.

1. The generator $S$ commutes with the density matrix $\rho_0$, in which case $S_{ph} = 0$ and the equation (3.14) is trivially satisfied. This case occurs when $S$ is a generator of a self-consistent symmetry.
2. $S$ does not commute with the density matrix $\rho_0$. Then the RPA matrix in Eq. (3.14) has an eigenvector $(S, -S^*)$ belonging to a zero eigenvalue. This occurs for every generator of a continuous broken symmetry. The eigenvector associated to this zero eigenvalue of the RPA matrix is called spurious mode, which is not associated with as intrinsic excitation of the system, but with a collective motion without restoring force.

In the particular case of the broken translational symmetry, the operator $S$ is the total momentum $\boldsymbol{P}$ of the system. In a bona fide self-consistent RPA the spurious state would appear as an eigenstate at zero energy, exhausting the whole strength of the operator $\boldsymbol{R}$ corresponding to a Galileian boost, with spherical components

$$R_\mu = \sqrt{\frac{4\pi}{3}}\frac{1}{\tilde{A}}i\sum_{i=1}^{A} r_i Y_{1\mu}(\hat{r}_i), \tag{3.15}$$

where $\tilde{A}$ is the effective particle number, which is equal to the actual particle number $A$ if the spurious state is exactly orthogonal to the excited states. However, in actual numerical implementations the spurious state is at low but not zero energy because of small numerical inaccuracies and therefore, strength associated with the operator $\boldsymbol{R}$ will be shared among the physical states. Moreover, the physical states should be orthogonal also to the states of the operator $\boldsymbol{P}$, conjugated to $\boldsymbol{R}$. The spherical components of the operator $\boldsymbol{P}$ are

$$P_\mu = \sum_{i=1}^{A} \nabla_\mu^{(r_i)}, \tag{3.16}$$

where the phases are chosen to have the canonical commutation relations $[R_\mu^\dagger, P_\mu] = i$.

Starting from the RPA states $|n'\rangle$, described by the vector $V^{n'} = {}^T(X^{n'}\, Y^{n'})$, we construct a new set of normalized states $|n\rangle$, with angular-momentum-coupled RPA amplitudes defined by

$$\begin{pmatrix} X^n \\ Y^n \end{pmatrix} = V^n = \alpha \left[ \begin{pmatrix} X^{n'} \\ Y^{n'} \end{pmatrix} - \lambda \begin{pmatrix} \mathcal{P} \\ -\mathcal{P}^* \end{pmatrix} - \mu \begin{pmatrix} \mathcal{R} \\ -\mathcal{R}^* \end{pmatrix} \right] = \alpha \left( V^{n'} - \lambda P - \mu R \right), \tag{3.17}$$

where $\mathcal{R}$ and $\mathcal{P}$ are the $ph$ matrix elements of the operators $R_\mu$ and $P_\mu$ and $\alpha$ is a normalization constant. Obviously, since $R_\mu$ is Hermitian and $P_\mu$ is anti-Hermitian, $\mathcal{R}^* = \mathcal{R}$ and $\mathcal{P}^* = -\mathcal{P}$. The two parameters $\lambda$ and $\mu$ can be determined by requiring the orthogonalization against the two zero-energy states, which translates into the two conditions

$$\sum_{ph} X^n_{\text{ph}} \mathcal{R}^*_{\text{ph}} + Y^n_{\text{ph}} \mathcal{R}_{\text{ph}} = R^\dagger \eta V^n \equiv 0 \tag{3.18a}$$

$$\sum_{ph} X^n_{\text{ph}} \mathcal{P}^*_{\text{ph}} + Y^n_{\text{ph}} \mathcal{P}_{\text{ph}} = P^\dagger \eta V^n \equiv 0. \tag{3.18b}$$

Inserting the definition of the new states (3.17) into Eq. (3.18) we obtain for the parameters $\lambda$ and $\mu$

$$\begin{matrix} \lambda = -iR^\dagger \eta V^n \\ \mu = iP^\dagger \eta V^n \end{matrix} \quad \Rightarrow \quad \begin{matrix} \lambda^* = iV^{n\dagger} \eta R \\ \mu^* = -iV^{n\dagger} \eta P \end{matrix} .$$

Specializing these relation in our practical cases, in which $V^n$, $R$ are real and $P$ is purely imaginary, we get

$$\begin{matrix} \lambda^* = -\lambda \\ \mu^* = \mu \end{matrix} .$$

The normalization constant $\alpha$ can be determined by requiring that the vectors $V^n$ are normalized:

$$\begin{aligned}
1 = V^{n\dagger}\eta V^n &= \alpha^2 \left(V^{n'\dagger} - \lambda^* P^\dagger - \mu^* R^\dagger\right) \eta \left(V^{n'} - \lambda P - \mu R\right) \\
&= \alpha^2 \Big(V^{n'\dagger}\eta V^{n'} - \lambda V^{n'\dagger}\eta P - \mu V^{n'\dagger}\eta R - \lambda^* P^\dagger \eta V^{n'} + \lambda^*\lambda P^\dagger \eta P \\
&\quad + \lambda^*\mu P^\dagger \eta R - \mu^* R^\dagger \eta V^{n'} + \mu^*\lambda R^\dagger \eta P + \mu^*\mu R^\dagger \eta R\Big) \\
&= \alpha^2 \left(1 - i\lambda\mu^* + i\mu\lambda^* + i\mu\lambda^* - i\mu\lambda^* - i\mu^*\lambda + i\mu^*\lambda\right) \\
&= \alpha^2 \left(1 - 2i\lambda\mu\right).
\end{aligned}$$

## 3.3 Transition operators, transition probability and transition densities

The vibrational states are excited under the action of an external field, which can be electromagnetic or hadronic.

In general, an isoscalar or isovector external field $F$ can be expanded in spherical harmonics and the $\lambda-$multipole of the field reads

$$\begin{aligned}
F^{(IS)}_{\lambda\mu} &= i^\lambda \sum_{i=1}^{A} f_\lambda(r_i) Y_{\lambda\mu}(\hat{r}_i), \\
F^{(IV)}_{\lambda\mu} &= i^\lambda \sum_{i=1}^{A} f_\lambda(r_i) Y_{\lambda\mu}(\hat{r}_i)\tau_z(i),
\end{aligned} \tag{3.19}$$

where the sums run over all nucleons and $\tau_z(i)$ is the $z-$component of the isospin operator of nucleon $i$. The phase $i^\lambda$ has been introduced in keeping with the expression for the wave functions (2.35).

The particular case of the electromagnetic excitation process, using either real photons or virtual photons (Coulomb excitation), is treated in textbooks (e.g. see Ref. [BW62; dSF74]). The multipole decomposition of the photon plane wave leads to terms which contain spherical Bessel functions $j_\lambda(qr)$, where $q$ is the photon momentum. If $E$ is the energy of the gamma ray in MeV and $R$ the nuclear radius in Fermi, we have

$$qR = \frac{ER}{197\,\text{fm MeV}}.$$

Thus for gamma-ray energies of the order of $\approx 10\,$MeV and for even large nuclei, $qR \ll 1$. Thus, the long wavelength limit of the spherical Bessel functions $j_\lambda(qr)$ can be used:

$$j_\lambda(qr) = \frac{(qr)^\lambda}{(2\lambda+1)!!}\left\{1 - \frac{(qr)^2}{2(2\lambda+3)} + O\left[(qr)^4\right]\right\}.$$

In this way, the electromagnetic operator becomes

$$F^{(\text{e.m.})}_{\lambda\mu} = i^\lambda \sum_{i=1}^{A} r_i^\lambda Y_{\lambda\mu}(\hat{r}_i)\tau_z(i). \tag{3.20}$$

More precisely, due to momentum conservation in a radiative transition [dSF74], neutron and protons acquire an effective charge. Therefore, the electromagnetic operator becomes

$$\begin{aligned} F^{(\text{e.m.})}_{\lambda\mu} &= i^{\lambda} \sum_{i=1}^{A} r_i^{\lambda} Y_{\lambda\mu}(\hat{r}_i) \left( \frac{(A-1)^{\lambda} + (-1)^{\lambda}(2Z-1)}{2A^{\lambda}} - \frac{(A-1)^{\lambda} - (-1)^{\lambda}}{2A^{\lambda}} \tau_z(i) \right) \\ &= i^{\lambda} \sum_{i=1}^{A} e_i^{eff} r_i^{\lambda} Y_{\lambda\mu}(\hat{r}_i), \end{aligned} \tag{3.21}$$

where

$$e_i^{\text{eff}} = \begin{cases} \left[ \left(1 - \dfrac{1}{A}\right)^{\lambda} + (-)^{\lambda} \dfrac{Z-1}{A^{\lambda}} \right] & \text{for protons} \\ Z\left( - \dfrac{1}{A}\right)^{\lambda} & \text{for neutrons} \end{cases} \tag{3.22}$$

Actually, the effective charges are substantially different from the bare charges only in the case of dipole transitions: e.g. for a quadrupole transition in $^{208}$Pb, $e_p^{\text{eff}} = 0.9923$ and $e_n^{\text{eff}} = 0.0019$.

In the case of hadron inelastic scattering, one arrives at similar excitation operators starting, e.g., from the Distorted Wave Born Approximation (DWBA) expression for the cross section. This cross section is in fact proportional to the square of the transition amplitude [HvW01]

$$T_{fi} = \int \mathrm{d}_3 R \chi_f^*(\boldsymbol{R}) \langle f | V(\boldsymbol{R}) | i \rangle \chi_i(\boldsymbol{R}), \tag{3.23}$$

where $\chi$ stands for the distorted waves describing the relative projectile-target motion, $\boldsymbol{R}$ is the distance between the corresponding centers of mass, and $V$ is the projectile-target interaction. Under the assumptions of (i) distorted waves replaced by plane waves, (ii) zero-range interaction of the form $V_0 \sum_i^A \delta(\boldsymbol{r}_i - \boldsymbol{R}) + V_1 \sum_i^A \delta(\boldsymbol{r}_i - \boldsymbol{R}) \boldsymbol{\tau}(i) \cdot \boldsymbol{T}$ where $\boldsymbol{T}$ is the projectile isospin, and (iii) small momentum transfer $\boldsymbol{q}$, the DWBA transition amplitude $T_{fi}$ for non-charge-changing transitions becomes proportional to the matrix elements of the isoscalar (IS) and isovector (IV) operators:

$$F^{(IS)}_{\lambda\mu} = i^{\lambda} \sum_{i=1}^{A} r_i^{\lambda} Y_{\lambda\mu}(\hat{r}_i), \tag{3.24}$$

$$F^{(IV)}_{\lambda\mu} = i^{\lambda} \sum_{i=1}^{A} r_i^{\lambda} Y_{\lambda\mu}(\hat{r}_i) \tau_z(i). \tag{3.25}$$

We do not consider the spin-dependent operators here because they will not be used in the following.

In the $\lambda = 0$ case the operator (3.24) is a constant, so we have to consider the next term in the expansion of the spherical Bessel function $j_0(qr)$, that is, the operator

$$F^{(IS)}_{00} = i^{\lambda} \sum_{i=1}^{A} r_i^2 Y_{00}. \tag{3.26}$$

Also in the dipole case, when $\lambda = 1$, the lowest order term of the isoscalar operator does not produce a physical excitation: the operator (3.24) induces simply a translation of the whole system, since it is proportional to the Galileian boost operator (3.15). The resulting isoscalar dipole operator is

$$F^{(IS)}_{1\mu} = i^{\lambda} \sum_{i=1}^{A} r_i^3 Y_{1\mu}(\hat{r}_i). \tag{3.27}$$

The spurious state must be subtracted from the vector $^T(X^n, Y^n)$ from the excited RPA states, as explained in section 3.2.

If only the strength function (see later in this section) is needed, the removal of the spurious state can be performed by modifying the dipole operator:

$$F^{(IS)}_{1\mu} = i^{\lambda} \sum_{i=1}^{A} \left( r_i^3 - \eta r_i \right) Y_{1\mu}(\hat{r}_i). \tag{3.28}$$

with $\eta = 5/3\langle r^2 \rangle$ [VS81].

For the isovector dipole excitation, the center of mass is subtracted as in Eq. (3.21) with $\lambda = 1$

$$F^{(IV)}_{1\mu} = i \sum_{i=1}^{A} r_i Y_{1\mu}(\hat{r}_i) \left( \frac{N-Z}{A} - \tau_z(i) \right) = i\frac{2N}{A} \sum_{p=1}^{Z} r_p Y_{1\mu}(\hat{r}_p) - i\frac{2Z}{A} \sum_{n=1}^{N} r_n Y_{1\mu}(\hat{r}_n). \tag{3.29}$$

For any of the operators (3.21), (3.24), or (3.25), the reduced transition probability is given by (see, e.g., Ref. [BM69])

$$B(\mathscr{T}\lambda; i \to f) = \frac{1}{2J_i + 1} |\langle f \| F_\lambda \| i \rangle|^2, \tag{3.30}$$

where $\langle f \| F_J \| i \rangle$ is the reduced matrix element of $F_{\lambda\mu}$ and $\mathscr{T}$ stands for $E$ (electromagnetic), $IS$ (isoscalar) or $IV$ (isovector). In the particular case of the excitation of a state $|\nu\rangle$, the matrix element of the operator is evaluated at the RPA level,

$$B(\mathscr{T}\lambda; 0 \to \nu) = \left| \sum_{ph} \left( X^{(\nu)\lambda}_{ph} + Y^{(\nu)\lambda}_{ph} \right) \langle p \| F_\lambda \| h \rangle \right|^2 = \left| \sum_{ph} A_{ph}(\mathscr{T}\lambda; 0 \to \nu) \right|^2 , \tag{3.31}$$

i.e., the transition strength is the square of the so-called transition amplitude $A_{ph}$. This quantity is important because allows one to determine the coherency (relative sign) and magnitude ($|A_{ph}(\mathscr{T}\lambda, 0 \to \nu)|$) of all the $ph$ contributions to the reduced transition probability. An RPA state is claimed to be a resonant excitation if the corresponding reduced amplitude is composed by several $ph$ excitations similar in magnitude and adding coherently. The explicit calculation of Eq. (3.31) can be found in the appendix A.

In order to have a qualitative estimate for the collectivity displayed by the different dipole responses, the reduced transition probabilities can be evaluated in single-particle units (s.p.u., or Weisskopf units [BM69]). Such unit is based on a macroscopic approach: we evaluate the average transition rate of a typical excitation in terms of the angular momentum carried by the probe and the radius of the nucleus under analysis; in this way, the result is nucleus-independent. In particular in the following we will be interested in the isovector and isoscalar dipole single-particle transitions. The isovector dipole response in Weisskopf units accounting for the center of mass correction can be computed as,

$$B_{\mathrm{W}}^{(IV)}(E1) = \frac{3^3 R^2}{4^3 \pi} \times \begin{cases} \left(\frac{N}{A}\right)^2 & \text{for protons} \\ \left(-\frac{Z}{A}\right)^2 & \text{for neutrons} \end{cases} \tag{3.32}$$

where the radius $R$ is taken to be $r_0 A^{1/3}$ and $r_0$ is the radius of the average sphere that one nucleon occupies, at the standard saturation density of 0.16 fm$^{-3}$ (that is, $r_0 = 1.14$ fm). For the isoscalar dipole case, once the spurious state has been subtracted from the operator, one finds

$$B_{\mathrm{W}}^{(IS)}(E1) = \frac{3}{4\pi}\left(\frac{1}{2}R^3 - \eta\frac{3}{4}R\right)^2 = \frac{3R^6}{4^3\pi} \tag{3.33}$$

with, again, $\eta = \frac{5}{3}\langle r^2\rangle$ and $\langle r^2\rangle = \frac{3}{5}R^2$. Such a unit allows us to account qualitatively for the nature of different excitations since a given RPA state will contribute with several single-particle units if it is collective. Moreover, it also enables the comparison between the results obtained for different nuclei.

A quantity of interest that represent the density variations associated with the transition from the ground state to the excited state, is the transition density. This is defined as the off-diagonal matrix elements of the density operator. Its integral with a multipole operator gives the corresponding transition amplitude of that operator. Thus the reduced transition probability becomes

$$B(\mathscr{T}\lambda; 0 \to \nu) = (2\lambda + 1)\left|\int \mathrm{d}r r^{\lambda+2}\delta\rho_\nu(r)\right|^2. \tag{3.34}$$

The spherical harmonic expansion of the transition density is given by

$$\delta\rho_\nu(\boldsymbol{r}) \equiv \langle\nu|\rho(\boldsymbol{r})|0\rangle = \delta\rho_\nu(r) Y^*_{\lambda\nu}(\hat{r}). \tag{3.35}$$

Using the helicity representation for the single-particle wave functions [BM69], the radial part of the transition density can be written as a function of the $X^\nu$ and $Y^\nu$ amplitudes of a given RPA state $|\nu\rangle$ [Ber80]

$$\delta\rho_\nu(r) = \frac{1}{\sqrt{2J+1}}\sum_{ph}\left(X^\nu_{ph} + Y^\nu_{ph}\right) \times \langle p\|Y_J\|h\rangle \frac{u_p(r)u_h(r)}{r^2}\ , \tag{3.36}$$

where $u_\alpha(r)$ is reduced radial wave function of the single-particle state $\alpha$. Note that the summations in the expression above can be done for neutrons or protons separately. This allows one to calculate the neutron and proton transition densities $\delta\rho^q_\nu(r)$ $(q = n, p)$ and define accordingly the isoscalar $(IS)$ and isovector $(IV)$ transition densities as

$$\begin{aligned} \delta\rho^{(IS)}_\nu(r) &\equiv \delta\rho^n_\nu(r) + \delta\rho^p_\nu(r) \quad \text{and} \\ \delta\rho^{(IV)}_\nu(r) &\equiv \delta\rho^n_\nu(r) - \delta\rho^p_\nu(r). \end{aligned} \tag{3.37}$$

The interest of the transition densities relies on the fact that their spatial shape reveal the nature of the excitations: volume or surface type, isoscalar or isovector, etc. More-over, they can be used as input in calculations of inelastic scattering cross sections, being thus liable to experimental investigations. It is possible create a new set of RPA states, identified with $\tilde{\nu}$, for which the corresponding neutron and proton transition densities are not contaminated by the spurious state. First, we impose that the translational operator which is proportional to the radial coordinate $r$ does not give any finite transition amplitude. This means

$$\int \mathrm{d}r\, r^2 r \left(\delta\rho^n_{\tilde{\nu}} + \delta\rho^p_{\tilde{\nu}}\right) = 0.$$

As a second condition, we impose on these new transition densities that the strength of the isovector dipole operator is not modified. That is, we write

$$\int \mathrm{d}r\, r^2 r \left(\frac{2Z}{A}\delta\rho^n_{\tilde{\nu}} - \frac{2N}{A}\delta\rho^p_{\tilde{\nu}}\right) = \int \mathrm{d}r r^2 r \left(\frac{2Z}{A}\delta\rho^n_{\nu} - \frac{2N}{A}\delta\rho^p_{\nu}\right).$$

By writing

$$\delta\rho^q_{\tilde{\nu}} = \delta\rho^q_{\nu} - \alpha^q \frac{\mathrm{d}\rho^q_{\mathrm{HF}}(r)}{\mathrm{d}r},$$

where $\rho^q_{\mathrm{HF}}(r)$ is the proton ($q = p$) or neutron ($q = n$) HF density, we find the following solution:

$$\alpha^n = \frac{2N}{A} \frac{\int \mathrm{d}r\, r^2 r \delta\rho_\nu}{\int \mathrm{d}r\, r^2 r \frac{\mathrm{d}\rho^n_{\mathrm{HF}}}{\mathrm{d}r}},$$
$$\alpha^p = \frac{2Z}{A} \frac{\int \mathrm{d}r\, r^2 r \delta\rho_\nu}{\int \mathrm{d}r\, r^2 r \frac{\mathrm{d}\rho^p_{\mathrm{HF}}}{\mathrm{d}r}}.$$

## 3.4 Sum rules

Given a one-body operator $F_{\lambda\mu}$, its strength function is defined as

$$S(E) = \sum_\nu |\langle\nu\|F_\lambda\|0\rangle|^2 \delta(E - E_\nu) \tag{3.38}$$

and it consists of a sum of delta functions. Its $k^{\mathrm{th}}$ moments are defined by

$$S_k = \sum_\nu (E_\nu - E_0)^k |\langle\nu|F|0\rangle|^2. \tag{3.39}$$

From the definition (3.30) of the reduced transition probability, it is apparent that the strength function can be rewritten as

$$S_k = \sum_\nu (E_\nu - E_0)^k B(\mathscr{T}\lambda; 0 \to \nu), \tag{3.40}$$

establishing a relation between the excitation probability $B(\mathscr{T}\lambda)$ and the EWSR.

Among them, the linear moment $S_1$ (known as Energy Weighted Sum Rule, or EWSR)

is important since the following equality holds

$$S_1 = \sum_\nu (E_\nu - E_0)|\langle\nu|F|0\rangle|^2 = \frac{1}{2}\langle 0|\,[F,[H,F]]\,|0\rangle. \qquad (3.41)$$

In particular, Thouless [Tho61] showed that the equality is fulfilled if the left-hand side is evaluated with RPA wave functions and energies and the right-hand side is calculated using the HF ground state wave function (if the calculation is self-consistent). Therefore $S_1$ can be used to test the validity of any numerical approximation.

For the isoscalar operator (3.24), the operator commutes with the potential [1] and one obtains

$$S_1 = \frac{\hbar^2}{2m}\lambda(2\lambda+1)^2\frac{A}{4\pi}\langle r^{2\lambda-2}\rangle. \qquad (3.42)$$

The above sum rule is model-independent and sometimes the right hand side is experimentally accessible (e.g., in the quadrupole case where $\langle r^2\rangle$ appears). In the monopole case, using the operator (3.26) the double commutator sum rule is

$$m_1 = \frac{\hbar^2}{m}\frac{A}{2\pi}\langle r^2\rangle. \qquad (3.43)$$

The sum rule associated with the isoscalar dipole operator (3.28) is

$$m_1 = \frac{\hbar^2}{2m}\frac{A}{4\pi}\left(33\langle r^4\rangle - 25\langle r^2\rangle^2\right), \qquad (3.44)$$

On the contrary, the isovector operator (3.29), depending on the isospin, does not commute with the nuclear potential, giving rise to a contribution to the EWSR in addition to the kinetic one:

$$S_1 = \frac{1}{2}\langle 0|\,[F,[H,F]]\,|0\rangle = \frac{1}{2}\langle 0|\,[F,[T,F]]\,|0\rangle(1+\kappa), \qquad (3.45)$$

where $\kappa$ is called enhancement factor. For the general expression for the isovector operator in Eq. (3.19), the enhancement factor is given by [Sil+06]

$$\kappa = \frac{2m}{\hbar^2}\frac{1}{A\langle g_\lambda\rangle}\left(t_1\left(1+\frac{x_1}{2}\right)+t_2\left(1+\frac{x_2}{2}\right)\right)\int \mathrm{d}_3 r\, g_\lambda(r)\rho_p(r)\rho_n(r), \qquad (3.46)$$

where

$$\langle g_\lambda\rangle = \frac{1}{A}\int\left[\left(\frac{\mathrm{d}f_\lambda}{\mathrm{d}r}\right)^2 + \lambda(\lambda+1)\left(\frac{f_\lambda}{r}\right)^2\right]\rho(r)\mathrm{d}_3 r.$$

The value of the enhancement factor has been the subject of many theoretical and experimental investigations: the experimental value is obtained integrating the photoabsorption cross section and is known to depend critically on the value of the energy up to which the integration is carried out. A value around 0.2 is currently considered a plausible estimate in the dipole case.

In the dipole case, when a Skyrme interaction is used, the sum rule (also known as

[1] Although the Skyrme interaction is velocity dependent, the expression can be shown to be valid due to the Galileian invariance [BL76].

Thomas-Reiche-Kuhn, or TRK, sum rule) reads

$$\begin{aligned} m_1 =& \frac{9}{\pi}\frac{\hbar^2}{2m}\frac{NZ}{A}\,(1+\kappa_D) \\ =& \frac{9}{\pi}\frac{\hbar^2}{2m}\frac{NZ}{A}\left(1+\frac{2m}{\hbar^2}\frac{4\pi A}{NZ}\left(t_1\left(1+\frac{x_1}{2}\right)+t_2\left(1+\frac{x_2}{2}\right)\right)\int \mathrm{d}r r^2\rho_p\rho_n\right). \end{aligned} \tag{3.47}$$

It is possible to define a set of energies

$$\mathscr{E}_k = \sqrt{\frac{S_k}{S_{k-2}}}$$

which characterizes the strength distribution: if it is sharply peaked, all the $\mathscr{E}_k$ coincide. The degree to which they are different reflects the width of the distribution. In particular $\mathscr{E}_1$ is called constrained energy and $\mathscr{E}_3$ scaling energy. The centroid energy on the other hand is defined as

$$\bar{E} = \frac{S_1}{S_0}$$

A wider treatment of the sum rules and their importance in the study of GRs can be found in [LS89].

## 3.5 The $\gamma$ decay to the ground state

We consider in this section the decay of an excited RPA state to the ground state. The $\gamma$ decay is the inverse process of the electromagnetic operator. For this reason the reduced transition probability can be obtained using the relation

$$(2J_i+1)B(\mathscr{T}\lambda; i\to f) = (2J_f+1)B(\mathscr{T}\lambda; f\to i). \tag{3.48}$$

In our case, the reduced transition probability becomes

$$B(E\lambda;\nu\to 0) = \frac{1}{2J_\nu+1}B(E\lambda;0\to\nu) = \frac{1}{2J_\nu+1}\left|\sum_{ph}\left(X_{ph}^{\nu\lambda}+Y_{ph}^{\nu\lambda}\right)\langle p\|F_\lambda^{e.m.}\|h\rangle\right|^2, \tag{3.49}$$

where $F_{\lambda\mu}^{e.m.}$ is the electromagnetic operator (3.21). Given the reduced transition probability, it is possible to define the $\gamma-$decay transition probability $T_\gamma(E\lambda;\nu\to 0)$ and the $\gamma-$decay width $\Gamma_\gamma(E\lambda;\nu\to 0)$, summed over the magnetic substates of the photon and of the final nuclear state:

$$T_\gamma(E\lambda; i\to f) = \frac{8\pi(\lambda+1)}{\lambda[(2\lambda+1)!!]^2}\left(\frac{E}{\hbar c}\right)^{2\lambda+1}B(E\lambda; i\to f) \tag{3.50}$$

$$\Gamma_\gamma(E\lambda; i\to f) = \frac{8\pi(\lambda+1)}{\lambda[(2\lambda+1)!!]^2}\left(\frac{E}{\hbar c}\right)^{2\lambda+1}B(E\lambda; i\to f), \tag{3.51}$$

where $E$ on the r.h.s. of the equations is the energy of the transition.

Chapter 4

# Equation of state of nuclear matter

In this chapter we are going to present the main feature of the nuclear matter. In particular, we will focus on the parameters of the equation of state which are important for the discussion of giant resonances (chapter 5) and our results (chapter 6)

Only recently, it has been possible to develop new experimental techniques to synthesize and perform measurements on radioactive nuclei. These new techniques have been or will be soon implemented in facilities such as HIRFL@CSR (China), Spiral2@GANIL (France), Spes@LNL (Italy), FAIR@GSI (Germany), FRIB@MSU (USA) and RIBF@RIKEN (Japan). At these facilities, it will be possible to extend the amount of known nuclides towards the boundaries of the nuclear chart. These nuclei are in general not present on Earth and, because of that, they are commonly known as exotic. They are usually short-lived, less bound and present a larger isospin asymmetry with respect to Earth-nuclei. Nowadays, some new phenomena have already been revealed in such systems, such as the modification of magic numbers or the appearance of a proton or neutron halo structure. Furthermore, observational data from astrophysical objects and processes, such as neutron stars, proto-neutron stars, astrophysical nucleosynthesis, cooling processes, supernova dynamics, etc., could be useful in constraining the nuclear equation of state (EoS) in density ranges and isospin asymmetries unreachable on Earth. This has stimulated much interest and a lot of activity in a new research direction in nuclear physics, namely isospin physics.

The simplest way to simulate the interior of a heavy nucleus is to consider a hypothetical system, called nuclear matter. By nuclear matter we mean an infinite system of nucleons interacting by their mutual nuclear forces and no electromagnetic interactions. It is characterized by its energy per particle $\frac{E}{A}$ as a function of density and other thermodynamic quantities, if appropriate (e.g. temperature). Such relation is known as the nuclear matter EoS. The translational invariance of the system facilitates theoretical calculations, since the wave functions are clearly plane waves. At the same time, adopting what is known as local density approximation, one can use the EoS directly in calculations of finite systems, e.g. in Thomas-Fermi calculations within the liquid drop model, where an appropriate energy functional is written in terms of the EoS.

The equation of state of isospin symmetric nuclear matter at zero temperature is given by $\frac{E}{A} = e(\rho)$, where $\rho$ is the density, defined in the usual way,

$$\rho = (2s+1)(2t+1)\frac{k_{\mathrm{F}}^3}{6\pi^2} = \frac{2k_{\mathrm{F}}^3}{3\pi^2},$$

where $(2s+1)$ and $(2t+1)$ are the spin and isospin degeneracy factors, each one equal to 2.

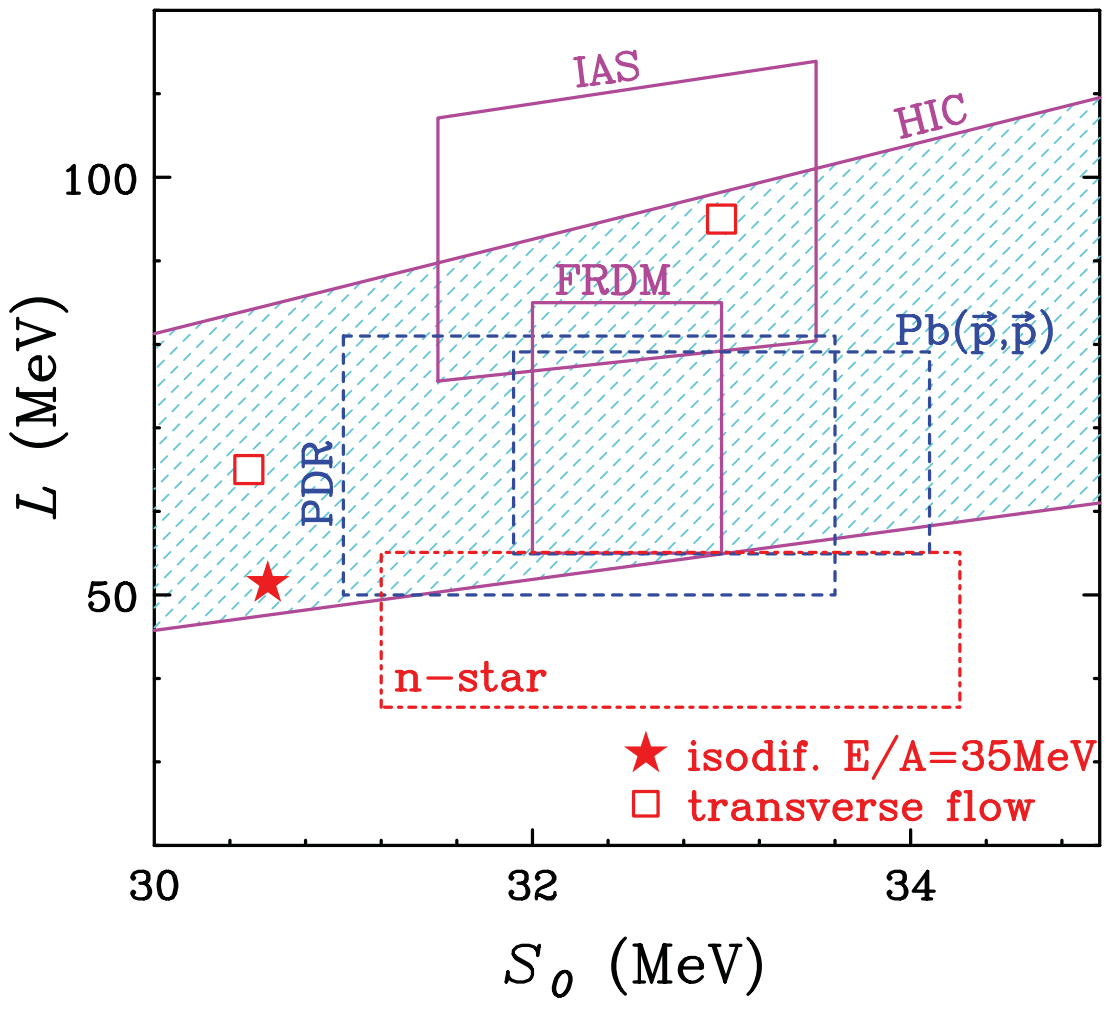

**Figure 4.1**
Constraints on $L$ and $J$ from different experimental results (figure taken from [Tsa+12], to which we address the reader for the details on the labels). Note that here the symmetry energy at saturation is named $S_0$.

The saturation density $\rho_\infty$ is obtained by requiring that the energy has a minimum

$$\left.\frac{\mathrm{d}e(\rho)}{\mathrm{d}\rho}\right|_{\rho=\rho_\infty} = 0.$$

The commonly accepted value for this quantity is $\rho_0 = 0.16\,\mathrm{fm}^{-3}$, to which corresponds a energy per particle of about $-16\,\mathrm{MeV}$.

Expanding the EoS around saturation, it is possible to define the compression modulus $K_\infty$

$$K_\infty = 9\rho_\infty^2 \left.\frac{\mathrm{d}^2 e(\rho)}{\mathrm{d}\rho^2}\right|_{\rho=\rho_\infty}.$$

A commonly accepted value for this quantity is $240 \pm 20\,\mathrm{MeV}$ [SKC06]. The compression modulus in related to compression modes in nuclei, e.g. the giant monopole resonance and the isoscalar dipole resonance, as we will recall in chapter 5. Moreover, it has an impact on the physics of supernovae explosions and the consequent formation of the neutron stars.

If now we consider the asymmetric nuclear matter, we can define the local asymmetry parameter $\delta = \frac{\rho_n - \rho_p}{\rho}$. In this way, we can expand the corresponding EoS as a function of $\delta$, obtaining

$$e(\rho,\delta) = e(\rho_\infty, \delta = 0) + S_2(\rho)\delta^2 + O(\delta^4). \tag{4.1}$$

In this equation, $e(\rho_\infty, \delta = 0)$ is the EoS for symmetric nuclear matter and $S_2(\rho)$ is called symmetry energy, defined as

$$S_2(\rho) = \frac{1}{2}\left.\frac{\partial^2 e(\rho,\delta)}{\partial \rho^2}\right|_{\delta=0}.$$

In Eq. (4.1), there are no odd-order $\delta$ terms due to the exchange symmetry between protons and neutrons (the charge symmetry of nuclear forces). Higher-order terms in $\delta$ are generally negligible for most purposes. For example, the magnitude of the $\delta^4$ term at $\rho_0$ has been estimated to be less than 1 MeV [Vid+09]. Actually, the presence of higher order terms in $\delta$ at supra-saturation densities can significantly modify the proton fraction and the cooling processes in neutron stars [LCK08]. Eq. (4.1) is considered to be valid only at small isospin asymmetries. However, many non-relativistic and relativistic calculations have shown that it is actually valid up to $\delta = 1$, at least for densities up to moderate values [BL91; Lee+98; Vid+09]. Furthermore, around saturation density $\rho_\infty$, the symmetry energy $S_2(\rho)$ can be expanded to second-order in density as $[\epsilon = (\rho - \rho_\infty)/\rho_\infty]$

$$S_2(\rho) = J + L\epsilon + \frac{K_{\rm sym}}{2}\epsilon^2.$$

Here, $J$ is the symmetry energy at saturation, while $L$ and $K_{\rm sym}$ are the slope and curvature parameters of the symmetry energy at saturation, i.e.,

$$J = S_2(\rho_\infty), \quad L = 3\rho_\infty \left.\frac{\partial S_2(\rho)}{\partial \rho}\right|_{\rho=\rho_\infty}, \text{ and } \quad K_{\rm sym} = 9\rho_\infty^2 \left.\frac{\partial^2 S_2(\rho)}{\partial \rho^2}\right|_{\rho=\rho_\infty}. \tag{4.2}$$

The parameters $L$ and $K_{\rm sym}$ characterize the density dependence of nuclear symmetry energy around saturation density, and thus provide important information on the behavior of the nuclear symmetry energy at both high and low densities. In particular, the slope parameter $L$ has been found to be correlated linearly with the neutron-skin thickness of heavy nuclei, and information on the slope parameter $L$ can thus in principle be obtained from the thickness of the neutron skin in heavy nuclei [Bro00; HP01; Die+03; CKL05; SL05; Cen+09; Roc+11; Tsa+12; Viñ+13]. Unfortunately, because of large uncertainties in the measured neutron skin thickness of heavy nuclei, this has so far not been possible. Moreover, the symmetry energy is important for the isovector excitation modes of nuclei, like the giant dipole resonance and the giant quadrupole resonance, as we will see in chapter 5 and 6, and on the composition of neutron stars, their cooling properties and the behavior of binary mergers, which are possible sources of gravitational waves. A systematic analysis of available results for $J$ and $L$ has been carried out in Refs. [Tsa+12; LL13]. In Fig. 4.1 we show a plot, taken from Ref. [Tsa+12], of $L$ as a function of $J$, in which a state-of-the-art summary of the constraints coming from experiments are introduced. In Fig. 4.2, which is taken from [Viñ+13], a summary of the experimental constrains on $L$ is provided. In principle, $K_{\rm sym}$ can be extracted experimentally studying the giant monopole resonance in neutron-rich nuclei. However, the large uncertainties on this value obtained from a systematic study of this resonance in different mass regions do not allow to distinguish between different theoretical predictions [LCK08].

The theoretical studies on the isospin-asymmetric nuclear matter were pioneered by Brueckner et al. [BCD68]. Since them, various approaches involving different physical approximations and numerical techniques have been developed to deal with the many-body problem of isospin-asymmetric nuclear matter. A recent review can be found in Ref. [LCK08]. These approaches can be roughly classified into three categories: the mi-croscopic many-body approach, the effective-field theory approach, and the phenomenological approach. In the first, bare nucleon-nucleon (NN) interactions, obtained from fitting experimental NN scattering phase shifts and deuteron properties, and empirical three-nucleon (3N) forces are used. Effective-field theory employs an effective interaction based on a perturbative expansion of the NN interaction or the nuclear mean-field within power-counting schemes, in which a separation of scales is introduced such that short-range correlations are separated from long- and intermediate-range parts of the NN interaction. At present, the effective-field theory approach in nuclear physics is based on either the density functional theory or the chiral perturbation theory. The phenomenological approach is based on effective density-dependent nuclear forces or effective Lagrangians. In these approaches, a number of parameters are adjusted to fit the properties of many finite nuclei and nuclear matter. This type of model mainly includes the relativistic mean-field (RMF) theory, and non-relativistic Hartree-Fock approaches using Skyrme or Gogny interaction.

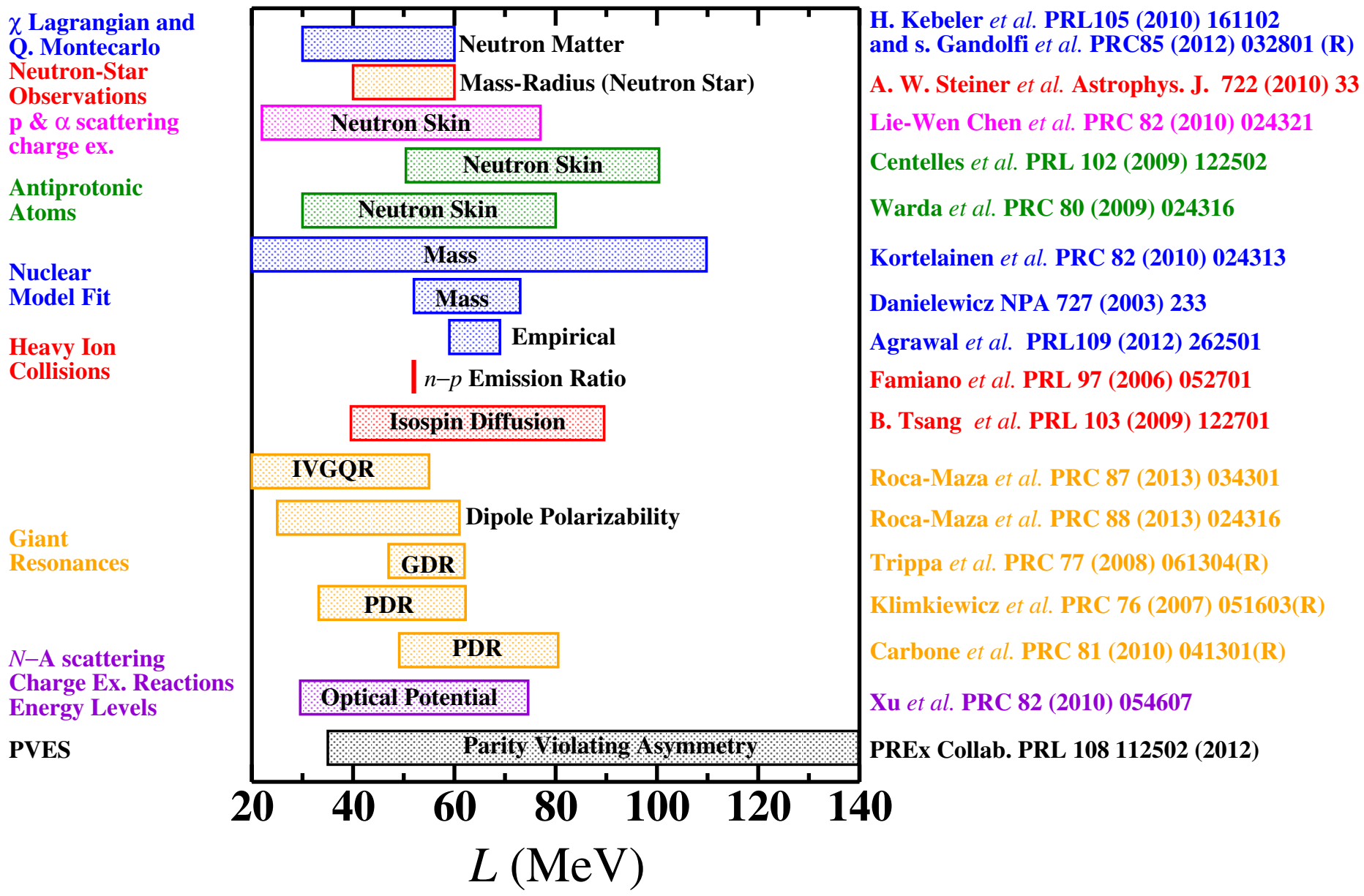

**Figure 4.2**
Experimental constraints on $L$ from different experiments (figure taken from [Viñ+13]). In the labels the type of experiment and the corresponding reference is listed.

In Fig. 4.3 (taken from Ref. [LCK08]) we show the symmetry energy as predicted by different Skyrme interactions (a) and different RMF models (b). Although all these predict a value for $J$ in the range of $26 - 44$ MeV, the predicted slope and curvature at saturation are very different. Therefore, the current knowledge about the EoS of asymmetric nuclear matter is still rather limited. In particular, the behavior of symmetry energy at supranormal densities, which is essential for understanding the properties of neutron stars, is one of the most uncertain among all properties of dense nuclear matter.

## 4.1 The symmetry energy and the neutron skin thickness

For the following discussion the relation between the neutron skin thickness of a nucleus and the $L$ parameter will be important. The neutron skin thickness of a nucleus is con-

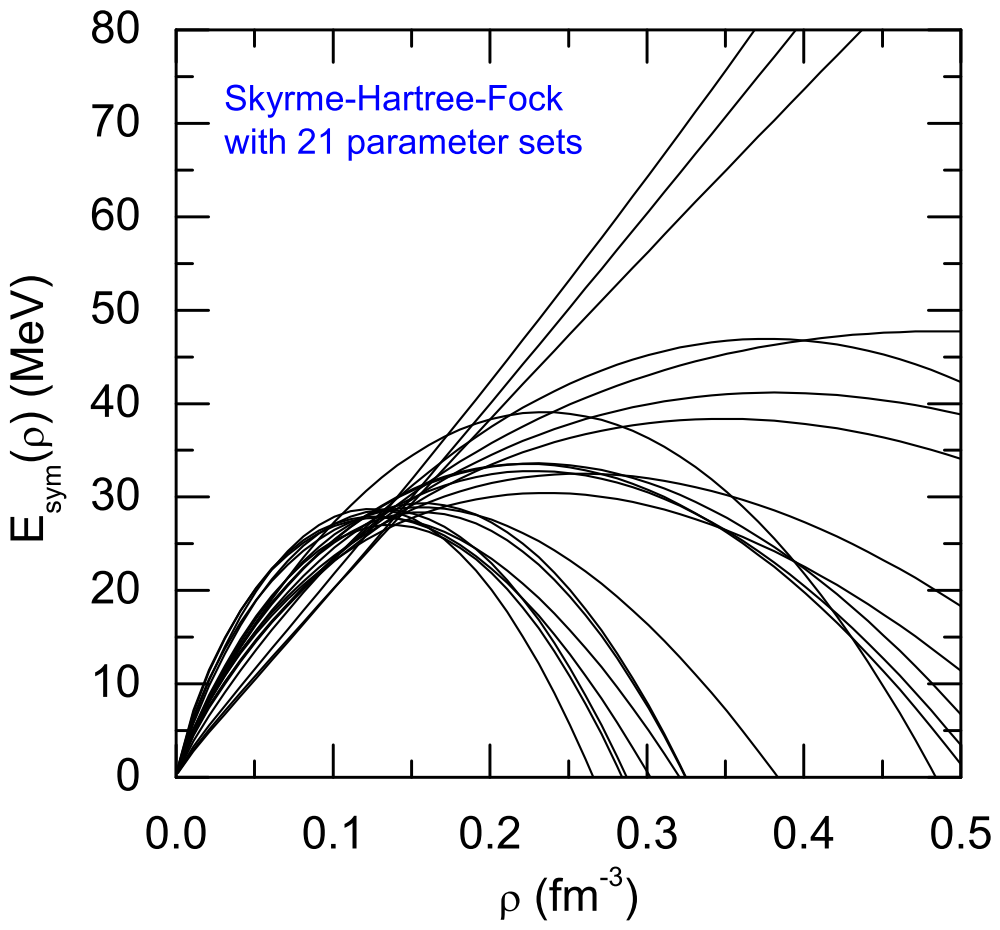

(A)

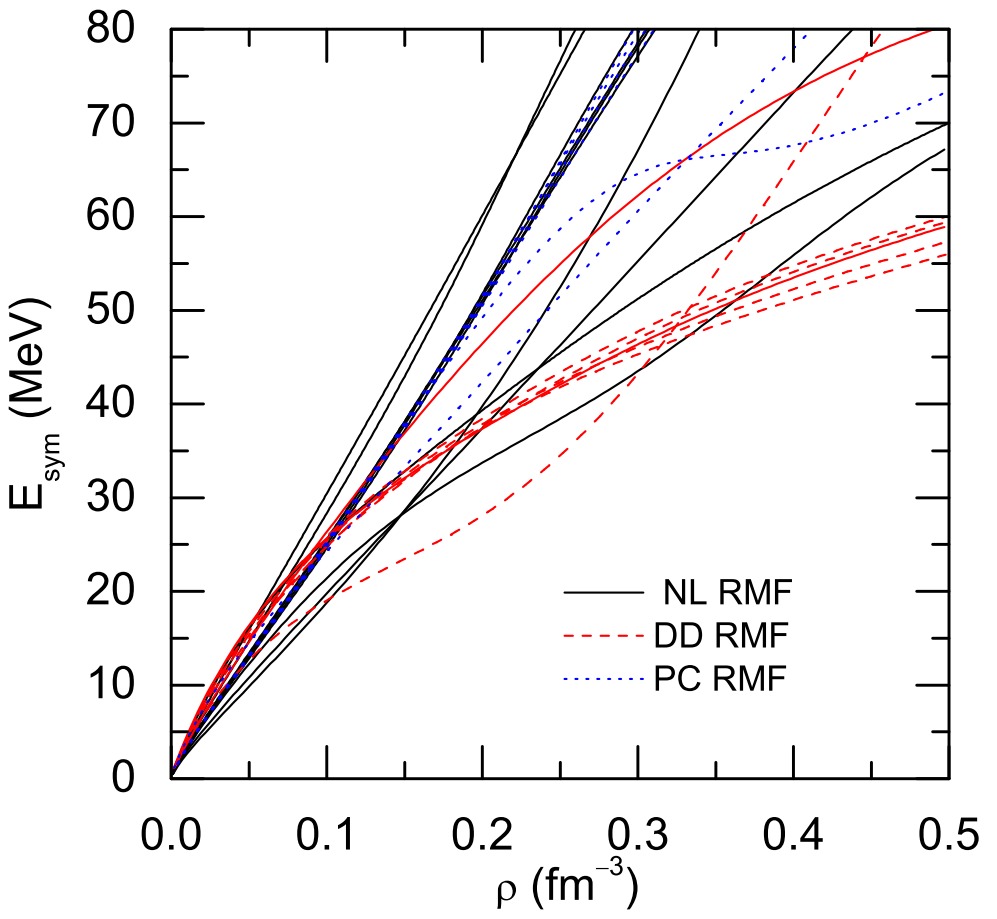

(B)

**Figure 4.3**
(A) Density dependence of the nuclear symmetry energy $S_2$ from Skyrme-Hartree-Fock with 21 sets of Skyrme interaction parameters. (B) Same as (A) from 23 different RMF model. These figures are taken from Ref. [LCK08], to which we address the reader for further details.

ventionally defined as the difference between the neutron and proton rms radii:

$$\Delta r_{np} = \sqrt{\langle r^2 \rangle_n} - \sqrt{\langle r^2 \rangle_p}. \tag{4.3}$$

As already recalled, a linear correlation between the neutron skin thickness of heavy nuclei (such as $^{208}$Pb) and the slope of the symmetry energy has been found. This correlation can be clearly seen in Fig. 4.4, taken from Ref. [Viñ+13], where Hartree-Fock or Hartree calculations of the neutron skin thickness of $^{208}$Pb with different Skyrme, Gogny and relativistic mean-field models is displayed as a function of $L$. The linear fit of the results gives

$$\Delta r_{np}\ (\text{fm}) = 0.101 + 0.00147L\ (\text{MeV}). \tag{4.4}$$

This correlation can be explained in the Droplet Model [Cen+09]. In this model, the neutron skin thickness can be written as

$$\Delta r_{np}^{\text{DM}} = \sqrt{\frac{3}{5}} \left[ t - e^2 Z/70J + \frac{5}{2r_0 A^{1/3}} (b_n^2 - b_p^2) \right].$$

The quantity $t$ is the distance between the neutron and proton mean surface locations. The correction $e^2 Z/70J$ is due to the Coulomb interaction, and $b_n$ and $b_p$ are the surface widths of the neutron and proton density profiles. Therefore, the neutron skin thickness in a heavy nucleus may be formed by two basic mechanisms. One of them is due to the separation between the neutron and proton mean surfaces, which is a bulk effect. The other one is a surface effect due to the fact that the widths of the neutron and proton

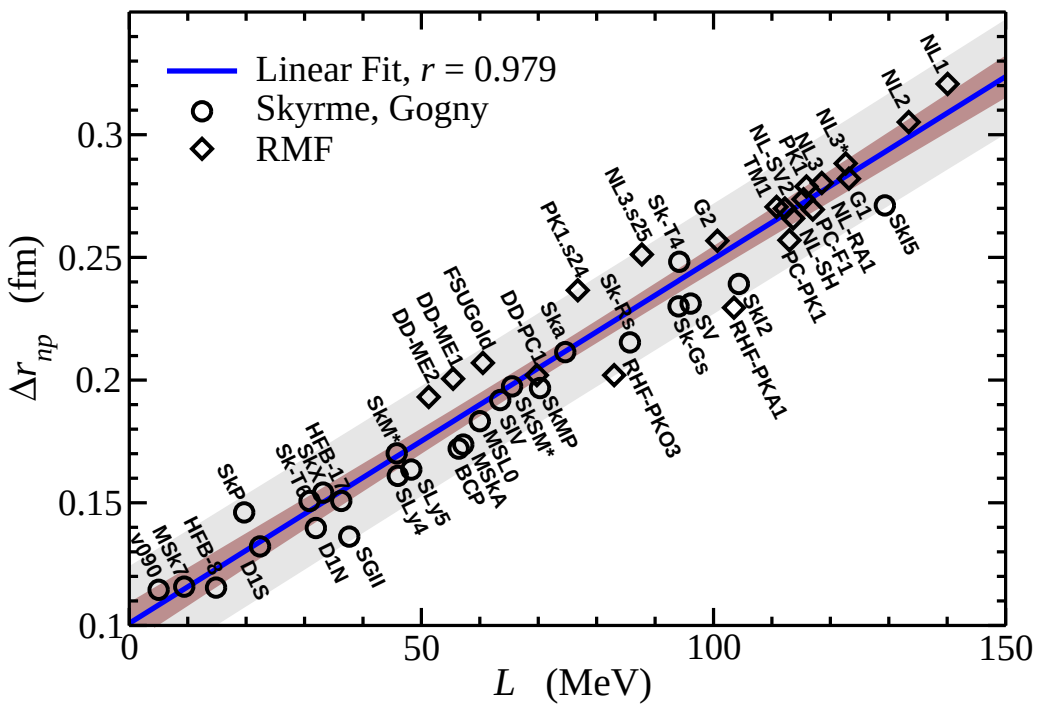

**Figure 4.4**
Neutron skin thickness $\Delta r_{np}$ of $^{208}$Pb against the slope parameter of the symmetry energy at saturation density $L$. The predictions of diverse nuclear energy density functionals (including Skyrme and Gogny forces and RMF) are shown.The thinner and thicker shadowed regions represent, respectively, the 95%-confidence band and the 95%-prediction band of the linear regression. Figure taken from Ref. [Viñ+13].

surfaces may be different. The quantity $t$ can be expressed as

$$t = \frac{2r_0}{3J} L \left(1 - \epsilon \frac{K_{\text{sym}}}{2L}\right) \epsilon A^{1/3} (I - I_C),$$

where $I = (N - Z)/A$. The corrective term with $K_{\text{sym}}$ has a limited influence on the result as far as $\epsilon$ is small.

Chapter 5

# Giant Resonances

The study of the properties of a physical system finds a powerful method in the analysis of the response of the system itself to a weak external perturbation. As it was pointed out in chapter 3, for the atomic nucleus the external probe can be hadronic or electromagnetic. In the energy range below 50 MeV the system can respond both through the excitation of relatively simple states involving only few (even one) particles, or exhibiting broad resonances. These are known as giant resonances (GRs) and correspond to a collective state involving many or all the particles in the nucleus [BBB98; HvW01].

A GR can be viewed as a high-frequency, damped, nearly harmonic density (or shape) vibration around the equilibrium configuration. Like any resonance, a GR is described by three observables: the excitation energy $E_x$, which is relatively easy to determine from the experimental data, the width $\Gamma$ and strength $S$, being these two more difficult to determine because in general in the experimental spectrum various resonances often overlap and are on top of a large continuum. From a microscopic point of view, the GRs can be described as RPA states (section 3.1), thus they can be viewed as a coherent superposition of particle-hole excitations. Because of that, in the analysis of GR's spectra, it is useful to make use of sum rules, to which section 3.4 is devoted.

Since GRs are excitations involving almost all the nucleons, it should be expected that their characteristic properties do not depend on the detailed microscopic structure of the nucleus but rather on its bulk structure. This implies that GRs should be present in all but very light nuclei and that their parameters (for a fixed multipolarity) vary smoothly as a function of the number of nucleons $A$. As a matter of fact, GRs currently provide the most reliable information on the properties of nuclear matter such as compression modulus and symmetry energy (see chapter 4). Moreover, the investigation of the strength distribution gives access to the study of the nuclear deformations in the ground state as well as to the shape evolution of nuclei as a function of spin and temperature of the system.

The first evidence of a giant-resonance phenomenon was obtained in 1937 by Bothe and Gentner [BG37], investigating the production of radioactivity induced by the bombardment of various targets with photons. They found that in some samples the experimental cross sections were two order of magnitude larger than expected and concluded that there exists a resonant absorption. A systematic study of resonance behavior began only in 1947 through the work of Baldwin and Klaiber, who observed strong resonance behavior in photo-absorption experiments at an excitation energy in the range 13 – 15 MeV in uranium [BK47]. This resonance can be interpreted as the excitation, by the electromagnetic dipole field of the photon, of a collective nuclear vibration in which all the protons in the nucleus move collectively against all the neutrons, resulting in a separation between the center of mass and the center of charge. A wider historical overview can be found in [HvW01].

## 5.1 Decay and damping of giant resonances

The study of the decay properties may have interesting implications: an insight into the mechanisms responsible for strong damping or the test for the microscopic description of the GR itself. The variety of mechanisms which can be responsible for the GR's relaxation can be grouped in two sets depending on whether the energy of the vibration escapes the system, or whether it is redistributed into other degrees of freedom within the system. The total width has four different contribution

$$\Gamma_{tot} = \Gamma_{Land} + \Gamma^{\uparrow} + \Gamma_{\gamma} + \Gamma^{\downarrow}$$

The first term, $\Gamma_{Land}$, corresponds to the "Landau damping", first introduced by Landau in the classic treatment of a plasma. Such spreading can occur if some non-collective $1p - 1h$ configurations with the same quantum number as the collective state have an energy close to that of this state. Thus, the collective states is strongly coupled to the $1p - 1h$ states and its strength is fragmented.

In general, GRs are well above the particle-decay threshold, therefore they acquire a width by emission of particles, deriving from the coupling to the continuum. This is called the escape width $\Gamma^{\uparrow}$. In heavy nuclei, neutron emission is predominant, the Coulomb barrier hindering proton tunneling. While in light nuclei is large, it becomes less important in medium-heavy nuclei, where typical values are of the order of 1 MeV. The width $\Gamma_{\gamma}$ is related to the photon emission and in general is only few per cent of the total width, being of the order of the keV or smaller.

The last contribution to the total width is the spreading width $\Gamma^{\downarrow}$. The GRs are located at high excitation energy where a high density of $2p - 2h$ configurations of the same spin and parity as the GR $1p - 1h$ configuration occurs. Thus, the $1p - 1h$ state first mixes with $2p - 2h$ states, which then dissolve into progressively more complicated $3p - 3h, \ldots$ states, till finally the energy has been spread over all degrees of freedom and a compound nucleus is formed. It turns out that the $\Gamma^{\downarrow}$ term gives the largest contribution to the total width: it describes primarily the coupling of the $1p - 1h$ state to $2p - 2h$ ones; each successive phase of the decay chain has a much smaller width. In principle, also the $2p - 2h, \ldots$ intermediate states can decay by particle or $\gamma$ emission: these decays can be accounted for by using a statistical model.

## 5.2 Classification of giant resonance modes

Giant resonances can be classified according to their multipolarity $L$, spin $S$ and isospin $T$ quantum number. Protons and neutrons can oscillate in phase or out of phase against each other: in the former case the vibration is called isoscalar, i.e. without a variation in isospin ($\Delta T = 0$), in the latter the vibration is called isovector ($\Delta T = 1$). As a rule, for the same multipole mode, the isovector one will be at a higher excitation energy than the isoscalar one since extra energy is required to separate the neutron and proton distributions. In $\Delta S = 1$ modes, or magnetic, nucleons with spin up ($\uparrow$) vibrate against nucleons with spin down ($\downarrow$), while if $\Delta S = 0$ (electric mode) spin states are in phase. Finally, giant resonances are classified depending on the angular momentum $\Delta L$ and the parity, $\pi = (-)^{\Delta L}$. A pictorial representation of the different possible modes is shown in Fig. 5.1

The qualitative features of giant resonances can be understood by considering a schematic shell model picture. In this model,the parity of the single-particle wave functions in subsequent shells $N, N+1, N+2, \ldots$ is alternating and their energy difference is given by $\Delta E = \Delta N \cdot 1\hbar\omega = \Delta N \cdot 41\, A^{-1/3}$ MeV. Because of parity conservation, odd $L$ transitions require $\Delta N = 1, 3, \ldots$, while even $L$ transitions $\Delta N = 0, 2, \ldots$ (Fig. 5.2).

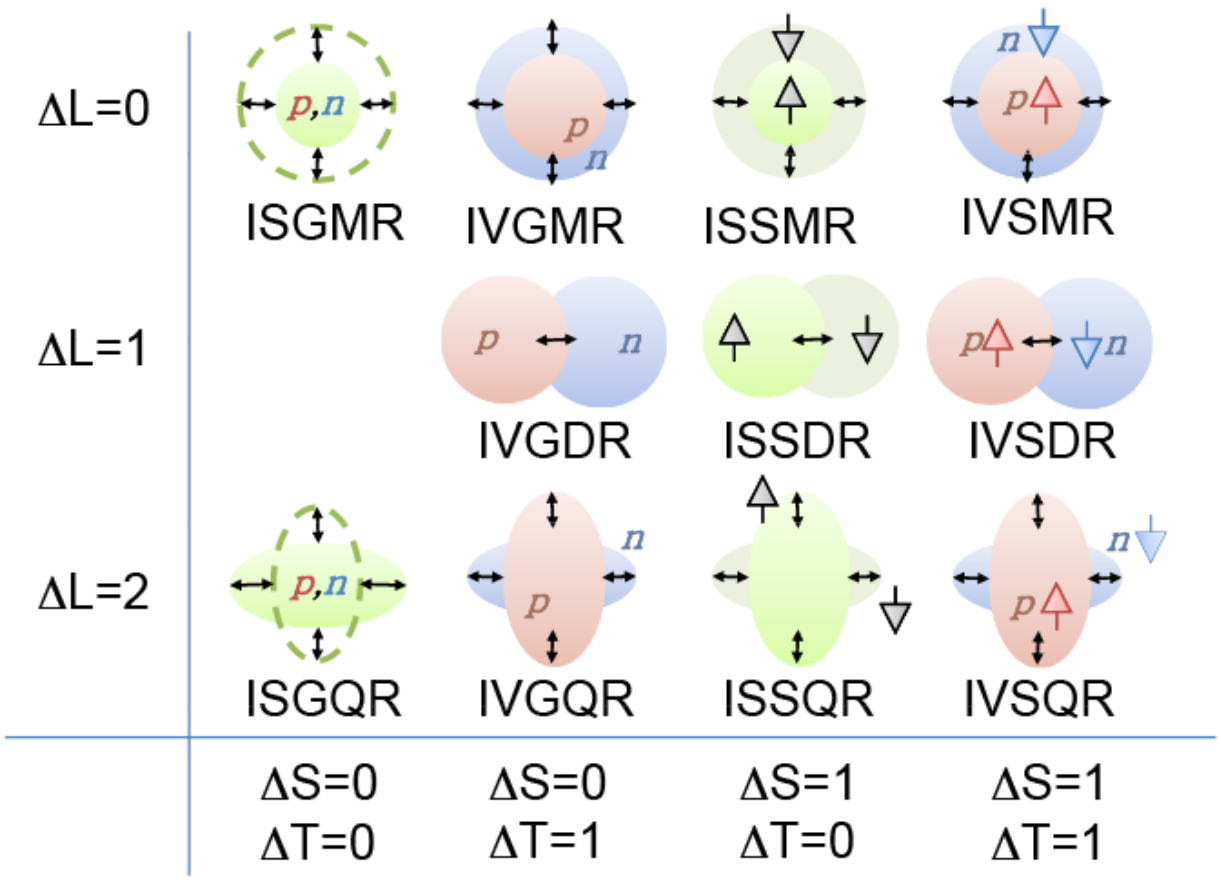

**Figure 5.1**
Pictorial representation of various collective modes.

In the following section we will give a brief description of the resonances of interest for this work.

### 5.2.1 The dipole spectrum

Being excited through photon scattering, the dipole spectrum has been studied since the early ages of nuclear physics and, up to now, it is by far the best known. In this section we will present the main feature of both isovector and isoscalar excitations.

#### The isovector giant dipole resonance (IVGDR)

As it was recalled in the introduction, the first GR discovered was the isovector dipole. For most nuclei with $A > 50$, the total absorption cross section in the $10 - 20$ MeV energy range can be fitted by a Lorentzian curve:

$$\sigma(E) = \frac{\sigma_m}{1 + \left[\left(E^2 - E_m^2\right)^2 / E^2\Gamma^2\right]}, \tag{5.1}$$

where the subscript $m$ refers to the peak cross section. The peak energy $E_m$ can be well reproduced by $E_m = 80\, A^{-1/3}$ MeV. The width $\Gamma$ is about $4 - 8$ MeV and it reaches its minimum values for closed shells nuclei. For axially symmetric deformed nuclei, the cross section is split into two parts, corresponding to an IVGDR vibration along or perpendicular to the symmetry axis [HvW01]. Moreover, in the last few years, the main properties of the GDR build on excited states have also been measured, exploring regions of excitation energies between 10 and 500 MeV and spins up to $60\hbar$ through both fusion reactions and inelastic scattering. These experiments have demonstrated that there is no significant shift of the centroid energy with either temperature or angular momentum, whereas the width increases both with excitation energy and spin, the latter becoming important only above $35\hbar$ [SB06].

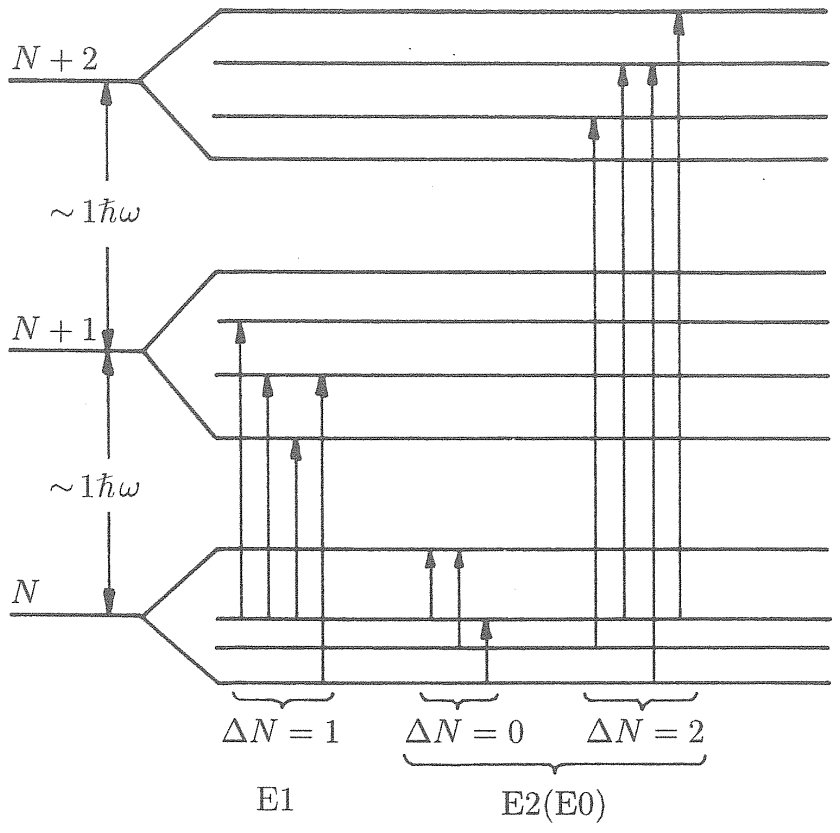

**Figure 5.2**
Schematic picture of of E1 and E2(E0) single-particle transitions between shell model spaces.

Even before the discovery of the resonance, Migdal theorized that the excitation energy for dipole absorption can be related to the symmetry energy $S_2(\rho)$ (see chapter 4). This correlation can be understood on the basis of the hydrodynamical model proposed by Lipparini and Stringari [LS89]. This model has been used in [TCV08] to write the constrained energy $E_{-1}$ as

$$\mathscr{E}_{-1} = \sqrt{\frac{3\hbar^2}{m\langle r^2\rangle}\frac{b_{vol}}{1+\frac{5}{3}\frac{b_{\rm surf}}{b_{\rm vol}}A^{-\frac{1}{3}}}(1+\kappa_D)} \approx \sqrt{\frac{6\hbar^2}{m\langle r^2\rangle}S_2(0.1\,{\rm fm}^{-3})(1+\kappa_D)},$$

where $S_2(0.1\,\mathrm{fm}^{-3})$ is the symmetry energy evaluated at $\rho = 0.1\,\mathrm{fm}^{-3}$ and $\kappa_D$ is the dipole enhancement factor. The strong resulting correlation has been used to extract a constraint on $S_2(0.1\,\mathrm{fm}^{-3})$ which, introducing an acceptable range for $\kappa_D$, is found to lie in the interval $24.1 \pm 0.8\,\mathrm{MeV}$.

### The isoscalar giant dipole resonance (ISGDR)

The isoscalar giant dipole resonance has remained, in many ways, one of the most elusive resonance for many years. In investigating this mode, one is looking for a "second-order" effect, since as it can be inferred by section 3.3, in the first order the isoscalar dipole excitation corresponds to a center-of-mass displacement. The isoscalar $E1$ strength displays a bimodal structure with two broad components: one in the region close to the IVGDR ($\approx 1\hbar\omega$, between 10 and 20 MeV) and the other at higher energy, close to the electric octupole resonance ($\approx 3\hbar\omega$, above 20 MeV).

Physically, the high energy component or ISGDR tends to correspond to a hydrodynamical density oscillation in which the volume of the nucleus remains constant and the state can be visualized in the form of a compression wave oscillating back and forth through the nucleus along a definite direction. This has been generally referred to as the "squeezing mode". Accordingly, its excitation energy can be evaluated in the liquid-drop

model to be [BM75]

$$E \approx 8\sqrt{K_{nm}}A^{-1/3}\,\text{MeV} = 124\,A^{-1/3}\,\text{MeV},$$

where $K_{nm}$ is the incompressibility of nuclear matter, assumed to be 240 MeV in the last estimate. This macroscopic scaling is relatively well obeyed by microscopic calculations [Paa+07]. On the other hand, the low energy part of the spectrum does not definitely scale with the nuclear incompressibility. Some authors have claimed that this low energy strength corresponds to a toroidal motion [Vre+02; Kva+11], however very few experimental information on this component is available.

From the experimental point of view, the first attempts to identifying the ISGDR were made from the beginning of the 80's, with contradictory results: indications of this resonance were reported in inelastic scattering measurements [Mor+80; Dja+82; Mor+83; Bon+84; Ada+86], while in other experiments the absence of the resonance was reported [Yam+81; Ber+86; McD+86]. The main issue is the superposition of the ISGDR and the high-energy octupole resonance, which lies in the same energy range; the expected angular distribution differ in any discernible manner only at very small angles ($\leq 5°$). The first measure around 0°was performed in 1997 [Dav+97] with an inelastic $\alpha$-scattering on $^{208}$Pb. Only later a wider systematic study has been performed on $A \geq 90$ nuclei [CLY01; Ito+03; Uch+04].

### The pygmy dipole resonance (PDR)

The fact that the observed cross section in the energy range in the IVGDR exhaust nearly 100 % of the sum rule for isovector dipole transition lead to the notation "pygmy" for the strength not attributed to the IVGDR. The first evidence of some enhancement in the strength around 5 − 7 MeV was found in the early 1960's by Bartholomew [Bar61], in a systematic study of $\gamma$-rays after thermal neutron capture in different nuclei. A first theoretical interpretation of this excitation mode was given by Mohan and collaborators in a three-fluid hydrodynamical model in 1971 [MDB71]. Here, the three fluids are the protons, the neutrons sitting in the same orbitals as the protons and the excess neutrons. This lead to two independent electric dipole resonances, one originating from the oscillation of all protons against all neutrons and an energetically lower lying mode where only the excess neutrons oscillate against a proton-neutron saturated core. Mohan estimated for $^{208}$Pb the former mode to be "more than two orders of magnitude stronger" than the latter one, which is roughly in agreement with our present-day experimental knowledge. The field received a boost with the advent of high-energy radioactive beams and the possibility to study their properties in reaction experiments in the 1980's, when abnormally large cross sections for nuclei located at the neutron drip line (in the lithium region) have been observed [Tan+85]. This enhancement can be explained because of the presence of weakly bound neutron forming a neutron halo: in such a situation, the dipole strength related to the excitation of the loosely bound halo neutrons is completely decoupled from the IVGDR and is located at the neutron separation threshold. A further acceleration both in the experimental and theoretical investigation has started in the nineties to understand the dipole response of neutron-rich nuclei in radioactive, as well as in stable nuclei. However, the dynamics of these low-lying dipole excitations has not been resolved yet. In particular, it is not clear whether some of these states correspond to a collective soft mode (i.e. with energy far below the giant resonance region) or they all simply result from incoherent single-particle excitations, contributing to the non-collective threshold strength. The main results of the extensive theoretical studies can be summarized as follows: (a) the dipole strength distributions in neutron-rich nuclei are more fragmented than in stable nuclei; (b) the centroids are calculated at significantly lower energies and (c) the ratio of neutron to proton particle-hole amplitudes of low-lying dipole states is much higher than in stable nuclei and, accordingly, the isoscalar (IS) transition densities do not vanish and isoscalar probes can excite these states. A recent review of the experimental findings related to this part of the spectrum can be found in Ref. [SAZ13], while for review on the theoretical description of this mode of excitation we address to Ref. [Paa+07].

From the theoretical point of view, such a mode also provides a unique test of the isospin-dependent components of effective nuclear interactions, which are particularly pronounced in nuclei with a large proton-neutron asymmetry.

Moreover, if the PDR is an oscillation of excess neutrons against an isospin saturated core, it is reasonable that the total strength of the PDR should be related to the thickness of the neutron-skin. Thus, the total strength provides an experimentally constrained approach to determine the neutron skin thickness of atomic nuclei, which is in turn connected with the the symmetry energy and its density dependence. In microscopic calculations, comparing different effective interactions, it was found that the centroid energy of the PDR is not sensitive to the density dependence of the symmetry energy, while the fraction of the EWSR exhausted by the PDR increases sharply with increasing neutron skin [Pie06; Kli+07; Car+10; Vre+12]. In particular, in [Car+10], a strong correlation has been found between the EWSR exhausted by the pygmy dipole resonance in $^{68}$Ni and $^{132}$Sn and the parameter $L$ in Skyrme-Hartree-Fock and relativistic mean-field models. This has allowed, via the experimental value of the EWSR, to constrain the values of $L$ in the range $64.6 \pm 15.7$ MeV, in agreement with the result coming from heavy-ion diffusion reactions. Besides, the linear correlation between $L$ and $\Delta r_{np}$, has allowed to constrain the neutron skin thickness of $^{68}$Ni, $^{132}$Sn and $^{208}$Pb, to, respectively, $0.200 \pm 0.015$ fm, $0.258 \pm 0.024$ fm and $0.194 \pm 0.024$ fm. A much different conclusion is drawn in [RN10], where the correlation between the properties of the PDR (centroid and ESWR) and the symmetry energy is claimed to be poor. The reason would be the unreal collective character of the PDR, whose strength would be caused only by shell effects around the Fermi surface acting on a single particle -hole pair.

Besides, the occurrence of low-lying dipole strength plays an important role in predictions of neutron-capture rates in the $r$-process nucleosynthesis, and consequently in the calculated elemental abundance distribution. Namely, although its transition strength is small compared with the total dipole strength, the low-lying collective dipole state located close to the neutron threshold can significantly enhance the radiative neutron-capture cross section on neutron-rich nuclei [GK02; GKS04].

### 5.2.2 The quadrupole spectrum

#### Isoscalar giant quadrupole resonance (ISGQR)

The ISGQR was first discovered in 1971-1972 in inelastic electron and proton collision [PW71; FT72; LB72]. The excitation energy is rather well described by $E_m = 64\,A^{-1/3}$ MeV: at this energy, $50 - 100\,\%$ of the EWSR is concentrated. The width is smaller in heavy nuclei than in light ones, and even in this case the widths reach a minimum for closed shell nuclei. In general the width is between 2 and 7 MeV.

In a simple approach in which the nucleus is described as a quantal harmonic oscillator (QHO), the excitation energy of the ISGQR is found to be proportional to the shell energy-gap $\hbar\omega_0 \sim 41\,A^{-1/3}$ MeV and, if the nuclear effective interaction is also velocity-dependent, to the nucleon effective mass, namely [BM75; Bla80]

$$E^{\mathrm{ISGQR}} = \sqrt{\frac{2m}{m^*}}\hbar\omega_0.$$

Comparison with the experimental findings allows us to put some constraints on the effective mass. For heavy nuclei, experimental data favor an effective mass close to 1, while in light nuclei a smaller effective mass seemed to be required. This may be an indication that the coupling with the collective modes, which is responsible for the enhancement of the effective mass at the Fermi surface in heavy nuclei (as we will see in section 7.3), may be less important in light nuclei.

**The isovector giant quadrupole resonance (IVGQR)**

The knowledge on the IVGQR is very limited because of lack of selective experimental probes that can excite this resonance. Moreover, it is a high-excitation energy mode, with large width and small excitation cross section, making it difficult to distinguish between the resonance and the underlying continuum. As a matter of fact, the results of experimental studies show a large spread in the reported parameters as a result of large backgrounds and model-dependent corrections [HvW01]. On the energy of the resonance there is reasonable agreement among different experiments, while the reported energy weighted sum rules are between 1.0 and 1.4 isovector quadrupole EWSRs and the reported widths vary between 3.5 and 10 MeV for nuclei having similar $A$, with large uncertainties in all cases. The accuracy in the experimental determination of the IVGQR has been considerably improved only recently [Hen+11], exploiting the use of intense, nearly monoenergetic, linearly polarized $\gamma$-ray beams in polarized Compton scattering. Also, a new measurement method consisting of making the ratio of cross sections perpendicular and parallel to the plane of polarization of the incident beam to locate the IVGQR is used.

Being a $2\hbar\omega$ excitation, it is expected to be located at an energy higher than $2\hbar\omega_0$. In the QHO approach, it can be found that the excitation energy depends on the symmetry potential $V_{\mathrm{sym}}$ [BM69] and on the effective mass [BM75]:

$$E^{\mathrm{IVGQR}} = 2\hbar\omega_0\sqrt{1 + \frac{5}{4}\frac{m^*\hbar^2}{2m}\frac{V_{\mathrm{sym}}\langle r^2\rangle}{(\hbar\omega_0)^2\langle r^4\rangle}}.$$

This estimate will be used in section 6.2 to extract information about the symmetry energy out of the excitation energy of the resonance.

Chapter 6

# Numerical Results

In this chapter we will present the results that we have obtained solving the SHF plus RPA equations [Col+13] in different circumstances. The chapter is organized as follows: in section 6.1 the results for the dipole spectrum are analyzed, and in section 6.2 we present the main features that we can extract for the isoscalar and isovector giant quadrupole resonances.

## 6.1 The dipole spectrum

In this section we present the main properties of the dipole spectrum, both isoscalar and isovector, focusing in particular on the low-energy part of it, where the so-called pygmy resonance has been attracting a lot interest for the last decade. In particular, as the collective nature of this structure has not been confirmed yet, we prefer the term pygmy dipole strength (PDS) rather than pygmy dipole resonance (PDR). Since the fraction of EWSR exhausted by this sector of the strength function was found to be related to the density dependence of the symmetry energy, in this work we have used three Skyrme interactions with different isovector properties, namely SGII ($L$ = 37.63 MeV), SLy5 ($L$ = 48.27 MeV) and SkI3 ($L$ = 100.52 MeV). We have analyzed three systems, $^{68}$Ni, $^{132}$Sn and $^{208}$Pb, chosen as representatives of different mass regions. We restrict ourselves to double closed shell nuclei in order to avoid pairing effects, which can in principle complicate the interpretation of our results. The low-energy dipole response of all studied nuclei has been measured [Rye+02; Adr+05; Wie+09].

As explained in section 3.2, the RPA calculation of dipole isoscalar response is affected by numerical approximations with the result that the spurious state does not lie at zero energy and thus it is not orthogonal to the other RPA states. We subtract the spurious state in two ways: by correcting the transition densities (see section 3.3) and by creating a new set of $X$ and $Y$ amplitudes (see section 3.2). However, since the spurious state lies at energies around (or smaller) 1 MeV, a small overlap with physical states is expected. As an example we consider $^{208}$Pb as a test nucleus, and the reliability of our method can be seen in Fig. 6.1 where we compare the strength function —calculated by convoluting the corresponding reduced transition probability with a Lorentzian of 1 MeV width— for the isoscalar dipole response predicted by the SLy5 interaction in four cases: in one case the spurious state has not been subtracted, in the second case the spurious state has been subtracted by correcting the isoscalar dipole operator (3.28), in the third case the spurious state has been subtracted from the transition densities, and in the last case the new set of $X$ and $Y$ amplitudes have been created. From this figure one clearly sees that the different prescriptions for correcting the spurious state are completely equivalent at the level of the strength function.

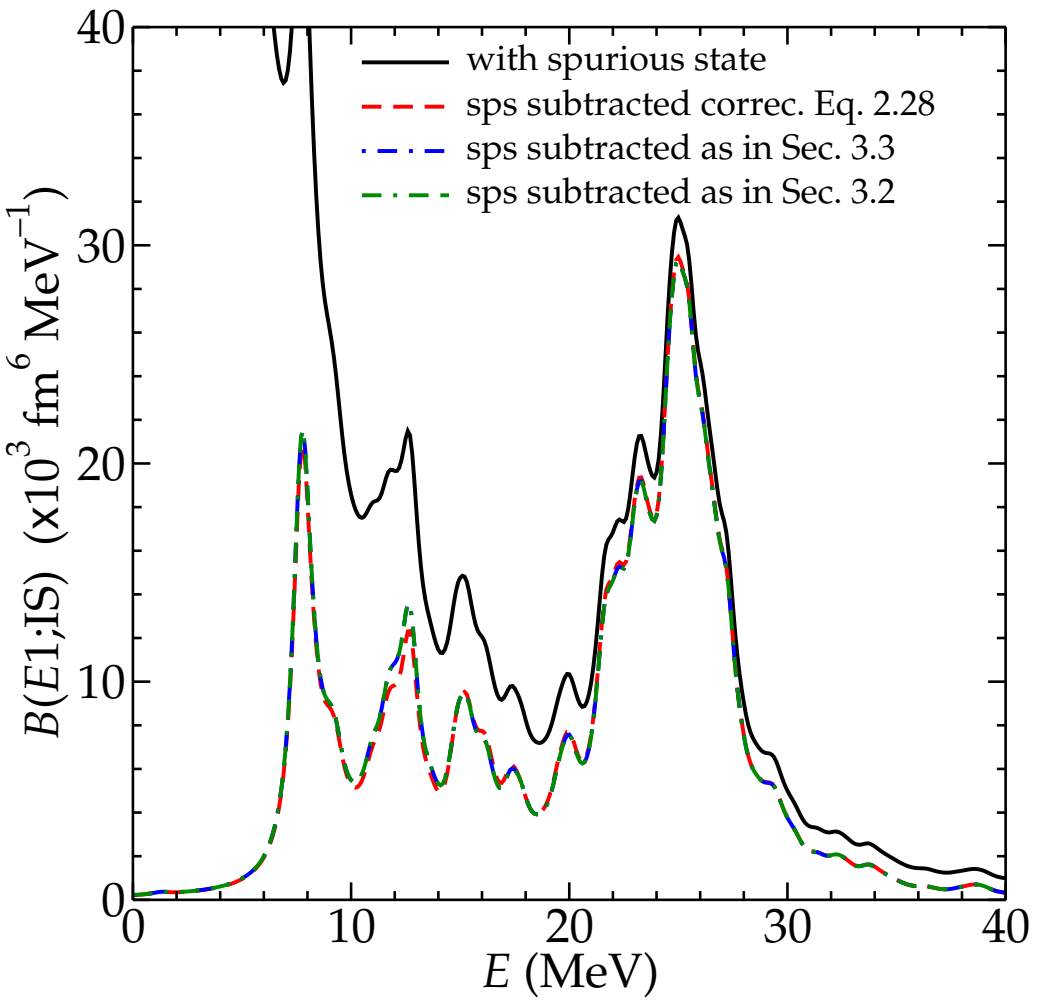

**Figure 6.1**
Isoscalar strength function in the case of the SLy5 interaction for $^{208}$Pb as a function of the excitation energy for four cases: (i) the spurious state has not been subtracted (solid line), (ii) the spurious state has been subtracted by correcting the isoscalar dipole operator (3.28) (dashed line), (iii) the spurious state has been subtracted as explained in section 3.3 (dot-dashed line), and (iv) the spurious state has been subtracted as explained in section 3.2.

### 6.1.1 Strength functions

In this subsection we analyze the isovector and isoscalar strength functions for the nuclei at hand and for the three interaction chosen. They have been calculated by convoluting the corresponding reduced transition probability with a Lorentzian of 1 MeV width.

We start analyzing the results for $^{208}$Pb. In Fig. 6.2(a) we show the strength function corresponding to the isovector dipole response as a function of the excitation energy. The inset displays in a larger scale the pygmy region. In Fig. 6.2(b), the same quantities are shown for the isoscalar dipole response as a function of the excitation energy. In both figures, the predictions of the three selected interactions are shown. The centroid energies of the PDS and the ISGDR as well as the energy peak of the IVGDR as predicted by the employed interactions ($E = 7.6 - 8.0$ MeV, $E = 20 - 21$ MeV and $E = 12 - 13$ MeV, respectively) fairly agree with the experimental data ($E = 7.37$ MeV within a window of $6 - 8$ MeV [Rye+02], $E = 20.1 - 20.5$ MeV [Gar99] and $E = 13.43$ MeV [BF75], respectively). In Table 6.1 the excitation energy and isoscalar and isovector reduced transition probabilities of the *RPA-pygmy* state —i.e. the RPA state which give rise to the largest peak in the PDS region— are detailed for all the studied nuclei as predicted by SGII, SLy5 and SkI3. We qualitatively observe that the low-energy peak found in the IV and IS dipole responses of $^{208}$Pb shows an increasing and outward trend with the excitation energy as the value of the parameter $L$ increases. This behavior is in agreement with Ref. [Car+10] where the energy weighted sum rule for the PDS was found to be linearly correlated with $L$ in mean-field models.

In the case of $^{132}$Sn and $^{68}$Ni, the strength functions for the dipole response are depicted in Fig. 6.3(a) and Fig. 6.4(a) (IV) and Fig. 6.3(b) and Fig. 6.4(b) (IS), respectively. Again, the predictions of SGII, SLy5 and SkI3 ($E = 8.5 - 9.2$ MeV for $^{132}$Sn and $E = 9.3 - 10.4$ MeV for $^{68}$Ni) are in rather good agreement with the experimental findings ($E = 9.1 - 10.5$ MeV for $^{132}$Sn [Adr+05] and $E = 11$ MeV and an energy width estimated to be less than 1 MeV for $^{68}$Ni [Wie+09]). Qualitatively in both nuclei, it seems again that the larger the value of $L$, the higher the values predicted for the excitation energy and the larger the different peaks arising in the low-energy region. In addition, we observe for all nuclei that the PDS is an order of magnitude smaller than the IVGDR and that its isoscalar counterpart is of the same order of magnitude than the corresponding ISGDR.

**Table 6.1**
Excitation energy $E$ and isoscalar ($\xi = IS$) and isovector ($\xi = IV$) reduced transition probabilities $B(E1;\xi)$ corresponding to the *RPA-pygmy* states of $^{68}$Ni, $^{132}$Sn and $^{208}$Pb as predicted by SGII, SLy5 and SkI3 interactions.

| | | $E$ [MeV] | $B(E1;IS)$ [$10^3$ fm$^6$] | $B(E1;IV)$ [fm$^2$] |
|---|---|---|---|---|
| | SGII | 9.77 | 1.9 | 1.4 |
| $^{68}$Ni | SLy5 | 9.30 | 1.7 | 0.8 |
| | SkI3 | 10.45 | 3.0 | 3.6 |
| | SGII | 8.52 | 3.3 | 1.2 |
| $^{132}$Sn | SLy5 | 8.64 | 10.0 | 1.6 |
| | SkI3 | 9.23 | 11.0 | 7.4 |
| | SGII | 7.61 | 17.0 | 2.9 |
| $^{208}$Pb | SLy5 | 7.74 | 28.0 | 2.8 |
| | SkI3 | 8.01 | 19.0 | 6.6 |

In Fig. 6.5, we focus on the relevant region for the study of PDS and we show the reduced transition probabilities in single-particle units Eq. (3.32)-(3.33). We display both the isovector (a) and isoscalar (b) dipole responses. We focus only on $^{132}$Sn, since also for the other nuclei we can draw similar conclusions. Our calculations predict an *RPA-pygmy* state characterized by $\approx$ 2-6 single-particle units: this result does not pin down clearly the collective nature of the state. As a reference, the most important RPA state in the IVGDR region contribute with about 30 single-particle units if the strength is fragmented, and with more than 60 if the strength is concentrated in one single peak. This is a clear indication of the collective nature of the IVGDR. On the contrary, from Fig. 6.5(b), we see that the RPA state leading to the pygmy peak is contributing with 15-20 single-particle units, very similar in magnitude to those displayed by the largest peak in the same isoscalar response at larger excitation energies. These large values indicate the collective character of the *RPA-pygmy* state when it is excited by an isoscalar probe.

Despite the fact that the reduced transition probabilities in s.p. units give a qualitative estimation of the collectivity, we can conclude that while the isoscalar dipole response seems to indicate that the *RPA-pygmy* state develops a certain amount of collectivity, the isovector response of the same excited state does not provide a clear trend: the collectivity displayed is very weak and depends on the used model.

### 6.1.2 Isospin nature of the low-energy RPA states

Given an excited RPA state, it can be said to be *purely* isovector if the transition densities of protons and neutrons have opposite sign, on the other hand it can be defined as *purely* isoscalar if they have the same sign. As isospin is not a good quantum number the most common situation corresponds to a mixture of a certain degree of isoscalarity and isovectoriality.

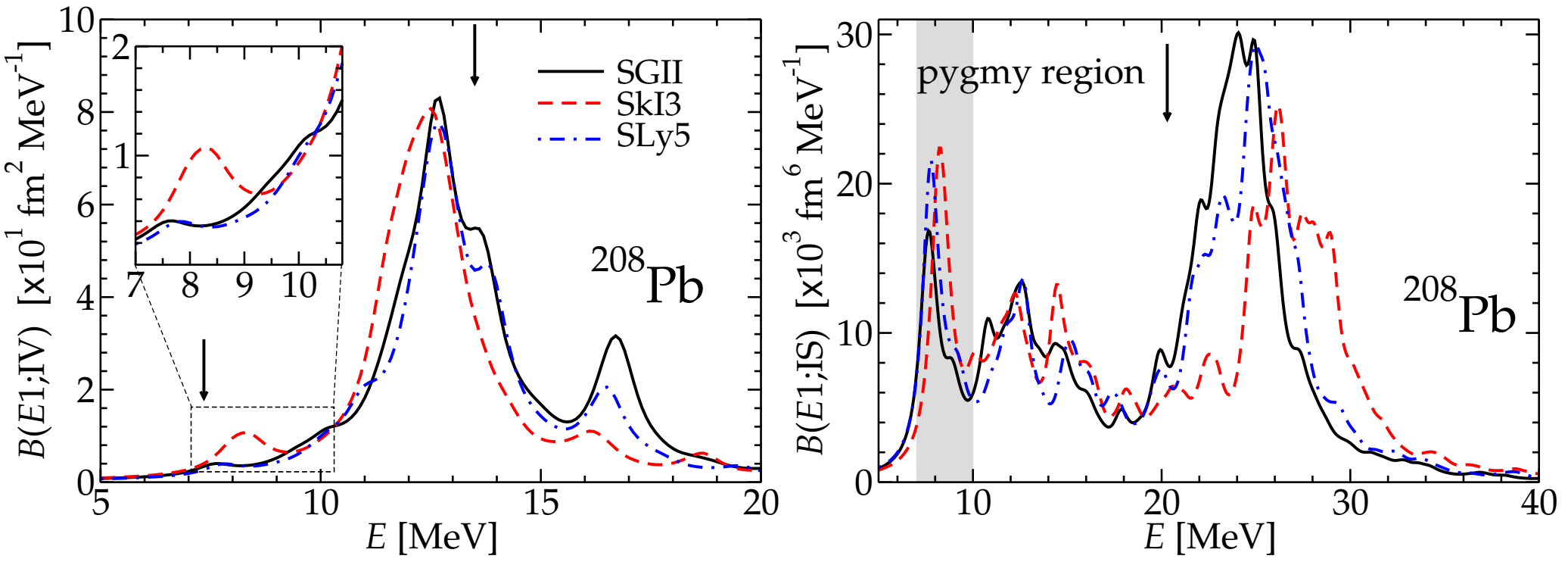

(A) (B)

**Figure 6.2**
Strength function corresponding to the isovector (A) and isoscalar (B) dipole response of $^{208}$Pb as a function of the excitation energy. The inset in (A) displays in a larger scale the pygmy region. In both figures the predictions of SGII, SLy5 and SkI3 are depicted. Black arrows indicate the experimental centroid energies for the PDS ($E$ = 7.37 MeV within a window of 6 − 8 MeV) [Rye+02], for the ISGDR ($E$ = 20.3 ± 2 MeV [Gar99]) and the energy peak for the IVGDR ($E$ = 13.43 MeV [BF75]).

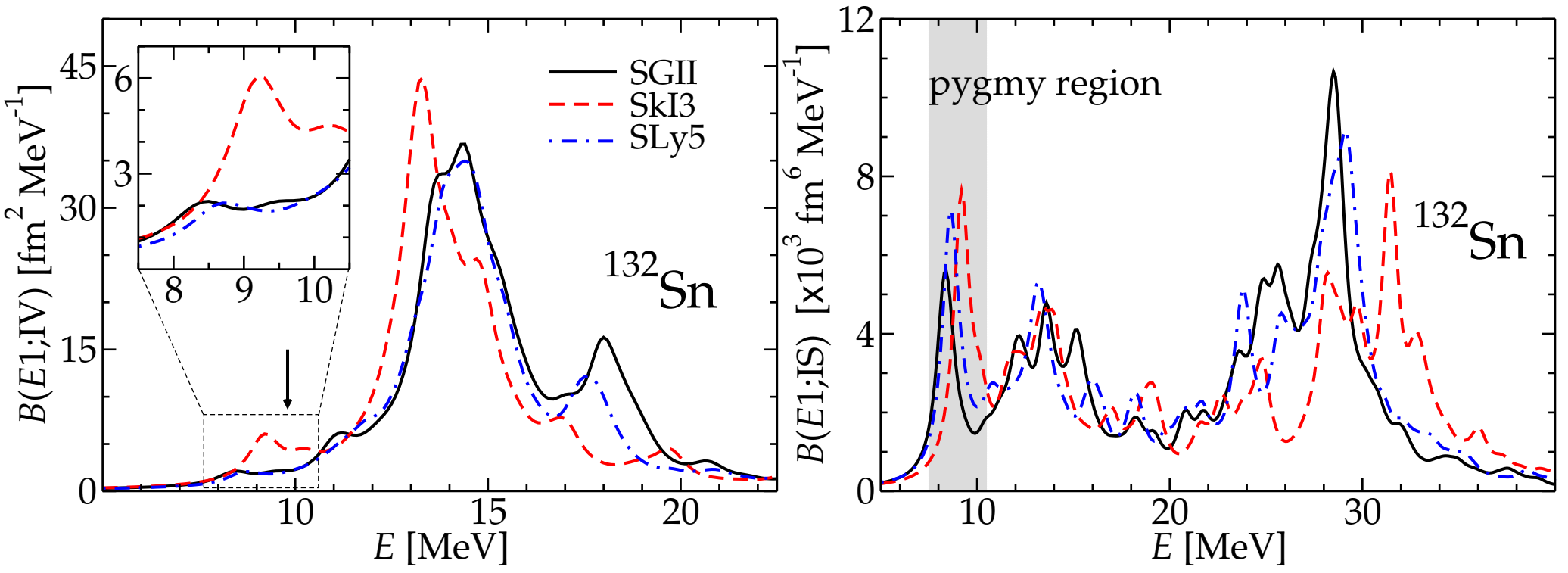

(A) (B)

**Figure 6.3**
Same as Fig. 6.2 for $^{132}$Sn. The experimental value for the peak energy of the PDS ($E$ = 9.8 ± 0.7 MeV) is indicated by a black arrow [Adr+05].

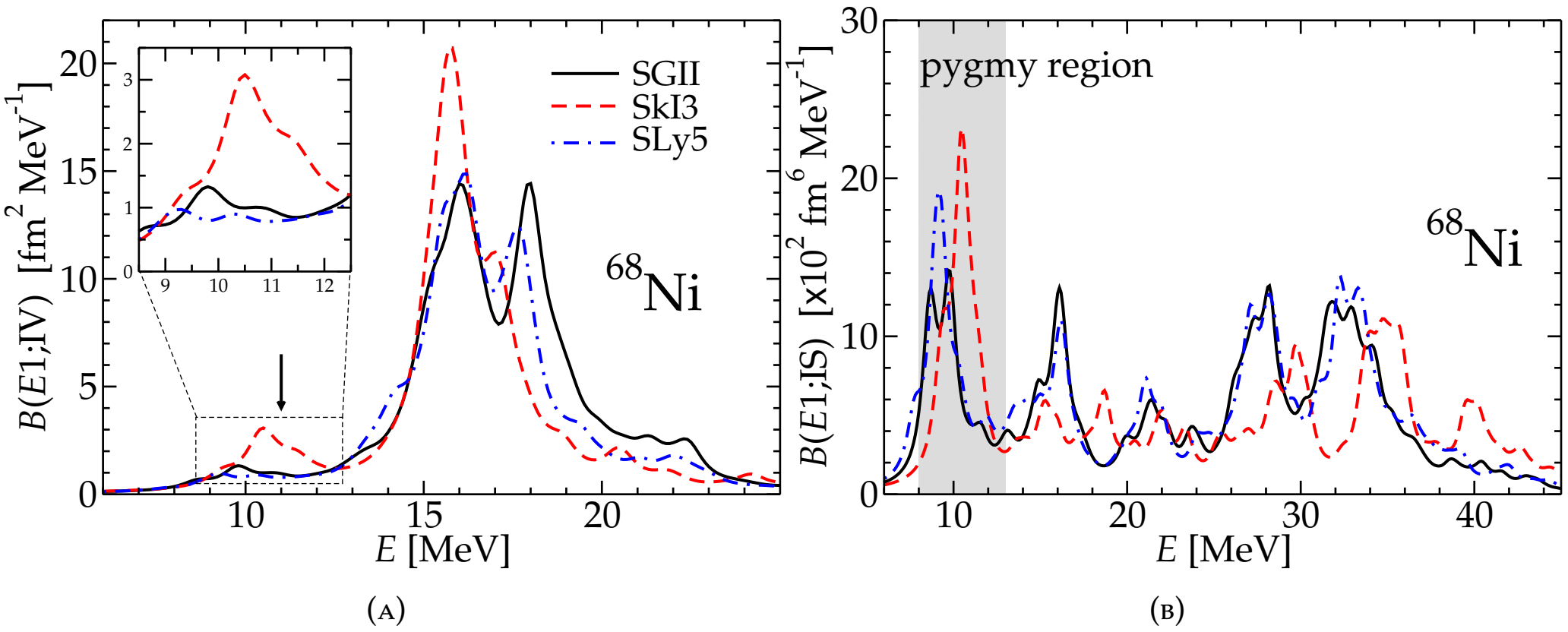

**Figure 6.4**
Same as Fig. 6.2 for $^{68}$Ni. The experimental value for the peak energy of the PDS ($E$ = 11 MeV and an energy width estimated to be less than 1 MeV) is indicated by a black arrow [Wie+09].

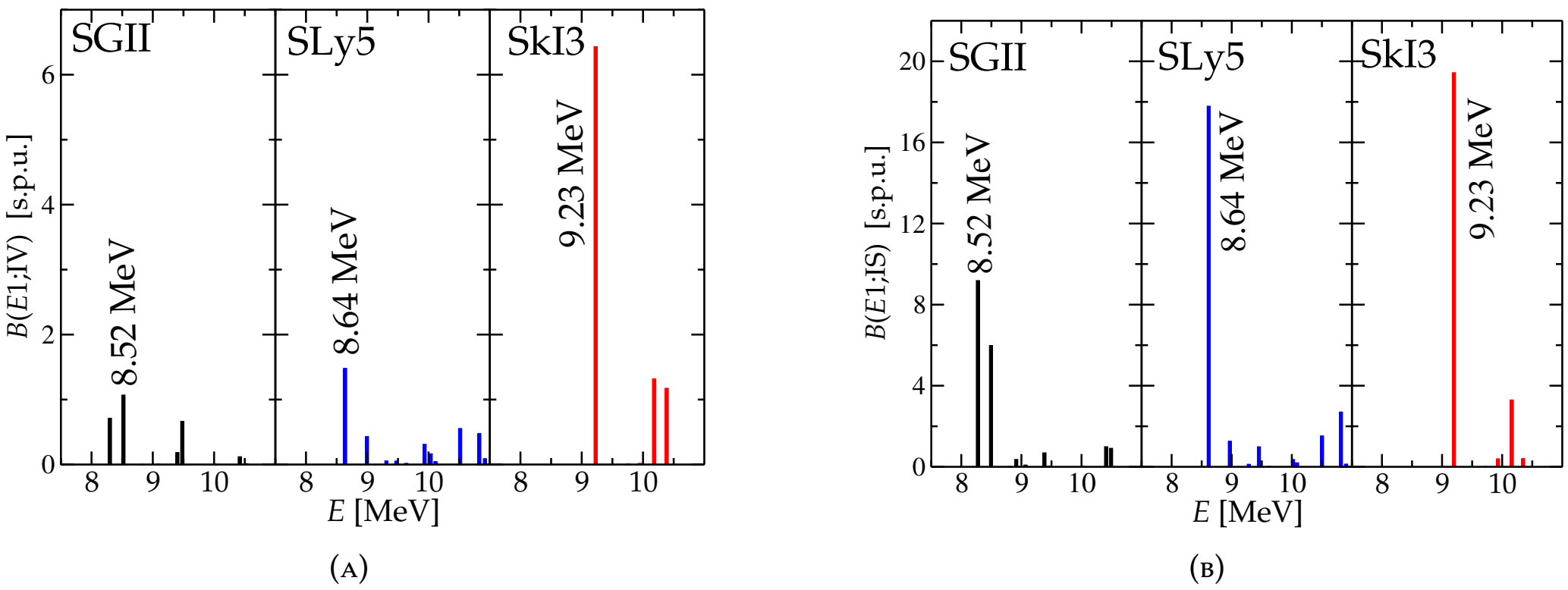

**Figure 6.5**
Reduced transition probabilities for the isovector (a) and isoscalar (b) dipole response, in the case of $^{132}$Sn in s.p. units, as a function of the excitation energy. Note that we only show the energy region relevant for our study of the *RPA-pygmy* state.

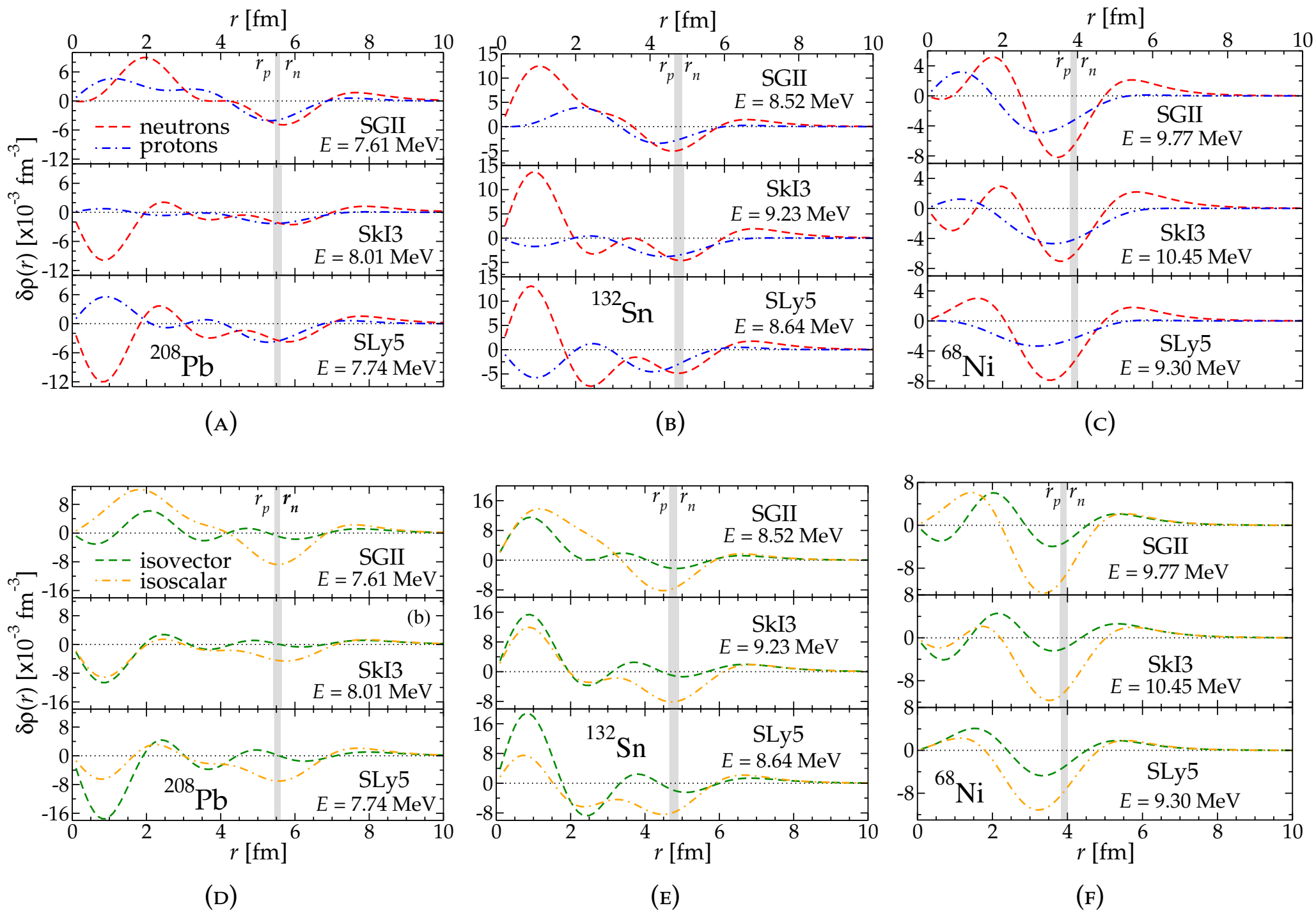

**Figure 6.6**
Neutron and proton [(a), (b) and (c)] and isoscalar and isovector [(d), (e) and (f)] transition densities of the *RPA-pygmy* state, as a function of the radial distance, for $^{208}$Pb [(a) and (d)], $^{132}$Sn [(b) and (e)] and $^{68}$Ni [(c) and (f)]. Proton ($r_p$) and neutron ($r_n$) rms radii are indicated for each interaction by the edges of the grey region.

The isoscalar or isovector nature of the low-energy RPA states has been already studied in Ref. [Paa+09] in the case of $^{140}$Ce. In that work, it was found that the low-lying dipole states of $^{140}$Ce are split into two groups depending on their isospin structure. More recently, similar conclusions were found in a study of the pygmy dipole strength in $^{124}$Sn [End+10], where it has been stated that the theoretical calculations were dominated by a low-lying isoscalar component basically due to oscillations of the neutron skin thickness of the nucleus under study. Both investigations were reported to be in qualitative agreement with the available experimental data. To study the isospin character of the PDS in the nuclei at hand, we first plot the neutron and proton, as well as the isoscalar and isovector transition densities corresponding to the *RPA-pygmy* state. We show the neutron and proton transition densities in Figs. 6.6(a)-(b)-(c), and the isoscalar and isovector transition densities in Figs. 6.6(d)-(e)-(f), respectively. The position of the proton ($r_p$) and neutron ($r_n$) rms radii corresponds to the edges of the grey region that defines in this way the neutron skin thickness.

For the case of $^{208}$Pb, neutrons and protons oscillate differently depending on the interaction but in all cases at the surface there is a predominant isoscalar character. On the contrary, the interior region is a mixture of isoscalar and isovector component, even if it is model dependent.

For $^{132}$Sn the situation is very similar to the one found in $^{208}$Pb. For $^{68}$Ni there is some differences: the behavior of the transition densities is predicted to be very similar within the studied models. Therefore, it is even more clear in this case that the interior of $^{68}$Ni is not dominated by isoscalar or isovector components. At the surface of the nucleus, the isoscalar part dominates but the isovector part is not negligible.

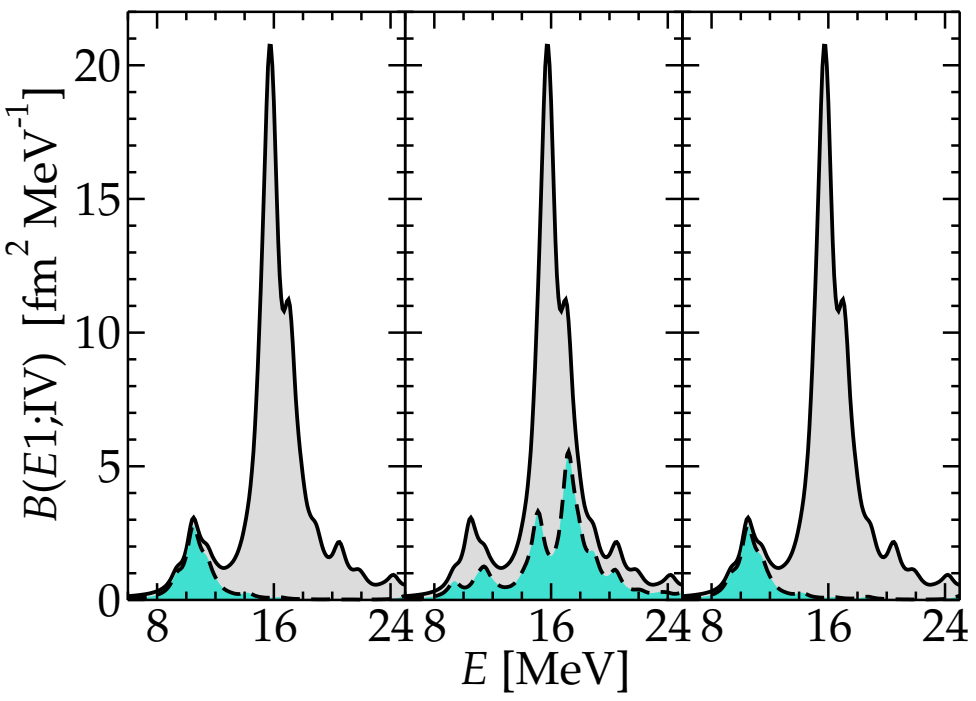

**Figure 6.7**
Strength function corresponding to the isovector dipole response of $^{68}$Ni as a function of the excitation energy (solid lines), and partial contribution due to those states which are at least 70% isoscalar (dashed line). See text for further explanations. The calculation is done using the SkI3 interaction.

It is also possible to use a local criterion to study quantitatively the isoscalar and isovector splitting of the *RPA-pygmy* state [Paa+09] based on the following. At each radial distance $r_i$, where $i = 1 \dots N$ at which the neutron and proton transition densities are calculated, we define that a certain RPA state is 70% isoscalar if at least the 70% of the calculated points fulfill the condition $|\delta\rho_\nu^{(IS)}(r)| > |\delta\rho_\nu^{(IV)}(r)|$. Accordingly, we can exploit this criterion to analyze the isoscalar or isovector nature of the *RPA-pygmy* state in different regions of the nucleus, imposing the above defined criteria of isoscalarity in two additional regions: one in the internal part of the nucleus, i.e. from 0 fm to $R/2$ and the other in external part of the nucleus, namely from $R/2$ to $R$, where $R = r_0 A^{1/3}$. We apply this criteria to all calculated excited states and plot in Fig. 6.7 their contribution to the isovector dipole strength function. We shows the strength function applying the criteria to those states that are 70% isoscalar in the region between 0 and $R$ (left panel in Fig. 6.7), then to those which are 70% isoscalar in the internal part of the nucleus (central panels in the same figure), and finally to those which are 70% isoscalar in the external part of the nucleus (right panels of the same figure). Since the results are not qualitatively different, we show here only the case of $^{68}$Ni as predicted by the SkI3 interaction. As a guidance, we also show the total isovector dipole strength function (solid line). From such a figure, it is evident that the RPA states which are mostly isoscalar in the whole region $[0, R]$ and in the external part of the nucleus $[R/2, R]$ are essentially the same ones since both give rise to almost the same contributions: most of the PDS and a small contribution to the rest of the strength function. On the contrary, those RPA states that are 70% isoscalar in the internal region of the nucleus [0,R/2], do not essentially contribute to the PDS.

Our results indicate that one is allowed to qualitatively distinguish the PDS from the IVGDR, and state that while the latter strength is basically isovector and involves the motion of mainly internal nucleons, the former is more isoscalar than isovector and involves the motion of external nucleons, that are mainly neutrons in a neutron rich nucleus.

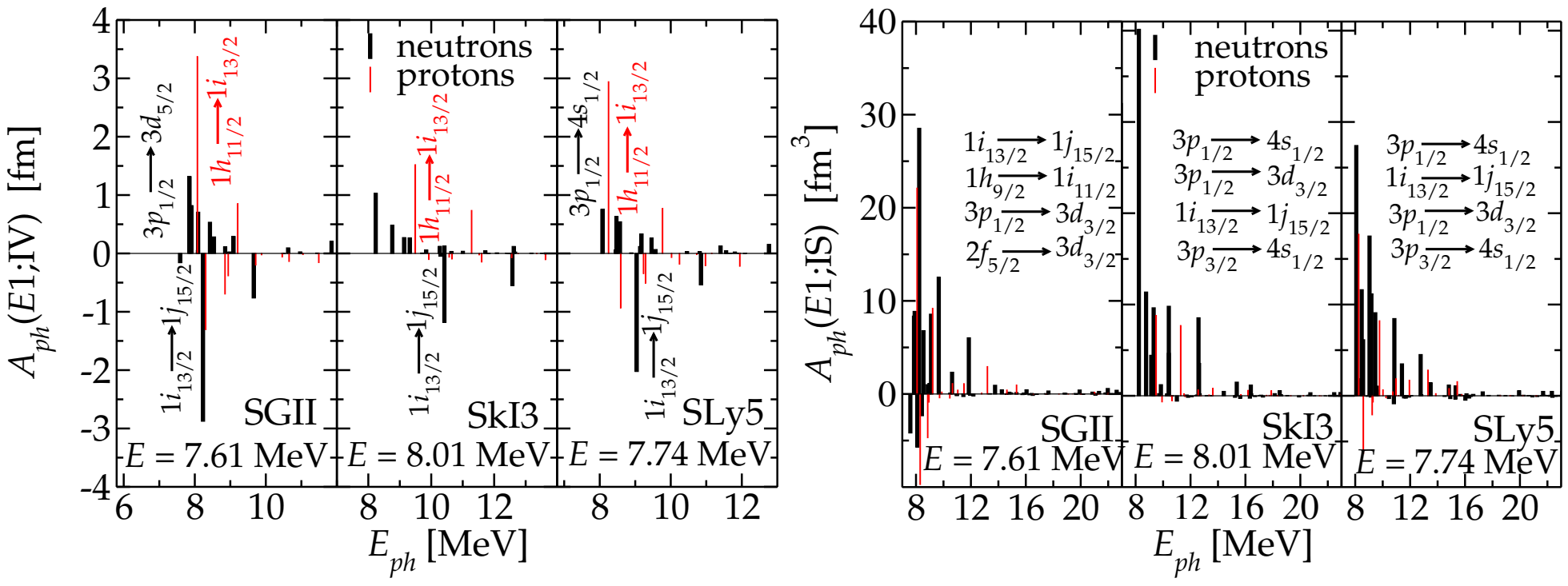

(a) (b)

**Figure 6.8**
*ph* contributions to the isovector reduced amplitude corresponding to the $^{208}$Pb *RPA-pygmy* state as a function of the *ph* excitation energy (a). Largest neutron *ph* contributions are also listed. Same as (a) but for the isoscalar reduced amplitude (b).

### 6.1.3 Relevant particle-hole excitations in the low-energy region

A given RPA state shows a collective character under the action of an external operator if there are many *ph* excitations providing non-negligible contributions that add coherently in Eq. (3.31). For these reasons, our purpose in this subsection is to analyze the contributions of the different *ph* excitations to the *RPA-pygmy* states depending on the (isoscalar or isovector) operator used to excite the nucleus. We show the results only for $^{208}$Pb, being the ones corresponding to the other nuclei similar.

We show in Fig. 6.8 all the neutron (black) and proton (red) *ph* contributions to the reduced amplitude $A^q_{ph}(E1;\xi)$ as a function of their excitation energy for the isovector and isoscalar dipole responses, respectively. Notice that not all contributions can be seen from these figures since most of them are very small.

It is evident from Fig. 6.8(a) that the contributions of the most relevant *ph* excitations to the isovector reduced amplitude are only a few in number and there is some amount of destructive interference. Opposite to that, it is also evident from Fig. 6.8(b) that the contributions of the most relevant *ph* excitations to the isoscalar reduced transition amplitude are basically coming from neutron transitions, and that most of them add coherently.

In particular, we generally find within all the employed models that the dynamics of the low-energy isoscalar dipole response of $^{208}$Pb seems to be governed by the excitations of the outermost neutrons, namely those that form the neutron skin of this nucleus. From the analysis of the *ph* contributions, we conclude that while the low-energy isoscalar dipole response of $^{208}$Pb arising form the *RPA-pygmy* state can be considered as a collective mode in all studied models, the PDS can not.

## 6.2 The quadrupole spectrum, the effective mass and the symmetry energy

In this section, the general features of the ISGQR and IVGQR are described. Moreover, we show our results for the dependence of the energy of the ISGQR from the effective mass and we present a macroscopic model, based on the quantal harmonic oscillator approach of Ref. [BM75], to highlight the relations between the energies of the ISGQR and the IVGQR, the effective mass and the symmetry energy at some subsaturation density. This analysis has been performed in $^{208}$Pb. The main results shown here were published in Ref. [Roc+13].

**Table 6.2**
Effective mass $m^*/m$ and neutron skin thickness $\Delta r_{np}$ for $^{208}$Pb predicted by the interactions used in the study of the IVGQR response.

| | $m^*/m$ | $\Delta r_{np}$ [fm] |
|---|---|---|
| SAMi | 0.68 | 0.147 |
| KDE | 0.76 | 0.155 |
| SkI3 | 0.58 | 0.227 |
| NL3 | – | 0.279 |
| DD-ME2 | – | 0.193 |

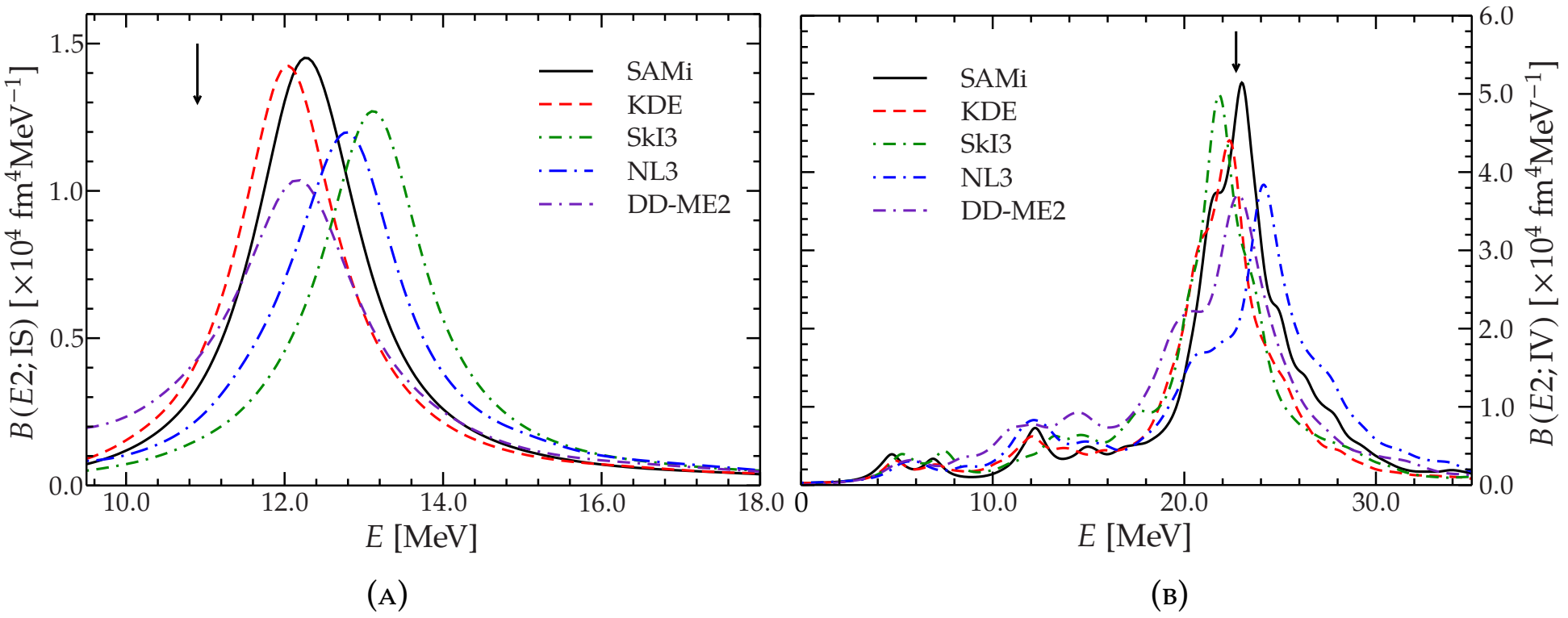

**Figure 6.9**
Isoscalar (a) and isovector (b) quadrupole strength functions. The strengths are calculated within the RPA for SAMi, KDE, SkI3, NL3 and DD-ME2. The experimental energies for the ISGQR (10.9 ± 0.1 MeV), and the IVGQR (22.7 ± 0.2 MeV) (weighted averages) listed in Table 6.3 are indicated by arrows.

In this analysis we have employed three Skyrme-type interactions, namely SAMi, KDE and SkI3, and two relativistic functionals, NL3 [LKR97] and DD-ME2 [Lal+05]. The Skyrme interactions have different effective masses $m^*/m$ and yield different values for the neutron skin thickness $\Delta r_{np}$ in $^{208}$Pb. The two covariant functionals are based on (i) finite-range meson exchange with non-linear self-interaction terms (NL3), and (ii) density-dependent meson-nucleon vertex functions (DD-ME2). Relativistic mean-field models are known to yield rather low values for the non-relativistic equivalent, or Schrödinger effective mass, typically around $0.6m$ at saturation density [JM89; LVM06]. The NL3 functional predicts values of the neutron skin that are considerably larger compared to non-relativistic functionals. The values for $m^*/m$ and for $\Delta r_{np}$ in $^{208}$Pb are listed in Table 6.2 for all the interaction used.

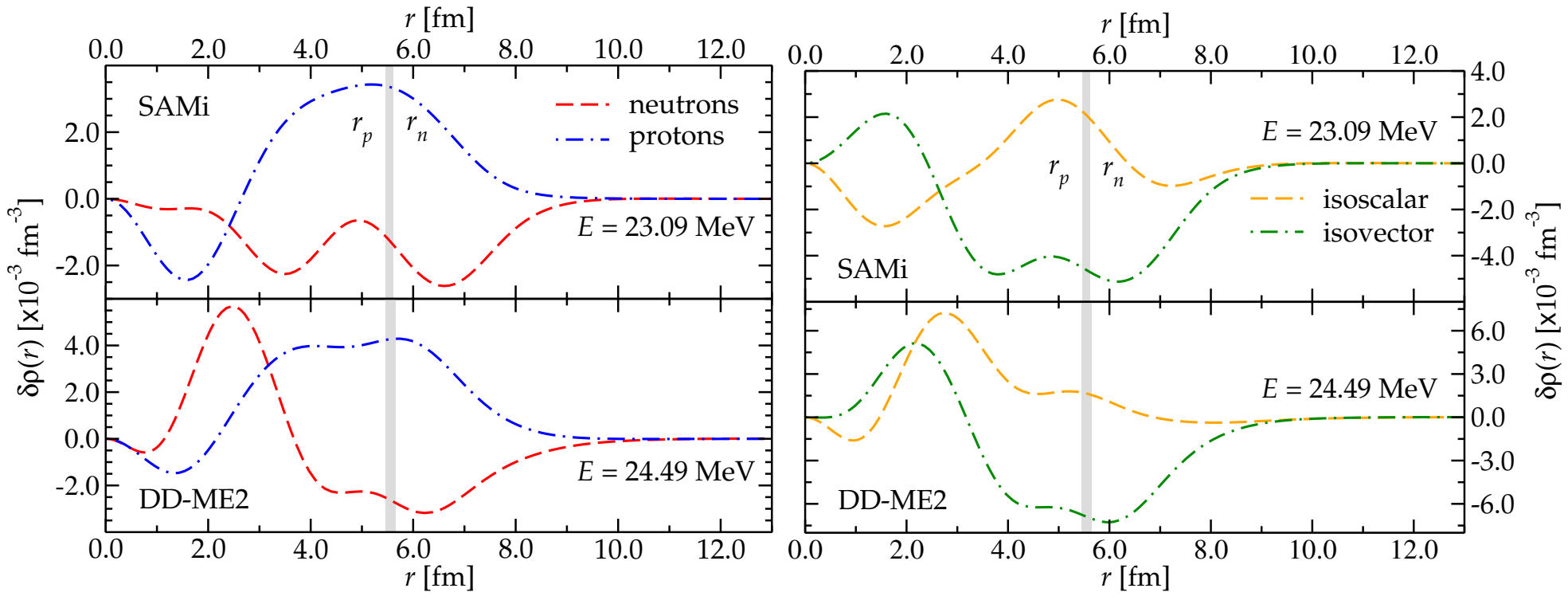

(A) (B)

**Figure 6.10**
Neutron and proton (A) and isoscalar and isovector (B) transition densities associated with the main peak of the isovector response. Only the predictions for the SAMi and DD-ME2 functionals are shown. The proton ($r_p$) and neutron ($r_n$) rms radii are indicated by the edges of the shaded region.

### 6.2.1 The strength function and the transition densities

In Fig. 6.9 the strength function for the isoscalar (A) and isovector (B) channels are reproduced. The RPA results are convoluted with Lorentzian functions, whose widths have been chosen in such a way that the total experimental ISGQR and IVGQR widths are reproduced in the corresponding medium and high energy regions, respectively.

The ISGQR peak is predicted at higher energy for all the interactions used, as should be expected from the values of the effective mass at saturation: as recalled in section 5.2.2 and later in this section, the energy of the ISGQR is inversely related to the square root of the effective mass.

The isovector spectrum shown in Fig. 6.9(B) consists of three distinct structures. The first one is the well known low-energy $2^+$ state at about 5 MeV. The second is the ISGQR that appears in the energy range between 10 and 15 MeV and, finally, the IVGQR located in the region above 20 MeV. The two lower structures arise because of isospin mixing in the RPA states and, therefore, these could be excited both by isoscalar and isovector probes. In the high-energy region all interactions predict the existence of a collective IVGQR peak. Our results are in good agreement with experimental findings, both for the excitation energy of the IVGQR and the fraction of the energy-weighted sum rule (EWSR). The measured fraction for the latter is 56% [Hen+11], whereas theoretical predictions range from 50% to 65%. Note that the EWSR fraction reported in Ref. [Hen+11] refers to the classical version of the sum rule, that is, without the multiplicative factor $(1+\kappa_Q)$, where $\kappa_Q$ is the isovector quadrupole enhancement factor. [1]

In Table 6.3 the experimental data for both isoscalar and isovector resonances are reported. Although the IVGQR in Ref. [Hen+11] was measured in $^{209}$Bi, calculations are carried out for $^{208}$Pb. The difference in energy of the nuclear response of $^{209}$Bi and $^{208}$Pb should scale with $A^{-1/3}$ [BM75], that is, it should be smaller than a few ‰. Another important reason for limiting the study to $^{208}$Pb is that it is a spherical double magic nucleus, and thus the dependence on the effective mass or the symmetry energy will not be screened by deformation or paring effects.

[1] From Eq. (3.46), $\kappa_Q = \frac{2m}{\hbar^2}\frac{4\pi}{A\langle r^2\rangle}\left(t_1\left(1+\frac{x_1}{2}\right)+t_2\left(1+\frac{x_2}{2}\right)\right)\int \mathrm{d}r r^4 \rho_p(r)\rho_n(r)$.

**Table 6.3**
Data for the IVGQR and ISGQR in $^{208}$Pb.

| | $E_x$ (MeV) | $\Gamma$ (MeV) | EWSR (%) | Reference |
|---|---|---|---|---|
| IVGQR | 24.3±0.4 | 4.5±0.5 | 140 | [Lei+81] |
| | 22.5 | 9 | 100 | [Sch+88] |
| | 20.2±0.5 | 5.5±0.5 | 140±30 | [DLA92] |
| | 23.0±0.2 | 3.9±0.9 | 56± 6 | [Hen+11] [1] |
| Weighted Average [2] | 22.7±0.2 | 4.8±0.3 | | |
| ISGQR | 10.60±0.25 | 2.8±0.25 | 100 | [Bue84] |
| | 11.0 ±0.2 | 2.7±0.3 | 105±25 | [You+81] |
| | 10.9 ±0.3 | 3.1±0.3 | 120-170 | [Bra85] |
| | 11.0 ±0.3 | 3.3±0.3 | 100-150 | [Bra85] |
| | 10.9 ±0.3 | 3.0±0.3 | 100±13 | [You+04] |
| Weighted Average | 10.9 ±0.1 | 3.0±0.1 | | |

[1] These experimental values are for $^{209}$Bi, and the EWSR corresponds to the classical value (see text)

[2] Weighted average of $O$ is defined in the standard way as $\bar{O} = \frac{\sum_{i=1}^{n} \omega_i O_i}{\sum_{i=1}^{n} \omega_i}$ where $\omega_i$ is defined as the inverse of the one standard deviation corresponding to the data point $O_i$. The standard deviation associated to $\bar{O}$ is calculated as $\sigma_{\bar{O}} = \left(\sum_{i=1}^{n} \omega_i^2\right)^{-1/2}$.

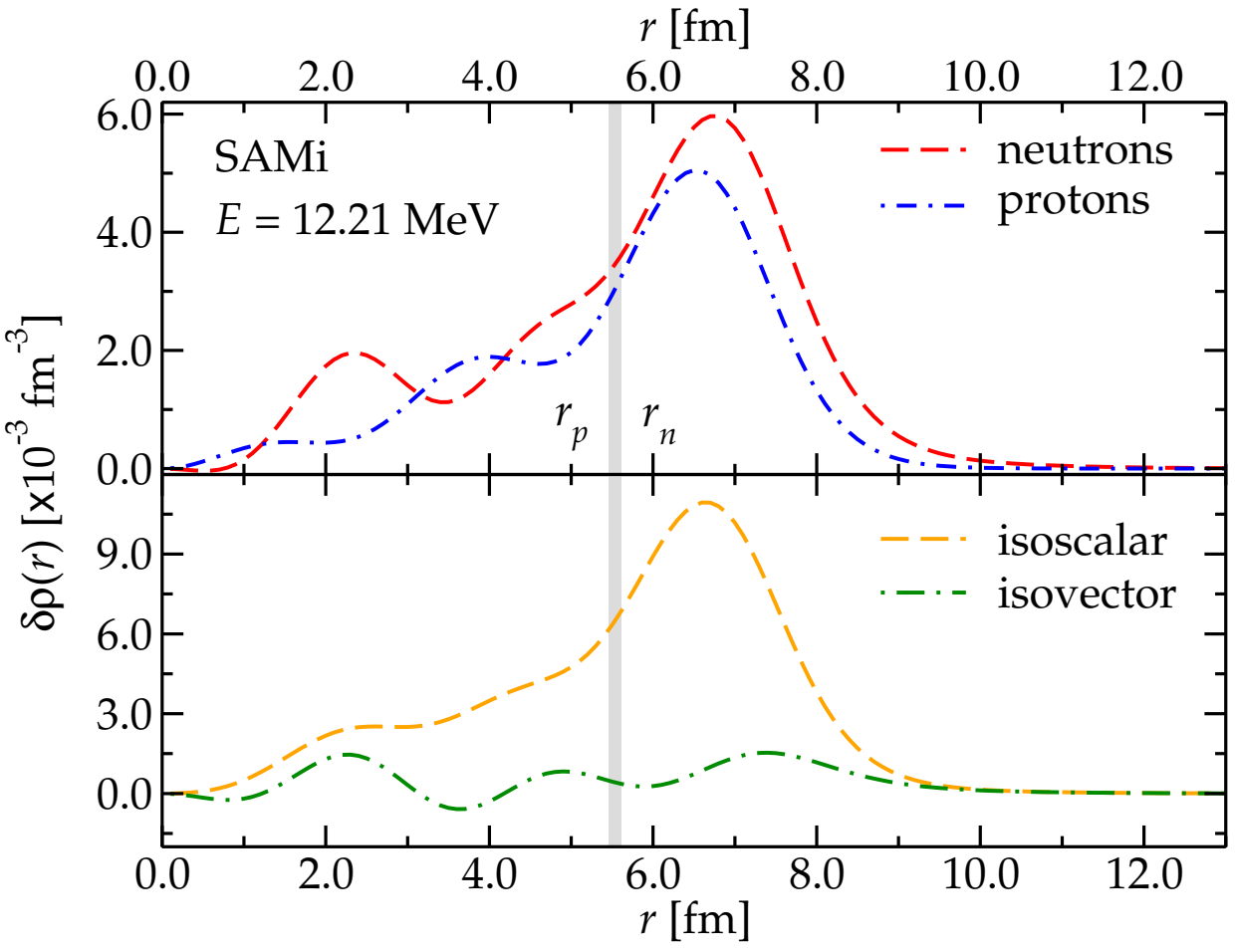

**Figure 6.11**
Neutron-proton and isoscalar-isovector transition densities associated with the main peak of the isoscalar response. Only the predictions for the SAMi functional are shown. The proton ($r_p$) and neutron ($r_n$) rms radii are indicated by the edges of the shaded region.

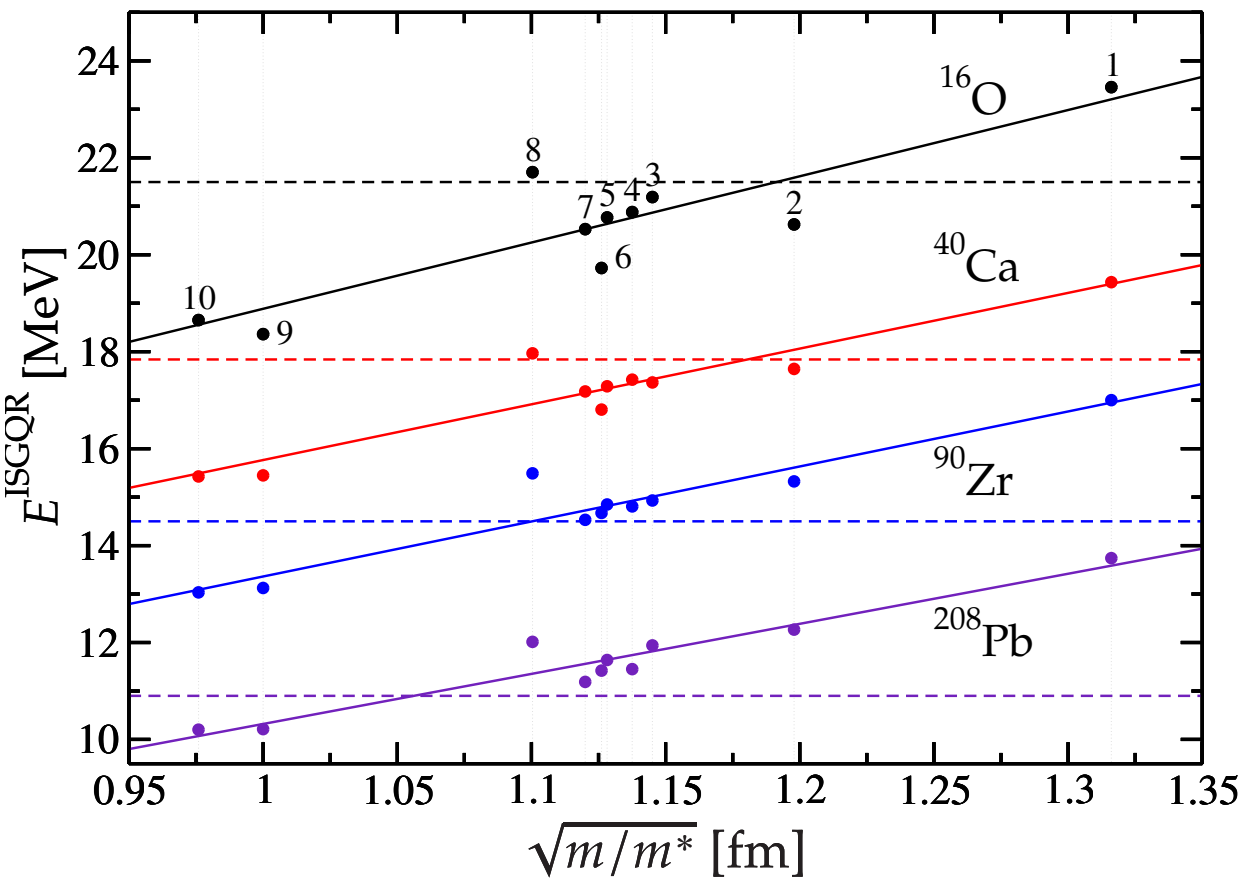

**FIGURE 6.12**
Excitation energy of the ISGQR as a function of $\sqrt{m/m^*}$, calculated with different Skyrme-type functionals, for the nuclei $^{16}$O, $^{40}$Ca, $^{48}$Ca, $^{90}$Zr, $^{208}$Pb. On the horizontal upper axis we display the corresponding values for $m^*/m$. The labels used to identify the interactions are listed in Table 6.4. The dashed lines indicate for each nucleus the experimental value, found in Ref. [Har+76], [YLC01], [Bro97] and [Mar07] for $^{16}$O, $^{40}$Ca, $^{90}$Zr and $^{208}$Pb.

In Fig. 6.10 we report the transition densities associated with the main peak of the isovector response. Fig. 6.10(A) displays the neutron and proton transition densities, and in Fig. 6.10(B) we plot the corresponding isoscalar and isovector transition densities calculated with the functionals SAMi and DD-ME2. The other functionals yield similar transition densities. The positions of the proton ($r_p$) and neutron ($r_n$) root mean square (rms) radii correspond to the edges of the shaded region that, in this way, denotes the neutron skin thickness calculated with a given functional. For all functionals and, in particular for those used in Fig. 6.10, protons and neutrons produce similar contributions but with opposite signs to the transition densities in the surface region. This shows that the excitation is predominantly isovector. In the bulk of the nucleus one finds a non-negligible isoscalar component, even when the state is mainly isovector. This is, in particular, the case for the relativistic DD-ME2 functional. For comparison, in Fig. 6.11 the transition densities for the main peak of the isoscalar response is shown. Only the predictions for the SAMi interaction are displayed, since in the other cases the results are similar. The isoscalar nature of this excitation is clear: protons and neutrons contribute in a similar way around the surface, resulting in a large isoscalar transition density.

### 6.2.2 Sensitivity to the effective mass and the symmetry energy

#### The macroscopic model for the excitation energy of the GQRs

As recalled in section 5.2.2, in a quantal harmonic oscillator (QHO) approach to the nuclear vibration with a velocity dependent interaction, the energy of the ISGQR would be related to the effective mass by

$$E^{\mathrm{ISGQR}} = \sqrt{\frac{2m}{m^*}}\hbar\omega_0, \tag{6.1}$$

where $\hbar\omega_0 \sim 41A^{-1/3}$.

**Table 6.4**
Interactions used in Fig. 6.12. In each entry, it is reported the label used to mark the interaction in the figure, the name of the parameterization and the corresponding value for the effective mass at saturation.

| Label | Interaction | $m^*/m$ | Label | Interaction | $m^*/m$ |
|---|---|---|---|---|---|
| 1 | SkI3 | 0.577 | 6 | SKM* | 0.789 |
| 2 | SLy5 | 0.697 | 7 | SK255 | 0.797 |
| 3 | SIII | 0.763 | 8 | LNS | 0.826 |
| 4 | SK272 | 0.773 | 9 | SkP | 1.000 |
| 5 | SGII | 0.786 | 10 | BSk1 | 1.050 |

The correlation between the ISGQR energy and the effective mass can studied employing a variety of Skyrme-type models, with different predictions for this quantity. Following the spirit of Fig. 22 of Ref. [Bla80], we plot in Fig. 6.12 the predictions produced by different Skyrme interactions for the energy of the main peak of the ISGQR as a function of $\sqrt{m/m^*}$, for different nuclei, namely $^{16}$O, $^{40}$Ca, $^{48}$Ca, $^{90}$Zr, $^{208}$Pb. The interactions used are listed in Table 6.4 with the corresponding predictions for the effective mass and the numerical label used in Fig. 6.12. The linear correlation is good for all the fits (the correlation coefficient $r$ is above 0.9 in all the case), even though it is worse for lighter systems. The experimental value for the ISGQR is also reported in the figure as dashed lines.

In the same QHO framework, it is possible to write for the energy of the IVGQR:

$$E^{\text{IVGQR}} = 2\sqrt{\frac{m^*}{m}(\hbar\omega_0)^2 + \frac{5}{4}\frac{\hbar^2}{2m}\frac{V_{\text{sym}}\langle r^2\rangle}{\langle r^4\rangle}}, \tag{6.2}$$

where $V_{\text{sym}}$ is the symmetry potential, which is proportional to the liquid drop model parameter $b_{\text{sym}}^{\text{pot}}$ [BM69, Eq. (2-28)]: $V_{\text{sym}} = 4b_{\text{sym}}^{\text{pot}}$. This is the potential part of the $b_{\text{sym}} = b_{\text{sym}}^{\text{pot}} + b_{\text{sym}}^{\text{kin}}$ coefficient, while the kinetic part can be approximated by $b_{\text{sym}}^{\text{kin}} = 2S^{\text{kin}}(\rho_\infty)$. In a non-relativistic approximation, this term can in turn be approximated with $\varepsilon_{\text{F}_\infty}/3$, where $\varepsilon_{\text{F}_\infty} = \hbar^2 k_{\text{F}_\infty}^2/2m \sim 37\,\text{MeV}$ is the Fermi energy for symmetric nuclear matter at saturation. The "standard" liquid drop parameter $a_{\text{sym}}^{\text{LDM}}$ is related to $b_{\text{sym}}$ by $b_{\text{sym}} \approx 2a_{\text{sym}}^{\text{LDM}}$ [BM69, Eq. (2-12)]. Actually, it is more appropriate to use the droplet model parameter $a_{\text{sym}}^{\text{DM}}$ (with which $a_{\text{sym}}^{\text{LDM}}$ can be identified) because it also takes into account the corrections due to the presence of a surface [MS69; MS74]. Besides, in Ref. [Cen+09] it has been demonstrated that the symmetry energy of a finite nucleus $a_{\text{sym}}^{\text{DM}}(A)$ equals the symmetry energy $S(\rho)$ of the infinite system at some sub-saturation density $\rho_A$, which is approximately $0.1\,\text{fm}^{-3}$ for the case of heavy nuclei such as $^{208}$Pb. Hence, one can rewrite the IVGQR energy as

$$\begin{aligned} E^{\text{IVGQR}} &\approx 2\left\{\frac{m}{m^*}(\hbar\omega_0)^2 + 6\frac{\varepsilon_{\text{F}_\infty}}{A^{2/3}}\left[S(\rho_A) - S^{\text{kin}}(\rho_\infty)\right]\right\}^{1/2} \\ &\approx 2\left[\frac{m}{m^*}(\hbar\omega_0)^2 + 6\frac{\varepsilon_{\text{F}_\infty}}{A^{2/3}}\left[S(\rho_A) - \frac{\varepsilon_{\text{F}_\infty}}{3}\right]\right\}^{1/2} \end{aligned} \tag{6.3}$$

$$\approx 2\left[\frac{\left(E^{\rm ISGQR}\right)^2}{2}+2\frac{\varepsilon^2_{\rm F_\infty}}{A^{2/3}}\left(\frac{3S(\rho_A)}{\varepsilon_{\rm F_\infty}}-1\right)\right]^{1/2},$$

where we have approximated the factor $\frac{14}{3}\left(\frac{8}{9\pi}\right)^{\frac{2}{3}}=2.0113$ by 2 on the r.h.s., considered $\langle r^n\rangle = 3r_0^nA^{n/3}/(n+3)$ where $r_0=[3/(4\pi\rho_\infty)]^{1/3}$, and used Eq. (6.1) in the last step.

From Eq. (6.3), some interesting features can be noted:

- $E^{\rm IVGQR}$ depends on the effective mass at saturation and, in addition, on the symmetry energy at some sub-saturation density $\rho_A$. In particular, $E^{\rm IVGQR}$ increases for decreasing values of $m^*$, and increasing values of $S(\rho_A)$.

- The larger the neutron skin thickness in a heavy nucleus such as $^{208}$Pb, the lower the excitation energy of the IVGQR. This characteristic can be understood as follows. If one expands $S(\rho)$ around the nuclear saturation density as $S(\rho)\approx J-L\epsilon$, where $\epsilon\equiv(\rho_\infty-\rho)/\rho$, it can explicitly be shown that at the sub-saturation density $\rho_A$, fixing $E^{\rm ISGQR}$ to the experimental value and for small variations of $J$, $E^{\rm IVGQR}$ decreases for increasing values of $L$, or, that is the same, for increasing value of the neutron skin thickness (see chapter 4).

The main novelty of this approach is the possibility to obtain the symmetry energy at sub-saturation density as a function of the energy of both the ISGQR and the IVGQR combined,

$$\tilde{S}(\rho_A)=\frac{A^{2/3}}{24\varepsilon_{\rm F_\infty}}\left[\left(E^{\rm IVGQR}\right)^2-2\left(E^{\rm ISGQR}\right)^2\right]+S^{\rm kin}(\rho_\infty) \quad (6.4a)$$

$$=\frac{\varepsilon_{\rm F_\infty}}{3}\left\{\frac{A^{2/3}}{8\varepsilon^2_{\rm F_\infty}}\left[\left(E^{\rm IVGQR}\right)^2-2\left(E^{\rm ISGQR}\right)^2\right]+1\right\}. \quad (6.4b)$$

By inserting the weighted averages of the experimental values of Table 6.3, and by using $\rho_{A=208}=0.1\,{\rm fm}^{-3}$, we find $\tilde{S}(0.1\,{\rm fm}^{-3})=23.3\pm0.6\,$MeV, in very good agreement with the estimate reported in Ref. [TCV08]: $23.3\,{\rm MeV}\leq S(0.1\,{\rm fm}^{-3})\leq 24.9\,$MeV. Note that the quoted error does not include an estimate of the theoretical uncertainty.

It is then possible to explicitly relate the excitation energies of the isoscalar and isovector GQRs with the neutron skin thickness of a heavy nucleus. For that, we use the DM expression for the neutron skin thickness that can be written as follows [Cen+09]

$$\frac{\Delta r_{np}-\Delta r_{np}^{\rm surf}}{\langle r^2\rangle^{1/2}}=\frac{2}{3}\left[1-\frac{S(\rho_A)}{J}\right](I-I_C)-\frac{2}{7}I_C, \quad (6.5)$$

where $I=(N-Z)/A$ is the relative neutron excess, $I_C=e^2Z/(20JR)$ and $\Delta r_{np}^{\rm surf}$ is the surface contribution to the neutron skin thickness[2] , which, for the case of $^{208}$Pb, has a value of $\approx 0.09\,$fm [Cen+10], if calculated with a large set of EDFs. Combining Eqs. (6.4)

---

[2] $\Delta r_{np}^{\rm surf}=\sqrt{3/5}[5(b_n^2-b_p^2)/(2R)]$, where $b_n$ and $b_p$ are the surface widths of the neutron and proton density profiles, respectively.

and (6.5) one finds,

$$\frac{\Delta r_{np} - \Delta r_{np}^{\rm surf}}{\langle r^2\rangle^{1/2}} = \frac{2}{3}\,(I - I_C)\left\{1 - \frac{\varepsilon_{\rm F_\infty}}{3J} - \frac{3}{7}\frac{I_C}{I - I_C} - \frac{A^{2/3}}{24\varepsilon_{\rm F_\infty}}\left[\frac{\left(E^{\rm IVGQR}\right)^2 - 2\left(E^{\rm ISGQR}\right)^2}{J}\right]\right\}. \tag{6.6}$$

This expression explicitly relates the neutron skin thickness of a heavy nucleus with the corresponding GQRs energies, and these can directly be determined in the experiment. Within our approach, only the parameter $J$ and $\Delta r_{np}^{\rm surf}$ contain a non-negligible uncertainty. The appropriate value of $J$ to be used in the expression above can be estimated from Fig. 4.1 of chapter 4 to be $J = 32 \pm 1$ MeV. For the case of $^{208}$Pb, $\Delta r_{np}^{\rm surf} = 0.09 \pm 0.01$ fm is consistent with the microscopic calculations of Ref. [Cen+10]. Using Eq. (6.6) and the data for the GQRs energies, we find $\Delta r_{np} = 0.22 \pm 0.02$ fm. This value is close to the upper limit derived from available estimates $\Delta r_{np} = 0.18 \pm 0.03$ fm [Tsa+12].

**Correlations using systematically varied interactions**

In order to make a deep study on these correlations, we have used families of functionals with systematically varied properties in the isoscalar and isovector channels. In particular, we have chosen the SAMi interaction because it has been fitted in order to improve the description of charge-exchange excitations without loss of quality in the description of other observables, such as masses, charge radii and non-charge exchange resonances. Using the fitting protocol described in the original reference [RCS12], we have first fixed the values of the nuclear incompressibility ($K_\infty = 245$ MeV) and the effective mass ($m^*/m = 0.675$), whereas the values of the symmetry energy at saturation ($J$) have been varied from 27 MeV (SAMi-J27) to 31 MeV (SAMi-J31) in steps of 1 MeV. Then, by fixing the values of $K_\infty = 245$ MeV, $J = 28$ MeV and $L = 44$ MeV, we have varied the effective mass from $m^*/m = 0.65$ (SAMi-m65) to 0.85 (SAMi-m85) in steps of 0.05. Among relativistic functional, we have adopted the set of interactions introduced in Ref. [VNR03], in which using the same fitting protocol of DD-ME2, the $J$ parameter was systematically varied from 30 MeV to 38 MeV in steps of 2 MeV (sets from DD-MEa to DD-MEe). In this way, we can be sure that differences in our results come only from the varied parameters, identifying possible correlations.

In Fig. 6.13 the excitation energy of the ISGQR as a function of $\sqrt{m/m^*}$ is represented, as calculated with the SAMi-m and SAMi-J families of functionals. The well known correlation between $E^{\rm ISGQR}$ and $\sqrt{m/m^*}$ is nicely reproduced. The variation of the peak energy for the SAMi-J family, although small, can be explained as follows: the neutron radius increases with $J$ [Fur02], and a larger size of the nucleus implies a lower ISGQR excitation energy.

Fig. 6.14 displays the predictions of the SAMi-m and SAMi-J families of functionals for the excitation energy of the IVGQR in $^{208}$Pb as a function of the effective mass (a) and neutron skin thickness (b), and those of the DD-ME family for the excitation energy of the IVGQR as a function of the neutron skin thickness (c). The two expected correlations between the $(E^{\rm IVGQR})^2$ and $m/m^*$ or $\Delta r_{np}$ are clearly displayed by the linear fits in Fig. 6.14(a)-(b) for the SAMi functionals. Also for the DD-ME family of functionals, the linear correlation is found between the square of the excitation energy of the IVGQR and the neutron skin thickness (Fig. 6.14(c)).

These results, as well as the macroscopic model of section 6.2.2, show that a measurement of the excitation energy of the IVGQR in $^{208}$Pb determines only a combination of the excitation energy of the ISGQR and $\Delta r_{np}$ (or the value of the slope of the symmetry energy at saturation $L$), but not their individual values (6.3).

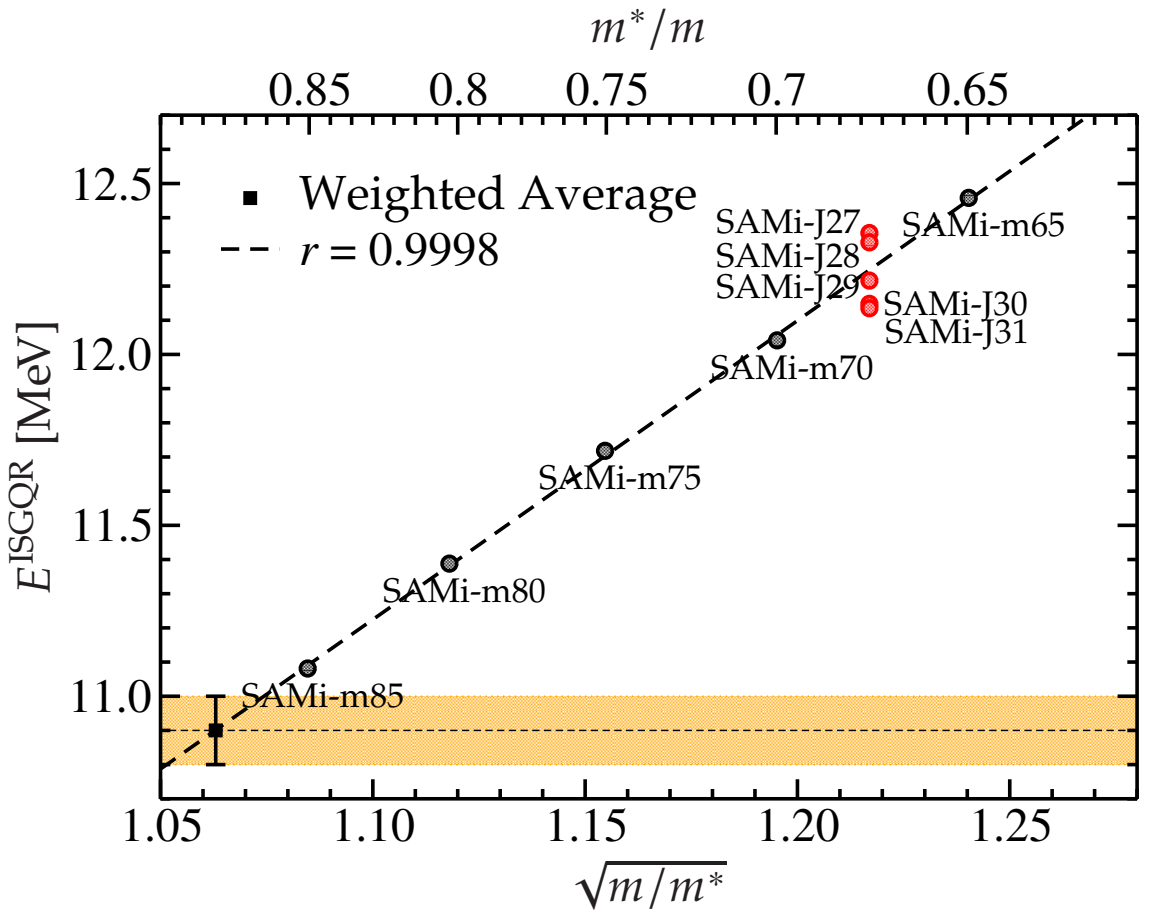

**Figure 6.13**
Excitation energy of the ISGQR as a function of $\sqrt{m/m^*}$, calculated with the SAMi-m and SAMi-J family of functionals. On the horizontal upper axis we display the corresponding values for $m^*/m$. The data from Table 6.3 are also included (square and shaded band).

Eq. (6.6) shows that $[(E^{\mathrm{IVGQR}})^2 - 2(E^{\mathrm{ISVGQR}})^2]/J$ is linearly correlated with $\Delta r_{np}$. This correlation is illustrated in Fig. 6.15 for the SAMi-J and DD-ME functionals. Both families show a high linear correlation ($r = 0.98$) between these two quantities, but predict different slopes. The slope obtained in the macroscopic model is: $\langle r^2\rangle^{1/2}(I - I_C)A^{2/3}/(36\varepsilon_{F_\infty})$, independent of $S^{\mathrm{kin}}(\rho_\infty)$. For $^{208}$Pb this yields 0.025 fmMeV$^{-1}$, in very good agreement with the value 0.027 fmMeV$^{-1}$ found for the SAMi family. The macroscopic formula obviously does not apply to the relativistic case since the slope for the DD-ME family of functionals is 0.057 fmMeV$^{-1}$.

Using the linear correlations shown in Fig. 6.15, the experimental values for $E^{\mathrm{IVGQR}} = 22.7 \pm 0.2$ MeV and $E^{\mathrm{IS}} = 10.9 \pm 0.1$ MeV from Table 6.3, and the value $J = 32 \pm 1$ MeV that yields $[(E^{\mathrm{IVGQR}})^2 - 2(E^{\mathrm{ISVGQR}})^2]/J = 8.7 \pm 0.4$ MeV, one finds $\Delta r_{np} = 0.14 \pm 0.03$ fm for the DDME family of functionals, and $\Delta r_{np} = 0.14 \pm 0.02$ fm for the SAMi-J functionals. The total range of allowed values 0.11 fm $\leq \Delta r_{np} \leq$ 0.17 fm is rather broad but in reasonable agreement with previous studies: $\Delta r_{np} = 0.18 \pm 0.03$ fm [Tsa+12], and $\Delta r_{np} = 0.188 \pm 0.014$ fm [ADS12]. In Fig. 6.17 some experimental constraints on the value of $\Delta r_{np}$ from different experiments are gathered. Finally, this result for the neutron skin thickness of $^{208}$Pb allows us to estimate the value of the slope of the symmetry energy at saturation for the DDME and SAMi-J families. Fig. 6.16 shows that this value is in the interval: 19 MeV $\leq L \leq$ 55 MeV. We note that the correlation coefficient is rather high and in agreement with those obtained in Refs. [Bro00; Cen+09; Roc+11; Tsa+12]. Our constraint on $L$ is also in agreement with previous estimates [Tsa+12; Viñ+12; LL13].

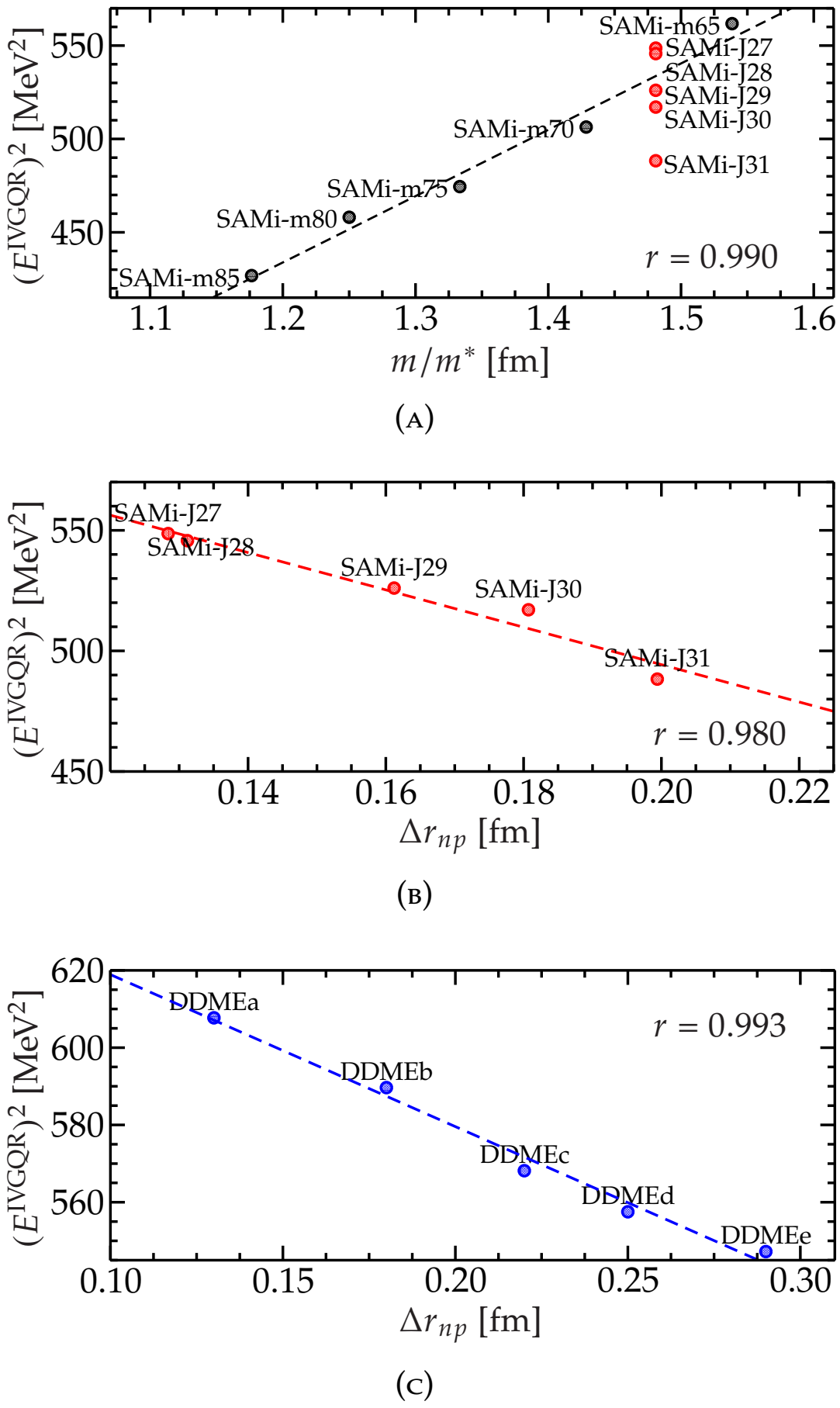

**Figure 6.14**
Square of the excitation energy of the IVGQR as a function of the effective mass (a), and the neutron skin thickness (b)-(b), predicted by the SAMi-m and SAMi-J ((a) and (b)), and DD-ME (c) families of energy density functionals.

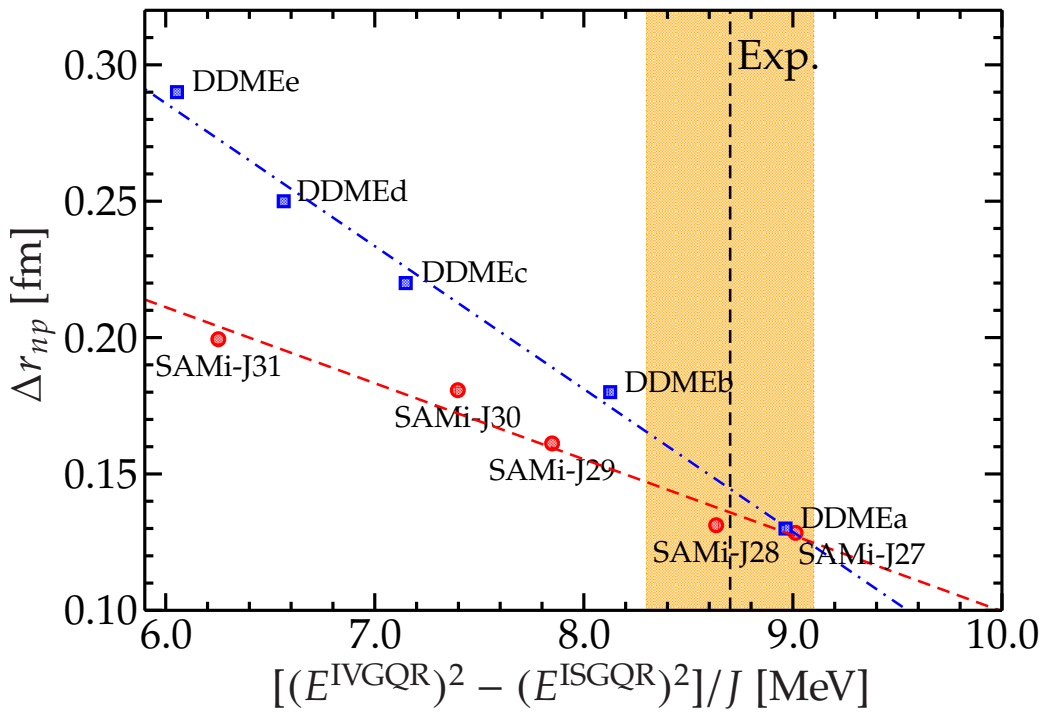

**Figure 6.15**
Values of $\Delta r_{np}$ as functions of $[(E^{\mathrm{IVGQR}})^2 - 2(E^{\mathrm{ISVGQR}})^2]/J$, calculated with the SAMi-J and DD-ME functionals. The dashed line and shaded band indicate the experimental value and corresponding uncertainty.

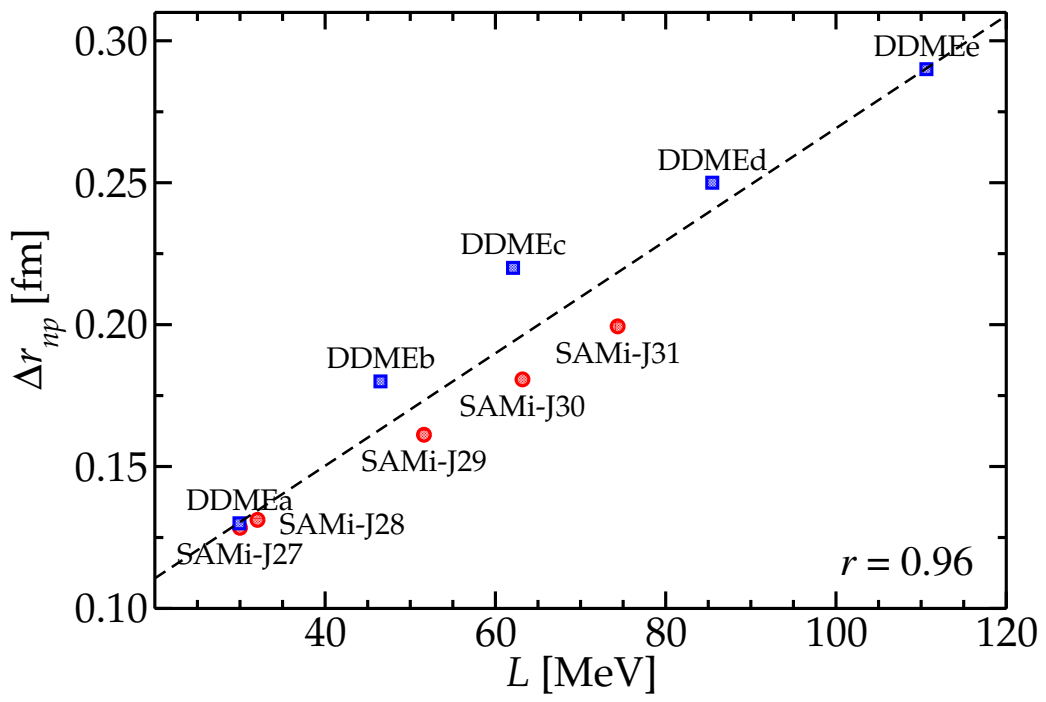

**Figure 6.16**
Neutron skin thickness $\Delta r_{np}$ of $^{208}$Pb as a function of the slope parameter of the symmetry energy at saturation density, for the two families of functionals: SAMi-J and DDME.

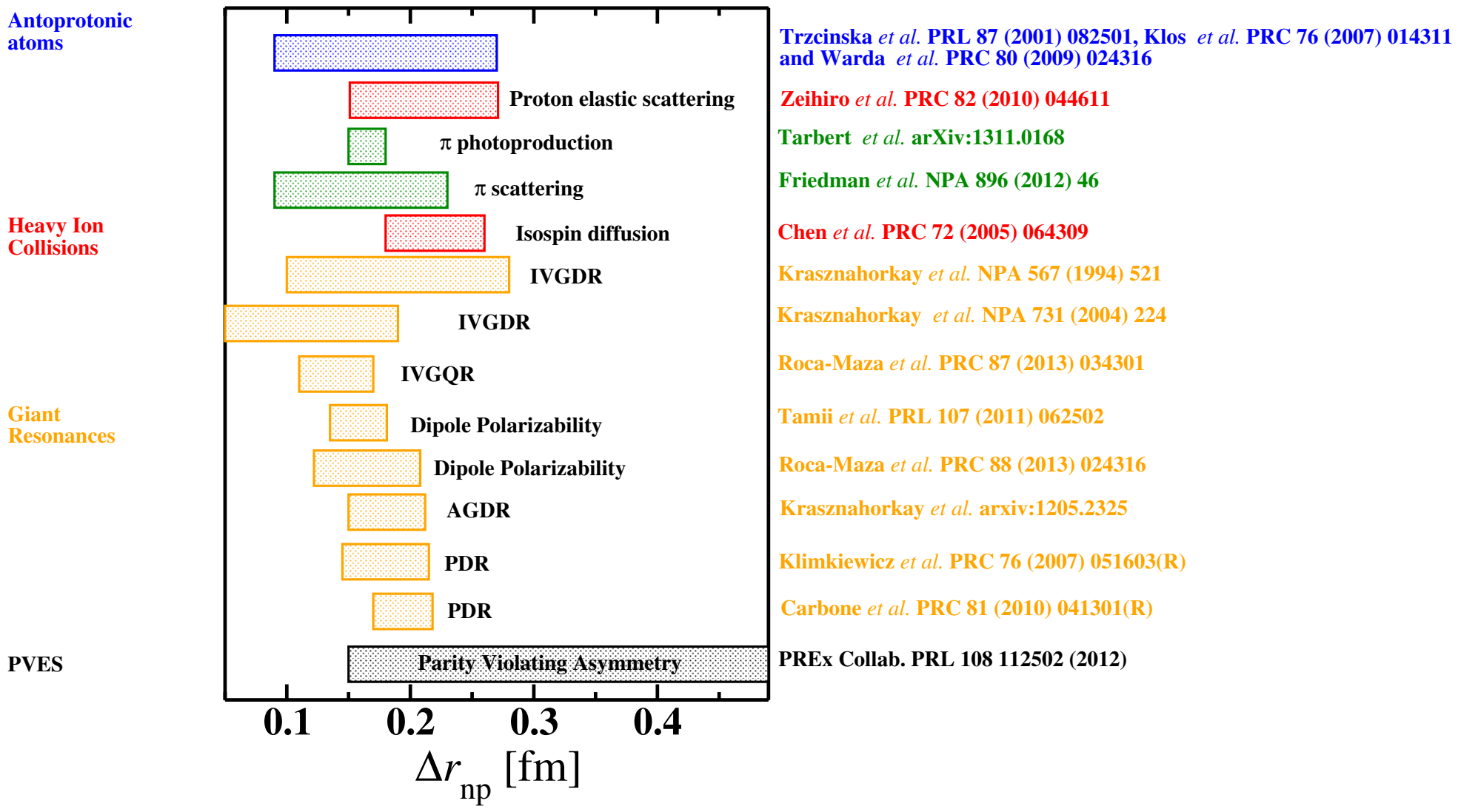

**Figure 6.17**
Experimental constraints on $\Delta r_{np}$ in $^{208}$Pb from different experiments.

# Part II

# Beyond the mean-field approach

Chapter 7

# Many-body approach to nuclear structure

As we have seen in the previous chapters, the description of the atomic nucleus as made of single-particles moving in an average potential is known to give good results for the bulk properties of the system. The potential emerging from SCMF calculations is static and in general non-local in space, or velocity-dependent. This feature can be translated in the definition of an effective mass, called $k$-mass, $m_k$. We have already encountered this quantity in Section 2.1.4, where we called it $m^*$. Nevertheless, SCMF methods have some important limitations. These models can be thought as based on the well-known Kohn-Sham theorem, thus the single-particle states are not strictly speaking within this framework. Moreover, properties such as the fragmentation of the states or their spreading widths cannot be reproduced. Moreover, they predict a lower density of states around the Fermi energy, causing a poor description of the low-lying collective states in the RPA approach.

Also, the lineshape of the strength functions and some decay properties of the GRs can be accounted for only including beyond RPA correlations, such as $2p-2h$ correlations. Some authors [CDK08; Kor+08; Zal+08] are currently trying to improve the accuracy of the present implementations aiming at functionals with so-called spectroscopic accuracy. In this case, the Kohn-Sham theorem applied to finite nuclei (see [Gir10] for a recent contribution about open problems on this issue) only guarantees that the energy of the lowest state with given quantum numbers can be exact if the functional is exact.

Another point of view, which is the one considered in this thesis, is the generalization of the concept of the shell model introducing a dynamical content [BV79; BV80; Mah+85; LR06; CSB10]. To this aim, the coupling between the single-particle motion and the collective modes is taken into account. This interweaving causes the single-particle potential to become energy dependent, and this fact can be characterized by the introduction of the so-called $\omega$-mass, $m_\omega$. Most experiments only probe the product of these two quantities, that is, the "total" effective mass $m^*$, defined as

$$\frac{m^*}{m} = \frac{m_k}{m}\frac{m_\omega}{m}. \tag{7.1}$$

The natural framework in which we can develop this extension of mean-field models is that of the standard many-body theory. We are not going to give here a complete description of this theory, for which we address the interested reader to standard textbooks like [BR86; FW03; DV08]. In the next sections we will introduce the main building blocks, which we will need in the following.

## 7.1 Green's functions

### Single-particle propagator

In a $N$-particle system, we consider the Hamiltonian $\mathcal{H}$ with the corresponding ground state eigenvalue $E_0^N$ and eigenvector $|\psi_0^N\rangle$. In the following we will always indicate with Greek letters the set of quantum numbers that completely characterize a state. The single-particle (sp) propagator, or Green's function, is defined as

$$G(\alpha,\beta;t,t') = -\frac{i}{\hbar}\langle\psi_0^N|\mathcal{T}\left[c_{\alpha_H}(t)c^\dagger_{\beta_H}(t')\right]|\psi_0^N\rangle. \tag{7.2}$$

The creation and annihilation operators are given in the Heisenberg picture by

$$c_{\alpha_H}(t) = e^{\frac{i}{\hbar}\mathcal{H}t}c_\alpha e^{-\frac{i}{\hbar}\mathcal{H}t}$$

respectively. The time ordering operator $\mathcal{T}$, defined as

$$\mathcal{T}\left[c_{\alpha_H}(t)c^\dagger_{\beta_H}(t')\right] = \theta(t-t')c_{\alpha_H}(t)c^\dagger_{\beta_H}(t') - \theta(t'-t)c^\dagger_{\beta_H}(t')c_{\alpha_H}(t),$$

puts operators with the later time to the left of operators at earlier times and includes a sign when a change of order is required (if we were considering bosons, no signs would be included).

The name "propagator" is then clear from the definition (7.2) of the Green's function: it corresponds to the probability amplitude for a process in which a particle is created in the state $\beta$ at the time $t'$ and is propagated to the state $\alpha$ at time $t$, where the particle is annihilated.

If the Hamiltonian of the system is not time-dependent, the Green's function depends only on the difference $t-t'$ and it can be written as

$$\begin{aligned} G(\alpha,\beta;t-t') = -\frac{i}{\hbar}\Bigg\{&\theta(t-t')\sum_m e^{\frac{i}{\hbar}(E_0^N-E_m^{N+1})(t-t')}\langle\psi_0^N|c_\alpha|\psi_m^{N+1}\rangle\langle\psi_m^{N+1}|c^\dagger_\beta|\psi_0^N\rangle \\ &-\theta(t'-t)\sum_n e^{-\frac{i}{\hbar}(E_0^N-E_n^{N-1})(t-t')}\langle\psi_0^N|c^\dagger_\beta|\psi_n^{N-1}\rangle\langle\psi_n^{N-1}|c_\alpha|\psi_0^N\rangle\Bigg\} \\ = G^+(\alpha,&\beta;t-t') + G^-(\alpha,\beta;t-t'), \end{aligned} \tag{7.3}$$

where it is clear that the sp propagator is composed of two parts: the first one is called the addition part, or alternatively, the "particle" or forward propagating part ($G^+$); the second term is likewise referred to as the removal, the "hole", or the backward propagating part ($G^-$).

Introducing the Fourier transform

$$\begin{aligned} G(\alpha,\beta;\omega) &= \int_{-\infty}^{+\infty} \mathrm{d}(t-t')\, e^{\frac{i}{\hbar}\omega(t-t')}G(\alpha,\beta;t-t') \\ &= \sum_m \frac{\langle\psi_0^N|c_\alpha|\psi_m^{N+1}\rangle\langle\psi_m^{N+1}|c^\dagger_\beta|\psi_0^N\rangle}{\omega-\left(E_m^{N+1}-E_0^N\right)+i\eta} + \sum_n \frac{\langle\psi_0^N|c^\dagger_\beta|\psi_n^{N-1}\rangle\langle\psi_n^{N-1}|c_\alpha|\psi_0^N\rangle}{\omega-\left(E_0^N-E_n^{N-1}\right)-i\eta} \\ &= G^+(\alpha,\beta;\omega) + G^-(\alpha,\beta;\omega) \end{aligned} \tag{7.4}$$

we obtain the so-called Lehmann representation [Leh54]. In the following we will always use the Lehmann representation for the propagators. In the particular case in which the Hamiltonian of the system can be diagonalized on a sp basis, the sp propagator (indicated by $G^{(0)}$) has a simpler Lehmann representation:

$$G^{(0)}(\alpha,\beta;\omega) = \delta_{\alpha,\beta}\left\{\frac{\theta\,(\varepsilon_\alpha - \varepsilon_{\mathrm{F}})}{\omega - \varepsilon_\alpha + i\eta} + \frac{\theta\,(\varepsilon_{\mathrm{F}} - \varepsilon_\alpha)}{\omega - \varepsilon_\alpha - i\eta}\right\} \tag{7.5}$$

For a finite system, all the information in the sp propagator can be related to experimental data. The poles contained in the first (second) energy denominators signal the location of the excited states in the $N + 1$ ($N - 1$)-particle systems with respect to the ground state of the $N$-particle system: this quantity is in principle measurable through stripping or pick-up experiments. The numerator determines the distribution of the transition strength from the ground state of the $N$-particle system to these states in the $N \pm 1$ systems, and this information is provided by the spectral function, related to the imaginary part of the sp propagator.

$$\begin{aligned}
\mathcal{S}_p(\alpha,\beta;\omega) &= -\frac{1}{\pi}\,\mathrm{Im}\,[G^+(\alpha,\beta;\omega)] \\
&= \sum_n \langle\psi_0^N|c_\alpha|\psi_m^{N+1}\rangle\langle\psi_m^{N+1}|c_\beta^\dagger|\psi_0^N\rangle\,\delta\left[\omega - \left(E_n^{N+1} - E_0^N\right)\right] \\
\mathcal{S}_h(\alpha,\beta;\omega) &= \frac{1}{\pi}\,\mathrm{Im}\,[G^-(\alpha,\beta;\omega)] \\
&= \sum_n \langle\psi_0^N|c_\beta^\dagger|\psi_n^{N-1}\rangle\langle\psi_n^{N-1}|c_\alpha|\psi_0^N\rangle\,\delta\left[\omega - \left(E_0^N - E_n^{N-1}\right)\right].
\end{aligned}$$

The diagonal part of the spectral function is interpreted as the probability density at energy $E$ for adding $[\mathcal{S}_p(\alpha;\omega)]$ or removing $[\mathcal{S}_h(\alpha;\omega)]$ a particle in the state $\alpha$, while leaving the remaining system at an energy $\pm\omega$ relative to the ground state $E_0^N$ of the system with $N$ particles. Using the spectral functions it is possible to evaluate the one-body density matrix elements as

$$\rho_{\alpha\beta} = \langle\psi_0^N|c_\beta^\dagger c_\alpha|\psi_0^N\rangle = \int_{-\infty}^{E_0^N - E_n^{N-1}} \mathrm{d}\omega\,\mathcal{S}_h(\beta,\alpha;\omega).$$

Accordingly, the matrix elements of a one-body operator $O$ are

$$\begin{aligned}
\langle\psi_0^N|O|\psi_0^N\rangle &= \sum_{\alpha\beta}\langle\alpha|O|\beta\rangle\langle\psi_0^N|c_\alpha^\dagger c_\beta|\psi_0^N\rangle = \sum_{\alpha\beta}\langle\alpha|O|\beta\rangle\rho_{\beta\alpha} \\
&= \sum_{\alpha\beta}\langle\alpha|O|\beta\rangle\int_{-\infty}^{E_0^N - E_n^{N-1}} \mathrm{d}\omega\,\mathcal{S}_h(\beta,\alpha;\omega).
\end{aligned}$$

The occupation number of a given sp particle state in the ground state can be obtained from the diagonal part of the density matrix:

$$n_\alpha = \langle\psi_0^N|c_\alpha^\dagger c_\alpha|\psi_0^N\rangle = \sum_n \left|\langle\psi_n^{N-1}|c_\alpha|\psi_0^N\rangle\right|^2 = \int_{-\infty}^{E_0^N - E_n^{N-1}} \mathrm{d}\omega\,\mathcal{S}_h(\alpha;\omega).$$

Correlations have the effect of reducing the occupation of the orbitals in the Fermi sea (which would be 1 for the unperturbed state) and to partially populate states that were originally empty. A measure of the emptiness of the orbital is obtained by integrating the particle spectral function

$$d_\alpha = \langle \psi_0^N | c_\alpha c_\alpha^\dagger | \psi_0^N \rangle = \sum_n \left| \langle \psi_n^{N+1} | c_\alpha^\dagger | \psi_0^N \rangle \right|^2 = \int_{E_n^{N+1} - E_0^N}^{+\infty} \mathrm{d}\omega \, \mathcal{S}_p(\alpha; \omega).$$

Using the fermion anti-commutation relations, one obtains the sum rule $n_\alpha + d_\alpha = 1$, valid for any orbital $\alpha$.This is an important result since it connects information on the removal and the addition of a particle to the system which refer to quite different experimental processes (viz. pick-up and stripping).

Moreover, also the ground-state energy of the $N$-particle system can be obtained by means of the Migdal-Galitski-Koltun (MGK) sum rule [GM58; Kol74], computing the expectation value of the potential energy, which is a two-body quantity, solely in terms of the one-body propagator:

$$E = \langle \psi_0^N | \mathcal{H} | \psi_0^N \rangle = \frac{1}{2} \int_{-\infty}^{E_0^N - E_n^{N-1}} \mathrm{d}\omega \sum_{\alpha\beta} \Big( \langle \alpha | T | \beta \rangle + \omega \delta_{\alpha\beta} \Big) \mathcal{S}_h(\beta, \alpha; \omega),$$

which is an exact result when only two-body forces are considered.

Exploiting the Wick theorem and the the time evolution in the interaction picture, it is possible to give a pictorial representation of the sp propagator in terms of the so-called Feynman or Goldstone diagrams. In these diagrams, the exact sp propagator is represented by two parallel arrowed lines, while the propagator for non-interacting particles is associated with a single arrowed line. The arrows indicate the flow of the energy inside the diagram. Eventually, the interaction vertices, i.e. the matrix elements of the interaction, are indicated with a dashed line. It may be possible to define a perturbative parameter and build a hierarchy of diagrams for the sp propagator in perturbation theory. If the interaction is considered small in comparison with the sp energies (as it is assumed in the standard formulation of the many-body theory), then its matrix elements can be taken as the perturbative parameter. In other cases, as we will see, some other choice should be made. To draw the $n^{\text{th}}$-order diagram of this expansion, first one has to identify $n$ interaction vertices, assigning them a time label. The difference between the Feynman and the Goldstone convention to appoint a arrowed line to a sp propagator consists in the way in which these interaction vertices are connected: in Goldstone diagram each arrowed line is either the propagator $G^+$ or the propagator $G^-$ of Eq. (7.3), while in Feynman diagrams each directed line indicate the Green's function $G$ as defined in Eq. (7.3). As a consequence two Goldstone diagrams are topologically identical if one can be distorted into the other without changing the time ordering of the interaction vertices, while the time ordering is not important in Feynman diagrams. Another possible phrasing is the following: in Goldstone diagrams the nature of the particle is clear, i.e. it is clear whether it is a particle or a hole, on the contrary, in Feynman diagrams the particle and hole parts of the propagator are summed.

The rules to translate the diagrams into formulae (and vice-versa) can be found for example in Ref. [DV08, Chap. 7-8].

It is also possible to organize the perturbation expansion in such a way that it automatically sums infinite sets of diagrams. From the diagrammatic representation of the sp propagator, we can infer that the exact sp propagator can be represented as in Fig. 7.1,

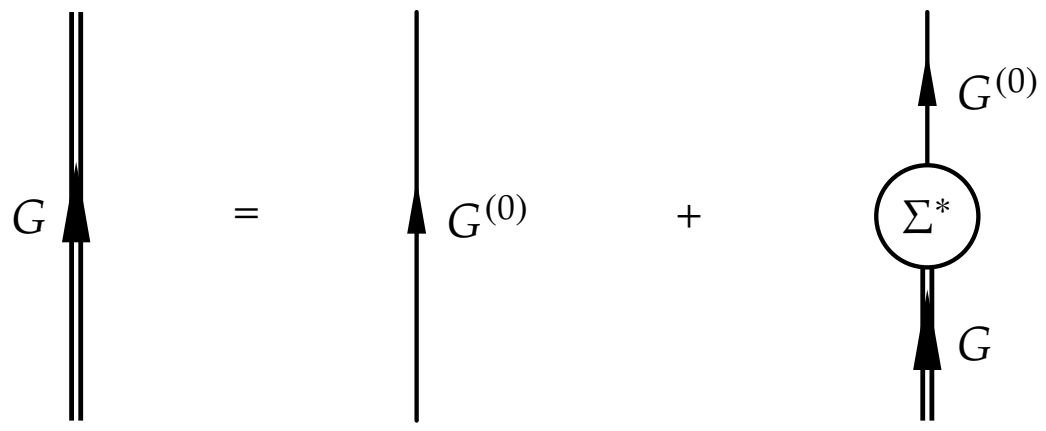

**Figure 7.1**
Diagrammatic representation of the sp propagator in terms of the irreducible self-energy $\Sigma^*$ and the non-interacting propagator $G^{(0)}$ representing Eq. (7.6).

which can be translated in the so-called Dyson equation for the sp propagator:

$$G(\alpha,\beta;\omega) = G^{(0)}(\alpha,\beta;\omega) + \sum_{\gamma\delta} G^{(0)}(\alpha,\gamma;\omega)\Sigma^*(\gamma,\delta;\omega)G(\delta,\beta;\omega). \qquad (7.6)$$

In this equation, the proper self-energy $\Sigma^*$ has been introduced, which is defined as the sum of all the irreducible contributions to the sp propagator that comes from the interaction between the particle and the medium. The word "irreducible" here means that the such diagrams do not contain two (or more) parts that are only connected by an unperturbed sp propagator $G^{(0)}$. Usually, to simplify matters, the first iteration step when solving self-consistently Eq. (7.6) is done explicitly by setting $G^{(0)} = G^{\mathrm{HF}}$, which is in turn obtained from Eq. (7.6) itself with the self-energy

$$\Sigma^*_{\mathrm{HF}}(\gamma,\delta;\omega) = -i\int \frac{d\omega'}{2\pi}\sum_{\mu\nu}\langle\gamma\mu|V|\delta\nu\rangle G^{\mathrm{HF}}(\nu,\mu;\omega'). \qquad (7.7)$$

With this choice, the HF propagator in the HF basis is diagonal

$$G^{\mathrm{HF}}(\alpha,\beta;\omega) = \delta_{\alpha,\beta}\left\{\frac{\theta\left(\varepsilon_\alpha-\varepsilon_{\mathrm{F}}\right)}{\omega-\varepsilon_\alpha+i\eta} + \frac{\theta\left(\varepsilon_{\mathrm{F}}-\varepsilon_\alpha\right)}{\omega-\varepsilon_\alpha-i\eta}\right\}. \qquad (7.8)$$

As expected from the fact that in the HF approximation a static (or instantaneous) potential is employed, the HF self-energy (7.7) do not depend on $\omega$.

In principle, the self-energy has non-diagonal contributions, even when evaluated with the diagonal HF sp propagator. However, in some cases, it is a good approximation to neglect the off-diagonal terms, e.g. in closed-shells nuclei off-diagonal elements would require mixing between major shells having a large energy separation. Within this approximation, the Dyson equation can have a simple algebraic solution

$$G(\alpha;\omega) = \frac{1}{\frac{1}{G^{\mathrm{HF}}(\alpha;\omega)} - \Sigma^*(\alpha;\omega)} = \frac{1}{\omega - \varepsilon_\alpha^{\mathrm{HF}} - \Sigma^*(\alpha;\omega)}.$$

Then, it is clear that the self-energy can be seen as an effective potential for the particles in the system, and accordingly the energy of the particle $\alpha$ can be written as

$$\varepsilon_\alpha = \varepsilon_\alpha^{\mathrm{HF}} + \Delta\varepsilon_\alpha = \varepsilon_\alpha^{\mathrm{HF}} + \mathrm{Re}\,\Sigma^*(\alpha;\omega=\varepsilon_\alpha).$$

Accordingly, the sp spectral function (e.g. for a hole state) can be written in the following

way:

$$\mathcal{S}_h(\omega) = \frac{1}{\pi}\,\mathrm{Im}\,G^-(\alpha,\beta;\omega) = \frac{1}{\pi}\,\mathrm{Im}\,\frac{1}{\omega - \varepsilon_\alpha^{\mathrm{HF}} - \Sigma^*(\alpha;\omega)}. \tag{7.9}$$

In order to take into account the fact that some couplings with some configuration are not considered explicitly in the self-energy, the spectral function is folded with a Lorentzian weight

$$\rho(E) = \frac{1}{\pi}\frac{\eta}{E^2 + \eta^2}.$$

It can be shown [BDL59] that averaging with this weight is equivalent to evaluating the spectral function at a complex energy $E + i\eta$, being $\eta$ the energy range over which the average is taken. The averaged spectral function can be expressed by means of a Lorentzian distribution with an energy dependent width,

$$\mathcal{S}_h(\omega) = \frac{1}{\pi}\frac{\mathrm{Im}\,\Sigma^*(\alpha;\omega)/2 + \eta}{[\omega - \varepsilon_\alpha^{\mathrm{HF}} - \mathrm{Re}\,\Sigma^*(\alpha;\omega)]^2 + [\mathrm{Im}\,\Sigma^*(\alpha;\omega)/2 + \eta]^2}$$

Let us consider for a moment an infinite system, where all the quantities of interest, such as the Green's function or the self-energy, can be written in the momentum space. From the dispersion relation of the particle

$$\varepsilon = \frac{k^2}{2m} + \Sigma^*(k;\omega = \varepsilon)$$

one can calculate the density of states

$$\begin{aligned}
\frac{\mathrm{d}\varepsilon}{\mathrm{d}k} &= \frac{k}{m} + \frac{\partial\Sigma^*(k;\varepsilon)}{\partial k} + \frac{\partial\Sigma^*(k;\varepsilon)}{\partial\varepsilon}\frac{\mathrm{d}\varepsilon}{\mathrm{d}k} \\
&= \frac{k}{m}\left(1 + \frac{m}{k}\frac{\partial\Sigma^*(k;\varepsilon)}{\partial k}\right)\left(1 - \frac{\partial\Sigma^*(k;\varepsilon)}{\partial\varepsilon}\right)^{-1}
\end{aligned}$$

In order to simplify this expression, it is useful to subsume the complicated effect of the medium into an effective mass. In particular, we can define

$$\begin{aligned}
\frac{m_k}{m} &= \left(1 + \frac{m}{k}\frac{\partial\Sigma^*(k;\varepsilon)}{\partial k}\right)^{-1} \\
\frac{m_\omega}{m} &= 1 - \frac{\partial\Sigma^*(k;\varepsilon)}{\partial\varepsilon} \\
\frac{m^*}{m} &= \frac{m_k}{m}\frac{m_\omega}{m}
\end{aligned}$$

### Polarization propagator

In order to treat the excited states in a $N$-particle system, from the propagator associated with a particle-hole excitations we can define the polarization propagator, with Lehmann representation (see Appendix B for a discussion on the particle/hole operators and the

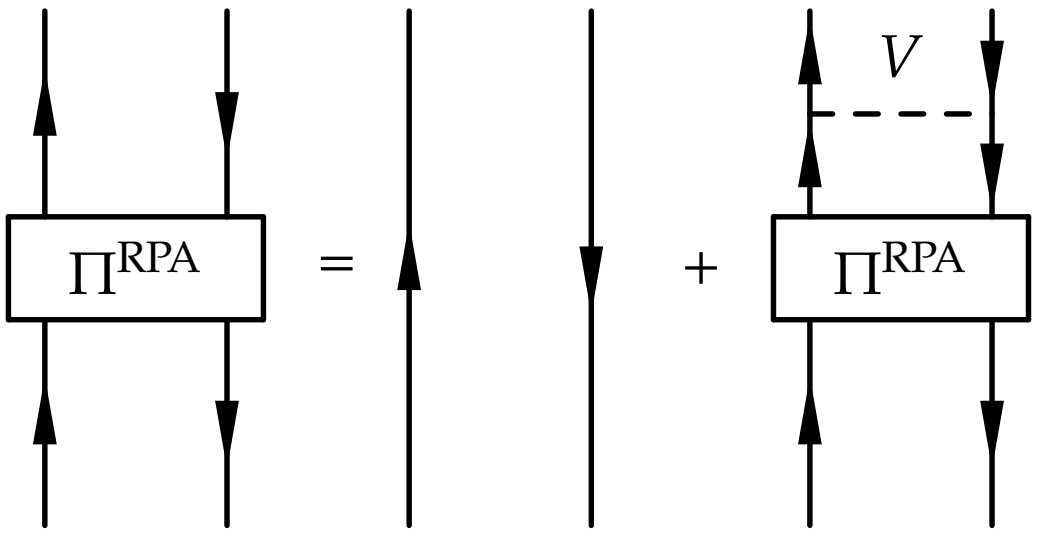

**Figure 7.2**
Diagrammatic representation the equation satisfied by the RPA polarization propagator.

particle-hole matrix element of the interaction)

$$\Pi(\alpha,\beta^{-1};\gamma,\delta^{-1};E) = \sum_{n\neq 0}\left\{\frac{\langle\psi_0^N|b_\beta a_\alpha|\psi_n^N\rangle\langle\psi_n^N|a_\gamma^\dagger b_\delta^\dagger|\psi_0^N\rangle}{E-(E_n^N-E_0^N)+i\eta} - \frac{\langle\psi_0^N|a_\gamma^\dagger b_\delta^\dagger|\psi_n^N\rangle\langle\psi_n^N|b_\beta a_\alpha|\psi_0^N\rangle}{E+(E_n^N-E_0^N)-i\eta}\right\}.$$

It is important to clarify that the indexes $\alpha,\beta,\gamma$ and $\delta$ can indicates states both above and below the Fermi level.

Also in this case, the denominator of the propagator incorporates the energy of the excited states of the $N$-particle system and the numerator contains the transition amplitudes connecting the ground state with the excited states. It can be shown that the polarization propagator satisfy an equation similar to the one for the sp propagator, represented in Fig. 7.2 and its first order approximation is the random phase approximation (RPA). Since in the following we will deal only with spherical system, it is useful to write the RPA polarization propagator in angular momentum coupled representation.

$$\Pi^{\text{RPA}}(a,b^{-1};c,d^{-1};J;E) = \sum_{n\neq 0}\left\{\frac{X_{ab}^{nJ}\left(X_{cd}^{nJ}\right)^*}{E-\varepsilon_{nJ}+i\eta} - \frac{\left(Y_{ab}^{nJ}\right)^* Y_{cd}^{nJ}}{E+\varepsilon_{nJ}-i\eta}\right\}, \tag{7.10}$$

where the sp quantum numbers takes on the form $\alpha = a, m_a$, $\varepsilon_{nJ} = E_{nJ}^N - E_0^N$ and

$$X_{ab}^{nJ} = \sum_{m_a m_b}(-)^{j_b-m_b}\langle j_a m_a j_b - m_b|JM\rangle\langle\psi_{nJM}^N|a_\alpha^\dagger b_{\bar\beta}^\dagger|\psi_0^N\rangle^*$$
$$Y_{ab}^{nJ} = (-)^{J+M+j_b-m_b}\sum_{m_a m_b}\langle j_a m_a j_b - m_b|JM\rangle\langle\psi_0^N|a_\alpha^\dagger b_{\bar\beta}^\dagger|\psi_{nJM}^N\rangle,$$

as in Eq. (3.7). A useful relation between $X$ and $Y$ amplitudes holds

$$Y_{ab}^{nJ} = (-)^{j_a-j_b+J}X_{ba}^{nJ}. \tag{7.11}$$

The angular-momentum-coupled version of the RPA propagator is expressed in terms of the uncoupled propagator by

$$\Pi^{\mathrm{RPA}}(a,b^{-1};c,d^{-1};J;E)=\sum_{\substack{m_a,m_b\\ m_c m_d}}(-)^{j_b-m_b+j_d-m_d}\langle j_a m_a j_b-m_b|JM\rangle\langle j_c m_c j_d-m_d|JM\rangle \times\Pi^{\mathrm{RPA}}(\alpha,\beta^{-1};\gamma,\delta^{-1};E). \tag{7.12}$$

As in the case of the sp propagator, the spectral function can be defined also for the polarization propagator (in the definition we use only the first part of the propagator which corresponds to positive energy $\varepsilon_{nJ}$)

$$\begin{aligned}\mathcal{S}_{\mathrm{RPA}}(a,b^{-1};c,d^{-1};J;E)&=-\frac{1}{\pi}\,\mathrm{Im}\,\Pi_{\mathrm{RPA}}(a,b^{-1};c,d^{-1};J;E)\\&=\sum_{n\neq0}X^{nJ}_{ab}\left(X^{nJ}_{cd}\right)^{*}\delta\left(E-\varepsilon_{nJ}\right)\end{aligned}$$

With this definition, if $F_J$ is a one-body excitation operator, the strength function of this operator (cf. Eq. (3.38)) becomes

$$\begin{aligned}S(E)&=\sum_{n\neq0}|\langle\psi^N_n\|F_J\|\psi^N_0\rangle|^2\delta(E-\varepsilon_{nJ})\\&=\sum_{abcd}\langle a\|F_J\|b\rangle\mathcal{S}_{\mathrm{RPA}}(a,b^{-1};c,d^{-1};J;E)\langle c\|F_J\|d\rangle.\end{aligned}$$

### Goldstone's Theorem

In order to calculate the energy of the system, the Goldstone theorem can be used [FW03]. Suppose that the Hamiltonian of the system is given by

$$\mathcal{H}=\mathcal{H}_0+V,$$

where $\mathcal{H}_0$ is diagonal and $V$ is the interaction. Let $|\phi_0\rangle$ be the ground state of $\mathcal{H}_0$, with energy $E_0$ and $|\psi_0\rangle$ the ground state of $\mathcal{H}$, with energy $E$. The energy shift of the ground state is

$$E-E_0=\langle\psi_0|V\sum_{n=1}^{\infty}\left(\frac{1}{E_0-\mathcal{H}_0}-V\right)^n|\psi_0\rangle_{\mathrm{connected}},$$

where the label "connected" means that the sum should comprise only connected diagrams. We can visualize these matrix elements in the following way: the operator $V$ acting on the state $|\phi_0\rangle$ creates two particles and two holes. This state then propagates with $(E_0-\mathcal{H}_0)^{-1}$, and the next $V$ can then create more particles and holes or scatter the existing particles or holes. The resulting intermediate state again propagates with $(E_0-\mathcal{H}_0)^{-1}$, and so on. The final $V$, must then return the system to the ground state $|\phi_0\rangle$. Goldstone's theorem is an exact restatement (to all orders) of the time-independent perturbation expression for the ground-state energy. This equivalence is readily verified in the first few terms by inserting a complete set of eigenstates of $\mathcal{H}_0$ between each interaction $V$,

$$E-E_0=\langle\psi_0|V|\psi_0\rangle+\sum_{n\neq0}\frac{\langle\psi_0|V|\psi_n\rangle\langle\psi_n|V|\psi_0\rangle}{E_0-E_n}+\dots \tag{7.13}$$

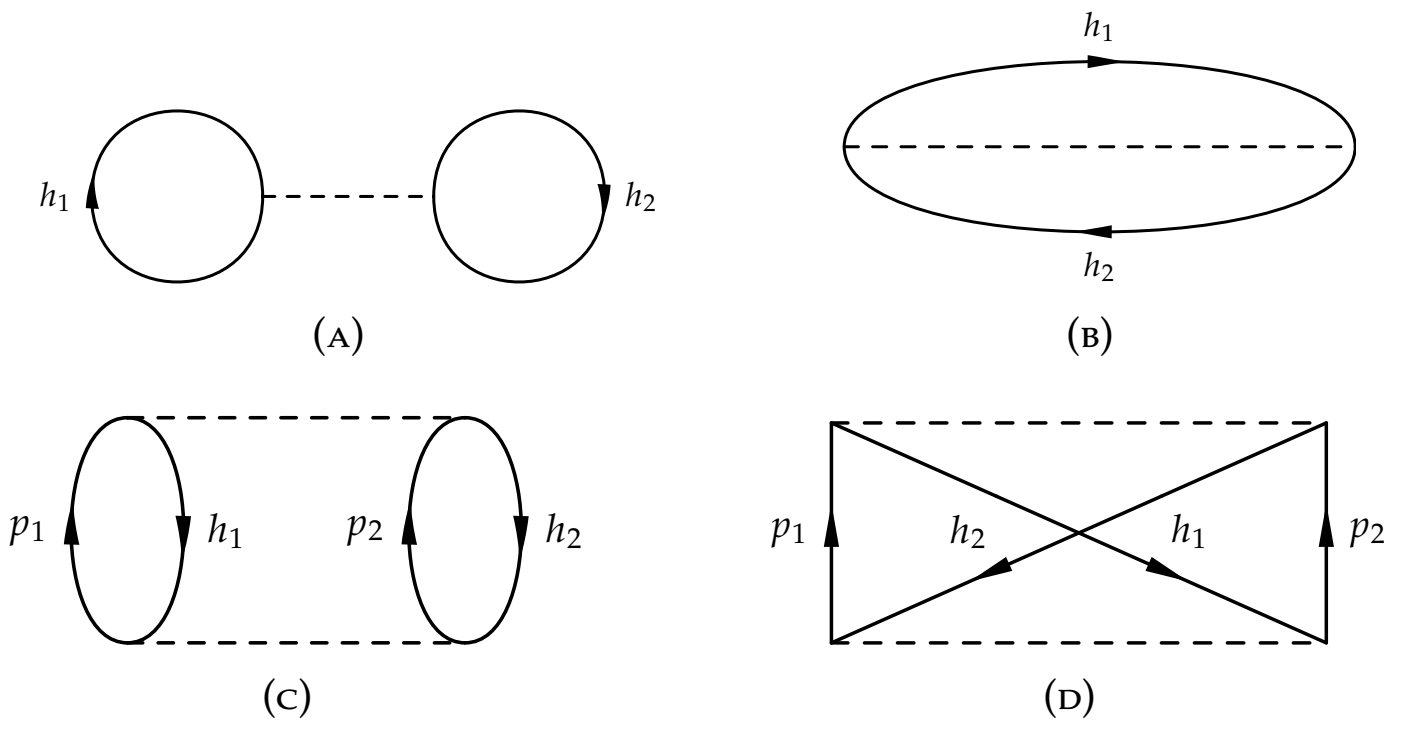

**FIGURE 7.3**
Direct and exchange contributions to the mean-field total energy (A) - (B) and second-order total energy (C)-(D). These are the diagrammatic representation of Eq. (7.13).

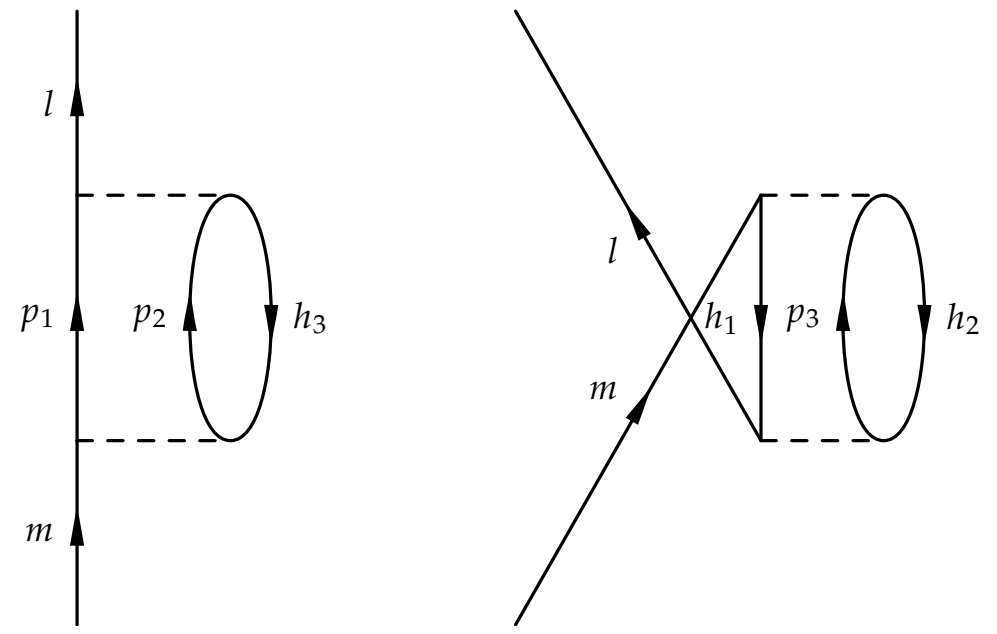

**FIGURE 7.4**
Second order contribution to the self-energy of the single-particle .

The corresponding Goldstone diagrams are shown in Fig. 7.3. The rules to evaluate these diagrams can be found in [Mat76].

### 7.1.1 Illustrative calculations

In this section, we will derive some diagrams which will be useful in the following.

#### The second order self-energy

The second order diagrams for the sp propagator are obtained by including the second order contribution to the self-energy. This contribution is depicted in Fig. 7.4 and it has the following explicit expression, in the particle-hole angular-momentum-coupled representation (the detailed calculation is carried out in Appendix C):

$$\Sigma^{(2)}(l,m;\omega) = \frac{1}{2}\left( \sum_{p_1,p_2,h_3} \frac{\langle lh_3|\bar{V}|p_1p_2\rangle\langle p_1p_2|\bar{V}|mh_3\rangle}{E-(\varepsilon_{p_1}+\varepsilon_{p_2}-\varepsilon_{h_3})+i\eta} + \sum_{h_1,h_2,p_3} \frac{\langle lp_3|\bar{V}|h_1h_2\rangle\langle h_1h_2|\bar{V}|mp_3\rangle}{E+(\varepsilon_{p_3}-\varepsilon_{h_1}-\varepsilon_{h_2})-i\eta} \right) \tag{7.14}$$

$$= \frac{1}{2}\sum_J \delta_{j_l,j_m}\frac{2J+1}{2j_m+1}\left(\sum_{p_1,p_2,h_3}\frac{V_J(p_1p_2lh_3)V_J(p_1p_2mh_3)}{E-(\varepsilon_{p_1}+\varepsilon_{p_2}-\varepsilon_{h_3})+i\eta}\right.$$
$$\left.+\sum_{h_1,h_2,p_3}\frac{V_J(h_1h_2lp_3)V_J(h_1h_2mp_3)}{E+(\varepsilon_{p_3}-\varepsilon_{h_1}-\varepsilon_{h_2})-i\eta}\right), \tag{7.15}$$

being $V_J$ the particle-hole angular-momentum-coupled matrix element

$$V_J(abcd) = \sum_{\{m\}}(-)^{j_b-m_b+j_c-m_c}\langle j_a m_a j_c - m_c|JM\rangle\langle j_d m_d j_b - m_b|JM\rangle\times$$
$$\times\langle j_a m_a j_b m_b|V|j_c m_c j_d m_d\rangle. \tag{7.16}$$

**Total energy of the system: the mean-field contribution**

The mean-field contribution to the total energy corresponds to the first term in the r.h.s of Eq. (7.13), represented in Fig. 7.3(A). It can be readily evaluated as

$$E^{\rm MF}_{dir} = (-1)^4\frac{1}{2}\sum_{\substack{h_1h_2\\ \varepsilon_{h_1,h_2}\leq\varepsilon_{\rm F}}}\langle h_1h_2|V|h_1h_2\rangle,$$

where the two minus signs come from the two hole lines and the other two from the two fermion loops, and the factor $\frac{1}{2}$ comes from the fact that the diagram is symmetric. The exchange diagram on the other hand reads (Fig. 7.3(B))

$$E^{\rm MF}_{ex} = (-)^3\frac{1}{2}\sum_{\substack{h_1h_2\\ \varepsilon_{h_1,h_2}\leq\varepsilon_{\rm F}}}\langle h_1h_2|V|h_2h_1\rangle,$$

where the two minus signs come from the two hole lines and the other from the fermion loop, and the factor $\frac{1}{2}$ comes from the fact that the diagram is symmetric.

**Total energy of the system: the second order contribution**

The diagrams that contribute at second order to the total energy are drawn in Fig. 7.3(C)-(D).

The direct contribution is given by

$$\Delta E^{(2)}_{dir} = (-)^4\frac{1}{2}\sum_{pp'hh'}\frac{\langle pp'|V|hh'\rangle\langle hh'|V|pp'\rangle}{\epsilon_h+\epsilon_{h'}-\epsilon_p-\epsilon_{p'}} = \frac{1}{2}\sum_{pp'hh'}\frac{\langle pp'|V|hh'\rangle\langle hh'|V|pp'\rangle}{\epsilon_h+\epsilon_{h'}-\epsilon_p-\epsilon_{p'}} \tag{7.17}$$

where the two minus signs come from the two hole lines and the other two from the two fermion loops. On the other hand, the exchange term is

$$\Delta E^{(2)}_{ex} = (-)^3\frac{1}{2}\sum_{pp'hh'}\frac{\langle pp'|V|h'h\rangle\langle hh'|V|pp'\rangle}{\epsilon_h+\epsilon_{h'}-\epsilon_p-\epsilon_{p'}} = -\frac{1}{2}\sum_{pp'hh'}\frac{\langle pp'|V|h'h)\langle hh'|V|pp'\rangle}{\epsilon_h+\epsilon_{h'}-\epsilon_p-\epsilon_{p'}} \tag{7.18}$$

If we wish to put together the direct and exchange terms, we obtain

$$\Delta E^{(2)} = \frac{1}{2} \sum_{pp'hh'} \frac{\langle pp'|V|hh'\rangle\langle hh'|\bar{V}|pp'\rangle}{\epsilon_h + \epsilon_{h'} - \epsilon_p - \epsilon_{p'}} \tag{7.19}$$

However, we can note that, given a two-body operator $V$,

$$\begin{aligned}
\sum_{\substack{ab\\cd}} \langle ab|\bar{V}|cd\rangle\langle cd|\bar{V}|ab\rangle &= \sum_{\substack{ab\\cd}} |\langle ab|\bar{V}|cd\rangle|^2 = \sum_{\substack{ab\\cd}} |\langle ab|V|cd\rangle - \langle ab|V|dc\rangle|^2 \\
&= \sum_{\substack{ab\\cd}} |\langle ab|V|cd\rangle|^2 + \sum_{\substack{ab\\cd}} |\langle ab|V|dc\rangle|^2 - 2\sum_{\substack{ab\\cd}} \langle ab|V|cd\rangle\langle ab|V|dc\rangle \\
&= \sum_{\substack{ab\\cd}} |\langle ab|V|cd\rangle|^2 + \sum_{\substack{ab\\cd}}^{c\leftrightarrow d} |\langle ab|V|cd\rangle|^2 - 2\sum_{\substack{ab\\cd}} \langle ab|V|cd\rangle\langle ab|V|dc\rangle \\
&= 2\sum_{\substack{ab\\cd}} \left(\langle ab|V|cd\rangle\langle cd|V|ab\rangle - \langle ab|V|cd\rangle\langle dc|V|ab\rangle\right).
\end{aligned}$$

Therefore,

$$\sum_{\substack{ab\\cd}} \langle ab|V|cd\rangle\langle ab|\bar{V}|cd\rangle = \frac{1}{2}\sum_{\substack{ab\\cd}} \langle ab|\bar{V}|cd\rangle\langle cd|\bar{V}|ab\rangle.$$

Then, Eq. (7.19) can be rewritten as

$$\Delta E^{(2)} = \frac{1}{4} \sum_{pp'hh'} \frac{\left|\langle pp'|\bar{V}|hh'\rangle\right|^2}{\epsilon_h + \epsilon_{h'} - \epsilon_p - \epsilon_{p'}} = \frac{1}{4} \sum_{\substack{pp',\\hh'J}} \frac{(2J+1)\,|V_J(pp'hh')|^2}{\epsilon_h + \epsilon_{h'} - \epsilon_p - \epsilon_{p'}}, \tag{7.20}$$

where the last has been obtained using the particle-hole angular-momentum-coupled matrix element (7.16).

## 7.2 Going beyond the mean-field approximation

As already stated, the mean-field approach shows some important limitations in the description of some observables. In order to improve on the description of collective states, a straightforward approach would be the diagonalization of the effective Hamiltonian in the configuration space which includes at least the $2p-2h$, or four-quasiparticle states. However, this route is usually not feasible because of the large number of $2p-2h$ configurations ($\approx 10^3 \div 10^4$ per MeV). One possible approximation is the so-called second RPA (SRPA), which is based on non-interacting $2p-2h$ configurations and only the interaction between $2p-2h$ and $1p-1h$ state is explicitly taken into account. Practical implementation of this framework have been restricted to relatively light nuclei, being often not self-consistent [Dro+90], although some recent calculations have been done including the same interaction both at HF-RPA and SRPA level (except for the Coulomb and spin-orbit interaction in the SRPA residual interaction) [GGC10; Gam+12]. However, in heavy nuclei it is impossible to solve the SRPA equations. Therefore, usually they are projected on the $1p-1h$ space, obtaining a matrix equation similar to Eq. (3.8).

Another possible way to improve the description of the system is to consider the next higher-order contribution to the self-energy of a particle in the medium. One route is to include in the sp propagator the particle-particle, hole-hole and particle-hole correlation, approximating the $n$-particle Green's function with sp and two-particle propagators. A recent review of these methods can be found in Refs. [MP00; DB04]

In other theoretical frameworks the interplay between single-particle and collective degrees of freedom are considered, in the so-called particle-vibration coupling (PVC), or particle-rotation coupling in deformed systems [BM75]. These frameworks are based on the fact that, in the particular case of the atomic nuclei, the excitation spectrum comprises in the same energetic region both kind of states. We can enumerate the following:

- In the effective theory of finite Fermi systems (ETFFS) [Kam+93; KST04] the $1p-1h$, complex $1p-1h\otimes 1$ phonon configurations, the single-particle continuum and ground-state correlations are taken into account.
- In the quasi-particle phonon model (QPM) [Sol92; BP99] the wave functions of the excited states are a combination of one-, two- and three QRPA phonon configurations.
- The nuclear field theory (NFT) is the framework which is used in this work, and will be widely discussed in the next section.

A similar implementation is also possible in the covariant realm [LR06]. In addition, the possibility of giving a formal theory of the PVC with density dependent effective forces starting from the many-body Hedine equations [FW03; DV08] is currently under investigation [Bal+13].

## 7.3 Nuclear Field Theory

Several evidences in the last decades [Bor+77] have shown that collective excitations should be considered as independent degrees of freedom with respect to single-particle ones. Apart from density vibrations, also rotational bands (in deformed nuclei), or pairing vibrations and rotations (the latter in superfluid nuclei), deriving from the coherent superposition of creation and annihilation of particle (or hole) pairs, are collective states. Therefore, as elementary modes of excitations we consider:

- single-particle states;
- density vibrations and rotations;
- pairing vibrations and rotations.

In this thesis we will deal only with doubly magic nuclei, thus we are not considering in the following discussion either pairing excitation, which have been estimated to give a small contribution [Ber+79], nor rotations. The nuclear bosonic fields are built out of the particle degrees of freedom, thus the chosen basis is non-orthogonal, overcomplete and contains states violating the Pauli principle.

The potential associated with the vibrations of the nuclear density gives rise to the so-called particle-vibration coupling (PVC). The resulting system is made up with coupled single-particle and collective degrees of freedom, whose interweaving can be described by a field theory made up with interacting fermions and bosons, named Nuclear Field Theory (NFT), which is widely discussed in Refs. [Bes+74; Bes+76a; Bes+76b; Bes+76c; Bro+76; Bor+77]. The total Hamiltonian in NFT is given by

$$\mathcal{H} = \mathcal{H}'_{sp} + \mathcal{H}_B + h, \tag{7.21}$$

where $\mathcal{H}'_{sp}$ and $\mathcal{H}_B$ are the uncoupled sp and collective Hamiltonians, respectively. The former must contain the Hartree-Fock contribution of the two-body interaction, while the latter comes from a RPA diagonalization with the residual interaction. The Hamiltonian $h$ contains the interaction between quasi-particles and vibrations. It can generate all different diagrams of perturbation theory (except those containing bubbles – see later in this section) and it is composed by

$$h = \mathcal{H}_{pv} + \mathcal{H}'_{tb}, \tag{7.22}$$

where $\mathcal{H}_{pv}$ and $\mathcal{H}'_{tb}$ are the particle-vibration and the two-body (or four-points) Hamiltonians, respectively. The two contributions to $h$ are depicted in Fig. 7.5. The overcompleteness implicit in the product basis is corrected through the perturbative treatment of $h$, while the violation of the Pauli principle is corrected by the fact that for a given diagram in which two fermion lines are simultaneously in the same single-particle state, there is another diagram in which the correspondent end points are interchanged.

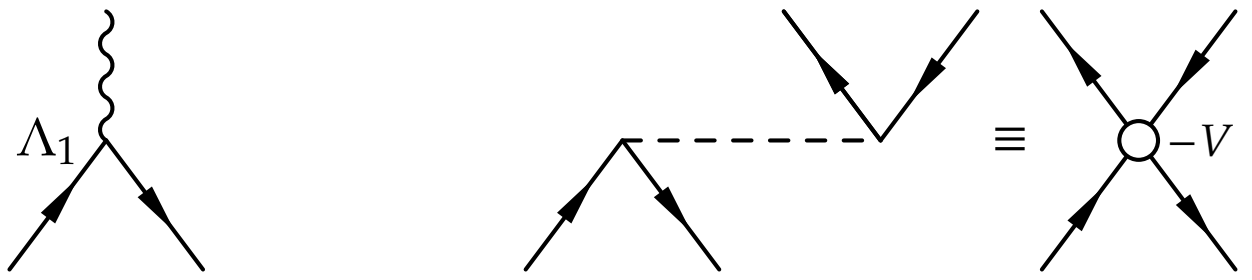

**Figure 7.5**
Graphical representation of $h$

**Rules for evaluating the diagrams**

Since a significant part of the original interaction has already been included in generating the collective mode, the rules for evaluating the possible diagrams involve a number of restrictions:

(1) In order to draw the diagrams the usual rules of many-body theory are used [DV08]. However, in initial and final states, proper diagrams do not involve any particle configuration that can be replaced by a combination of collective modes. This restriction does not apply to internal lines of these diagrams.

(2) The vertices are allowed to act in all orders to generate the different diagrams of perturbation theory. In particular, the particle-hole interaction should be used if the PVC vertex is handled.

(3) Unlinked diagrams are neglected.

(4) The internal lines of diagrams are restricted by the exclusion of diagrams in which a particle-hole pair is created and subsequently annihilated without having participated in any interactions (bubble).

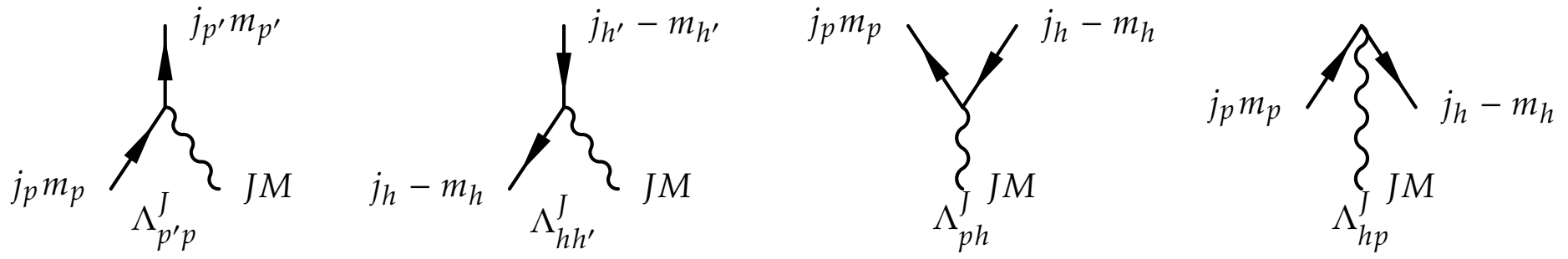

**Figure 7.6**
The four particle-vibration coupling vertices.

(5) In intermediate states, the requirement of antisymmetry between fermion lines and of symmetry between phonon ones is neglected. However, there is a factor $(n!)^{-\frac{1}{2}}$ for each subset of $n$ identical phonons in initial and final states.

(6) The numerator is given by the product of the vertices appearing in the diagram.

(7) The denominators of each diagrams are calculated according to Releigh - Schrödinger, Brillouin-Wigner (see for examples [GRH91]) or Bloch-Horowitz, [DL71] perturbation theory.

In order to explain the basics of the NFT, we present in the appendix D a schematic two-level model with a schematic particle-hole interaction. Rules (6)-(7) come from the fact that in the diagrams sp and polarization propagators should be included and integration on the energy of intermediate states should be carried out. An example of the application of these rules is shown in Appendix E, where the lowest-order self-energy of a particle (or a hole) at PVC level is evaluated both using the sp and polarization propagators and using rules (6)-(7).

Except for the pioneering work of Ref. [BV79; BV80] in which a simplified Skyrme interaction (including only its velocity independent part) was used in the PVC vertex, many other calculations have been done so far with phenomenological inputs and uncontrolled approximations (for a review, see Ref. [Mah+85]). Only recently a fully self-consistent version of the PVC has been proposed in Ref. [CSB10] in the Skyrme framework. In that paper, the lowest-order self-energy for sp particles at PVC level has been calculated for $^{40}$Ca and $^{208}$Pb, finding corrections of the order of MeV and hundreds of keV, respectively, going in the direction of making the agreement with the experiment better.

In general, for the reduced matrix element of the interaction, at the vertex (see Fig. 7.6) we obtain

$$\langle a||V||b,nJ\rangle = \sqrt{2J+1}\sum_{ph} X^{nJ}_{ph} V_J(ahbp) + (-)^{j_h-j_p+J} Y^{nJ}_{ph} V_J(apbh). \tag{7.23}$$

The detailed derivation of the reduced matrix element (7.23) is provided in the appendix of Ref. [CSB10] and appendix F. It should be noted that the result ensues from the extension to a microscopic interaction of the original schematic PVC vertex $\Lambda_\nu$ of appendix D. The matrix element $\langle a,nJ||V||b\rangle$, where the phonon is in the final state, can be easily obtained by the addition of a phase [BM75]

$$\langle b,nJ||V||a\rangle = (-)^{j_a-j_b+J}\langle a||V||b,nJ\rangle, \tag{7.24}$$

and for the explicit derivation of this matrix element we address the reader to the Appendix F.

The four possible particle-phonon vertices are depicted in Fig. 7.6, depending on the fermionic state involved being a particle or a hole. We will work out the expressions for the particle-vibration coupling vertices $\Lambda^{nJ}_{ab}$ (in keeping with the original model exposed in appendix D) from the interaction $V$.

$$\begin{aligned}\Lambda^{nJ}_{p'p} &= \langle j_{p'} m_{p'} | V | j_p m_p, nJM \rangle = \\ &= \frac{(-)^{j_p - m_p}}{\sqrt{2J+1}} \langle j_{p'} m_{p'} j_p - m_p | JM \rangle \langle p' \| V \| p, nJ \rangle \\ &= (-)^{j_{p'} - m_{p'}} \begin{pmatrix} j_{p'} & j_p & J \\ m_{p'} & -m_p & -M \end{pmatrix} \langle p' \| V \| p, nJ \rangle \end{aligned} \tag{7.25a}$$

$$\begin{aligned}\Lambda^{nJ}_{hh'} &= \langle (j_{h'} m_{h'})^{-1} | V | (j_h m_h)^{-1}, nJM \rangle \\ &= (-)^{j_{h'} + m_{h'} + j_h + m_h + 1} \langle j_h - m_h | V | j_{h'} - m_{h'}, nJM \rangle \\ &= \frac{(-)^{j_{h'} + m_{h'} + J + M}}{\sqrt{2J+1}} \langle j_{h'} m_{h'} j_h - m_h | JM \rangle \langle h \| V \| h', nJ \rangle \\ &= (-)^{j_{h'} + m_{h'}} \begin{pmatrix} j_{h'} & j_h & J \\ m_{h'} & -m_h & -M \end{pmatrix} \langle h \| V \| h', nJ \rangle \end{aligned} \tag{7.25b}$$

$$\begin{aligned}\Lambda^{nJ}_{ph} &= \langle j_p m_p (j_h m_h)^{-1} | V | nJM \rangle = (-)^{j_h - m_h} \langle j_p m_p | V | j_h - m_h, nJM \rangle \\ &= \frac{-1}{\sqrt{2J+1}} \langle j_p m_p j_h m_h | JM \rangle \langle p \| V \| h, nJ \rangle \\ &= (-)^{J+M} \begin{pmatrix} j_h & j_p & J \\ m_h & m_p & -M \end{pmatrix} \langle p \| V \| h, nJ \rangle \end{aligned} \tag{7.25c}$$

$$\begin{aligned}\Lambda^{nJ}_{hp} &= \langle 0 | V | j_p m_p (j_h m_h)^{-1}, nJM \rangle = (-)^{j_h - m_h} \langle j_h - m_h | V | j_p m_p, nJM \rangle \\ &= \frac{(-)^{J+M}}{\sqrt{2J+1}} \langle j_h m_h j_p m_p | J - M \rangle \langle h \| V \| p, nJ \rangle \\ &= \begin{pmatrix} j_h & j_p & J \\ m_h & m_p & M \end{pmatrix} \langle h \| V \| p, nJ \rangle . \end{aligned} \tag{7.25d}$$

It is important to note that Eq. (7.25a) and Eq. (7.25b) are different only for a sign (having identified $h$ and $h'$ as $p$ and $p'$, respectively), as we should expect from the correlations between the particle and the hole existing in collective vibrations [BBB83].

We indicate the vertices in which the phonon is created, instead of annihilated, with the label $\Lambda'^{nJ}_{ba}$. These vertices can be obtained from $\Lambda^{nJ}_{ab}$ with the addition of the phase factor $(-)^{j_a - j_b + J}$ and the use of the matrix element (7.24). Moreover, note that the $X$ and $Y$ RPA amplitudes are related to the vertices $\Lambda_{ph}$ and $\Lambda_{hp}$

$$X^{nJ}_{ph} = -\frac{1}{\sqrt{2J+1}} \frac{\langle p \| V \| h, nJ \rangle}{E_{nJ} - \varepsilon_p + \varepsilon_h + i\eta} \tag{7.26a}$$

$$Y^{nJ}_{ph} = -\frac{1}{\sqrt{2J+1}} \frac{\langle h \| V \| p, nJ \rangle}{E_{nJ} + \varepsilon_p - \varepsilon_h + i\eta}, \tag{7.26b}$$

as it can be inferred from the RPA equations (3.8).

Chapter 8

# The width of giant resonances

In this chapter we apply the nuclear field theory, expounded in the previous chapter, to the case of the width of giant resonances. It has been known for several decades that coupling with low-lying vibrations is the main source of the GR width [BBB83]. In this work we concentrate on the spreading width of GRs (Section 8.1), which was considered in Ref. [BB81] with the use of a phenomenological separable force, and on the width associated with the $\gamma$ decay of GRs (Section 8.2.1), which was as well handled by using a separable force in Ref. [BBB84]. The particle decay was described in a microscopic, though not completely self-consistent, model using a Skyrme interaction some years ago in Refs. [CVB94; CVS95].

## 8.1 The lineshape functions of giant resonances

One of the most important observables that can be described using a PVC approach is the strength function of giant resonances. The strength function has already been defined in Eq. (3.38) and in Eq. (7.1) it has been re-expressed in terms of the spectral function of the RPA polarization propagator $S_{\text{RPA}}$. As in the case of the single-particle spectral function, the RPA spectral function can be expressed at a beyond mean-field level taking into account the contributions coming from the two-body particle-hole self-energy. The diagrams that contribute to the p-h self-energy are depicted in Fig. 8.1.

We indicate by $\Delta E_{\text{GR}}$ and $\Gamma_{\text{GR}}$ the real and imaginary part of the p-h self-energy, respectively. In this way, the probability of finding the GR state per unit energy can be written as

$$P(E) = \frac{1}{\pi} \frac{\Gamma_{\text{GR}}(E) + \frac{\eta}{2}}{(E - E_{\text{GR}} - \Delta E_{\text{GR}}(E))^2 + \left(\Gamma_{\text{GR}}(E) + \frac{\eta}{2}\right)^2}. \tag{8.1}$$

The parameter $\eta$ corresponds to the energy interval over which averages are taken and represents, in an approximate way, the coupling of the intermediate states to more complicated configurations. Figures 8.1(a) through (d) correspond to the self-energy of the particle (or the hole), i.e., the processes in which the particle or the hole reabsorbs the intermediate excitation $\lambda$, while Figs. 8.1(e) – (h) are vertex corrections which describe the process in which the phonon is exchanged between particle and hole. If $J$ or $\lambda$ is a density oscillation (as it is in this case), the latter contributions have opposite sign with respect to the former one, regardless of the spin and isospin character of $\lambda$ or $J$, respectively [BBB98].

The eight diagrams in Figs. 8.1(a) – (h) are evaluated by the following expressions:

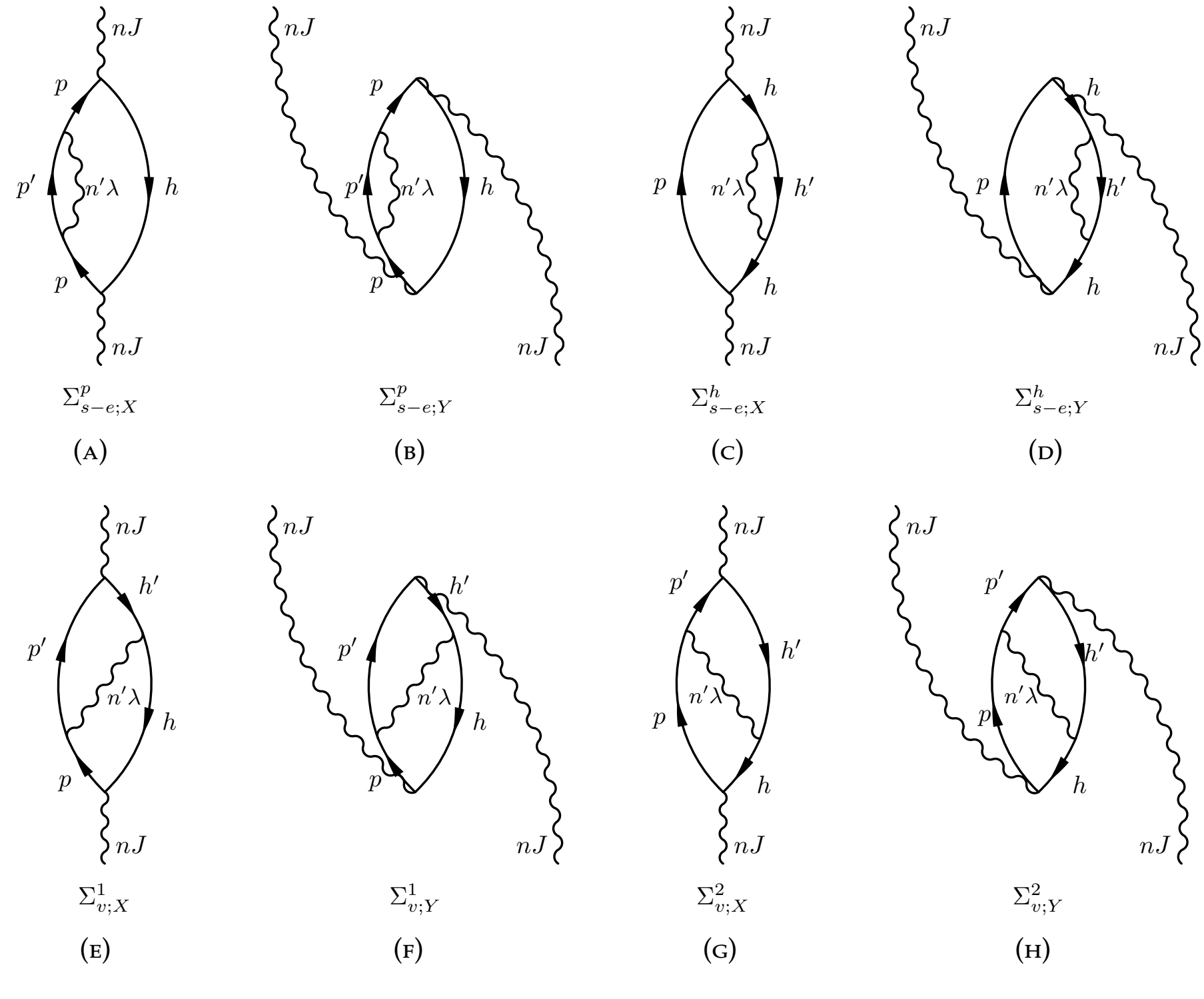

**Figure 8.1**
NFT diagrams contributing to the strength function of the giant resonance.

$$\Sigma^{p}_{s-e;X}(GR, E_{nJ}) = \sum_{\substack{pp'h \\ n'\lambda}} \frac{1}{(2j_p+1)} \frac{\left(X^{nJ}_{ph}\right)^2 |\langle p\|V\|p', n'\lambda\rangle|^2}{E_{nJ} - E_{n'\lambda} - \epsilon_{p'h} + i\eta} \tag{8.2.a}$$

$$\Sigma^{p}_{s-e;Y}(GR, E_{nJ}) = \sum_{\substack{pp'h \\ n'\lambda}} \frac{-1}{(2j_p+1)} \frac{\left(Y^{nJ}_{ph}\right)^2 |\langle p\|V\|p', n'\lambda\rangle|^2}{E_n J + E_{n'\lambda} + \epsilon_{p'h} + i\eta} \tag{8.2.b}$$

$$\Sigma^{h}_{s-e;X}(GR, E_{nJ}) = \sum_{\substack{phh' \\ n'\lambda}} \frac{1}{(2j_h+1)} \frac{\left(X^{nJ}_{ph}\right)^2 |\langle h'\|V\|h, n'\lambda\rangle|^2}{E_{nJ} - E_{n'\lambda} - \epsilon_{ph'} + i\eta} \tag{8.2.c}$$

$$\Sigma^{h}_{s-e;Y}(GR, E_{nJ}) = \sum_{\substack{phh'' \\ n'\lambda}} \frac{-1}{(2j_h+1)} \frac{\left(Y^{nJ}_{ph}\right)^2 |\langle h'\|V\|h, n'\lambda\rangle|^2}{E_{nJ} + E_{n'\lambda} + \epsilon_{ph'} + i\eta} \tag{8.2.d}$$

$$\Sigma^{1}_{v;X}(GR, E_{nJ}) = \sum_{\substack{pp' \\ hh' \\ n'\lambda}} (-)^{J+\lambda+j_p-j_{h'}} \begin{Bmatrix} j_h & j_p & J \\ j_{p'} & j_{h'} & \lambda \end{Bmatrix} \frac{X^{nJ}_{ph} X^{nJ}_{p'h'} \langle h\|V\|h', n'\lambda\rangle \langle p\|V\|p', n'\lambda\rangle}{E_{nJ} - E_{n'\lambda} - \epsilon_{p'h} + i\eta} \tag{8.2.e}$$

$$\Sigma^{1}_{v;Y}(GR, E_{nJ}) = \sum_{\substack{pp' \\ hh' \\ n'\lambda}} (-)^{J+\lambda+j_p+j_{h'}} \begin{Bmatrix} j_h & j_p & J \\ j_{p'} & j_{h'} & \lambda \end{Bmatrix} \frac{Y^{nJ}_{ph} Y^{nJ}_{p'h'} \langle h\|V\|h', n'\lambda\rangle \langle p\|V\|p', n'\lambda\rangle}{E_{nJ} + E_{n'\lambda} + \epsilon_{p'h} + i\eta} \tag{8.2.f}$$

$$\Sigma^{2}_{v;X}(GR, E_{nJ}) = \sum_{\substack{pp'' \\ hh' \\ n'\lambda}} (-)^{J+\lambda+j_{p'}-j_h} \begin{Bmatrix} j_h & j_p & J \\ j_{p'} & j_{h'} & \lambda \end{Bmatrix} \frac{X^{nJ}_{ph} X^{nJ}_{p'h'} \langle p'\|V\|p, n'\lambda\rangle \langle h'\|V\|h, n'\lambda\rangle}{E_{nJ} - E_{n'\lambda} - \epsilon_{ph'} + i\eta} \tag{8.2.g}$$

$$\Sigma^{2}_{v;Y}(GR, E_{nJ}) = \sum_{\substack{pp' \\ hh' \\ n'\lambda}} (-)^{J+\lambda+j_{p'}-j_h} \begin{Bmatrix} j_h & j_p & J \\ j_{p'} & j_{h'} & \lambda \end{Bmatrix} \frac{Y^{nJ}_{ph} Y^{nJ}_{p'h'} \langle p'\|V\|p, n'\lambda\rangle \langle h'\|V\|h, n'\lambda\rangle}{E_{nJ} + E_{n'\lambda} + \epsilon_{ph'} + i\eta}. \tag{8.2.h}$$

The calculation of the first and fifth one, corresponding to Fig. 8.1(a) and 8.1(e), is detailed in Appendix E.

## 8.2 The $\gamma$ decay of GRs

In the past, the study of the $\gamma$ decay of GRs has played a less central role, because this decay gives the smallest contribution to the width (or, that is the same, the less probable decay branch) – of the order of $\approx 10^{-3}$ of the total width. Nonetheless, the study of the $\gamma$ decay has been considered a valuable tool for about 30 years [Bee+89; Bee+90]. In these works, the fact that $\gamma$ decay can be a sensitive probe of the excited multipolarity, and that $\gamma$-ejectile coincidence measurements can improve the extraction of the properties of GRs, has been thoroughly discussed.

Eventually, we should point out that in the following only the direct decay of GRs will be considered. To allow a complete comparison with the experiment, also the $\gamma$ decay from the compound nucleus (formed because of the coupling between the GR to $1p-1h, 2p-2h, \dots, np-nh, \dots$ configurations) should be taken into account. This last contribution can be described by means of a statistical model [Bee+85; BBB98].

### 8.2.1 The $\gamma$ decay to low-lying states

While the decay to the GS, being a mean-field observable, has already been presented in section 3.5, the decay to low-lying states lies outside the mean-field framework. The reason is that, by construction, RPA is an appropriate theory to describe the transition amplitudes between states that differ only by one vibrational (phonon) state. For other processes, like the one at hand, the extension to a treatment beyond RPA is mandatory. For this reason, in this work we consider all the lowest-order contributions to the $\gamma$ decay between two different phonons. This amounts to writing and evaluating all lowest-order perturbative diagrams involving single-particle states and phonon states, which can lead from the initial state to the final state by the action of the external electromagnetic field.

The perturbative diagrams associated with the $\lambda$-pole decay of the initial RPA state $|nJ\rangle$ (at energy $E_{nJ}$) to the final state $|n'J'\rangle$ (at energy $E_{n'J'}$) are shown in Fig. 8.2, and the way to evaluate the first 8.2(a), the fifth 8.2(e) and the ninth 8.2(i) is sketched in the Appendix E. The resulting analytic expressions read

$$\langle n'J'\,||Q_\lambda||\,nJ\rangle_{(\mathrm{A})} = \sum_{pp'h}(-)^{J+\lambda+j_{p'}+j_h}\begin{Bmatrix} J & \lambda & J' \\ j_{p'} & j_h & j_p \end{Bmatrix} X^{nJ}_{ph}X^{n'J'}_{p'h}Q^{\lambda pol}_{p'p}\,, \tag{8.3.a}$$

$$\langle n'J'\,||Q_\lambda||\,nJ\rangle_{(\mathrm{B})} = \sum_{pp'h}(-)^{J'+j_{p'}+j_h}\begin{Bmatrix} J & \lambda & J' \\ j_{p'} & j_h & j_p \end{Bmatrix} Y^{nJ}_{ph}Y^{n'J'}_{p'h}Q^{\lambda pol}_{pp'}\,, \tag{8.3.b}$$

$$\langle n'J'\,||Q_\lambda||\,nJ\rangle_{(\mathrm{C})} = \sum_{hh'p}(-)^{J'+j_p-j_{h'}}\begin{Bmatrix} J & \lambda & J' \\ j_{h'} & j_p & j_h \end{Bmatrix} X^{nJ}_{ph}X^{n'J'}_{ph'}Q^{\lambda pol}_{hh'}\,, \tag{8.3.c}$$

$$\langle n'J'\,||Q_\lambda||\,nJ\rangle_{(\mathrm{D})} = \sum_{hh'p}(-)^{J+\lambda+j_p-j_{h'}}\begin{Bmatrix} J & \lambda & J' \\ j_{h'} & j_p & j_h \end{Bmatrix} Y^{nJ}_{ph}Y^{n'J'}_{ph'}Q^{\lambda pol}_{h'h}\,, \tag{8.3.d}$$

$$\langle n'J'\,||Q_\lambda||\,nJ\rangle_{(\mathrm{E})} = \sum_{pp'h}(-)^{J'+j_p-j_{p'}}\begin{Bmatrix} J & \lambda & J' \\ j_{p'} & j_p & j_h \end{Bmatrix} \frac{X^{nJ}_{ph}\langle p||V||p', n'J'\rangle Q^{\lambda pol}_{hp'}}{E_{nJ}-E_{n'J'}-\epsilon_{p'h}+i\eta}\,, \tag{8.3.e}$$

$$\langle n'J'\,||Q_\lambda||\,nJ\rangle_{(\mathrm{F})} = \sum_{pp'h}(-)^{J+\lambda+j_p-j_{p'}}\begin{Bmatrix} J & \lambda & J' \\ j_{p'} & j_p & j_h \end{Bmatrix} \frac{Y^{nJ}_{ph}\langle p'||V||p, n'J'\rangle Q^{\lambda pol}_{p'h}}{E_{nJ}-E_{n'J'}+\epsilon_{p'h}-i\eta}\,, \tag{8.3.f}$$

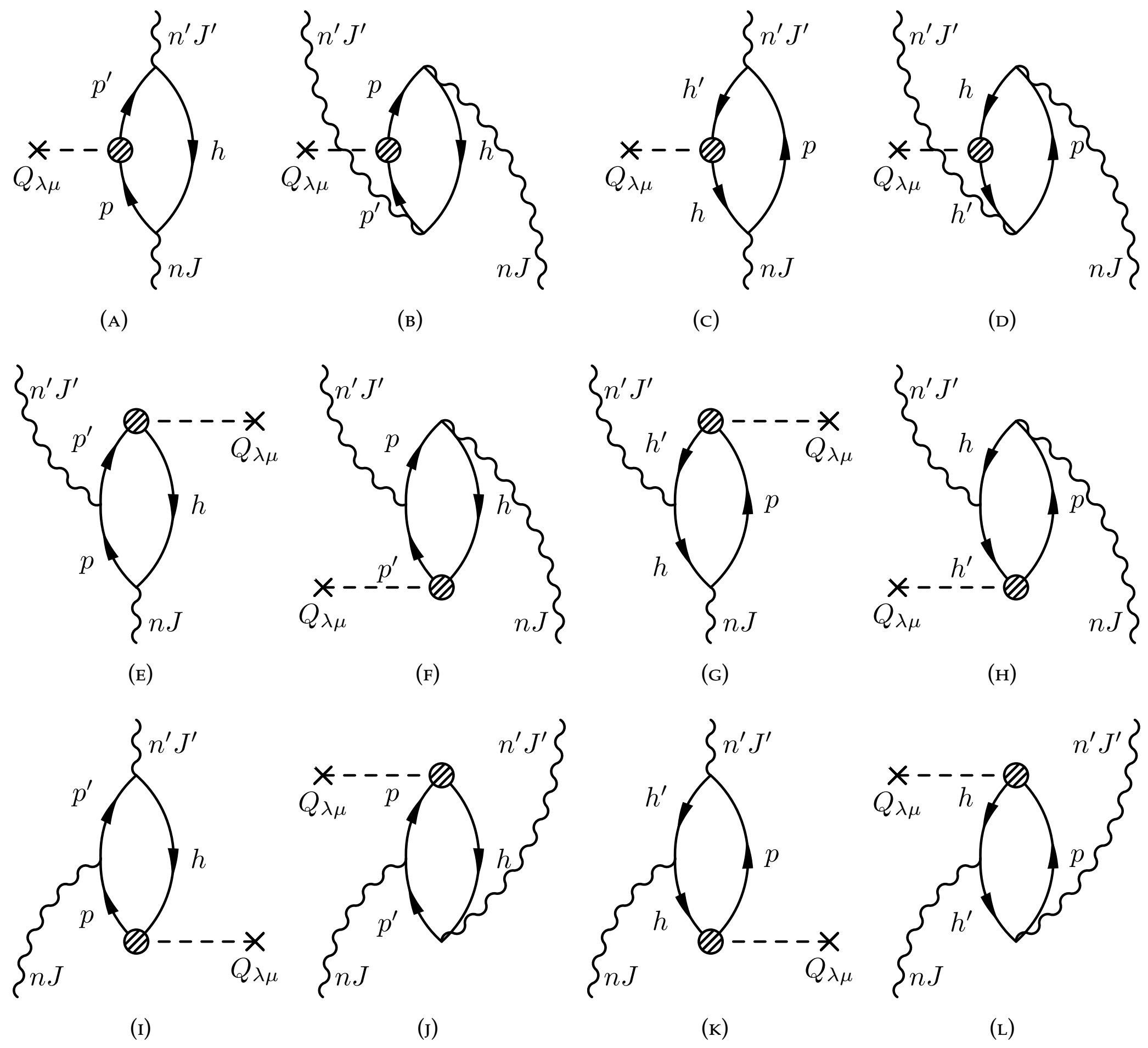

**Figure 8.2**
NFT diagrams contributing to the decay of the $|nJ\rangle$ state to the $|n'J'\rangle$ state.

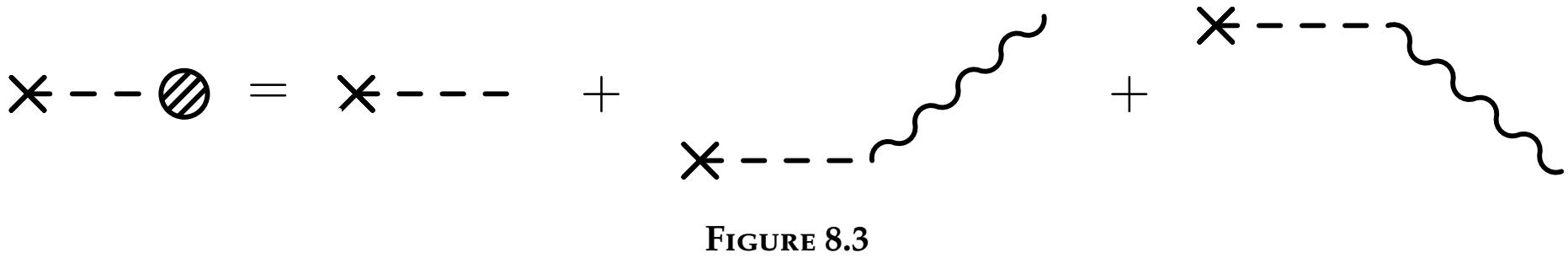

**Figure 8.3**
Polarization contribution to the operator $Q_{\lambda\mu}$

$$\langle n'J' \| Q_\lambda \| nJ\rangle_{(G)} = \sum_{hh'p} (-)^{J+\lambda+j_h+j_{h'}} \begin{Bmatrix} J & \lambda & J' \\ j_{h'} & j_h & j_p \end{Bmatrix} \frac{X^{nJ}_{ph} \langle h' \| V \| h, n'J'\rangle Q^{\lambda pol}_{h'p}}{E_{nJ} - E_{n'J'} - \epsilon_{ph'} + i\eta}, \quad (8.3.G)$$

$$\langle n'J' \| Q_\lambda \| nJ\rangle_{(H)} = \sum_{hh'p} (-)^{J'+j_h+j_{h'}} \begin{Bmatrix} J & \lambda & J' \\ j_{h'} & j_h & j_p \end{Bmatrix} \frac{Y^{nJ}_{ph} \langle h \| V \| h', n'J'\rangle Q^{\lambda pol}_{ph'}}{E_{nJ} - E_{n'J'} + \epsilon_{ph'} - i\eta}, \quad (8.3.H)$$

$$\langle n'J' \| Q_\lambda \| nJ\rangle_{(I)} = \sum_{pp'h} (-)^{J'+j_h+j_{p'}} \begin{Bmatrix} J & \lambda & J' \\ j_h & j_{p'} & j_p \end{Bmatrix} \frac{X^{n'J'}_{p'h} \langle p' \| V \| p, nJ\rangle Q^{\lambda pol}_{ph}}{E_{nJ} - E_{n'J'} + \epsilon_{ph} + i\eta}, \quad (8.3.I)$$

$$\langle n'J' \| Q_\lambda \| nJ\rangle_{(J)} = \sum_{pp'h} (-)^{J+\lambda+j_{p'}+j_h} \begin{Bmatrix} J & \lambda & J' \\ j_h & j_{p'} & j_p \end{Bmatrix} \frac{Y^{n'J'}_{p'h} \langle p \| V \| p', nJ\rangle Q^{\lambda pol}_{hp}}{E_{nJ} - E_{n'J'} - \epsilon_{ph} + i\eta}, \quad (8.3.J)$$

$$\langle n'J' \| Q_\lambda \| nJ\rangle_{(K)} = \sum_{hh'p} (-)^{J+\lambda+j_{p'}-j_h} \begin{Bmatrix} J & \lambda & J' \\ j_p & j_{h'} & j_h \end{Bmatrix} \frac{X^{n'J'}_{ph'} \langle h \| V \| h', nJ\rangle Q^{\lambda pol}_{ph}}{E_{nJ} - E_{n'J'} + \epsilon_{ph} + i\eta}, \quad (8.3.K)$$

$$\langle n'J' \| Q_\lambda \| nJ\rangle_{(L)} = \sum_{hh'p} (-)^{J'+j_p-j_{h'}} \begin{Bmatrix} J & \lambda & J' \\ j_p & j_{h'} & j_h \end{Bmatrix} \frac{Y^{n'J'}_{ph'} \langle h' \| V \| h, nJ\rangle Q^{\lambda pol}_{hp}}{E_{nJ} - E_{n'J'} - \epsilon_{ph} + i\eta}. \quad (8.3.L)$$

In these equations $\epsilon_{ph}$ is equal to the difference of the Hartree-Fock (HF) sp energies $\epsilon_p - \epsilon_h$. In all the energy denominators we include finite imaginary parts $\eta$ to take into account the coupling to more complicated configurations not included in the model space, as in the case of the spreading width of GRs.

In all the above equations, the matrix elements of the operator $Q_\lambda$ include the contribution from the nuclear polarization (consequently they carry the label $pol$). They read

$$Q^{\lambda pol}_{ij} = \langle i \| Q_\lambda \| j\rangle + \sum_{n''} \frac{1}{\sqrt{2\lambda+1}} \left[ \frac{\langle 0 \| Q_\lambda \| n''\lambda\rangle \langle i, n''\lambda \| V \| j\rangle}{(E_{nJ} - E_{n'J'}) - E_{n''\lambda} + i\eta} - \frac{\langle i \| V \| j, n''\lambda\rangle \langle n''\lambda \| Q_\lambda \| 0\rangle}{(E_{nJ} - E_{n'J'}) + E_{n''\lambda} + i\eta} \right], \quad (8.4)$$

where $|n''\lambda\rangle$ are the RPA states having multipolarity $\lambda$ (and lying at energy $E_{n''\lambda}$), while the bare operator $Q_\lambda$ has been defined in Eq. 3.21. The polarization contribution, that is, the second and third term in the latter equation, has the effect of screening partially the external field. In a diagrammatic way, the bare and the polarization contributions to Eq. (8.4)) are drawn in Fig. 8.3.

It should be noted that the diagrams of Fig. 8.2 are related two by two by particle-hole conjugation, so that Fig. 8.2(A) is the opposite of Fig. 8.2(D) after the substitutions $h' \to p'$ and $h \leftrightarrow p$, and the same holds for the pairs in Figs. 8.2(B) through 8.2(C), 8.2(E) through 8.2(H), 8.2(F) through 8.2(G), 8.2(I) through 8.2(L) and 8.2(J) through 8.2(K).

Chapter 9

# Numerical results

In this chapter we present the results obtained for the spreading width and the $\gamma$-decay width of GRs. The former is computed for the ISGQR and IVGQR in $^{208}$Pb, while the latter is computed for the $\gamma$ decay of the ISGQR in $^{208}$Pb and $^{90}$Zr to the GS and to the first collective octupole state. The interactions used in the calculations will be detailed in the following.

The HF equations are solved in a radial mesh that extends up to 20 fm for $^{208}$Pb and 18 fm for $^{90}$Zr, with a radial step of 0.1 fm in both cases. Once the HF solution is found, the RPA equations are solved in the usual matrix formulation. More information on our RPA implementation can be found in Ref. [Col+13]. The RPA model space consists of all the occupied states and all the unoccupied states lying below a cutoff energy $E_p^{\max}$ equal to 50 and 70 MeV for the ISGQR and IVGQR in $^{208}$Pb, respectively, while for the ISGQR in $^{90}$Zr the maximum particle energy is 40 MeV. The states at positive energy are obtained by setting the system in a box, that is, the continuum is discretized. These states have increasing values of the radial quantum number $n$, and are calculated for those values of $l$ and $j$ that are allowed by the selection rules. With this choice of the model space the energy-weighted sum rules (EWSRs) satisfy the double commutator values at the level of about 99 %; moreover, the energy and the fraction of EWSR of the states that are relevant for the following discussion are well converged.

In all the calculations, collective states up to 30 MeV are used as intermediate states and a lower cutoff of 5 % on the EWSR of the intermediate states is introduced. We need a lower cutoff on the collectivity of the intermediate states for at least two reasons: first, RPA is known to be not reliable for non-collective states, and second, introducing them will oblige us to take into account the issue of the Pauli principle correction. Although there is not a general recipe to choose the lower bound of the isovector and isoscalar EWSR, we take 5 % in keeping with several previous works (e.g., Ref. [CSB10]).

**Table 9.1**
Spreading width and energy centroid of the ISGQR and IVGQR as produced by the model explained in the main text. The weighted averages of experimental results and listed in Table 6.3 are also included for comparison.

| | $E$ [MeV] | $E_{\text{exp}}$ [MeV] | $\Gamma^{\downarrow}$ [MeV] | $\Gamma^{\downarrow}_{\text{exp}}$ [MeV] |
|---|---|---|---|---|
| ISGQR | 11.3 | 10.9 ± 0.1 | 2.3 | 3.0 ± 0.1 |
| IVGQR | 22.0 | 22.7 ± 0.2 | 4.0 | 4.8 ± 0.3 |

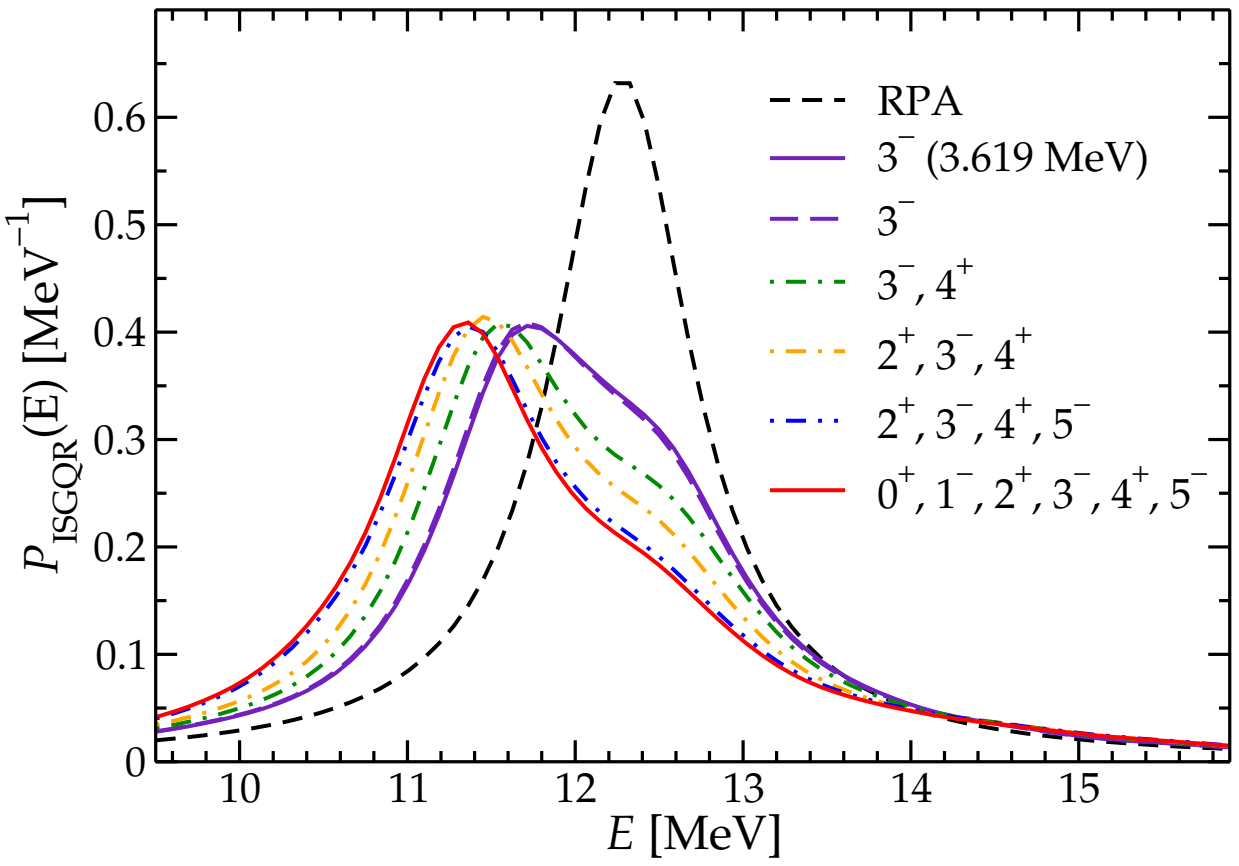

**Figure 9.1**
Probability $P$ per unit energy to find the ISGQR at an energy $E$ in $^{208}$Pb. Each line corresponds to the probability obtained when the phonons listed in the legend are used as intermediate states. The label RPA (black dashed line) refers to the RPA result, in which none of the diagrams in Fig. 8.1 are taken into account, but a Lorentzian averaging with functions having 1 MeV width is introduced. The interaction used is SLy5. The violet solid line is the probability we get when only the low-lying $3^-$ phonon at 3.61 MeV is considered as intermediate state.

## 9.1 The lineshape functions of giant resonances

As already mentioned, we have computed the contributions of section 8.1 in the case of the isoscalar and isovector GQR in $^{208}$Pb. The results has been published in Refs. [BCB12; Roc+13]. The interaction used are SLy5 and SAMi for the ISGQR and IVGQR, respectively. The results for the probability of finding the GRs, calculated by including in the diagrams an increasing number of intermediate phonons, are collected in Fig. 9.1 - 9.2. In our calculation we set the imaginary part of the denominator in Eq. (8.2) at 1 MeV.

In Table 9.1 we summarized the results obtained for the ISGQR and IVGQR. In both cases, phonons with multipolarity $\lambda$ ranging from 0 to 5 and with natural parity $(-1)^\lambda$ are considered.

For the ISGQR, the most important contribution to the spreading width $\Gamma^\downarrow$ of the resonance is given by the low-lying $3^-$ state, while the other phonons do contribute basically only to the energy shift. We obtain eventually a spreading width of the order of 2.3 MeV and the energy centroid of the resonance is shifted down, as compared to the RPA value, to 11.3 MeV. These results are in good agreement with the experimental findings that give a spreading width of 3.0 ± 0.1 MeV (see Table 9.1).

In the IVGQR case, the single RPA state splits into two components: the peak at higher energy is barely affected by increasing the number of intermediate phonons, whereas the one at lower energy broadens and is shifted downwards as the number of phonons increases. The energy centroid is shifted by 1 MeV, from 23 MeV for the RPA to 22 MeV. The spreading width is 4.0 MeV. Both results are in reasonable agreement with the experimental results listed in Table 9.1.

## 9.2 The $\gamma$ decay of GRs

In this section we will present the results obtained for the $\gamma$ decay of giant resonances to the ground state (GS) as well as to low-lying states of the system. In particular, we will focus on the decay of the ISGQR in $^{208}$Pb and $^{90}$Zr to the first $3^-$ collective state. For the first nucleus, there exists an experimental result, given in Ref. [Bee+89]. The results shown in this section have been the subject of Ref. [BCB12].

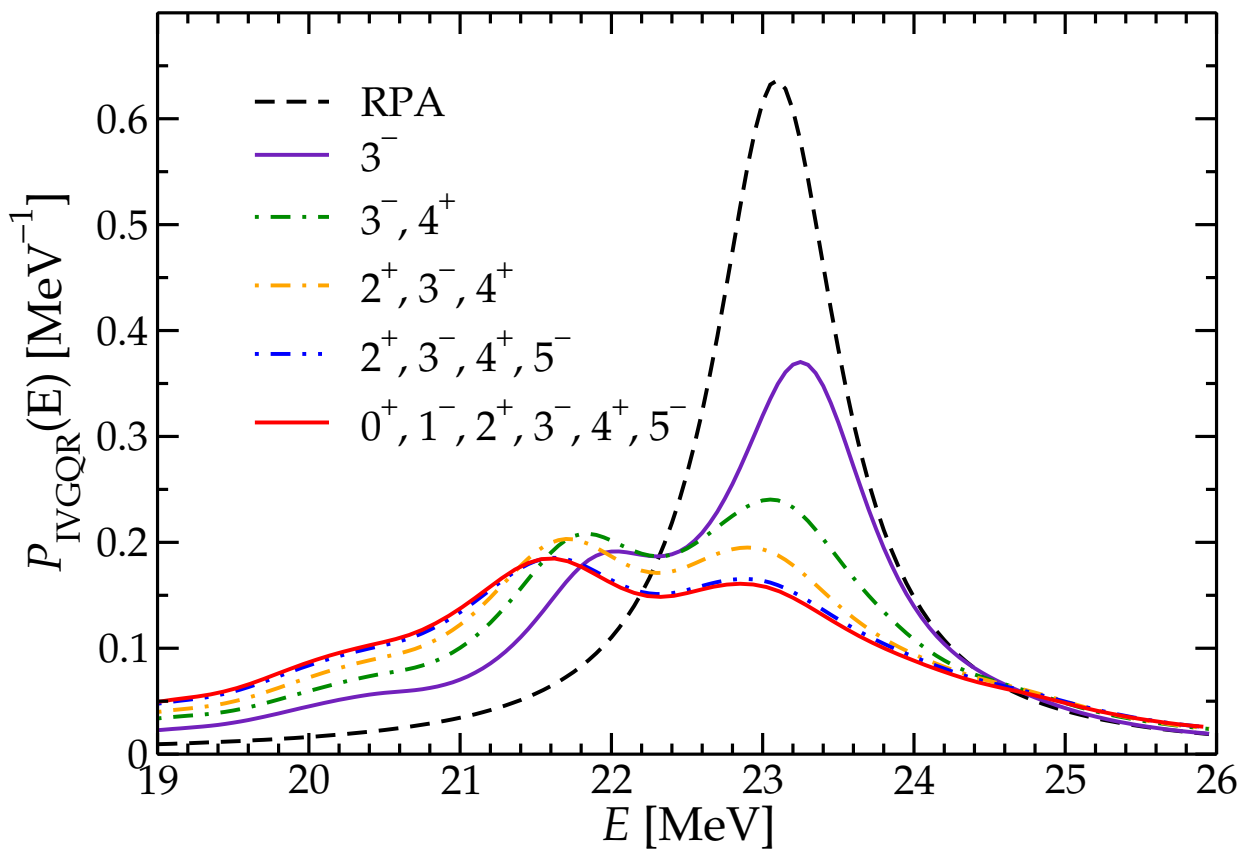

**FIGURE 9.2**
The same as Fig. 9.1 for the IVGQR. The interaction used is SAMi.

We use different Skyrme parameterization, namely SLy5, SGII, SkP and LNS. We use these particular sets for the following reasons:

**SLy5** It is a modern interaction which provides successful results for many observables. It was fitted also to the equation of state of neutron matter, so it is suitable for the description of neutron-rich nuclei.

**SGII** It is a standard Skyrme interaction which has been used for decades with reliable results.

**SkP** It has the peculiarity that the effective mass of the nucleons at saturation is equal to the bare mass, at saturation density. This fact has an effect on the energy of the energy of the ISGQR, as discussed in Section 5.2.2.

**LNS** Even though in general it does not reproduce the experimental findings at the same level of accuracy of other parameterizations, we use it because it is fitted to the equation of state of nuclear matter at Brueckner-Hartree-Fock level.

### 9.2.1 The $\gamma$ decay to the ground state

Even if the decay to the ground state is a mean-field observable, as it can be treated at RPA level, we put it in this section in order to make a comparison with the decay to low-lying states. In Table 9.2, we group the results obtained for the decay to the ground state. For completeness, in the same table the previous theoretical values found in literature [BBB84; Bee+85; Spe+85] are listed as well. In Ref. [BBB84], the PVC with a separable force [BB81] is used to evaluate the reduced transition probability and the decay width. In Ref. [Spe+85], the theory of finite Fermi systems [SWW77] is implemented with a separable interaction to obtain the decay width. In Ref. [Bee+85], eventually, the value is estimated from the empirical energies and fraction of EWSR.

In general, our calculations reproduce the experiment quite well, without any parameter adjustment. They tend at the same time to overestimate the decay width: this is true in particular for SLy5. However, even in this worst case, our result lies within 2.5$\sigma$ from the experimental value. This discrepancy is basically due to the fact that the energy of the GR do not fit accurately the experimental findings. As a matter of fact, in Eq. (3.51) the energy of the transition is raised to the fifth power: consequently, an increase of the energy by 1 MeV produces an increase of the $\gamma$-decay width by about 50 % (at 10 MeV). To substantiate this point, in the fourth column of Table 9.2 we report the values obtained in $^{208}$Pb for the decay width after having rescaled the ISGQR energy to the experimental value. The energies of the resonances can be found in Table 9.3. We are not giving a detailed description the strength functions and of the transition densities of the resonances, since it would be beyond the scope of this work.

**Table 9.2**
Energy $E$ of the ISGQR and $\gamma$-decay width associated with its transition to the ground-state. The first four rows correspond to the present RPA calculations performed with different Skyrme parameter sets, for the two nuclei at hand. In this case, for $^{208}$Pb we show both the bare $\Gamma_\gamma$ from Eq. (3.51) as well as the renormalized value which is discussed in the main text. The next three rows report the results of previous theoretical calculations [BBB84; Bee+85; Spe+85] for $^{208}$Pb. In the last row the experimental value for $^{208}$Pb from Ref. [Bee+89] is displayed.

| | $^{208}$Pb | | | $^{90}$Zr | |
|---|---|---|---|---|---|
| | $E$ [MeV] | $\Gamma_\gamma$ [eV] | $\Gamma_\gamma^{ren}$ [eV] | $E$ [MeV] | $\Gamma_\gamma$ [eV] |
| SLy5 | 12.28 | 231.54 | 127.58 | 15.33 | 211.77 |
| SGII | 11.72 | 163.22 | 113.57 | 14.90 | 182.03 |
| SkP | 10.28 | 119.18 | 159.72 | 13.09 | 107.27 |
| LNS | 12.10 | 176.57 | 104.74 | 15.48 | 182.71 |
| Ref. [Bee+85] | 11.20 | 175 | | | – |
| Ref. [BBB84] | 11.20 | 142 | | | – |
| Ref. [Spe+85] | 10.60 | 112 | | | – |
| Ref. [Bee+89] | 10.60 | 130±40 | | | – |

**Table 9.3**
Main properties of the ISGQR in $^{208}$Pb (second and third columns) and in $^{90}$Zr (fourth and fifth columns). The label Exp. indicates the corresponding experimental values (the italic number after the value is the experimental error on the last significant figure), these values are from Refs. [Mar07] and [Bro97].

| | $^{208}$Pb | | $^{90}$Zr | |
|---|---|---|---|---|
| | $E$ [MeV] | EWSR [%] | $E$ [MeV ] | EWSR [%] |
| SLy5 | 12.28 | 69.27 | 15.33 | 80.50 |
| SGII | 11.72 | 72.31 | 14.90 | 81.04 |
| SkP | 10.28 | 81.79 | 13.09 | 81.05 |
| LNS | 12.10 | 66.98 | 15.48 | 75.48 |
| Exp. | 10.9 *0.1* | 90 *20* | 14.5 *0.3* | 54 *15* |

We can conclude that, since for all the interactions the experimental value of the ground-state decay width can be obtained simply by scaling the energy to the experimental value, it means that this kind of measurement is not particularly able to discriminate between models more than usual integral properties.

### 9.2.2 The $\gamma$ decay to low-lying states

As mentioned above, the model has been applied to the decay of the ISGQR to the low-lying octupole state in $^{208}$Pb and in $^{90}$Zr. Table 9.4 collects our results in these two cases, together with values of previous theoretical estimation [BBB84; Spe+85] for $^{208}$Pb and the experimental result from Ref. [Bee+89] in the same nucleus. The previous theoretical predictions are coming from different models: in Ref. [Spe+85], the theory of finite Fermi systems [SWW77] with a phenomenological interaction is used to calculate the decay width, while in Ref. [BBB84] the decay width is obtained by means of the NFT, with a separable interaction at the PVC vertex. On the other hand, the experimental findings confirm that the decay to low-lying states is a really touchy observable and the only indication that it gives us is that the $\Gamma_\gamma(\text{ISGQR} \to 3^-)$ is only a few percent of $\Gamma_\gamma(\text{ISGQR} \to \text{g.s.})$. As can be seen from Table 9.4, qualitatively all the interactions are able to reproduce this feature. However, only two interactions, viz. SLy5 and SkP, can reasonably reproduce the experimental value for the decay width.

In principle, all the multipolarities allowed by angular momentum selection rules for the electromagnetic transitions should be considered. In practical calculations, however, we consider only $E1$ transitions since the higher multipolarities are quenched (see Eq. (3.51))

The value for the imaginary part $\eta$ in the polarization is 1 MeV for $^{208}$Pb and 1.75 MeV for $^{90}$Zr. The reasons for this choices will be clarified in the following.

In order to better understand the role of the several factors in Eqs. (8.3), we analyzed the physical inputs in great detail in the case of $^{208}$Pb. In Table 9.5, we collect, for the four interaction used, the contribution of the factors included in Eqs. (8.3). Similar factors from Ref. [BBB84] are provided as well. The decay width that is obtained considering a typical particle-hole transition is of the order of $\approx$ keV, as it can be qualitatively estimated by means of the Weisskopf reduced transition probability [BM69]. The label *recoupling coefficient* indicates the quenching deriving from the angular momentum coupling coefficients. Then, because of the isovector nature of the operator (3.21), diagrams involving protons and neutrons have opposite sign and partially cancel each other (the corresponding factor is called *$\pi$-$\nu$ cancellation*). Moreover, diagrams in which the operator acts on a particle line must have an opposite sign to the ones in which it acts on a hole line, reflecting the correlations between particles and holes in vibrations [BBB98]. Accordingly, the label associated with this contribution is *p-h cancellation*. Eventually, the polarization contribution (8.4), deriving from the screening of the external field by the mediation of the giant dipole resonance, represent a further and more important quenching of the original decay width, giving then a final width of the order of electronvolts.

**Table 9.4**
Decay width to the low-lying $3^-$ for the interactions used, calculated including beyond RPA contributions for the two nuclei $^{208}$Pb and $^{90}$Zr. In particular, for $^{208}$Pb the results from Ref. [Spe+85] and Ref. [BBB84] are also listed and in the last row, and the experimental value from Ref. [Bee+89] is provided as well.

| | $^{208}$Pb | | $^{90}$Zr | |
|---|---|---|---|---|
| Interaction | $E_{\text{trans}}$ [MeV] | $\Gamma_\gamma$ [eV] | $E_{\text{trans}}$ [MeV] | $\Gamma_\gamma$ [eV] |
| SLy5 | 8.66 | 3.39 | 12.51 | 5.81 |
| SGII | 8.58 | 29.18 | 12.16 | 50.58 |
| SkP | 6.99 | 8.34 | 10.42 | 5.14 |
| LNS | 8.90 | 39.87 | 12.72 | 16.95 |
| Ref. [BBB84] | 8.59 | 3.5 | | – |
| Ref. [Spe+85] | 7.99 | 4 | | – |
| Ref. [Bee+89] | 7.99 | 5±5 | | – |

**Table 9.5**
The various quenching factors that combine to produce the decay width $\Gamma_\gamma$ from a typical particle-hole dipole transition, for $^{208}$Pb. The decay width reported here refers to a cutoff of 5 % on the percentage of isovector EWSR for the dipole states. The same quantities from [BBB84] are displayed.

| | SLy5 | SGII | SkP | LNS | Ref. [BBB84] |
|---|---|---|---|---|---|
| $p-h$ transition [eV] | $10^3$ | $10^3$ | $10^3$ | $10^3$ | $10^3$ |
| Recoupling coefficient | 3 | 3 | 3 | 3 | 3 |
| $\pi-\nu$ cancellation | 5 | 4 | 3–4 | 4 | 4 |
| $p-h$ cancellation | 3–4 | 2–3 | 2–3 | 3–4 | 2–3 |
| Polarization | 6 | 3 | 7–8 | 4 | 15 |
| $\Gamma_\gamma$ [eV] | 3.39 | 29.18 | 8.34 | 39.87 | 3.50 |

**Table 9.6**
Same as Table 9.5 but for $^{90}$Zr.

| | SLy5 | SGII | SkP | LNS |
|---|---|---|---|---|
| $p-h$ transition [eV] | $10^3$ | $10^3$ | $10^3$ | $10^3$ |
| Recoupling coefficient | 3 | 3 | 3 | 3 |
| $\pi-\nu$ cancellation | 4–5 | 3–4 | 2 | 10 |
| $p-h$ cancellation | 4 | 3–4 | 3–4 | 4 |
| Polarization | 10 | 1.5 | 18 | 1 |
| $\Gamma_\gamma$ [eV] | 5.81 | 50.58 | 5.14 | 16.95 |

**Table 9.7**
The effect on the $\gamma$-decay width $\Gamma_\gamma$ of the width $\Gamma_D$ of the intermediate dipole states. All the states exhausting the EWSR for more than 5 % are considered. The decay width is an almost monotonically non-decreasing function of this parameter, as expected from the Bohr-Mottelson model [BM75].

| | $\Gamma_D$ [MeV] | 0.01 | 0.1 | 0.5 | 1.0 | 2.0 | 3.0 | 4.0 | 5.0 |
|---|---|---|---|---|---|---|---|---|---|
| $\Gamma_\gamma$ [eV] | SLy5 | 2.34 | 2.35 | 2.46 | 2.69 | 3.39 | 4.32 | 5.34 | 6.36 |
| | SGII | 27.28 | 27.19 | 27.00 | 27.25 | 29.18 | 32.68 | 37.21 | 42.25 |
| | SkP | 7.35 | 7.35 | 7.40 | 7.59 | 8.34 | 9.54 | 11.08 | 12.86 |
| | LNS | 37.93 | 37.82 | 37.55 | 37.76 | 39.87 | 43.84 | 49.03 | 54.86 |

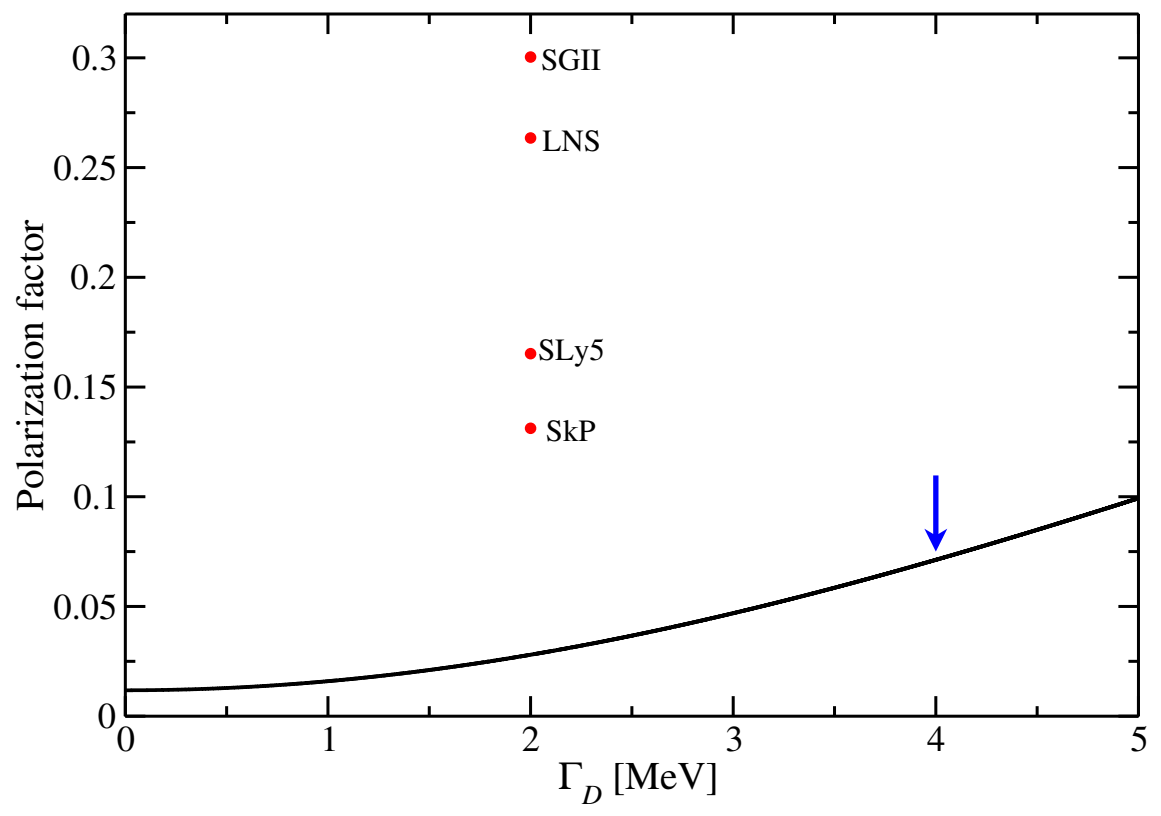

**Figure 9.3**
Polarization contribution in the separable framework as a function of the parameter $\Gamma_D$ (solid line). The arrow indicates the value used in Ref. [BBB84]. The points are the analogous factors obtained within our model in which all the dipole states having a fraction of EWSR larger than 5 % are considered and each one is given a width $\Gamma_D = 2\,\text{MeV}$.

A similar analysis carried out on $^{90}$Zr would produce similar results: the polarization of the nuclear medium plays a crucial role.

Being the polarization the major quenching effect, we studied which parameters can affect this contribution. In Table 9.7, the variation of the $\gamma$ decay width $\Gamma_\gamma$ with the parameter $\eta = \frac{\Gamma_D}{2}$ that appears in Eq. (8.4) as an imaginary part of the energy denominator, is discussed. If only a single dipole intermediate state is considered, as in Ref. [BBB84], this parameter should be set equal to the IVGDR width ($\sim 4\,\text{MeV}$); since in our model, the dipole strength is fragmented, we should take a smaller value and we give here the trend of the decay width as a function of this parameter. As indicated by the plot in Fig. 9.3, the polarization factor (and consequently the decay width) should be monotonically non-decreasing when $\Gamma_D$ increases and reaches a roughly constant value as $\Gamma_D$ goes to zero. In the same plot, the points represent the polarization factors that we obtain using the value 2 MeV for the parameter $\Gamma_D$, but including all the dipole states having a fraction of EWSR larger than 5 %. This value has been chosen in order give a width of the RPA dipole states, each convoluted with a Lorentzian of width equal to $\Gamma_D$, similar to the experimental IVGDR width. The polarization that we get is then consistent with the one of Bohr-Mottelson model [BM75], indicated with the arrow in Fig. 9.3.

# Part III

# Beyond mean-field theories and divergences

Chapter 10

# Divergences in beyond mean-field models

We have seen in part II that the introduction of beyond mean-field correlations is mandatory to describe some important observables. The main limitation of our approach concerns the use of an effective interaction, rather than a bare nucleon-nucleon force. As a matter of fact, the effective interactions are originally designed for SCMF calculations: they contain some parameters which are fitted at mean-field level to reproduce some experimental results. Therefore, these parameters take into account also beyond mean-field correlations in an effective and uncontrolled way. The explicit computation of beyond mean-field correlations with these effective interaction may produce some double-counting. For this reason, the refitting of the effective Hamiltonian should be envisaged. Regarding this issue, a possible alternative solution to the refit of the interaction has been proposed by V. I. Tselyaev in Ref. [Tse07]: in order to avoid double-counting in the calculation of the beyond mean-field p-h self-energy, he states that it is enough to subtract from the self-energy the contribution coming from the self-energy evaluated at zero energy. This subtraction should remove the "static" self-energy, which is the one effectively incorporated in the effective interaction. We should point out that this recipe, although reasonable, do not have any grounds in the underlying many-body theory.

Moreover, if zero-range effective interactions are used in beyond mean-field theories, divergences arise. We should note that this problem affects not only the Skyrme interaction which, as it was recalled in chapter 1, is entirely built with delta functions and is the one interesting for this work, but also the Gogny interaction, in which the density dependent term is zero-range, and point coupling Lagrangians used in relativistic mean-field models. The presence of a divergence can be understood with a simple consideration: if the interaction is of contact type in the coordinate space, it is a constant in the momentum domain (connected to the coordinate one by the Fourier transform); as a consequence it would allow the transfer of arbitrarily high momentum (or energy, which is the same for this qualitative argument). This is of course unphysical. Usually, beyond mean-field calculations are done by truncating the model space in some arbitrary way and this is by far an unsatisfactory procedure.

The issue of divergences has been attacked successfully in Refs. [Mog+10; Mog+12a]. In these works, the total energy up to the second order in perturbation theory has been considered for nuclear matter with different level of proton-neutron asymmetry with the Skyrme interaction. In order to avoid the divergences, a cutoff $\Lambda$ on the transferred momentum between particles is introduced so that all the divergent integrals on intermediate states become finite. In other words, the states with momentum greater than the cutoff $\Lambda$ are discarded from the theory. In order to take into account the rejected states, for each value of the cutoff $\Lambda$ the parameters of the interaction are to be re-fitted to a reasonable equation of state, chosen to be the mean-field EoS for the given Skyrme interaction (see section 10.1).

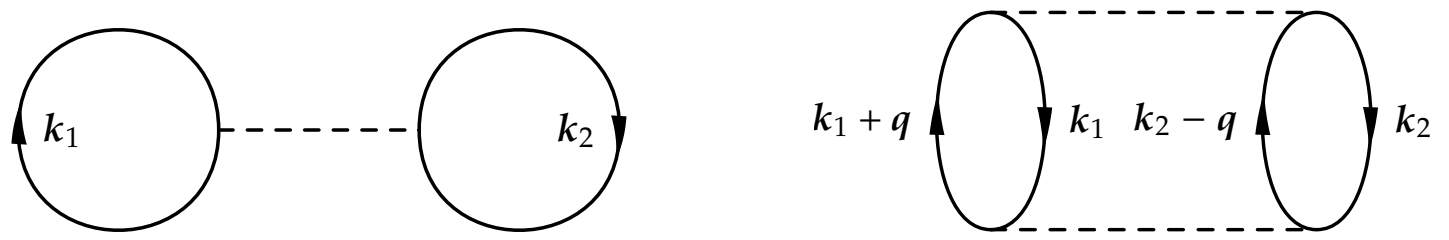

**Figure 10.1**
Direct contributions to the mean-field total energy (A) and second-order total energy (B). The exchange contributions are not drawn since they are proportional to the corresponding direct ones.

In this part of the thesis, we would like to investigate whether it is possible to use the interaction fitted in nuclear matter to perform calculation in nuclei, just performing a minor adjustment the parameters. In our explorative approach, we are going to use a simplified Skyrme interaction, in which only the $t_0$, $t_3$ and $\alpha$ parameters are non-zero, and we are considering only even-even isospin-symmetric nuclei. Since the finite system is not translationally invariant, the transferred momentum is not a well-defined quantum number; therefore we cannot use a cutoff on this quantity. In order to make a connection between infinite matter and nuclei, inspired by Ref. [CTvB13] we want to use the relative momentum between the nucleons, defined as the Fourier conjugate variable of the relative coordinate. In this respect, we need two cutoffs, for the initial two-particle state and for the final one. Note that this is in agreement with the idea that the Skyrme interaction is a low-relative-momentum polynomial expansion [Sky59a]. In order to investigate the relation between the cutoff on the transferred momentum and the ones on the relative momenta, the total energy of symmetric nuclear matter has to be computed in the center of mass and relative motion coordinate system (see section 10.2).

While in nuclear matter the implementation of the cutoffs on the relative momenta in the matrix elements of the interaction is trivial, being the wave functions plane waves, in a finite system it requires some intermediate steps:

- a manipulation of the form of the interaction, which consequently acquires a finite range (see section 10.3);
- the transformation of the two-particle states needed to compute the matrix elements of the interaction to the center of mass and relative motion coordinate system. This procedure can be performed analytically only it the sp wave functions are written on a harmonic oscillator basis, by exploiting the Brody-Moshinsky transformations [Law80; BM96; Kam+01] (see section 10.4).

## 10.1 Divergences in the nuclear matter EoS

The problem of divergences has been attacked in Refs. [Mog+10; Mog+12a]. In these works, the EoS of nuclear matter with different level of proton-neutron asymmetry has been considered with the Skyrme interaction. Nuclear matter has been chosen because the nucleon wave functions are plane-waves and the EoS can be written analytically as a function of the parameters of the Skyrme interaction. In Ref. [Mog+10] the SkP interaction was used: this is because for the SkP interaction, being the effective mass equal to the bare mass, the EoS is written in terms of only the $t_0$, $t_3$ and $\alpha$ parameters. The energy per particle of the system is computed up to second-order (see Fig. 10.1) and a cutoff $\Lambda$ is included in the transferred momentum $\boldsymbol{q}$. The three interaction parameters are fitted for each value of the cutoff $\Lambda$ in such a way that the resulting energy per particle equals the mean-field EoS. In Ref. [Mog+12a] the same procedure was applied to the EoS obtained with the SLy5 interaction, considering also the momentum dependent part of the Skyrme force.

**Table 10.1**
Parameter sets (named $\mathrm{SkP}_\Lambda$) obtained in the fits associated with different values of the cutoff $\Lambda$ compared with the original set SkP, labelled with SkP (first line) [Mog].

| | $t_0$ | $t_3$ | $\alpha$ | | $t_0$ | $t_3$ | $\alpha$ |
|---|---|---|---|---|---|---|---|
| SkP | −2931.70 | 18709.00 | 1/6 | | | | |
| $\mathrm{SkP}_{0.1}$ | −2937.45 | 18758.12 | 0.16674 | $\mathrm{SkP}_{1.9}$ | −649.68 | 7431.97 | 1.13340 |
| $\mathrm{SkP}_{0.2}$ | −2931.54 | 18723.70 | 0.16713 | $\mathrm{SkP}_{2.0}$ | −618.70 | 7062.93 | 1.16305 |
| $\mathrm{SkP}_{0.3}$ | −2906.45 | 18577.75 | 0.16881 | $\mathrm{SkP}_{2.1}$ | −593.41 | 6596.73 | 1.16744 |
| $\mathrm{SkP}_{0.4}$ | −2842.25 | 18204.63 | 0.17328 | $\mathrm{SkP}_{2.2}$ | −573.43 | 6052.99 | 1.14457 |
| $\mathrm{SkP}_{0.5}$ | −2719.66 | 17494.17 | 0.18249 | $\mathrm{SkP}_{2.3}$ | −558.79 | 5469.05 | 1.09369 |
| $\mathrm{SkP}_{0.6}$ | −2531.08 | 16406.95 | 0.19873 | $\mathrm{SkP}_{2.4}$ | −549.99 | 4892.54 | 1.01547 |
| $\mathrm{SkP}_{0.7}$ | −2288.58 | 15022.28 | 0.22432 | $\mathrm{SkP}_{2.5}$ | −548.24 | 4374.67 | 0.91252 |
| $\mathrm{SkP}_{0.8}$ | −2020.60 | 13517.37 | 0.26140 | $\mathrm{SkP}_{3.0}$ | −544.99 | 3624.67 | 0.66267 |
| $\mathrm{SkP}_{0.9}$ | −1758.46 | 12085.78 | 0.31144 | $\mathrm{SkP}_{3.5}$ | −514.79 | 3386.33 | 0.62361 |
| $\mathrm{SkP}_{1.0}$ | −1524.15 | 10862.96 | 0.37503 | $\mathrm{SkP}_{4.0}$ | −489.40 | 3180.44 | 0.59654 |
| $\mathrm{SkP}_{1.1}$ | −1326.93 | 9904.53 | 0.45153 | $\mathrm{SkP}_{5.0}$ | −448.19 | 2858.89 | 0.56329 |
| $\mathrm{SkP}_{1.2}$ | −1166.61 | 9204.84 | 0.53904 | $\mathrm{SkP}_{8.0}$ | −368.24 | 2279.23 | 0.52259 |
| $\mathrm{SkP}_{1.3}$ | −1038.29 | 8724.34 | 0.63454 | $\mathrm{SkP}_{10.0}$ | −334.14 | 2045.30 | 0.51106 |
| $\mathrm{SkP}_{1.4}$ | −935.83 | 8409.46 | 0.73409 | $\mathrm{SkP}_{20.0}$ | −244.47 | 1457.29 | 0.49046 |
| $\mathrm{SkP}_{1.5}$ | −853.56 | 8203.44 | 0.83327 | $\mathrm{SkP}_{40.0}$ | −176.94 | 1035.19 | 0.48119 |
| $\mathrm{SkP}_{1.6}$ | −786.87 | 8050.45 | 0.92736 | $\mathrm{SkP}_{60.0}$ | −145.95 | 846.69 | 0.47822 |
| $\mathrm{SkP}_{1.7}$ | −732.24 | 7899.09 | 1.01172 | $\mathrm{SkP}_{80.0}$ | −127.16 | 733.92 | 0.47675 |
| $\mathrm{SkP}_{1.8}$ | −687.11 | 7704.42 | 1.08184 | $\mathrm{SkP}_{100.0}$ | −114.20 | 656.81 | 0.47589 |

Since in the following we will use a simplified Skyrme interaction with only the $t_0$, $t_3$ and $\alpha$ parameters, we now focus on the work done in nuclear matter with the SkP interaction in symmetric nuclear matter. In this case the interaction can be written as

$$v(\boldsymbol{r}_1, \boldsymbol{r}_2) = \left[t_0 + \frac{1}{6} t_3 \rho\left(\frac{\boldsymbol{r}_1 + \boldsymbol{r}_2}{2}\right)\right] \delta_3\left(\boldsymbol{r}_1 - \boldsymbol{r}_2\right) = g(\boldsymbol{r}_1, \boldsymbol{r}_2)\, \delta_3\left(\boldsymbol{r}_1 - \boldsymbol{r}_2\right), \tag{10.1}$$

and it has constant matrix elements $v = \frac{g}{\Omega}$ (being $\Omega$ the quantization box) in the momentum space. In this simple case, the divergence can be clearly seen with a simple power counting argument, since the second order contribution to the energy reads

$$\Delta E = -\frac{6m\Omega^3 v^2}{(2\pi)^9 \hbar^2} \int_{k_1, k_2 < k_\mathrm{F}, |\boldsymbol{k}_1 + \boldsymbol{q}|, |\boldsymbol{k}_2 - \boldsymbol{q}| > k_\mathrm{F}} \frac{\mathrm{d}_3 q\, \mathrm{d}^3 k_1 \mathrm{d}^3 k_2}{q^2 + \boldsymbol{q} \cdot (\boldsymbol{k}_1 - \boldsymbol{k}_2)} \sim \int \frac{\mathrm{d}_3 q}{q^2}. \tag{10.2}$$

The divergence has been cured with the introduction of the cutoff $\Lambda$ as the upper limit on the integral on $q$. It can also be shown that this divergence is linear in $\Lambda$, if $\Lambda$ is large enough [Mog+12b].

The refitting procedure has been successful in producing for each cutoff a new set of parameters. We list in Table 10.1 the new parameters for different cutoffs.

## 10.2 The nuclear matter EoS revised

The use of the momentum transfer as a (continuous) quantum number is motivated by the fact that nuclear matter is translationally invariant. As already recalled, however, this quantity cannot be used in finite systems, which is not invariant under space translations. Instead, we want to use the relative momentum between the nucleons, defined as the Fourier conjugate variable of the relative coordinate. Therefore, we should assess the relation between introducing a cutoff on the transferred momentum versus one on the relative momentum. To do that, we compute here the mean-field and second order contribution to the total energy in symmetric nuclear matter writing the equation in the center of mass and relative motion coordinate system. At variance with the work done in [Mog+10], since a relative momentum can be defined also at mean-field level, we have to include a cutoff in the Hartree-Fock total energy.

The Hartree-Fock potential energy contribution to the energy per particle is

$$\begin{aligned}\frac{E}{A} &= \frac{d\Omega^2}{(2\pi)^6}\frac{1}{\rho\Omega}\int_{k_1,k_2<k_F} \mathrm{d}_3k_1\mathrm{d}_3k_2\, v(\boldsymbol{k}_1,\boldsymbol{k}_2,\boldsymbol{k}_1,\boldsymbol{k}_2) = \frac{d}{(2\pi)^6}\frac{g}{\rho}\int_{k_1,k_2<k_F}\mathrm{d}_3k_1\mathrm{d}_3k_2 = \frac{dg}{\rho(2\pi)^6}\frac{(4\pi)^2}{3^2}k_F^6\\ &= \frac{3}{8}g\rho,\end{aligned} \tag{10.3}$$

where $d = (n^2 - n)/2$, $n$ being the level degeneracy (4 in the case of symmetric nuclear matter).

We introduce the following new set of variables

$$\begin{pmatrix}\boldsymbol{k}\\ \boldsymbol{k}''\end{pmatrix} = \begin{pmatrix}\frac{1}{\sqrt{2}} & -\frac{1}{\sqrt{2}}\\ \frac{1}{\sqrt{2}} & \frac{1}{\sqrt{2}}\end{pmatrix}\begin{pmatrix}\boldsymbol{k}_1\\ \boldsymbol{k}_2\end{pmatrix} \quad\Rightarrow\quad \begin{pmatrix}\boldsymbol{k}_1\\ \boldsymbol{k}_2\end{pmatrix} = \begin{pmatrix}\frac{1}{\sqrt{2}} & \frac{1}{\sqrt{2}}\\ -\frac{1}{\sqrt{2}} & \frac{1}{\sqrt{2}}\end{pmatrix}\begin{pmatrix}\boldsymbol{k}\\ \boldsymbol{k}''\end{pmatrix}. \tag{10.4}$$

By using transformation (10.4), Eq. (10.3) becomes

$$\begin{aligned}\frac{E}{A} &= \frac{dgk_F^6}{\rho(2\pi)^6}\int_{|k''+k|,|k''-k|<\sqrt{2}}\mathrm{d}_3k\mathrm{d}_3k''\\ &= 8\frac{dgk_F^6}{\rho(2\pi)^6}\int \mathrm{d}_3\tilde{k}\mathrm{d}_3\tilde{k}''\,\theta\left(1-\left|\tilde{k}''+\tilde{k}\right|\right)\theta\left(1-\left|\tilde{k}''-\tilde{k}\right|\right)\theta(1-\tilde{k}).\end{aligned} \tag{10.5}$$

where $\tilde{k} = \frac{k}{\sqrt{2}}$ and $\tilde{k}'' = \frac{k''}{\sqrt{2}}$ in units of $k_\mathrm{F}$. We should compute first the integral on $\tilde{\boldsymbol{k}}''$, because we would like to be able to introduce later a cutoff on the relative momentum $\tilde{\boldsymbol{k}}$. The first integral is computed in Appendix G, following Ref. [FW03, pg. 28], with $d = 2k$ and $r = 1$.

$$\begin{aligned}\mathcal{I} &= \int \mathrm{d}_3k''\,\theta\left(1-\left|\tilde{k}''+\tilde{k}\right|\right)\theta\left(1-\left|\tilde{k}''-\tilde{k}\right|\right) = \int \mathrm{d}_3\tilde{k}''\,\theta\left(1-\left|\tilde{k}''+\tilde{k}\right|\right)\theta\left(1-\left|\tilde{k}''-\tilde{k}\right|\right)\\ &= \frac{4\pi}{3}\left(1-\frac{3}{2}\tilde{k}+\frac{1}{2}\tilde{k}^3\right).\end{aligned} \tag{10.6}$$

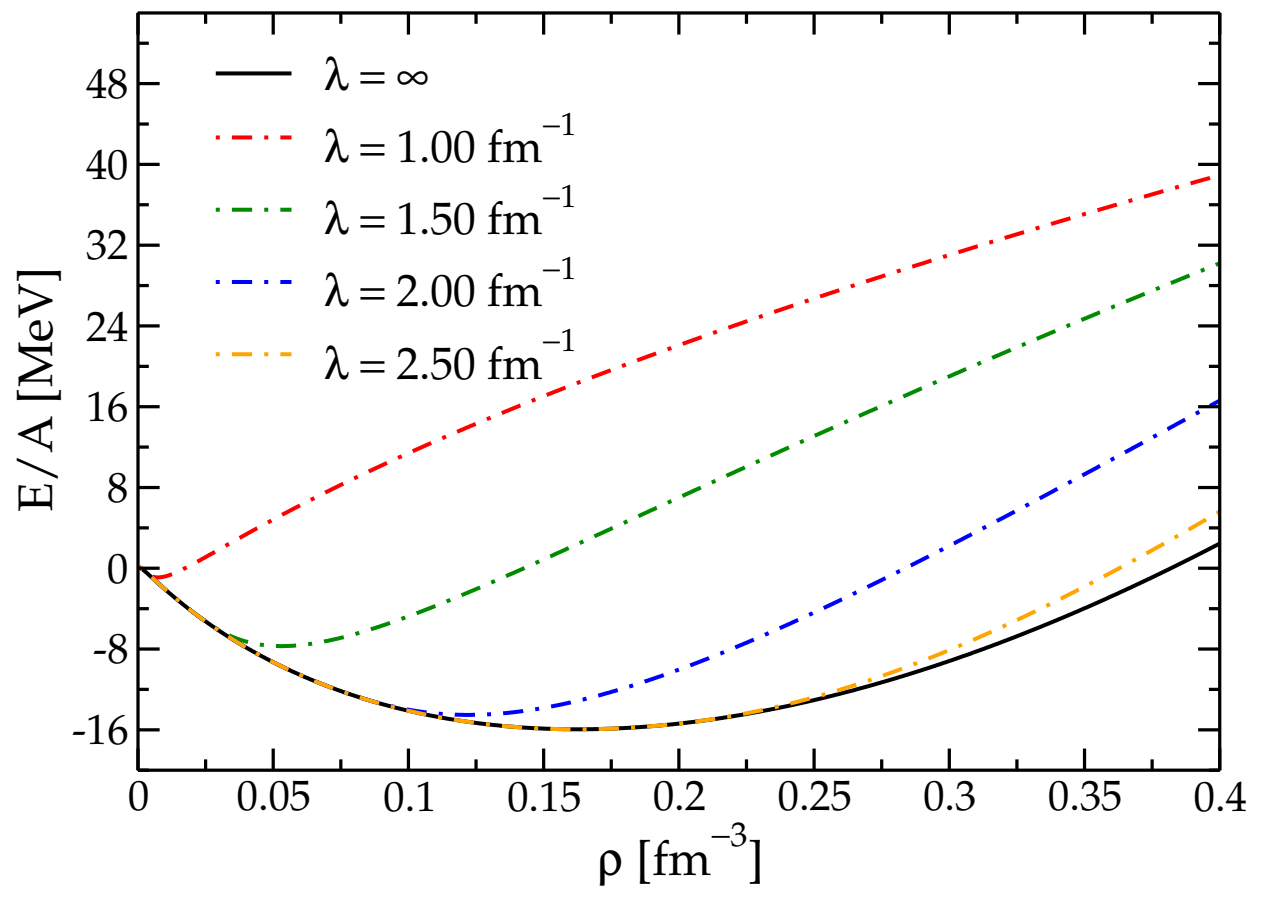

**Figure 10.2**
Energy per particle for different values of the cutoff $\lambda$ on the relative momentum $\boldsymbol{k}$.

Therefore, inserting Eq. (10.6) in Eq. (10.5) the energy per particle turns out to be

$$\frac{E}{A} = 8\frac{dgk_F^6}{\rho(2\pi)^6}\frac{4\pi}{3}\int \mathrm{d}_3\tilde{k}\left(1-\frac{3}{2}\tilde{k}+\frac{1}{2}\tilde{k}^3\right)\theta(1-\tilde{k}) = \frac{3}{8}g\rho. \tag{10.7}$$

If we want to introduce a cutoff $\lambda$ on the relative momentum $\boldsymbol{k}$, we just have to add a factor $\theta\left(\frac{\lambda}{\sqrt{2}k_F}-\tilde{k}\right)$. Then, Eq. (10.7) becomes

$$\begin{aligned}\frac{E}{A} &= \frac{dgk_F^6}{\rho(2\pi)^6}8\frac{4\pi}{3}\int \mathrm{d}_3\tilde{k}\left(1-\frac{3}{2}\tilde{k}+\frac{1}{2}\tilde{k}^3\right)\theta(1-\tilde{k})\theta\left(\frac{\lambda}{\sqrt{2}k_F}-\tilde{k}\right)\\ &= \frac{dgk_F^6}{\rho(2\pi)^6}8\frac{(4\pi)^2}{3}\int_0^\beta \mathrm{d}\tilde{k}\,\tilde{k}^2\left(1-\frac{3}{2}\tilde{k}+\frac{1}{2}\tilde{k}^3\right)\\ &= \frac{3}{8}g\rho\left(8\beta^3-9\beta^4+2\beta^6\right),\end{aligned} \tag{10.8}$$

where $\beta = \min\{1, \frac{\lambda}{\sqrt{2}k_F}\}$. Note that if $\lambda > \sqrt{2}k_F$, then $\beta = 1$ and we get Eq. (10.7) back. In Fig. 10.2 Eq. (10.8) is reproduced for different values of the cutoff $\lambda$.

Analogously, for the second order contribution to the total energy, we can rewrite Eq. (10.2) in terms of the relative momenta. We want to introduce the following new set of variables:

$$\begin{pmatrix}\boldsymbol{k}\\ \boldsymbol{k}'\\ \boldsymbol{k}''\end{pmatrix} = \begin{pmatrix}\frac{1}{\sqrt{2}} & -\frac{1}{\sqrt{2}} & 0\\ \frac{1}{\sqrt{2}} & -\frac{1}{\sqrt{2}} & \sqrt{2}\\ \frac{1}{\sqrt{2}} & \frac{1}{\sqrt{2}} & 0\end{pmatrix}\begin{pmatrix}\boldsymbol{k}_1\\ \boldsymbol{k}_2\\ \boldsymbol{q}\end{pmatrix} \quad\Rightarrow\quad \begin{pmatrix}\boldsymbol{k}_1\\ \boldsymbol{k}_2\\ \boldsymbol{q}\end{pmatrix} = \begin{pmatrix}\frac{1}{\sqrt{2}} & 0 & \frac{1}{\sqrt{2}}\\ -\frac{1}{\sqrt{2}} & 0 & \frac{1}{\sqrt{2}}\\ -\frac{1}{\sqrt{2}} & \frac{1}{\sqrt{2}} & 0\end{pmatrix}\begin{pmatrix}\boldsymbol{k}\\ \boldsymbol{k}'\\ \boldsymbol{k}''\end{pmatrix}. \tag{10.9}$$

The determinant for the Jacobian matrix of this transformation is 1. With this transfor-mation, the denominator in Eq. (10.2) reads

$$\begin{aligned}\epsilon_{k_1}+\epsilon_{k_2}-\epsilon_{k_1+q}-\epsilon_{k_2-q} &= \frac{\hbar^2}{2m}\left[\left(\frac{\boldsymbol{k}''+\boldsymbol{k}}{\sqrt{2}}\right)^2+\left(\frac{\boldsymbol{k}''-\boldsymbol{k}}{\sqrt{2}}\right)^2-\left(\frac{\boldsymbol{k}''+\boldsymbol{k}'}{\sqrt{2}}\right)^2-\left(\frac{\boldsymbol{k}''-\boldsymbol{k}'}{\sqrt{2}}\right)^2\right]\\ &= -\frac{\hbar^2}{2m}(k'^2-k^2).\end{aligned} \tag{10.10}$$

Then, Eq. (10.2) becomes, expressing all the wave vectors in units of $\sqrt{2}k_F$,

$$\frac{\Delta E}{A}=\chi(\rho)\frac{\sqrt{2}}{4\pi^3}\int_{\mathcal{D}(\boldsymbol{k},\boldsymbol{k}',\boldsymbol{k}'')}\mathrm{d}_3k\mathrm{d}_3k'\mathrm{d}_3k''\,\frac{1}{k'^2-k^2}, \tag{10.11}$$

where

$$\begin{aligned}\mathcal{D}(\boldsymbol{k},\boldsymbol{k}',\boldsymbol{k}'') \equiv \Big\{&\boldsymbol{k},\boldsymbol{k}',\boldsymbol{k}''\in\mathbb{R}^3:\,k\le 1, k''\le 1,\\ &(|\boldsymbol{k}''+\boldsymbol{k}|<1\cap|\boldsymbol{k}''-\boldsymbol{k}|<1)\cup(|\boldsymbol{k}''+\boldsymbol{k}'|>1\cap|\boldsymbol{k}''-\boldsymbol{k}'|>1)\Big\}\end{aligned}$$

and

$$\chi(\rho)=-\frac{3}{4\pi^6}\frac{mk_F^7g^2}{\hbar^2\rho}.$$

In order to make a qualitative comparison between the results of Ref. [Mog+10] and the present approach, we limit ourselves to the case in which the cutoff on relative momenta is larger than $2\sqrt{2}k_{\mathrm{F}}$, being the calculations less cumbersome. The complete calculation in this case is performed in appendix H. If $\lambda$ is the cutoff in units of $\sqrt{2}k_{\mathrm{F}}$, the second order contribution to the total energy becomes

$$\begin{aligned}\frac{\Delta E}{A}=\chi(\rho)\Big\{&-\frac{11}{105}+\frac{2}{105}\ln 2+\frac{2}{35}\lambda-\frac{11}{35}\lambda^3-\frac{2}{21}\lambda^5-\left(\frac{4\lambda^5}{5}-\frac{4\lambda^7}{21}\right)\ln(\lambda)\\ &+\left(\frac{1}{35}-\frac{\lambda^4}{3}+\frac{2\lambda^5}{5}-\frac{2\lambda^7}{21}\right)\ln(\lambda-1)\\ &-\left(\frac{1}{35}-\frac{\lambda^4}{3}-\frac{2\lambda^5}{5}+\frac{2\lambda^7}{21}\right)\ln(\lambda+1)\Big\}.\end{aligned} \tag{10.12}$$

We note that the part that is independent on $\lambda$ is equal to the one in Ref. [Mog+10]. We want now to study the behavior for $\lambda \gg 1$. In this case, Eq. (10.12) can be rewritten as

$$\begin{aligned}\frac{\Delta E}{A}=\chi(\rho)\Big[&-\frac{11}{105}+\frac{2}{105}\ln 2+\frac{2}{35}\lambda-\frac{11}{35}\lambda^3-\frac{2}{21}\lambda^5\\ &-\left(\frac{1}{35}-\frac{\lambda^4}{3}+\frac{2\lambda^5}{5}-\frac{2\lambda^7}{21}\right)\left(\frac{1}{\lambda}+\frac{1}{2\lambda^2}+\frac{1}{3\lambda^3}+\frac{1}{4\lambda^4}+\frac{1}{5\lambda^5}+\frac{1}{6\lambda^6}\right)\\ &-\left(\frac{1}{35}-\frac{\lambda^4}{3}-\frac{2\lambda^5}{5}+\frac{2\lambda^7}{21}\right)\left(\frac{1}{\lambda}-\frac{1}{2\lambda^2}+\frac{1}{3\lambda^3}-\frac{1}{4\lambda^4}+\frac{1}{5\lambda^5}-\frac{1}{6\lambda^6}\right)\Big]\\ =\chi(\rho)&\left(-\frac{11}{105}+\frac{2}{105}\ln 2+\frac{\lambda}{9}\right).\end{aligned} \tag{10.13}$$

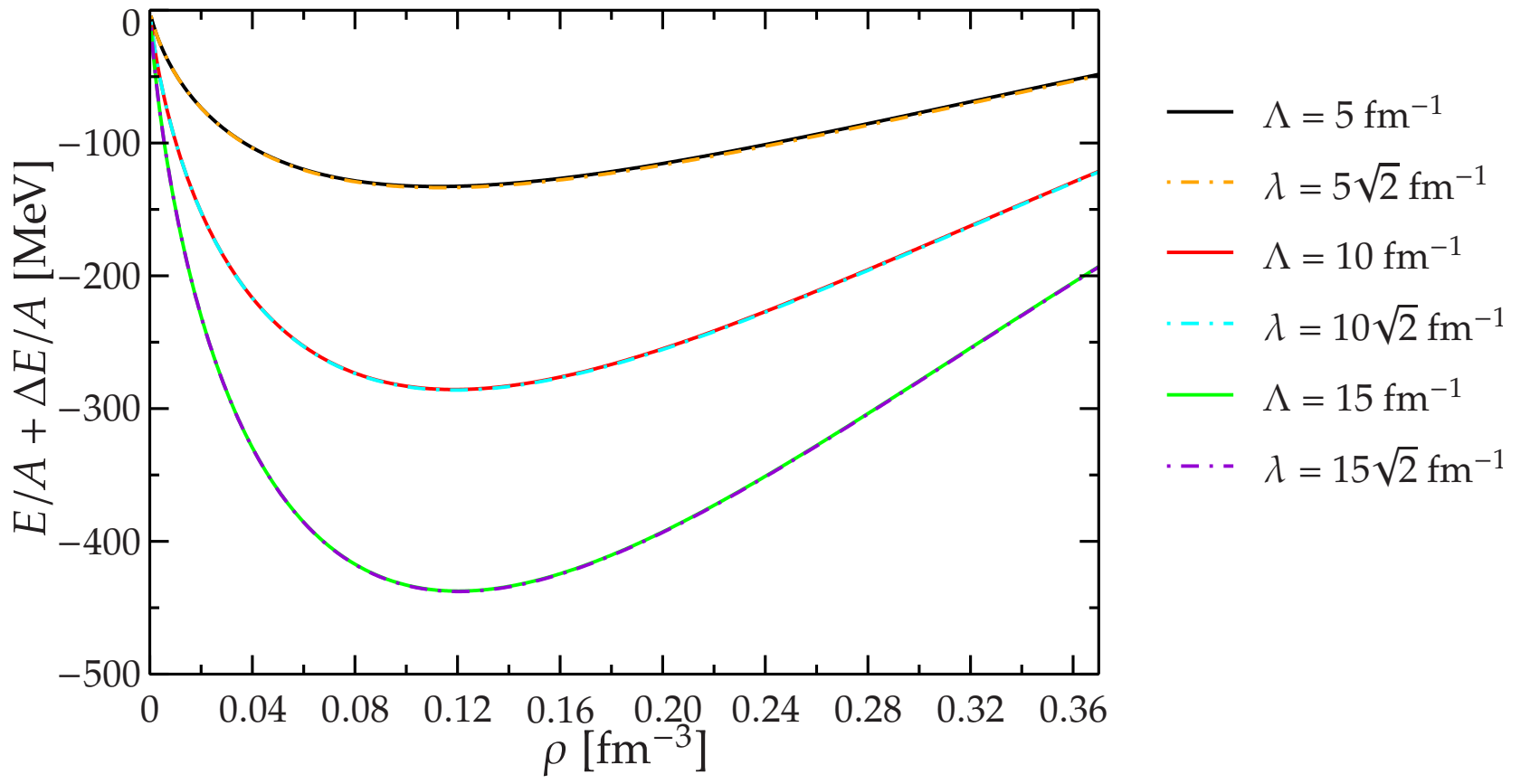

**Figure 10.3**
Total energy at second order level for symmetric nuclear matter using SkP. Different cutoffs are used. In particular the cutoff $\Lambda$ refers to the approach of Ref. [Mog+10] while the cutoff $\lambda$ refers to the present approach.

Then, also the asymptotic behavior is equal to the one of Ref. [Mog+10].

In Fig. 10.3, it is depicted the total energy (mean-field plus second order) calculated using Eq. (10.12) and Eq. (8) of Ref. [Mog+10], for different values of the cutoff. The two calculations are almost indistinguishable when $\lambda = \sqrt{2}\Lambda$.

## 10.3 The renormalized interaction

The fact that in nuclear matter, due to translational invariance, it is possible to work entirely in the momentum domain has greatly simplified the calculations in section 10.2. However, since for a finite system only the space domain can be considered, the introduction of cutoffs on relative momenta requires the modification of the interaction in coordinate space, which eventually acquires a finite range. In order to perform this reformulation, we need to write the velocity independent part of the Skyrme force as a non-local interaction:

$$V(\boldsymbol{r}_1', \boldsymbol{r}_2', \boldsymbol{r}_1, \boldsymbol{r}_2) = g\left(\frac{\boldsymbol{r}_1 + \boldsymbol{r}_2}{2}\right)\delta_3(\boldsymbol{r}_1 - \boldsymbol{r}_2)\delta_3(\boldsymbol{r}_1 - \boldsymbol{r}_1')\delta_3(\boldsymbol{r}_2 - \boldsymbol{r}_2') \tag{10.14}$$

We want to write the interaction in the center of mass and relative motion frame of reference, in order to have the possibility of introducing the relative momentum. The change of the variables, having the determinant of Jacobian matrix equal to 1, is the following

$$\begin{pmatrix} \boldsymbol{r}^{(\prime)} \\ \boldsymbol{R}^{(\prime)} \end{pmatrix} = \begin{pmatrix} \frac{1}{\sqrt{2}} & -\frac{1}{\sqrt{2}} \\ \frac{1}{\sqrt{2}} & \frac{1}{\sqrt{2}} \end{pmatrix} \begin{pmatrix} \boldsymbol{r}_1^{(\prime)} \\ \boldsymbol{r}_2^{(\prime)} \end{pmatrix} \quad \Rightarrow \quad \begin{pmatrix} \boldsymbol{r}_1^{(\prime)} \\ \boldsymbol{r}_2^{(\prime)} \end{pmatrix} = \begin{pmatrix} \frac{1}{\sqrt{2}} & \frac{1}{\sqrt{2}} \\ -\frac{1}{\sqrt{2}} & \frac{1}{\sqrt{2}} \end{pmatrix} \begin{pmatrix} \boldsymbol{r}^{(\prime)} \\ \boldsymbol{R}^{(\prime)} \end{pmatrix}. \tag{10.15}$$

In this way, the interaction (10.14) can be written as

$$
\begin{aligned}
V(\mathbf{r}',\mathbf{R}',\mathbf{r},\mathbf{R}) &= \frac{\sqrt{2}}{4} g\left(\frac{\mathbf{R}}{\sqrt{2}}\right)\delta_3(\mathbf{r})\delta_3(\mathbf{r}')\delta_3(\mathbf{R}-\mathbf{R}') \\
&= \frac{\sqrt{2}}{4} g\left(\frac{\mathbf{R}}{\sqrt{2}}\right) v(\mathbf{r}',\mathbf{r})\delta_3(\mathbf{R}-\mathbf{R}');
\end{aligned}
\tag{10.16}
$$

where $g(\mathbf{R}) = t_0 + \frac{t_3}{6}\,[\rho\,(\mathbf{R})]^{\alpha}$.

Computing the Fourier transform of Eq. (10.16), we get

$$
\begin{aligned}
v\,(\mathbf{k}_3,\mathbf{k}_4,\mathbf{k}_1,\mathbf{k}_2) = &\frac{\sqrt{2}}{4}\frac{1}{\Omega}\int \mathrm{d}_3 R \mathrm{d}_3 R'\, \mathrm{e}^{-i\frac{\mathbf{k}_3+\mathbf{k}_4}{\sqrt{2}}\cdot\mathbf{R}'} g\left(\frac{\mathbf{R}}{\sqrt{2}}\right)\delta_3\,(\mathbf{R}-\mathbf{R}')\,\mathrm{e}^{i\frac{\mathbf{k}_1+\mathbf{k}_2}{\sqrt{2}}\cdot\mathbf{R}} \\
&\frac{1}{\Omega}\int \mathrm{d}_3 r \mathrm{d}_3 r'\, \mathrm{e}^{-i\frac{\mathbf{k}_3-\mathbf{k}_4}{\sqrt{2}}\cdot\mathbf{r}'} v(\mathbf{r}',\mathbf{r})\mathrm{e}^{i\frac{\mathbf{k}_1-\mathbf{k}_2}{\sqrt{2}}\cdot\mathbf{r}} \\
= &\frac{\sqrt{2}}{4}\frac{1}{\Omega^2}\int \mathrm{d}_3 R\, \mathrm{e}^{i\frac{\mathbf{k}_1+\mathbf{k}_2-\mathbf{k}_3-\mathbf{k}_4}{\sqrt{2}}\cdot\mathbf{R}} g\left(\frac{\mathbf{R}}{\sqrt{2}}\right).
\end{aligned}
\tag{10.17}
$$

We can write $\mathbf{k} \equiv \frac{\mathbf{k}_1-\mathbf{k}_2}{\sqrt{2}}$ and $\mathbf{k}' \equiv \frac{\mathbf{k}_3-\mathbf{k}_4}{\sqrt{2}}$, so that

$$
v(\mathbf{k}',\mathbf{k}) = \frac{1}{\Omega}\int \mathrm{d}_3 r \mathrm{d}_3 r'\, \mathrm{e}^{-i\mathbf{k}'\cdot\mathbf{r}'} v(\mathbf{r}',\mathbf{r})\mathrm{e}^{i\mathbf{k}\cdot\mathbf{r}} = \frac{1}{\Omega},
\tag{10.18}
$$

If we want to introduce a cutoff on initial and final relative momenta, we compute the Fourier anti-transform multiplying Eq. (10.18) times two step functions $\theta(\Lambda-k)\theta(\Lambda'-k')$. Here we want to distinguish the two cutoffs in order to be as much general as possible.

$$
\begin{aligned}
v^{\Lambda\Lambda'}(\mathbf{r}',\mathbf{r}) =& \frac{1}{\Omega}\int \mathrm{d}_3 k \mathrm{d}_3 k'\, \mathrm{e}^{i\mathbf{k}'\cdot\mathbf{r}'} v\,(\mathbf{k}',\mathbf{k})\,\theta(\Lambda-k)\theta(\Lambda'-k')\mathrm{e}^{-i\mathbf{k}\cdot\mathbf{r}} \\
=& \frac{(4\pi)^2}{\Omega^2}\left(\sum_{lm}(-i)^l\int_0^{\Lambda}\mathrm{d}k\,k^2 j_l(kr)\int \mathrm{d}\hat{k} Y_{lm}(\hat{k})Y^*_{lm}(\hat{r})\right) \\
&\left(\sum_{l'm'}(i)^{l'}\int_0^{\Lambda'}\mathrm{d}k'\,k'^2 j_{l'}(k'r')\int \mathrm{d}\hat{k}' Y_{l'm'}(\hat{k}')Y^*_{l'm'}(\hat{r}')\right) \\
=& \frac{16\pi^2}{\Omega^2}\int_0^{\Lambda}\mathrm{d}k k^2 j_0(kr)\int_0^{\Lambda'}\mathrm{d}k' k'^2 j_0(k'r') \\
=& \frac{(4\pi)^2}{\Omega^2}\frac{\Omega^2}{(2\pi)^6}\frac{\Lambda^2\Lambda'^2}{rr'} j_1(r\Lambda)j_1(r'\Lambda') \\
=& \frac{1}{4\pi^4}\frac{\Lambda^2\Lambda'^2}{rr'} j_1(r\Lambda)j_1(r'\Lambda') \\
&\xrightarrow[\substack{\Lambda\to+\infty \\ \Lambda'\to+\infty}]{} \frac{16\pi^2}{\Omega^2}\frac{\Omega^2}{16\pi^2}\delta_3(\mathbf{r})\delta_3(\mathbf{r}') = \delta_3(\mathbf{r})\delta_3(\mathbf{r}')
\end{aligned}
\tag{10.19}
$$

by using Eq. (3.5) of Ref. [Meh11]. This $v^{\Lambda\Lambda'}(\mathbf{r}',\mathbf{r})$ is used instead of $v(\mathbf{r}',\mathbf{r})$ in the evaluation of the matrix elements of the interaction (10.16).

## 10.4 The transformation to relative and center of mass wave functions

Since we are interested in the introduction of a cutoff on the relative momenta, we have to write the two-particle states used to compute the matrix elements of the interaction in the center of mass and relative motion coordinate system. In a finite system, this transformation can be performed analytically only for two particles in a harmonic oscillator potential. For this reason, it is mandatory to expand the sp wave functions on a harmonic oscillator basis.

Consider a two-particle state, $|n_a l_a j_a m_a, n_b l_b j_b m_b\rangle$, in a harmonic oscillator potential. We can write the wave function of the single-particle state as

$$|nljm\tau\rangle = \psi^{m_\tau}_{nljm}(\boldsymbol{r}) = i^l R_{nl}(\beta r)\left[Y_l(\hat{r}) \otimes \chi_{\frac{1}{2}}\right]_{jm} \xi_\tau, \tag{10.20}$$

where $\beta^2 = \frac{m\omega}{\hbar}$.

We want to couple the two particles to the total angular momentum $JM$:

$$|n_a l_a j_a m_a \tau_a, n_b l_b j_b m_b \tau_b\rangle = \sum_{JM_J} \langle j_a m_a j_b m_b | JM_J\rangle |(n_a l_a j_a \tau_a, n_b l_b j_b \tau_b) JM_J\rangle \tag{10.21}$$

In order to separate the two-particle wave function into the relative motion wave function and the center of mass wave function, it is necessary to switch from the $j$–$j$ coupling scheme to the $L$–$S$ coupling one.

$$\begin{aligned}|n_a l_a j_a m_a \tau_a, n_b l_b j_b m_b \tau_b\rangle = \sum_{\substack{JM_J\\ \Lambda\Sigma}} \hat{\Lambda}\hat{\Sigma}\hat{j}_a\hat{j}_b \langle j_a m_a j_b m_b | JM_J\rangle \begin{Bmatrix} l_a & l_b & \Lambda \\ \frac{1}{2} & \frac{1}{2} & \Sigma \\ j_a & j_b & J \end{Bmatrix} \\ |[n_a n_b, (l_a, l_b)\Lambda, (\frac{1}{2}, \frac{1}{2})\Sigma, \tau_a \tau_b] JM_J\rangle,\end{aligned} \tag{10.22}$$

and writing explicitly the state $|[n_a n_b, (l_a, l_b)\Lambda, (\frac{1}{2}, \frac{1}{2})\Sigma, \tau_a \tau_b] JM_J\rangle$ we get (in the following, $\hat{j}$ is a shorthand notation for $\sqrt{2j+1}$)

$$\begin{aligned}|n_a l_a j_a m_a \tau_a, n_b l_b j_b m_b \tau_b\rangle &= \sum_{\substack{JM_J\\ \Lambda\Sigma}} \hat{\Lambda}\hat{\Sigma}\hat{j}_a\hat{j}_b \langle j_a m_a j_b m_b | JM_J\rangle \begin{Bmatrix} l_a & l_b & \Lambda \\ \frac{1}{2} & \frac{1}{2} & \Sigma \\ j_a & j_b & J \end{Bmatrix} \\ &\qquad |[n_a n_b, (l_a, l_b)\Lambda, (\frac{1}{2}, \frac{1}{2})\Sigma, \tau_a \tau_b] JM_J\rangle \\ &= \sum_{\substack{JM_J\\ \Lambda\Sigma}} i^{l_a+l_b} \hat{\Lambda}\hat{\Sigma}\hat{j}_a\hat{j}_b \langle j_a m_a j_b m_b | JM_J\rangle \begin{Bmatrix} l_a & l_b & \Lambda \\ \frac{1}{2} & \frac{1}{2} & \Sigma \\ j_a & j_b & J \end{Bmatrix} \\ &\qquad \sum_{M_\Lambda M_\Sigma} \langle \Lambda M_\Lambda \Sigma M_\Sigma | JM_J\rangle R_{n_a l_a}(\beta r_a) R_{n_b l_b}(\beta r_b) \\ &\qquad \left[Y_{l_a} \otimes Y_{l_b}\right]_{\Lambda M_\Lambda} \left[\chi_{\frac{1}{2}}(1) \otimes \chi_{\frac{1}{2}}(2)\right]_{\Sigma M_\Sigma} \xi_{\tau_a}(1)\xi_{\tau_b}(2)\end{aligned}$$

$$= \sum_{\substack{JM_J \\ \Lambda\Sigma}} i^{l_a+l_b} \hat{\Lambda}\hat{\Sigma}\hat{j}_a\hat{j}_b \langle j_a m_a j_b m_b | J M_J \rangle \begin{Bmatrix} l_a & l_b & \Lambda \\ \frac{1}{2} & \frac{1}{2} & \Sigma \\ j_a & j_b & J \end{Bmatrix}$$
$$\sum_{M_\Lambda M_\Sigma} \langle \Lambda M_\Lambda \Sigma M_\Sigma | J M_J \rangle R_{n_a l_a}(\beta r_a) R_{n_b l_b}(\beta r_b)$$
$$\sum_{m_{l_a} m_{l_b}} \langle l_a m_{l_a} l_b m_{l_b} | \Lambda M_\Lambda \rangle Y_{l_a m_{l_a}} Y_{l_b m_{l_b}}$$
$$\sum_{\sigma_a \sigma_b} \langle \frac{1}{2}\sigma_a \frac{1}{2}\sigma_b | \Sigma M_\Sigma \rangle \chi_{\sigma_a}(1)\chi_{\sigma_b}(2)\xi_{\tau_a}(1)\xi_{\tau_b}(2)$$
$$= \sum_{\substack{JM_J \\ \Lambda\Sigma}} i^{l_a+l_b} \hat{\Lambda}\hat{\Sigma}\hat{j}_a\hat{j}_b \langle j_a m_a j_b m_b | J M_J \rangle \begin{Bmatrix} l_a & l_b & \Lambda \\ \frac{1}{2} & \frac{1}{2} & \Sigma \\ j_a & j_b & J \end{Bmatrix}$$
$$\sum_{M_\Lambda M_\Sigma} \langle \Lambda M_\Lambda \Sigma M_\Sigma | J M_J \rangle$$
$$\sum_{nlNL} M_\Lambda(NLnl; n_a l_a n_b l_b) R_{nl}(\beta r) R_{NL}(\beta R) \tag{10.23}$$
$$\sum_{M_l M_L} \langle L M_L l M_l | \Lambda M_\Lambda \rangle Y_{l M_l}(\hat{r}) Y_{L M_L}(\hat{R})$$
$$\sum_{\sigma_a \sigma_b} \langle \frac{1}{2}\sigma_a \frac{1}{2}\sigma_b | \Sigma M_\Sigma \rangle \chi_{\sigma_a}(1)\chi_{\sigma_b}(2)\xi_{\tau_a}(1)\xi_{\tau_b}(2)$$

$$= \sum_{J\Lambda\Sigma} \sum_{\substack{M_\Lambda M_\Sigma M_J \\ M_l M_L}} \sum_{\sigma_a \sigma_b} i^{l_a+l_b} (-)^{j_a - j_b + \Lambda + M_\Lambda + \Sigma + M_\Sigma + L + l} \hat{J}^2 \hat{\Lambda}^2 \hat{\Sigma}^2 \hat{j}_a \hat{j}_b \begin{Bmatrix} l_a & l_b & \Lambda \\ \frac{1}{2} & \frac{1}{2} & \Sigma \\ j_a & j_b & J \end{Bmatrix}$$
$$\begin{pmatrix} j_a & j_b & J \\ m_a & m_b & -M_J \end{pmatrix} \begin{pmatrix} \Lambda & \Sigma & J \\ M_\Lambda & M_\Sigma & -M_J \end{pmatrix} \begin{pmatrix} \frac{1}{2} & \frac{1}{2} & \Sigma \\ \sigma_a & \sigma_b & -M_\Sigma \end{pmatrix} \begin{pmatrix} L & l & \Lambda \\ M_L & M_l & -M_\Lambda \end{pmatrix} \tag{10.24}$$
$$\sum_{nlNL} M_\Lambda(NLnl; n_a l_a n_b l_b) R_{nl}(\beta r) R_{NL}(\beta R) Y_{l M_l}(\hat{r}) Y_{L M_L}(\hat{R})$$
$$\chi_{\sigma_a}(1)\chi_{\sigma_b}(2)\xi_{\tau_a}(1)\xi_{\tau_b}(2),$$

where we use the notation $\boldsymbol{r} = \frac{1}{\sqrt{2}}(\boldsymbol{r}_a - \boldsymbol{r}_b)$ and $\boldsymbol{R} = \frac{1}{\sqrt{2}}(\boldsymbol{r}_a + \boldsymbol{r}_b)$. The quantities $M_\Lambda$ are the Brody-Moshinsky coefficients [Law80; BM96; Kam+01], which allow us to write the wave functions to the center of mass and relative motion coordinate system.

## 10.5 The matrix element of the interaction

In this section we want to compute the matrix elements of the Skyrme interaction, written as in Eq. (10.14), between the two-particle states (10.24). Obtaining the actual matrix elements between HF sp states is as simple as multiplying by the corresponding harmonic oscillator expansion coefficients. We are interested in the antisymmetrized interaction $\bar{V} = V(1 - P_M P_\sigma P_\tau)$, where $P_M$ is the Majorana exchange operator, $P_\sigma = \frac{1+\sigma(1)\sigma(2)}{2}$ and $P_\tau = \frac{1+\tau(1)\tau(2)}{2}$ are the spin and isospin exchange operators. Since we are dealing with a delta interaction (acting only in S waves), $P_M = 1$.

$$\bar{V} = V(1 - P_M P_\sigma P_\tau) = V\left(\frac{3}{4} - \frac{1}{4}\sigma(1)\sigma(2) - \frac{1}{4}\tau(1)\tau(2) - \frac{1}{4}\sigma(1)\sigma(2)\tau(1)\tau(2)\right). \quad (10.25)$$

First, we compute the matrix elements not performing the transformation to center of mass and relative coordinate system. The matrix element is derived in detail in appendix I. We indicate with the labels 0, $\sigma$, $\tau$ and $\sigma\tau$ the scalar-isoscalar, vector-isoscalar, scalar-isovector and vector-isovector matrix elements of the interaction, respectively. We use the notation:

$$\mathscr{F}(\tau) = \begin{cases} \delta_{\tau_a\tau_c}\delta_{\tau_b\tau_d} & \text{for } \langle ab|\bar{V}|cd\rangle_{0,\sigma} \\ \sum_\mu (-)^{1+\tau_a+\tau_b+\mu}\begin{pmatrix} \frac{1}{2} & 1 & \frac{1}{2} \\ \tau_c & \mu & -\tau_a \end{pmatrix}\begin{pmatrix} \frac{1}{2} & 1 & \frac{1}{2} \\ \tau_d & -\mu & -\tau_b \end{pmatrix} & \text{for } \langle ab|\bar{V}|cd\rangle_{\tau,\sigma\tau} \end{cases}$$

$$\mathscr{G}(\Sigma) = \begin{cases} 1 & \text{for } \langle ab|\bar{V}|cd\rangle_{0,\tau} \\ (-)^{1+\Sigma}\begin{Bmatrix} \frac{1}{2} & \frac{1}{2} & \Sigma \\ \frac{1}{2} & \frac{1}{2} & 1 \end{Bmatrix} & \text{for } \langle ab|\bar{V}|cd\rangle_{\sigma,\sigma\tau} \end{cases}$$

$$\mathscr{N} = \begin{cases} \frac{3}{4} & \text{for } \langle ab|\bar{V}|cd\rangle_0 \\ -\frac{3}{2} & \text{for } \langle ab|\bar{V}|cd\rangle_{\sigma,\tau} \\ -9 & \text{for } \langle ab|\bar{V}|cd\rangle_{\sigma\tau} \end{cases}$$

$$\begin{aligned} \langle ab|\bar{V}|cd\rangle_{0,\sigma,\tau,\sigma\tau} = & \mathscr{N}\mathscr{F}(\tau)\sum_{JM_J}\sum_{\Sigma\Lambda} i^{-l_a-l_b+l_c+l_d}\hat{J}^2\hat{\Lambda}^2\hat{\Sigma}^2\frac{\hat{j}_a\hat{j}_b\hat{j}_c\hat{j}_d\hat{l}_a\hat{l}_b\hat{l}_c\hat{l}_d}{4\pi}(-)^{l_a-l_b+l_c-l_d} \\ & \mathscr{G}(\Sigma)\begin{Bmatrix} l_a & l_b & \Lambda \\ \frac{1}{2} & \frac{1}{2} & \Sigma \\ j_a & j_b & J \end{Bmatrix}\begin{Bmatrix} l_c & l_d & \Lambda \\ \frac{1}{2} & \frac{1}{2} & \Sigma \\ j_c & j_d & J \end{Bmatrix}\begin{pmatrix} l_a & l_b & \Lambda \\ 0 & 0 & 0 \end{pmatrix}\begin{pmatrix} l_c & l_d & \Lambda \\ 0 & 0 & 0 \end{pmatrix} \\ & (-)^{j_a-j_b+j_c-j_d}\begin{pmatrix} j_a & j_b & J \\ m_a & m_b & -M_J \end{pmatrix}\begin{pmatrix} j_c & j_d & J \\ m_c & m_d & -M_J \end{pmatrix} \\ & \int \mathrm{d}r\, r^2 R_{n_al_a}(\beta r)R_{n_bl_b}(\beta r)g(r)R_{n_cl_c}(\beta r)R_{n_dl_d}(\beta r) \end{aligned} \quad (10.26)$$

We are interested in the particle-particle angular-momentum-coupled matrix element. In general,

$$\begin{aligned} \langle(\alpha\beta)JM_J|\bar{V}|(\gamma\delta)JM_J\rangle &= \sum_{\substack{m_\alpha m_\beta \\ m_\gamma m_\delta}}\langle j_\alpha m_\alpha j_\beta m_\beta|JM_J\rangle\langle j_\gamma m_\gamma j_\delta m_\delta|JM_J\rangle\langle\alpha\beta|\bar{V}|\gamma\delta\rangle \\ &= \sum_{\substack{m_\alpha m_\beta \\ m_\gamma m_\delta}}(-)^{j_\alpha-j_\beta+j_\gamma-j_\delta}\hat{J}^2\begin{pmatrix} j_\alpha & j_\beta & J \\ m_\alpha & m_\beta & -M_J \end{pmatrix}\begin{pmatrix} j_\gamma & j_\delta & J \\ m_\gamma & m_\delta & -M_J \end{pmatrix}\langle\alpha\beta|\bar{V}|\gamma\delta\rangle \end{aligned} \quad (10.27)$$

Thus, the four terms in Eq. (10.26) become

$$\begin{aligned}
&\langle (n_a l_a j_a \tau_a, n_b l_b j_b \tau_b) J M_J | \bar{V} | (n_c l_c j_c \tau_c, n_d l_d j_d \tau_d) J M_J \rangle_{0,\sigma,\tau,\sigma\tau} = \\
&= \mathcal{N} \mathcal{F}(\tau) \sum_{\Sigma\Lambda} i^{-l_a - l_b + l_c + l_d} \hat{\Lambda}^2 \hat{\Sigma}^2 \frac{\hat{j}_a \hat{j}_b \hat{j}_c \hat{j}_d \hat{l}_a \hat{l}_b \hat{l}_c \hat{l}_d}{4\pi} (-)^{l_a - l_b + l_c - l_d} \mathcal{G}(\Sigma) \\
&\begin{Bmatrix} l_a & l_b & \Lambda \\ \frac{1}{2} & \frac{1}{2} & \Sigma \\ j_a & j_b & J \end{Bmatrix} \begin{Bmatrix} l_c & l_d & \Lambda \\ \frac{1}{2} & \frac{1}{2} & \Sigma \\ j_c & j_d & J \end{Bmatrix} \begin{pmatrix} l_a & l_b & \Lambda \\ 0 & 0 & 0 \end{pmatrix} \begin{pmatrix} l_c & l_d & \Lambda \\ 0 & 0 & 0 \end{pmatrix} \\
&\int \mathrm{d}r\, r^2 R_{n_a l_a}(\beta r) R_{n_b l_b}(\beta r) g(r) R_{n_c l_c}(\beta r) R_{n_d l_d}(\beta r)
\end{aligned} \tag{10.28}$$

## 10.6 The matrix element of the interaction in center of mass and relative coordinates

We can now compute the matrix elements of the interaction in the center of mass and relative motion coordinate system. In this case, the Majorana operator is non-trivial, since the interaction can act not only in S waves. The exchange operator acts on a two-particle state (10.24) in the following way.

$$\begin{aligned}
&|n_b l_b j_b m_b \tau_b, n_a l_a j_a m_a \tau_a\rangle = P_M P_\sigma P_\tau |n_a l_a j_a m_a \tau_a, n_b l_b j_b m_b \tau_b\rangle \\
&= i^{l_a + l_b} \hat{j}_a \hat{j}_b \sum_{JM_J} \sum_{\Lambda M_\Lambda} \sum_{\Sigma M_\Sigma} \hat{\Lambda} \hat{\Sigma} \langle j_b m_b j_a m_a | J M_J \rangle \begin{Bmatrix} l_b & l_a & \Lambda \\ \frac{1}{2} & \frac{1}{2} & \Sigma \\ j_b & j_a & J \end{Bmatrix} \langle \Lambda M_\Lambda \Sigma M_\Sigma | J M_J \rangle \\
&\quad R_{n_b l_b}(\beta r_1) R_{n_a l_a}(\beta r_2) \left[ Y_{l_b} \otimes Y_{l_a} \right]_{\Lambda M_\Lambda} \left[ \chi_{\frac{1}{2}}(1) \otimes \chi_{\frac{1}{2}}(2) \right]_{\Sigma M_\Sigma} \xi_{\tau_b}(1) \xi_{\tau_a}(2) \\
&= i^{l_a + l_b} \hat{j}_a \hat{j}_b \sum_{JM_J} \sum_{\Lambda M_\Lambda} \sum_{\Sigma M_\Sigma} \hat{\Lambda} \hat{\Sigma} \langle j_a m_a j_b m_b | J M_J \rangle \begin{Bmatrix} l_a & l_b & \Lambda \\ \frac{1}{2} & \frac{1}{2} & \Sigma \\ j_a & j_b & J \end{Bmatrix} \\
&\quad \langle \Lambda M_\Lambda \Sigma M_\Sigma | J M_J \rangle R_{n_b l_b}(\beta r_1) R_{n_a l_a}(\beta r_2) (-1)^{l_a + l_b + \Lambda} \left[ Y_{l_b} \otimes Y_{l_a} \right]_{\Lambda M_\Lambda} \\
&\quad \sum_{\sigma_a \sigma_b} \langle \frac{1}{2} \sigma_a \frac{1}{2} \sigma_b | \Sigma M_\Sigma \rangle P_\sigma \left[ \chi_{\sigma_a}(1) \chi_{\sigma_b}(2) \right] P_\tau \left[ \xi_{\tau_a}(1) \xi_{\tau_b}(2) \right] \\
&= i^{l_a + l_b} \hat{j}_a \hat{j}_b \sum_{JM_J} \sum_{\Lambda M_\Lambda} \sum_{\Sigma M_\Sigma} \hat{\Lambda} \hat{\Sigma} \langle j_a m_a j_b m_b | J M_J \rangle \begin{Bmatrix} l_a & l_b & \Lambda \\ \frac{1}{2} & \frac{1}{2} & \Sigma \\ j_a & j_b & J \end{Bmatrix} \\
&\quad \langle \Lambda M_\Lambda \Sigma M_\Sigma | J M_J \rangle \sum_{NLnl} M_\Lambda(NLnl; n_b l_b n_a l_a) R_{nl}(\beta r) R_{NL}(\beta R) \\
&\quad (-1)^{l_a + l_b + \Lambda} \sum_{M_L M_l} \langle L M_L l M_l | \Lambda M_\Lambda \rangle Y_{l M_l}(\hat{r}) Y_{L M_L}(\hat{R}) \\
&\quad \sum_{\sigma_a \sigma_b} \langle \frac{1}{2} \sigma_a \frac{1}{2} \sigma_b | \Sigma M_\Sigma \rangle P_\sigma \left[ \chi_{\sigma_a}(1) \chi_{\sigma_b}(2) \right] P_\tau \left[ \xi_{\tau_a}(1) \xi_{\tau_b}(2) \right]
\end{aligned}$$

$$
= i^{l_a+l_b} \hat{j}_a \hat{j}_b \sum_{JM_J} \sum_{\Lambda M_\Lambda} \sum_{\Sigma M_\Sigma} \sum_{LM_L} \sum_{lM_l} \sum_{Nn} \sum_{\sigma_a\sigma_b} (-1)^{l_a+l_b-L} \hat{J}^2 \hat{\Lambda}^2 \hat{\Sigma}^2 \begin{Bmatrix} l_a & l_b & \Lambda \\ \frac{1}{2} & \frac{1}{2} & \Sigma \\ j_a & j_b & J \end{Bmatrix}
$$
$$
(-1)^{j_a-j_b+\Lambda+M_\Lambda+\Sigma+M_\Sigma+L+l} M_\Lambda(NLnl; n_a l_a n_b l_b)
$$
$$
R_{nl}(\beta r) R_{NL}(\beta R) Y_{lM_l}(\hat{r}) Y_{LM_L}(\hat{R})
$$
$$
\begin{pmatrix} j_a & j_b & J \\ m_a & m_b & -M_J \end{pmatrix} \begin{pmatrix} \Lambda & \Sigma & J \\ M_\Lambda & M_\Sigma & -M_J \end{pmatrix} \begin{pmatrix} \frac{1}{2} & \frac{1}{2} & \Sigma \\ \sigma_a & \sigma_b & -M_\Sigma \end{pmatrix} \begin{pmatrix} L & l & \Lambda \\ M_L & M_l & -M_\Lambda \end{pmatrix}
$$
$$
P_\sigma \left[ \chi_{\sigma_a}(1) \chi_{\sigma_b}(2) \right] P_\tau \left[ \xi_{\tau_a}(1) \xi_{\tau_b}(2) \right],
$$

that is equal to Eq. (10.24) except for the $P_\sigma$, $P_\tau$ operators and the phase factor $(-1)^{l_a+l_b-L}$.

The matrix element of the interaction is computed in detail in appendix I. We indicate with the labels 0, $\sigma$, $\tau$ and $\sigma\tau$ the scalar-isoscalar, vector-isoscalar, scalar-isovector and vector-isovector matrix elements of the interaction, respectively.

In the following, we use also the notation

$$
\mathscr{F}(\tau) = \begin{cases} \delta_{\tau_a\tau_c} \delta_{\tau_b\tau_d} & \text{for } \langle ab|\bar{V}|cd\rangle_{0,\sigma} \\ \sum_\mu (-)^{1+\tau_a+\tau_b+\mu} \begin{pmatrix} \frac{1}{2} & 1 & \frac{1}{2} \\ \tau_c & \mu & -\tau_a \end{pmatrix} \begin{pmatrix} \frac{1}{2} & 1 & \frac{1}{2} \\ \tau_d & -\mu & -\tau_b \end{pmatrix} & \text{for } \langle ab|\bar{V}|cd\rangle_{\tau,\sigma\tau} \end{cases}
$$
$$
\mathscr{G}(\Sigma) = \begin{cases} 1 & \text{for } \langle ab|\bar{V}|cd\rangle_{0,\tau} \\ (-)^{1+\Sigma} \begin{Bmatrix} \frac{1}{2} & \frac{1}{2} & \Sigma \\ \frac{1}{2} & \frac{1}{2} & 1 \end{Bmatrix} & \text{for } \langle ab|\bar{V}|cd\rangle_{\sigma,\sigma\tau} \end{cases}
$$
$$
\mathscr{N} = \begin{cases} 1 & \text{for } \langle ab|\bar{V}|cd\rangle_{0} \\ -\frac{3}{2} & \text{for } \langle ab|\bar{V}|cd\rangle_{\sigma,\tau} \\ -9 & \text{for } \langle ab|\bar{V}|cd\rangle_{\sigma\tau} \end{cases}
$$
$$
\mathscr{M}(L_i) = \begin{cases} \left(1 - \frac{1}{4}(-1)^{l_c+l_d-L_i}\right) & \text{for } \langle ab|\bar{V}|cd\rangle_{0} \\ (-1)^{l_c+l_d-L_i} & \text{for } \langle ab|\bar{V}|cd\rangle_{\sigma,\tau,\sigma\tau} \end{cases}
$$

$$
\langle ab|\bar{V}|cd\rangle_{0,\sigma,\tau,\sigma\tau} =
$$
$$
\mathscr{N}\mathscr{F}(\tau) \sum_{JM_J} \sum_{\substack{\Lambda\Sigma \\ Ll}} i^{-l_a-l_b+l_c+l_d} (-)^{j_a-j_b+j_c-j_d} (-)^l \mathscr{G}(\Sigma) \begin{Bmatrix} l_a & l_b & \Lambda \\ \frac{1}{2} & \frac{1}{2} & \Sigma \\ j_a & j_b & J \end{Bmatrix} \begin{Bmatrix} l_c & l_d & \Lambda \\ \frac{1}{2} & \frac{1}{2} & \Sigma \\ j_c & j_d & J \end{Bmatrix} \quad (10.29)
$$
$$
\mathscr{M}(L) \frac{\hat{J}^2 \hat{\Lambda}^2 \hat{\Sigma}^2 \hat{j}_a \hat{j}_b \hat{j}_c \hat{j}_d}{\hat{l}} \begin{pmatrix} j_a & j_b & J \\ m_a & m_b & -M_J \end{pmatrix} \begin{pmatrix} j_c & j_d & J \\ m_c & m_d & -M_J \end{pmatrix}
$$
$$
\sum_{\substack{n_i N_i \\ n_f N_f}} M_\Lambda(N_f L n_f l; n_a l_a n_b l_b) M_\Lambda(N_i L n_i l; n_c l_c n_d l_d)
$$
$$
\int dR R^2 R_{N_f L}(\sqrt{2}\beta R) g(R) R_{N_i L}(\sqrt{2}\beta R) \int dr dr' \, r^2 r'^2 R_{n_f l}(\beta r') v_{lm}(r', r) R_{n_i l}(\beta r)
$$

We are interested in the particle-particle angular-momentum-coupled matrix element defined in Eq. (10.27). Thus, the four terms in Eq. (10.29) become

$$\langle (n_a l_a j_a \tau_a, n_b l_b j_b \tau_b) J M_J | \bar{V} | (n_c l_c j_c \tau_c, n_d l_d j_d \tau_d) J M_J \rangle_{0,\sigma,\tau,\sigma\tau} =$$
$$= \mathcal{N} \mathcal{F}(\tau) \sum_{\substack{\Lambda\Sigma \\ Ll}} i^{-l_a - l_b + l_c + l_d} (-)^l \mathcal{M}(L) \mathcal{G}(\Sigma) \begin{Bmatrix} l_a & l_b & \Lambda \\ \frac{1}{2} & \frac{1}{2} & \Sigma \\ j_a & j_b & J \end{Bmatrix} \begin{Bmatrix} l_c & l_d & \Lambda \\ \frac{1}{2} & \frac{1}{2} & \Sigma \\ j_c & j_d & J \end{Bmatrix}$$
$$\frac{\hat{\Lambda}^2 \hat{\Sigma}^2 \hat{j}_a \hat{j}_b \hat{j}_c \hat{j}_d}{\hat{l}} \sum_{\substack{n_i N_i \\ n_f N_f}} M_\Lambda(N_f L n_f l; n_a l_a n_b l_b) M_\Lambda(N_i L n_i l; n_c l_c n_d l_d) \tag{10.30}$$
$$\int \mathrm{d}R R^2 R_{N_f L}(\sqrt{2}\beta R) g(R) R_{N_i L}(\sqrt{2}\beta R)$$
$$\int \mathrm{d}r \mathrm{d}r' \, r^2 r'^2 R_{n_f l}(\beta r') v_{lm}(r', r) R_{n_i l}(\beta r)$$

### 10.6.1 The matrix element with the Skyrme interaction

In the case of the standard Skyrme interaction, the $(lm)$−coefficients (see appendix J) are

$$v_{lm}(r', r) = \frac{(-)^l}{\hat{l}} \sum_m \int \mathrm{d}\hat{r} \mathrm{d}\hat{r}' \, \frac{\delta(r)\delta(\theta)\delta(\varphi)}{r^2 \sin\theta} \frac{\delta(r')\delta(\theta')\delta(\varphi')}{r'^2 \sin\theta'} Y^*_{lm}(\hat{r}) Y_{lm}(\hat{r}')$$
$$= \frac{(-)^l}{\hat{l}} \frac{\delta(r)}{r^2} \frac{\delta(r')}{r'^2} \sum_m \frac{\hat{l}}{\sqrt{4\pi}} \frac{\hat{l}}{\sqrt{4\pi}} \delta_{m,0} \tag{10.31}$$
$$= \frac{(-)^l \hat{l}}{4\pi} \frac{\delta(r)}{r^2} \frac{\delta(r')}{r'^2}$$

Let us consider the integral on the relative momentum in Eq. (10.30). We can introduce the Skyrme interaction (10.19).

$$(-)^l \int \mathrm{d}r \mathrm{d}r' \, r^2 r'^2 R_{n_f l}(\beta r') v_{lm}(r', r) R_{n_i l}(\beta r) = \frac{\hat{l}}{4\pi} \int \mathrm{d}r \, R_{n_i l}(\beta r) \delta(r) \int \mathrm{d}r' \, R_{n_f l}(\beta r') \delta(r')$$
$$= \frac{\hat{l}}{4\pi} R_{n_i 0}(0) R_{n_f 0}(0)$$

Since $l = 0$, i.e. the interaction acts only in S waves, the phase factor $(-1)^{l_c + l_d - L} = 1$ from the property of the Brody-Moshinsky coefficients $(-1)^{l_1 + l_2} = (-1)^{L+l}$. Accordingly, the usual coefficients for central, spin, isospin and spin-isospin matrix elements of the interaction. It is worthwhile to redefine the coefficient $\mathcal{N}$ in this way:

$$\mathcal{N} = \begin{cases} \frac{3}{4} & \text{for } \langle ab | \bar{V} | cd \rangle_0 \\ -\frac{3}{2} & \text{for } \langle ab | \bar{V} | cd \rangle_{\sigma,\tau} \\ -9 & \text{for } \langle ab | \bar{V} | cd \rangle_{\sigma\tau} \end{cases}$$

Then the matrix element reads

$$\langle (n_a l_a j_a \tau_a, n_b l_b j_b \tau_b) J M_J | \bar{V} | (n_c l_c j_c \tau_c, n_d l_d j_d \tau_d) J M_J \rangle_{0,\sigma,\tau,\sigma\tau} =$$

$$
\begin{aligned}
=&\mathcal{N}\mathcal{F}(\tau)\sum_{\Sigma L} i^{-l_a-l_b+l_c+l_d}\frac{\hat{L}^2\hat{\Sigma}^2\hat{j}_a\hat{j}_b\hat{j}_c\hat{j}_d}{4\pi}\mathcal{G}(\Sigma)\begin{Bmatrix} l_a & l_b & L \\ \frac{1}{2} & \frac{1}{2} & \Sigma \\ j_a & j_b & J\end{Bmatrix}\begin{Bmatrix} l_c & l_d & L \\ \frac{1}{2} & \frac{1}{2} & \Sigma \\ j_c & j_d & J\end{Bmatrix}\\
&\sum_{\substack{n_iN_i\\ n_fN_f}} M_L(N_fLn_f0;n_al_an_bl_b)M_L(N_iLn_i0;n_cl_cn_dl_d) \qquad (10.32)\\
&R_{n_i0}(0)R_{n_f0}(0)\int \mathrm{d}RR^2R_{N_fL}(\sqrt{2}\beta R)g(R)R_{N_iL}(\sqrt{2}\beta R)
\end{aligned}
$$

### 10.6.2 The matrix element with the renormalized interaction

For the renormalized interaction the $(lm)-$coefficients (see appendix J) are

$$
\begin{aligned}
v_{lm}^{\lambda\lambda'}(r',r) =&\frac{1}{4\pi^4}\frac{\lambda^2\lambda'^2}{rr'}j_1(\lambda r)j_1(\lambda' r')\frac{(-)^l}{\hat{l}}\sum_m\int \mathrm{d}\hat{r}'\mathrm{d}\hat{r}\, Y^*_{lm}(\hat{r})Y_{lm}(\hat{r}')\\
=&\frac{1}{4\pi^4}\frac{\lambda^2\lambda'^2}{rr'}j_1(\lambda r)j_1(\lambda' r')\frac{(-)^l}{\hat{l}}\sum_m 4\pi\delta_{l,0}\delta_{m,0} \qquad (10.33)\\
=&\frac{1}{4\pi^3}\frac{\lambda^2\lambda'^2}{rr'}j_1(\lambda r)j_1(\lambda' r')\delta_{l,0}
\end{aligned}
$$

Let us consider the integral on the relative momentum in Eq. (10.30). We can introduce the renormalized interaction (10.19).

$$
\begin{aligned}
(-)^l\int \mathrm{d}r\mathrm{d}r'\, r^2r'^2R_{n_fl}(\beta r')v_{lm}(r',r)R_{n_il}(\beta r) =&\\
=\frac{\lambda^2\lambda'^2\delta_{l,0}}{\pi^3}&\int \mathrm{d}r\, r^2R_{n_il}(\beta r)\frac{j_1(r\lambda)}{r}\int \mathrm{d}r'\, r'^2R_{n_fl}(\beta r')\frac{j_1(r'\lambda')}{r'}\\
=\frac{\lambda^2\lambda'^2\delta_{l,0}}{\pi^3}&\int \mathrm{d}r\, rR_{n_i0}(\beta r)j_1(r\lambda)\int \mathrm{d}r'\, r'R_{n_f0}(\beta r')j_1(r'\lambda')
\end{aligned}
$$

Since $l = 0$, i.e. the interaction acts only in S waves, the phase factor $(-1)^{l_c+l_d-L} = 1$ from the property of the Brody-Moshinsky coefficients $(-1)^{l_1+l_2} = (-1)^{L+l}$. According to this, the usual coefficients for central, spin, isospin and spin-isospin matrix elements of the interaction. It is worthwhile to redefine the coefficient $\mathcal{N}$ in this way:

$$
\mathcal{N} = \begin{cases} \frac{3}{4} & \text{for } \langle ab|\bar{V}|cd\rangle_0 \\ -\frac{3}{2} & \text{for } \langle ab|\bar{V}|cd\rangle_{\sigma,\tau} \\ -9 & \text{for } \langle ab|\bar{V}|cd\rangle_{\sigma\tau}\end{cases}
$$

Then the matrix element reads

$$
\begin{aligned}
\langle(n_al_aj_a\tau_a,n_bl_bj_b\tau_b)JM_J|\bar{V}|(n_cl_cj_c\tau_c,n_dl_dj_d\tau_d)JM_J\rangle_{0,\sigma,\tau,\sigma\tau} =&\\
=\mathcal{N}\mathcal{F}(\tau)\sum_{\Sigma L} i^{-l_a-l_b+l_c+l_d}\hat{L}^2\hat{\Sigma}^2\hat{j}_a\hat{j}_b\hat{j}_c\hat{j}_d\mathcal{G}(\Sigma)&\begin{Bmatrix} l_a & l_b & L \\ \frac{1}{2} & \frac{1}{2} & \Sigma \\ j_a & j_b & J\end{Bmatrix}\begin{Bmatrix} l_c & l_d & L \\ \frac{1}{2} & \frac{1}{2} & \Sigma \\ j_c & j_d & J\end{Bmatrix}
\end{aligned}
$$

$$\frac{\lambda^2\lambda'^2}{\pi^3}\sum_{\substack{n_iN_i\\n_fN_f}} M_L(N_fLn_f0;n_al_an_bl_b)M_L(N_iLn_i0;n_cl_cn_dl_d) \tag{10.34}$$
$$\int dRR^2R_{N_fL}(\sqrt{2}\beta R)g(R)R_{N_iL}(\sqrt{2}\beta R)$$
$$\int dr\, rR_{n_i0}(\beta r)j_1(r\lambda)\int dr'\, r'R_{n_f0}(\beta r')j_1(r'\lambda')$$

Note that Eq. (10.34) can be obtained from Eq. (10.32) by making the substitution:

$$\frac{1}{4\pi}R_{n_i0}(0)R_{n_f0}(0)\rightarrow\frac{\lambda^2\lambda'^2}{\pi^3}\int dr\, rR_{n_i0}(\beta r)j_1(r\lambda)\int dr'\, r'R_{n_f0}(\beta r')j_1(r'\lambda') \tag{10.35}$$

## 10.7 Matrix elements of the interaction for Hartree-Fock

From the discussion of section 10.2 it follows that it is important to introduce a cutoff also at mean-field level. We re-write here the matrix element of the HF Hamiltonian, as discussed in section 2.1.3

$$h_{ab}^{(\alpha)} = t_{ab} \tag{10.36}$$
$$+\sum_{\substack{\beta\\ \varepsilon_\beta\le\varepsilon_F}}\sum_{cd} c^*_{\beta,c}\langle ac|\bar{V}|bd\rangle c_{\beta,d} \tag{10.37}$$
$$+\frac{1}{2}\sum_{\substack{\beta\gamma\\ \varepsilon_{\beta,\gamma}\le\varepsilon_F}}\sum_{\substack{cd\\ ef}} c^*_{\beta,c}c^*_{\gamma,d}\langle cd|\frac{\partial\bar{V}}{\partial c^*_{\alpha,a}}|ef\rangle c_{\beta,e}c_{\gamma,f}, \tag{10.38}$$

where the Greek letters represent the set of quantum number which identify a sp state and the Latin letters indicate the harmonic oscillator basis quantum number. The explicit expression for the term (10.36) can be easily found in Ref. [Ber72]. The second term (10.37) can be written using Eq. (10.34). Nevertheless, the expression can be further simplified because of two simple considerations [VB72]:

- we are dealing with even-even nuclei, thus the matrix elements of the operator $\sigma(1)\sigma(2)$ vanishes;
- there is no charge mixing of the HF states, so the isospin exchange operator $P_\tau$ reduces to a Kronecker delta.

With these simplifications and by using the orthogonality relations for the $9-j$ symbol [BS94], we get

$$\sum_{\substack{\beta\\ \varepsilon_\beta\le\varepsilon_F}}\sum_{cd} c^*_{\beta,c}\langle ac|\bar{V}|bd\rangle c_{\beta,d} = \sum_{\substack{\beta\\ \varepsilon_\beta\le\varepsilon_F}}\sum_{cd}\sum_{J}\frac{\hat{J}^2}{\hat{j}^2_\alpha}c^*_{\beta,c}\langle (ac)JM|\bar{V}|(bd)JM\rangle c_{\beta,d}$$
$$=\sum_{\substack{\beta\\ \varepsilon_\beta\le\varepsilon_F}}\sum_{cd} c^*_{\beta,c}c_{\beta,d}\sum_{L}\sum_{\substack{n_iN_i\\ n_fN_f}}\frac{\hat{L}^2\hat{j}^2_\beta}{\hat{l}^2_\alpha\hat{l}^2_\beta}M_L(N_fLn_f0;al_\alpha cl_\beta)M_L(N_iLn_i0;bl_\alpha dl_\beta)$$

$$\times \left(1 - \frac{1}{2}\delta_{q_\alpha,q_\beta}\right) \int \mathrm{d}RR^2 R_{N_f L}(\sqrt{2}\beta R) g(R) R_{N_i L}(\sqrt{2}\beta R) \tag{10.39}$$
$$\times \frac{\lambda^2 \lambda'^2}{\pi^3} \int \mathrm{d}r\, r R_{n_i 0}(\beta r) j_1(r\lambda) \int \mathrm{d}r'\, r' R_{n_f 0}(\beta r') j_1(r'\lambda').$$

Following the same strategy, the last term (10.38), which is the rearrangement term, can be written as

$$\frac{1}{2} \sum_{\substack{\beta\gamma \\ \varepsilon_{\beta,\gamma} \leq \varepsilon_F}} \sum_{\substack{cd \\ ef}} c^*_{\beta,c} c^*_{\gamma,d} \langle cd | \frac{\partial \bar{V}}{\partial c^*_{\alpha,a}} | ef \rangle c_{\beta,e} c_{\gamma,f}$$
$$= \frac{1}{2} \sum_{\substack{\beta\gamma \\ \varepsilon_{\beta,\gamma} \leq \varepsilon_F}} \sum_{\substack{cd \\ ef}} c^*_{\beta,c} c^*_{\gamma,d} c_{\beta,e} c_{\gamma,f} \tag{10.40}$$
$$\times \sum_L \sum_{\substack{n_i N_i \\ n_f N_f}} \frac{\hat{L}^2 \hat{j}^2_\gamma \hat{j}^2_\beta}{\hat{l}^2_\gamma \hat{l}^2_\beta} M_L(N_f L n_f 0; c l_\beta d l_\gamma) M_L(N_i L n_i 0; e l_\beta f l_\gamma)$$
$$\times \left(1 - \frac{1}{2}\delta_{q_\gamma,q_\beta}\right) \int \mathrm{d}RR^2 R_{N_f L}(\sqrt{2}\beta R) g'(R) R_{N_i L}(\sqrt{2}\beta R) \tag{10.41}$$
$$\times \frac{\lambda^2 \lambda'^2}{\pi^3} \int \mathrm{d}r\, r R_{n_i 0}(\beta r) j_1(r\lambda) \int \mathrm{d}r'\, r' R_{n_f 0}(\beta r') j_1(r'\lambda'),$$

where $g'(R) = \frac{t_3 \alpha}{24\pi} R_{a l_\alpha}(\beta R) R_{b l_\alpha}(\beta R) \rho^{\alpha-1}(R)$.

Chapter 11

# Numerical results

In this chapter we present our results for the total energy of $^{16}$O, computed including the contributions up to second order, already discussed in section 7.1.1. This attempt can be considered as a natural continuation of the work done in nuclear matter in Refs. [Mog+10; Mog+12a]. A simplified Skyrme interaction is used, namely, the velocity independent part of the SkP parameterization (in the following we refer to it simply as SkP), without the Coulomb, the spin-orbit and without any tensor interaction.

From the discussion in chapter 10, we expect that the total energy diverges: the features of the divergence are analyzed in section 11.3. In order to avoid this behavior, we want to introduce a cutoff on the relative momenta of the particles included in the model space as explained in chapter 10. Therefore, it is more convenient to write the second order contribution to the total energy with particle-particle, rather than particle-hole, angular-momentum-coupled matrix elements. The former is related to the latter by a purely geometric transformation, known as Pandya relation (see e.g. Refs. [BM69])

$$\langle (ab)J|V|(cd)J\rangle = \sum_{J'}(2J'+1)(-)^{j_c+j_d+J}\begin{Bmatrix} j_a & j_c & J' \\ j_d & j_b & J \end{Bmatrix} V_{J'}(abcd).$$

Therefore, Eq. (7.20) becomes

$$\Delta E = \frac{1}{4}\sum_{\substack{pp' \\ hh'}J} \frac{(2J+1)\,|\langle (pp')J|V|(hh')J\rangle|^2}{\epsilon_h+\epsilon_{h'}-\epsilon_p-\epsilon_{p'}}. \tag{11.1}$$

The divergence is caused by the fact that there is no upper limit in the energy of the particle states included in the model space used for the calculation. Instead of putting a somewhat arbitrary cutoff on the energy of the particles in the model space, we want to introduce a cutoff on the relative momentum of the particles in the initial and final states of the matrix elements of the interaction, applying the same procedure used in infinite matter (see chapter 10) and compute the second order total energy for different model spaces. We recall here that the cutoff $\Lambda$ considered when treating the divergence in nuclear matter is related with the cutoff $\lambda$ on the relative momentum of the particles used in finite nuclei by $\lambda = \sqrt{2}\Lambda$ (see section 10.2).

Our final goal is to perform a calculation of the total energy of the system including contributions up to second order beyond the mean-field using the interactions SkP$_\Lambda$, fitted in nuclear matter and listed in Table 10.1, in order to check whether the refitted interactions are able to cure or at least reduce the divergence of this observable. If these refitted interaction were successfully usable in finite systems, we would obtain a value for the second order total energy equal to the one obtained at mean-field level.

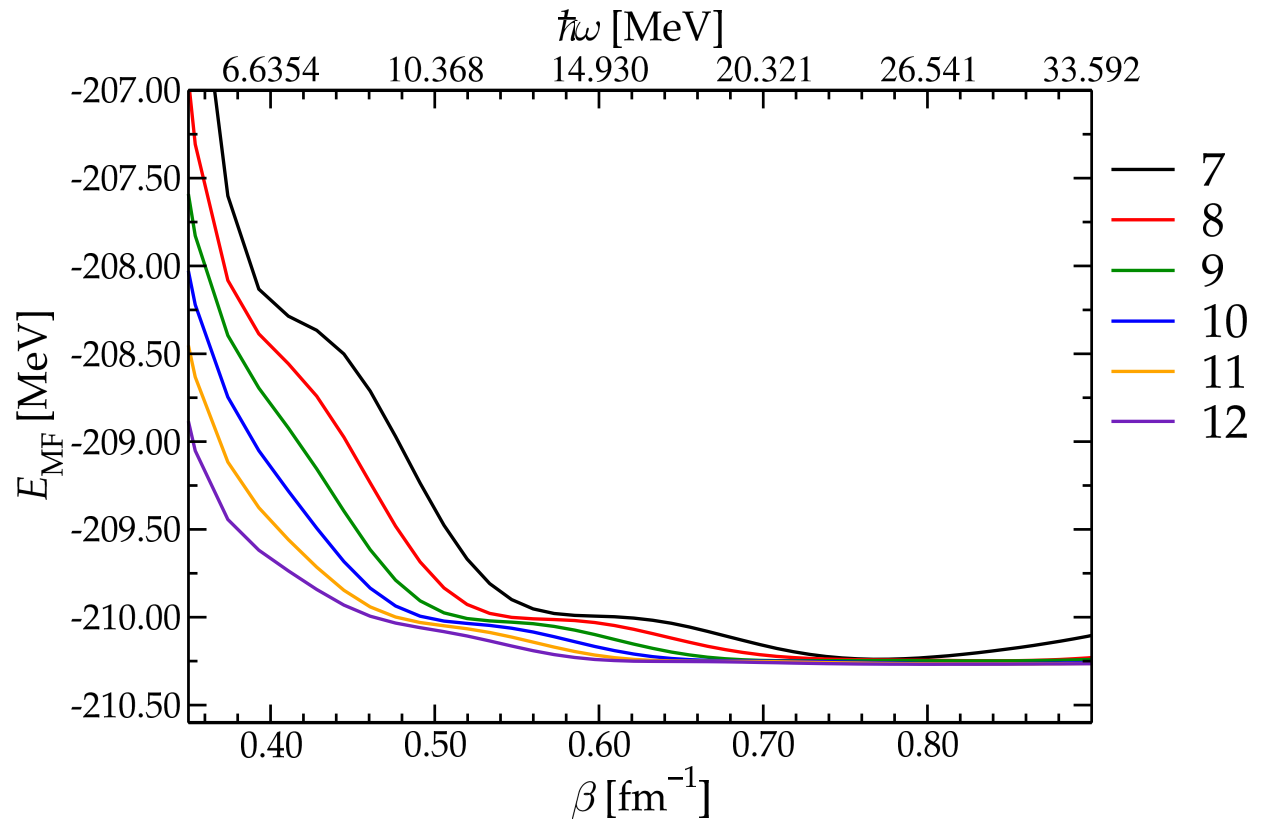

**Figure 11.1**
Mean-field total energy $E_{\mathrm{MF}}$ as a function of the harmonic oscillator parameter $\beta$ (lower axis) or the shell gap $\hbar\omega$ (upper axis). Each line represents values obtained with a different maximum value of the harmonic oscillator principal number $n_{\max}$, as reported in the legend.

Following the discussion in section 10.2, it is clear that the cutoff on the relative momenta of the particles should be introduced at mean-field as well as beyond mean-field level, thus the practical calculations are based on the following steps:

1. Solution of the HF equations on a harmonic oscillator basis. The dimension of the basis will be discussed in the following. In the matrix elements of the interaction, the cutoff is included.

2. Computation of the second order contribution (11.1), introducing in the matrix elements the same cutoff used at mean-field level.

We set the same cutoff $\lambda$ on relative momenta on initial and final states of the matrix elements of the interaction. Note that this calculations are self-consistent, in the sense that the same interaction is used at mean-field and beyond mean-field level. However, we do not solve iteratively the Dyson equation for the sp propagator considering the Hartree-Fock and the second order self-energy at the same level: we add the second order correction on top of Hartree-Fock, thus in a perturbative way. The consequences of this will be discussed in section 11.5.

## 11.1 The harmonic oscillator basis parameters

The first step of our approach is the solution of the HF equations on a harmonic oscillator basis, which depends on two parameters: the harmonic oscillator parameter $\beta = \sqrt{\frac{m\omega}{\hbar}}$, related to the shell gap $\hbar\omega$; and $n_{\max}$, the maximum number of the harmonic oscillator principal quantum number used in the expansion. Fig. 11.1 gives an illustration of the variation of the mean-field total energy when the two parameters of the basis are varied and no cutoff is introduced. We recall that the total energy of $^{16}$O is 127.619 MeV [Wan+12], even if we know that we cannot converge to this value since only a part of the Skyrme interaction is used in the present model and we do not want to refit it at this stage.

The energy decreases when $n_{\max}$ increases, which reflects the fact that in this way the variational space is enlarged. The exact solution is not expected to be reached unless $n_{\max}$ becomes infinite, but it is well known that a relatively small basis ($n_{\text{max}} \approx 10$) allows a fairly good description of the ground state wave functions for nuclei as heavy as $^{208}$Pb [BG77]. For a given value of $n_{\text{max}}$ the energy exhibits in general a minimum when $\beta$ varies. As $n_{\text{max}}$ increases, the energy becomes more and more independent on the choice for $\beta$. In general, a reasonable value for $\beta$ is the one for which the harmonic oscillator potential resembles a standard Woods-Saxon potential.

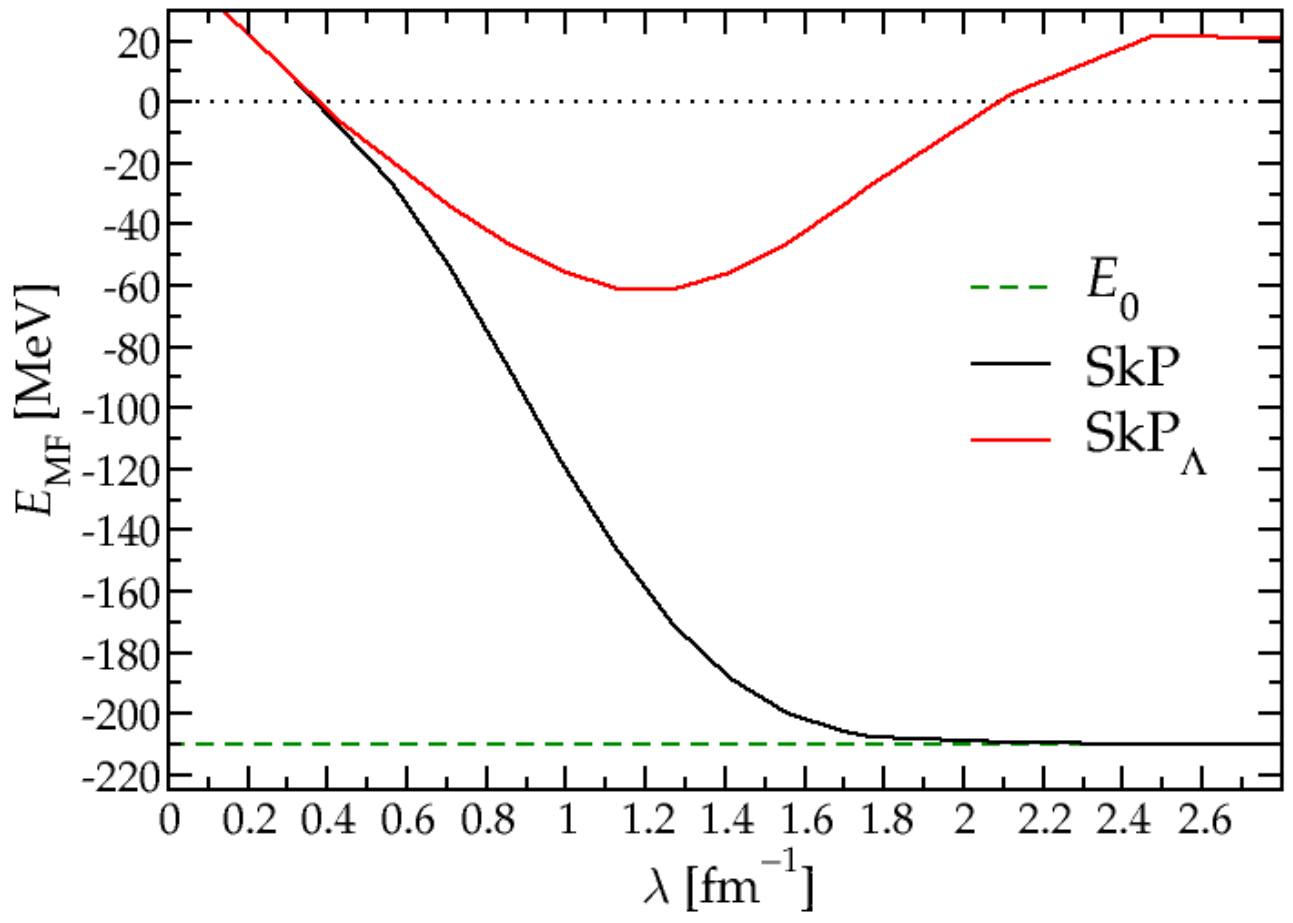

**Figure 11.2**
Mean-field total energy $E_{\text{MF}}$ as a function of the cutoff $\lambda$. The black curve represent the HF total energy obtained with the SkP interaction and the cutoff $\lambda$ is introduced. The red curve shows the same quantity, computed with the renormalized SkP$_\Lambda$ interactions. The green dashed curve represents as a reference the value of the HF total energy $E_0$ obtained with the SkP interaction and no cutoff included, while the black dotted curve identifies the zero of the binding energy.

Since we want to use the same basis for the calculations with cutoffs up to 3 fm$^{-1}$, we use a larger value for $n_{\text{max}}$ ($n_{\text{max}} = 10$) and a smaller one for $\beta$ ($\beta = 0.5\,\text{fm}^{-1}$) than the ones which give a deeper minimum in the total energy (the optimal $\beta$ would be around 0.75 fm$^{-1}$ and $n_{\text{max}} = 8$ would be enough), with an error of about 300 keV. The reason will become clear in the following.

## 11.2 Mean-field energy and density

In Fig. 11.2 we show the results for the mean-field total energy. In particular the dashed line corresponds to the value of the HF total energy $E_0$ obtained with the SkP interaction when no cutoff is included. The black solid line represent the total energy of the system when the SkP interaction is used and a cutoff $\lambda$ is introduced in the matrix elements of the interaction. As expected, for large enough cutoffs the energy converges to $E_0$. This happen when $\lambda \gtrapprox \sqrt{2}k_F \approx 1.8\,\text{fm}^{-1}$, in agreement with what was found in section 10.2. On the other hand, for small values of the cutoff the energy of the system drops and the system becomes more dilute, e.g. if $\lambda = 0.2\,\text{fm}^{-1}$ the mean square radius of the nucleus is around 10 fm. A qualitative explanation may be the following: the introduction of a cutoff $\lambda$ in the momentum space on the relative momentum between two particles implies that, in the coordinate space, the two particles cannot get closer than $\approx \frac{1}{\lambda}$. If $\lambda \sim 0.2 - 0.4\,\text{fm}^{-1}$, then $\frac{1}{\lambda} \sim 2.5 - 6.2\,\text{fm}$ and the two particles can feel only the tail of the mutual interaction.

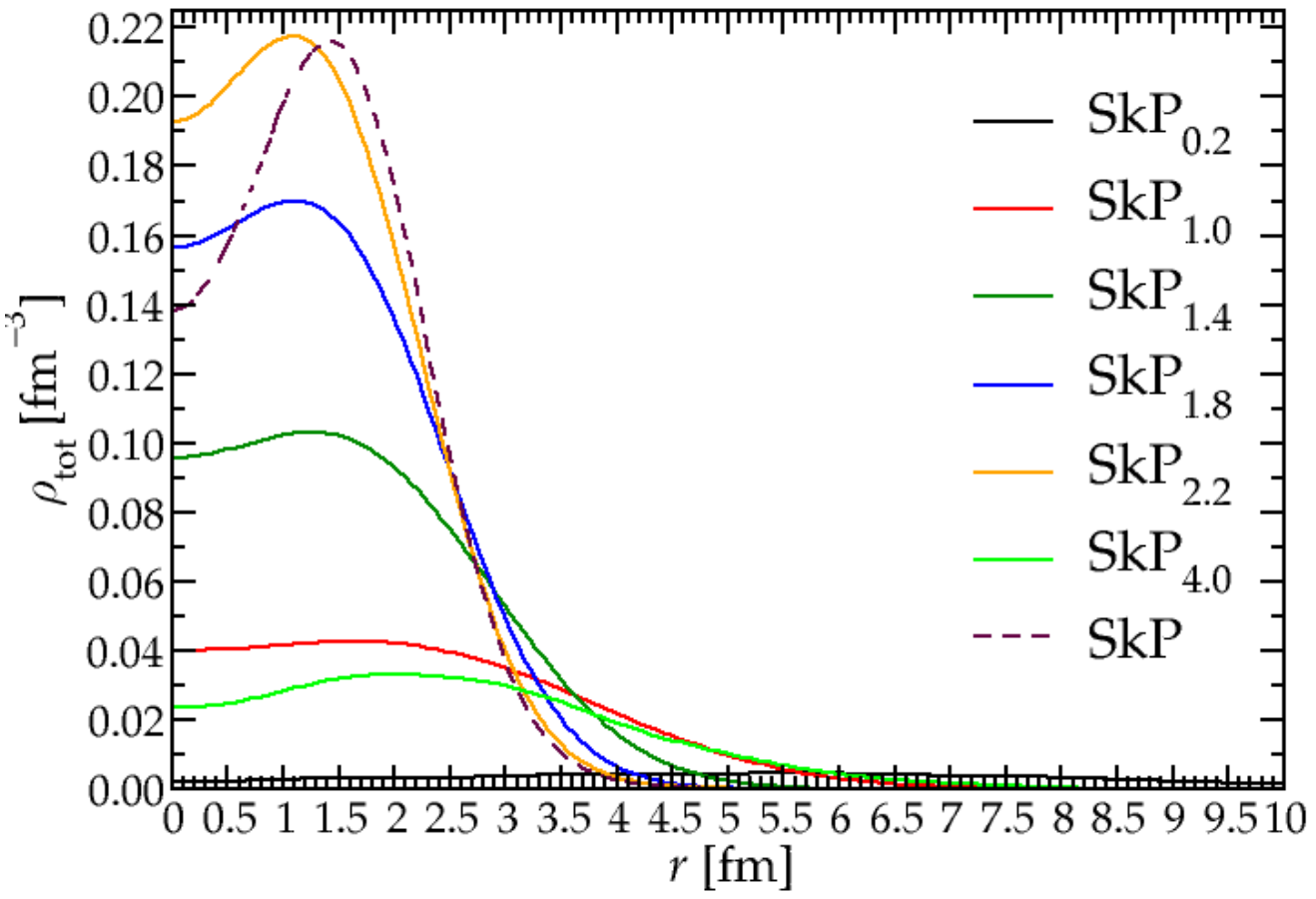

**Figure 11.3**
Density profiles obtained with the $SkP_\Lambda$ interactions. For small and large values of the cutoff the density indicates that the system is dilute.

The red curve in Fig. 11.2 represents the total energy of the system when the $SkP_\Lambda$ interactions are used for each cutoff. For small values of the cutoff the energy does not change because the parameters of the corresponding $SkP_\Lambda$ are barely modified. At the opposite extreme, for large values of the cutoff, the fact the particles can be packed in a smaller volume is balanced by the fact that their mutual interaction becomes weaker and the net effect is that the system is less bound. This behavior is reflected in the densities, which are represented in Fig. 11.3. The curves corresponding to large and small cutoffs describe a dilute system, while for intermediate values of the cutoff the density profile does not qualitatively differ from the one obtained with a standard Skyrme interaction. For this reason we expect that our model can give reasonable results only for intermediate values of the cutoff, where the densities do not differ too much from the density obtained with the original SkP interaction.

## 11.3 Divergence of the second order energy as a function of the model space

In this section we want to explore the dependence of the second order contribution to the total energy on the maximal energy $\varepsilon_p^{\max}$ of the particles in the model space. Since we do not expect the qualitative trends to be modified by the method used in the computation, we solve the usual SHF equations in a box using the code of Ref. [Col+13] and we compute the second order contribution to the total energy on top of it. In Fig. 11.4, we plot the second order contribution to the total energy as a function of the particle model space. As expected, the second order energy diverges and the divergence seems to be linear in this energy range. Incidentally, we can also analyze the dependence on the di-mension of the quantization box used. Each curve corresponds to different values of the radius of the box, from 8 to 30 fm in steps of 2 fm: all the curves lie on top of each others. The radius of the box does not qualitatively change the trend of $\Delta E$ as a function of the particle energy cutoff (quantitatively, the variations are below few percents). This result can be explained as follows. Let us fix the model space. Enlarging the quantization box, the number of states below the energy cutoff increases. Moreover, while the hole wave functions are basically the same since they are governed by the potential well, the particle wave functions are stretched outside the potential well. Because of that, the matrix elements of the interaction should decrease because the overlap between hole and particle wave functions decreases.

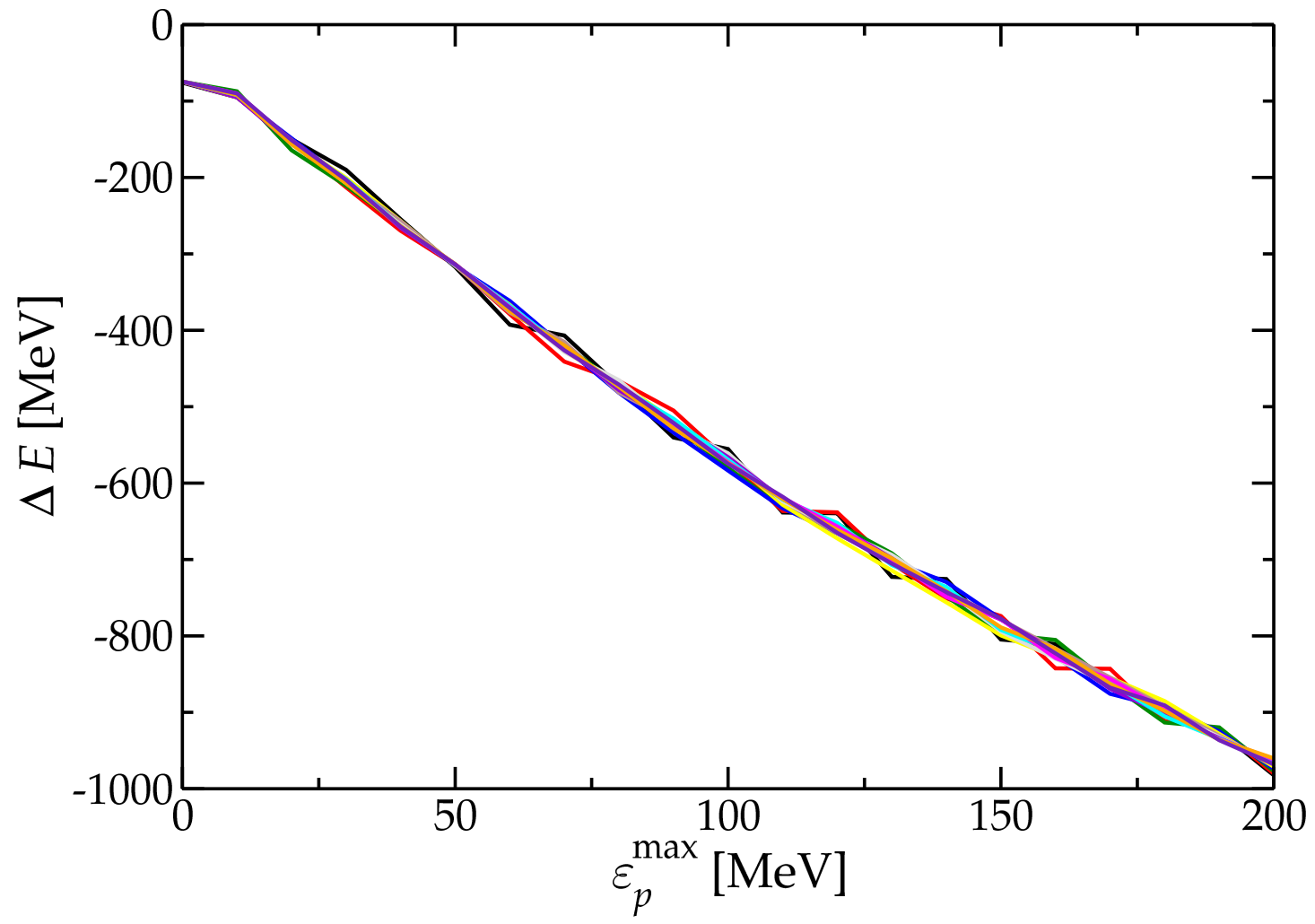

**Figure 11.4**
Second order contribution to the total energy $\Delta E$ as a function of the maximum energy $\varepsilon_p^{max}$ of the particles in the model space. Different colors correspond to different values of the quantization box radius from 8 to 30 fm in steps of 2 fm.

## 11.4 The total energy at second order

In Fig. 11.5 the total energy up to second order (i.e. $E_{\text{MF}}+\Delta E$) is drawn as a function of the maximum energy $\varepsilon_p^{max}$ for the particles, for different values of the cutoff $\lambda$ on the relative momenta. In the plot, the dashed line represent the mean-field total energy $E_0$, computed with the SkP interaction. If the interactions, renormalized in the infinite system, were successfully usable in nuclei, all the curves would collapse onto the dashed line for values of $\varepsilon_p^{max}$ larger than some critical energy. Although this does not happen in our case, we can identify a window in $\varepsilon_p^{max}$ (above 80 MeV) and $\lambda$ (between 0.99 and 1.77 fm$^{-1}$) in which the total energy of the system converge to a given value with an error of about 10−15%. From Fig. 11.6 we can reach the same conclusion. The above-mentioned window here is recognized by the fact that for energies larger than 80 MeV all the curves lie on top of each other.

Figs 11.7-11.8 gathers all the information together in a 3D plot (Fig 11.7) and a density plot (Fig 11.8): the fact that the total energy converges when $\varepsilon_p^{max} \gtrapprox 80$ MeV is represented in Fig 11.7 by the dark valley and by the dark area in Fig 11.8. A further fine tuning of the parameters of the interaction could be eventually performed to enlarge this valley to

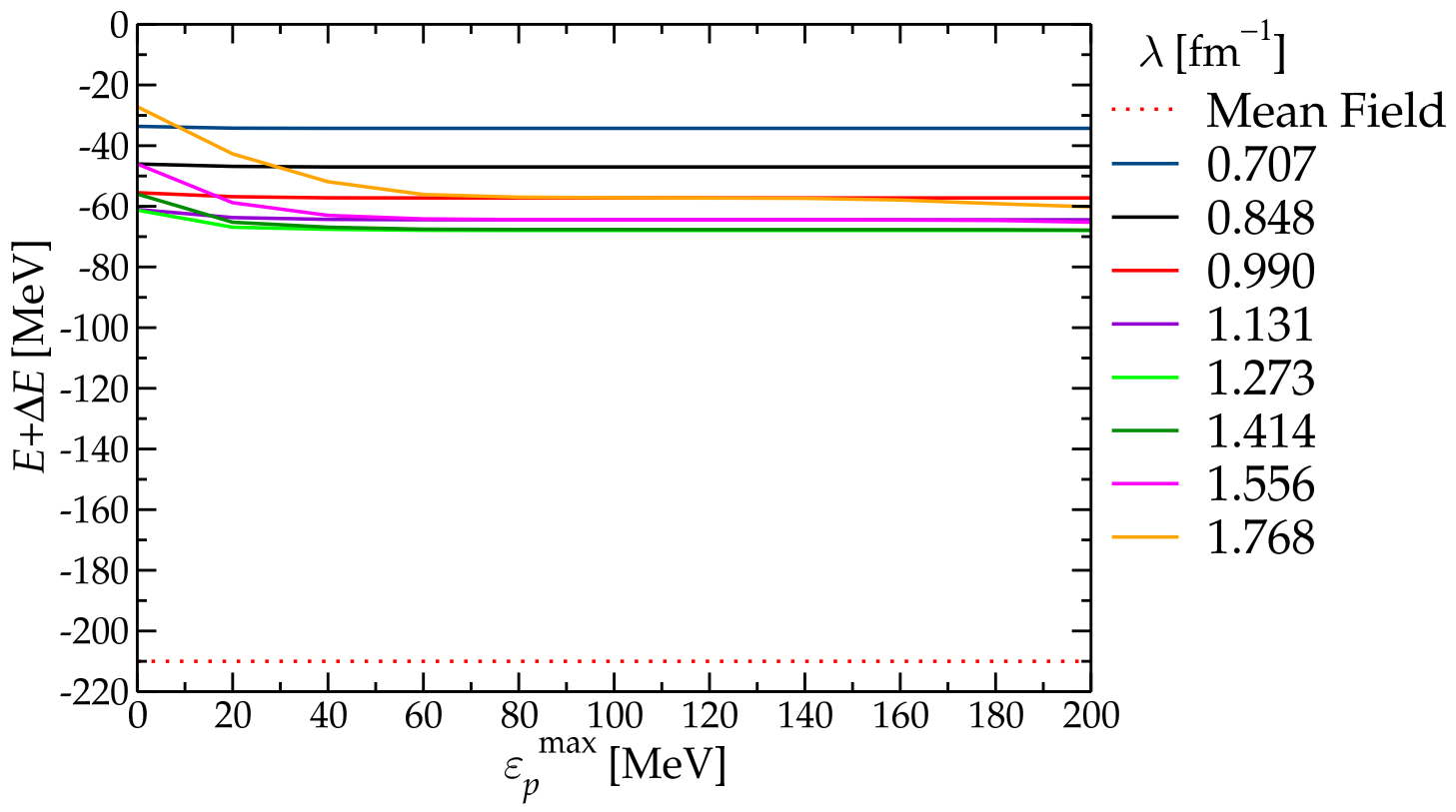

**Figure 11.5**
Total energy total energy $E_{MF} + \Delta E$ as a function of the energy cutoff $\varepsilon_p^{max}$ of the particles in the model space, computed with the refitted interaction, for different values of the cutoff $\lambda$. The red dashed line indicates the mean-field energy computed with the $SkP_\infty$ interaction.

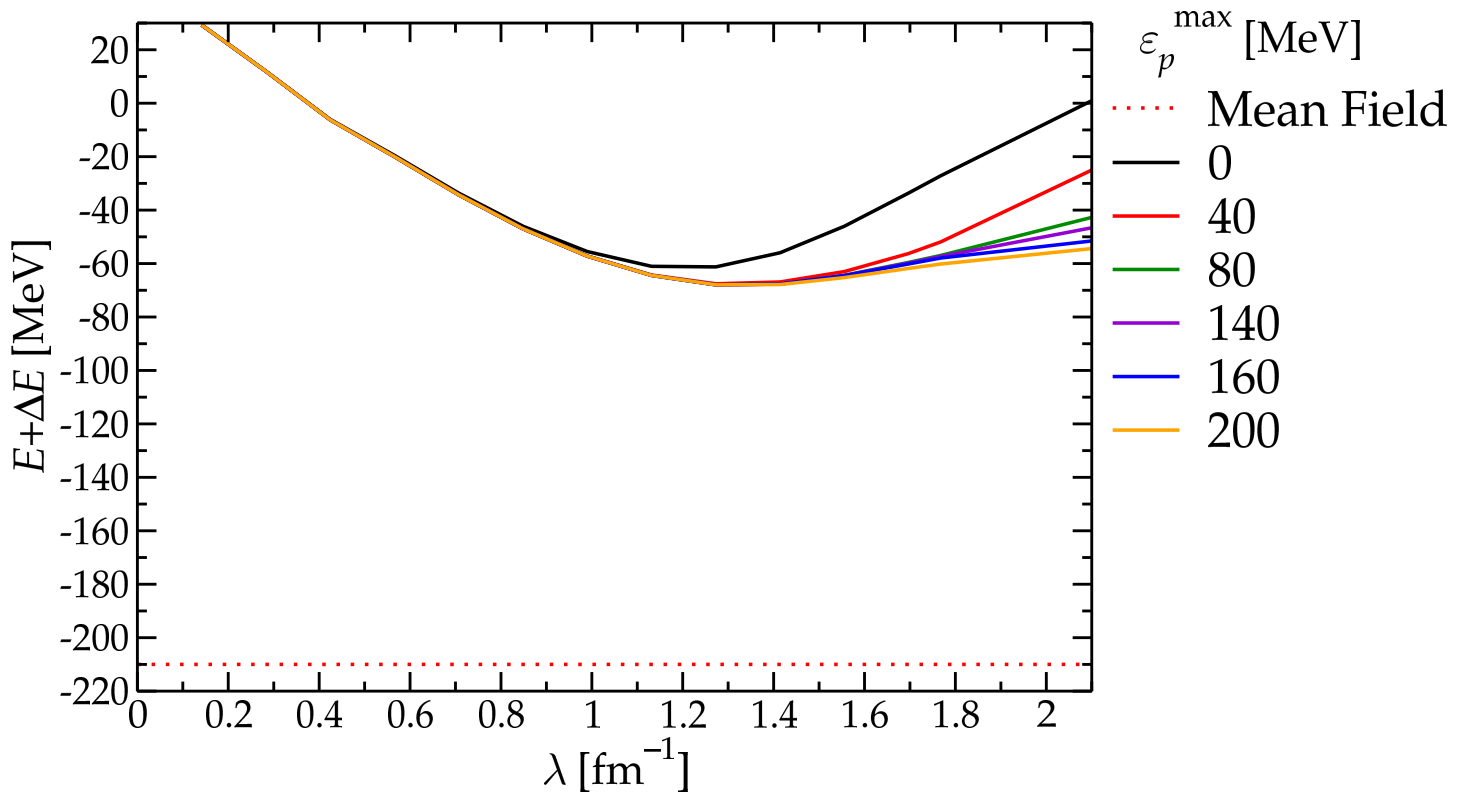

**Figure 11.6**
Total energy total energy $E_{MF} + \Delta E$ as a function of the cutoff $\lambda$, computed with the refitted interaction, for different values of the cutoff on the maximum energy of the particles in the model space.

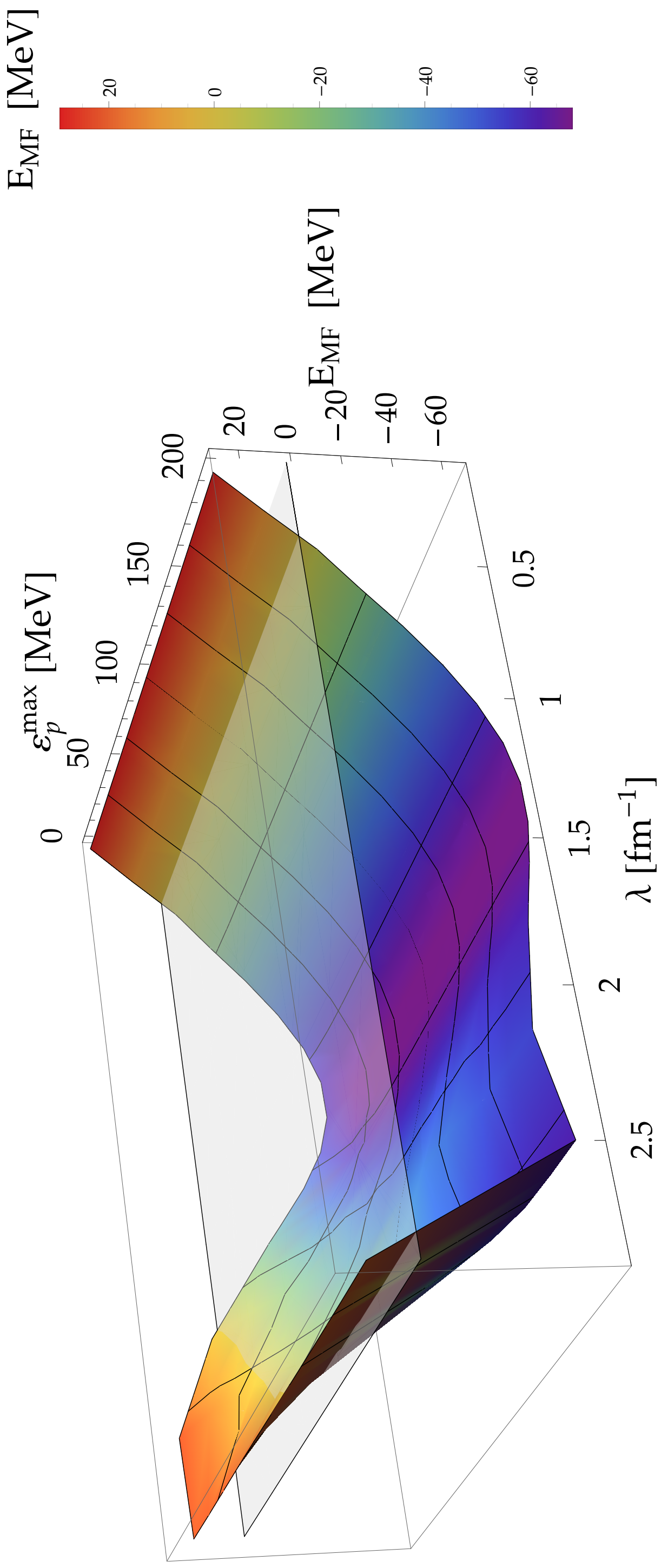

**FIGURE 11.7**
Total energy $E_{\mathrm{MF}} + \Delta E$ as a function of the cutoff $\lambda$ and the energy cutoff $\varepsilon_p^{\max}$ of the particles in the model space, computed with the refitted interaction.

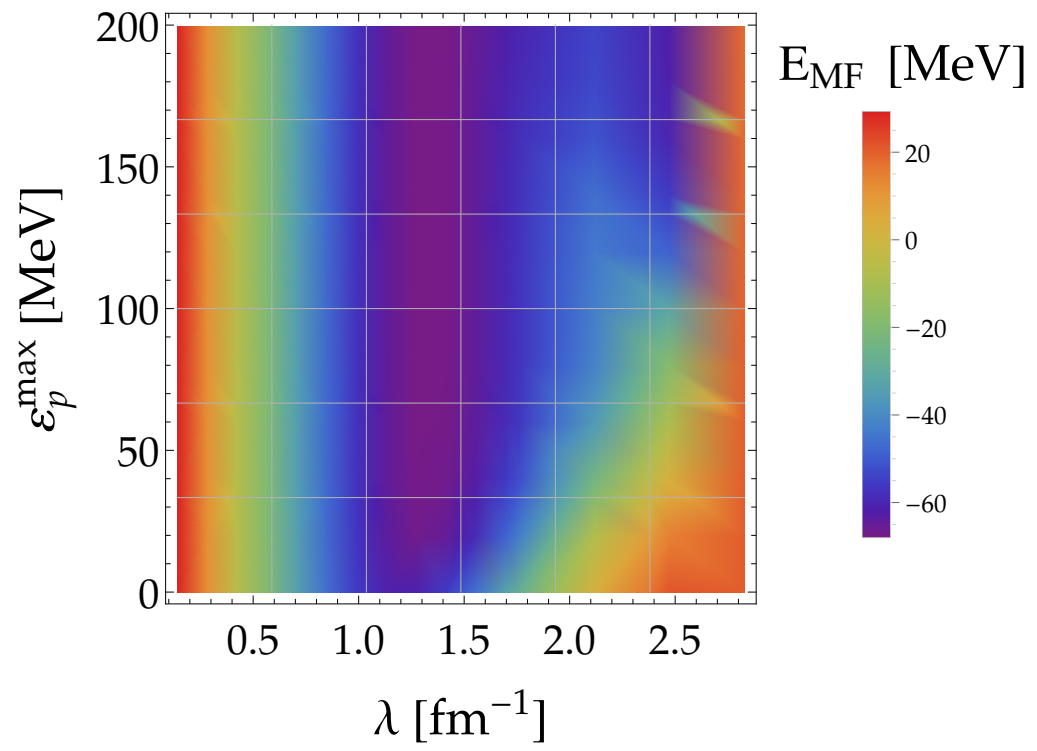

**Figure 11.8**
Density plot of total energy $E_{\mathrm{MF}}+\Delta E$ as a function of the cutoff $\lambda$ and the energy cutoff $\varepsilon_p^{\max}$ of the particles in the model space, computed with the refitted interaction. Lower values of the total energy correspond to the darker regions.

neighbouring values of the cutoff and, more importantly, to re-obtain the value $E_0$ for the total energy up to second order.

## 11.5 Limitations of the model

The reason why the energy of the system at second order level with different SkP$_\Lambda$ interactions (corresponding to different cutoffs $\lambda = \sqrt{2}\Lambda$) differs from the Hartree-Fock energy $E_0$ obtained with the SkP interaction are examined in this short section. At least two possible causes can be identified.

First, in our model a simplified Skyrme interaction is used, in which only the $t_0$ and $t_3$ parts are considered, without Coulomb and spin-orbit interactions. Therefore, the effects of the presence of nuclear surface, which plays a crucial role in many phenomena, are not taken into account properly, due to the lack of the velocity-dependent terms. Although we expect the Coulomb interaction not to change qualitatively the results presented here, from Ref. [Mog+12a], in which the problem of divergences has been successfully attacked in nuclear matter using the complete Skyrme interaction, it is clear that the velocity-dependent and the spin-orbit terms of the Skyrme could change deeply our results. However, in any case, a further adjustment of the parameters of the interactions is mandatory in order to take into account the fact that the nucleus is a finite system.

Moreover, we treat the problem in a perturbative way, adding the second order correction to the total energy on top of a Hartree-Fock diagonalization. For some region of the cutoff, namely $\lambda < 0.5\,\mathrm{fm}^{-1}$ and $\lambda > 2.1\,\mathrm{fm}^{-1}$, the biding energy of the system turns to positive values when the SkP$_\Lambda$ interactions are used: this is a clear evidence that the problem cannot be attacked in a perturbative way in that region. The reason is that the interaction for those values of the cutoff is too weak and the system resembles a gas of nucleons. However, also for intermediate values of the cutoff, the fact that the density profiles derived from the HF procedure with the SkP$_\Lambda$ interactions do not differ too much from the density obtained with the original SkP interaction is not enough in general to claim that the problem can be attacked in a perturbative way. For these reasons, at least for cutoffs $\lambda < 0.5\,\mathrm{fm}^{-1}$ and $\lambda > 2.1\,\mathrm{fm}^{-1}$, a completely self-consistent solution of the Dyson equation for the sp propagator considering together the HF and the second order self-energy is compulsory.

# Conclusions and outlook

In this thesis, our intention has been twofold: on the one side, in part I, to analyze in the standard fully self-consistent Skyrme-Hartree-Fock plus RPA approach the features of some nuclear excitations which have been experimentally studied lately; on the other side, in parts II-III, to implement a beyond mean-field microscopic consistent framework, based on the particle-vibration coupling idea, to compute some exclusive and inclusive properties of giant resonances. This last part has led us to face the problem of the divergences that rise when zero-range effective interactions, like the Skyrme force, are used in beyond mean-field theories.

In part I, some results at mean-field level have been presented. In particular, we focused on the pygmy dipole strength and on the IVGQR.

The low-lying part of the isovector and isoscalar dipole spectrum has been studied in detail in $^{68}$Ni, $^{132}$Sn and $^{208}$Pb, removing in an accurate way the spurious state, with three Skyrme interactions having different isovector properties. In both the isoscalar and isovector channel, the pygmy dipole strength is more prominent and lies at a higher energy in the case of the SkI3 interaction, which is the one with higher value for the $L$ parameter, i.e. the slope of the symmetry energy at saturation density. While in the isovector channel the strength in the pygmy peak is one order of magnitude smaller than the corresponding IVGDR, in the isoscalar spectrum, the strength of the low-lying peak is comparable to that of the ISGDR. The study of the transition densities clarifies some features of the pygmy state: around the surface it is mainly isoscalar rather than isovector; the neutron contribution has a tail that extends beyond the surface and eventually, the behavior in the interior part of the nucleus (which is more difficult to probe in experiments) is more complicated. This picture is confirmed by the study of the isovector response function in which only the RPA states that are 70% isoscalar in the outermost part of the nucleus are considered. A first qualitative indication of the collectivity comes from the strength in single-particle units: the isovector strength is between 2 and 6 sp units, being not enough to completely characterize the nature of the state; on the other hand 15-20 sp units contribute to the isoscalar strength. The study of the particle-hole excitations contributing to the RPA peak gives a deeper insight into the collectivity of the state. The most important contributions come from the neutrons in the outermost orbitals and, although more or less equal in number in the isovector and isoscalar case, only in the latter almost all the excitations sum up coherently. Then, we have been able to conclude that in our model the pygmy strength has a collective nature in the isoscalar channel, being less collective in the isovector one.

The study of the IVGQR in $^{208}$Pb, triggered by some recent results obtained at the HI$\vec{\gamma}$S facility, has been performed with both Skyrme interactions and covariant Lagrangians.

All the models predict for the resonance a predominant isovector nature at the surface and a non-negligible isoscalar contribution in the interior of the nucleus. A macroscopic model for the excitation energy of the IVGQR has been proposed. According to this, the energy of the IVGQR is related to the effective mass and to the symmetry energy at the subsaturation density of 0.1 fm$^{-3}$. Since it is well known (and it has been confirmed also in this work) that the excitation energy of the ISGQR is related to the effective mass, it is possible to express the symmetry energy at subsaturation density as a function of the excitation energy of both isoscalar and isovector GQRs, which are purely experimental observables. Using a Skyrme (SAMi) and a covariant (DD-ME) family of systematically varied functionals, a strong correlation between $[(E^{\mathrm{IVGQR}})^2 - 2(E^{\mathrm{ISGQR}})^2]/J$ and $\Delta r_{np}$ is found for the two families of EDFs considered in this work. Although quite interesting, this result has some degree of model dependence. This means that data on the excitation energy of the ISGQR and the IVGQR may be used to determine the neutron skin thickness of a heavy nucleus and the slope of the symmetry energy at saturation. With this approach we have obtained for the neutron skin thickness of $^{208}$Pb the value $\Delta r_{np} = 0.14 \pm 0.03$ fm, and for the slope parameter of the symmetry energy $L = 37 \pm 18$ MeV. These values are compatible with previous estimates [Tsa+12; LL13].

In part II, we have presented a microscopic particle-vibration coupling model using the Skyrme interaction which has allowed us to implement and compute two important observables related to giant resonances: the spreading width and the $\gamma$-decay width. In our model, therefore, there is no parameter to be adjusted to experimental data, once the Skyrme functional has been chosen. This is the first time in which such microscopic approach is fully fledged including also the velocity dependent term of the Skyrme interaction to obtain both an exclusive property (the spreading width) and an inclusive one (the $\gamma$ decay).

In particular, we have computed the lineshape of the ISGQR and IVGQR in $^{208}$Pb, obtaining that the energy centroid of the resonance is shifted downwards towards the experimental energy in both cases and that the spreading widths are consistent with the experimental findings: for the energy centroid our theoretical results are $E_{\mathrm{ISGQR}} = 11.3$ MeV and $E_{\mathrm{IVGQR}} = 22.0$ MeV, while for the spreading width $\Gamma^{\downarrow}_{\mathrm{ISGQR}} = 2.3$ MeV and $\Gamma^{\downarrow}_{\mathrm{IVGQR}} = 4.0$ MeV.

The $\gamma$-decay width is a more exclusive observable which was studied in the past decades using only phenomenological models. In this work, we have treated the ground-state decay within the fully self-consistent RPA and the decay to low-lying collective vibrations at the lowest contributing order of perturbation theory beyond RPA. We have applied our model to the decay of the ISGQR in $^{208}$Pb and $^{90}$Zr into the ground state and the first low-lying octupole vibration. In particular, in $^{208}$Pb, in the case of the ground-state decay, our outcomes are consistent the experimental findings. In particular, all the Skyrme parameterizations give a $\gamma$-decay width to the ground state of the order of hundreds of electronvolts, though, at the same time, they tend to overestimate it: these discrepancies are due to the fact that the energy of the resonance does not completely agree with the experimental data. For this reason, we have concluded that the $\gamma$ decay to the ground state is not so able to discriminate between different models, at least not more than any other inclusive observables (as energy and strength). However, the decay to low-lying collective states is more sensitive to the interaction used. As a matter of fact, although all the interactions agree that the decay width is only few percent of the ground state decay width, as the experiment indicates, only two interactions (namely SLy5 and SkP) manage to achieve a decay width of few electronvolts, consistently with the experimental finding. Actually, it should be recognized that it is just remarkable that Skyrme interactions can reproduce the order of magnitude (few eV) of this exclusive observable, in keeping with the fact that these functionals are fitted to reproduce basically macroscopic properties of nuclei at the scale of hundreds of keV or MeV. In particular, the description of the dipole spectrum is a crucial point because small differences in the strength of the dipole states, introduced as intermediate states, change significantly the polarization of the nuclear medium. For $^{90}$Zr, the general conclusion is similar: the $\gamma$ decay to low-lying collective states seems to be a good observable to test the quality of different Skyrme models, being very sensitive to the description of the polarization of the nuclear medium.

This model can be applied to the $\gamma$ decay of GRs of other multipolarities as far as doubly closed shell nuclei are considered. However, a further improvement of the model may be envisaged for applications to open shell nuclei, which require the inclusion of the effects of pairing [Sch+13].

Eventually, in part III of this thesis we have reported on the present attempts to overcome the main limitation of the results of part II, viz. the employment of interactions fitted at mean-field level in a higher order framework. As a matter of fact, fitting the interaction at mean-field level would include in an uncontrolled way the beyond mean-field correlations, possibly introducing a double counting when those correlations are explicitly introduced. The other main issue of the aforementioned results is the fact that when zero-range interactions are used in a beyond mean-field framework, divergences arise. This problem has been solved often in an unsatisfactory way, including a somewhat arbitrary cutoff in the model space. The fit of an interaction at beyond mean-field level which can properly renormalize the divergences including a cutoff has been successfully done in nuclear matter [Mog+10; Mog+12a]. In this work, we have presented a possible way to connect the uniform system to nuclei. Introducing a cutoff $\lambda$ on relative momenta (defined as $\frac{k_1-k_2}{\sqrt{2}}$) of the initial and final states entering the matrix elements of the interaction, it is possible to use for applications to finite nuclei the interactions refitted in nuclear matter for each value of $\lambda$.

To test our model we have computed the total energy of $^{16}$O up to second order in perturbation theory, with a simplified Skyrme interaction. We have found a window in $\lambda$ (between 0.99 and 1.77 fm$^{-1}$) and in the energy cutoff imposed on the particle states in the model space (above 80 MeV), in which the total energy of the system converges, when refitted interactions are used. If the refitted interactions could be used in nuclei without any further parameter adjustment, we would expect to find the same value for the energy up to second order and at mean-field level. Nevertheless, in our case the second order total energy is larger than the one obtained in Hartree-Fock. A fine tuning of the parameters of the interactions should be actually envisaged because it is well known that the surface of a finite system plays a crucial role. As a matter of fact, our promising result is so far obtained by using a simplified Skyrme force that does not include the velocity-dependent terms.

The main limitations and possible future improvements of this model has been addressed already in section 11.5. Here we add only the fact that the problem of divergences of the second order contribution to the total energy has been chosen as a springboard for the problem of divergences which rise when particle-vibration coupling corrections are included in the model. The renormalization of these last correlations would be of great importance to obtain a successful interaction, fitted at PVC level, that hopefully can better describe many experimental findings and improve the predictive power of the nuclear mean-field theory. Indeed, generally speaking, the problems associated to the renormalizability of zero-range forces, and the possibility of their refit, are still to a large extent unexplored [MGvK14]; thus this thesis constitutes a step forward in a basically uninvestigated territory.

# Appendices

APPENDIX A

# Matrix element of an excitation operator in RPA

In this appendix we would like to derive the expression (3.31) for the matrix element of a transition operator at RPA level.

Let $F_{\lambda\mu}$ be a generic one-body operator $F_{\lambda\mu} = \sum_{\alpha\beta} F^{\lambda\mu}_{\alpha\beta} c^\dagger_\alpha c_\beta$. We recall here the expression for a generic RPA state in the angular-momentum-coupled formalism:

$$|mJM\rangle = \Gamma^\dagger_m (JM)\, |0\rangle = \sum_{ph} \left[ X^{mJ}_{ph} A^\dagger_{ph} (JM) - Y^{mJ}_{ph} A_{ph} \left(\widetilde{JM}\right)\right] |0\rangle,$$

with

$$\begin{aligned} A^\dagger_{ph} (JM) &= \sum_{m_p m_h} (-)^{j_h - m_h} \langle j_p m_p j_h - m_h | JM \rangle a^\dagger_{j_p m_p} a_{j_h m_h} \\ A_{ph} \left(\widetilde{JM}\right) &= \sum_{m_p m_h} (-)^{J+M+j_h - m_h} \langle j_p m_p j_h - m_h | J - M \rangle a^\dagger_{j_h m_h} a_{j_p m_p} \end{aligned}$$

We have to make the following approximation, due to the fact we are using RPA to treat the collective phonons

$$a = \langle mJM | F_{\lambda\mu} | 0 \rangle \simeq \langle 0 | \left[ \Gamma_m (JM)\, , F_{\lambda\mu} \right] | 0 \rangle$$

where the commutator is

$$\left[ \Gamma_m (JM)\, , F_{\lambda\mu} \right] = \sum_{ph} X^{mJ}_{ph} \left[ A_{ph} (JM)\, , F_{\lambda\mu} \right] - Y^{mJ}_{ph} \left[ A^\dagger_{ph} \left(\widetilde{JM}\right), F_{\lambda\mu} \right] = a_1 + a_2$$

These commutators must be evaluated at the same level of approximation as in the RPA, that is using the quasi-boson commutation relation, reported in Eq. (3.6). The first commutator reads

$$\begin{aligned} \left[ A_{ph} (JM)\, , F_{\lambda\mu} \right] &= \sum_{\substack{m_p m_h \\ \alpha\beta}} (-)^{j_h - m_h} \langle j_p m_p j_h - m_h | JM \rangle F^{\lambda\mu}_{\alpha\beta} \left[ a^\dagger_{j_h m_h} a_{j_p m_p}, a^\dagger_\alpha a_\beta \right] \\ &= \sum_{m_p m_h} (-)^{j_h - m_h} \langle j_p m_p j_h - m_h | JM \rangle F^{\lambda\mu}_{ph} \end{aligned}$$

while the second one is

$$\left[A^{\dagger}_{ph}\left(\widetilde{JM}\right), F_{\lambda\mu}\right] = \sum_{\substack{m_p m_h \\ \alpha\beta}} (-)^{J+M+j_h-m_h}\langle j_p m_p j_h - m_h | J - M\rangle F^{\lambda\mu}_{\alpha\beta}\left[a^{\dagger}_{j_p m_p} a_{j_h m_h}, a^{\dagger}_{\alpha} a_{\beta}\right]$$

$$= -\sum_{m_p m_h} (-)^{J+M+j_h-m_h}\langle j_p m_p j_h - m_h | J - M\rangle F^{\lambda\mu}_{hp}$$

In both these expressions the labels $h$ and $p$ stand for $j_h m_h$ and $j_p m_p$ respectively.

We are interested in the reduced matrix element, that can be evaluated through the Wigner-Eckart theorem

$$\langle mJM|F_{\lambda\mu}|0\rangle = \frac{1}{\sqrt{2J+1}}\langle 00\lambda\mu|JM\rangle\langle mJ||F_{\lambda}||0\rangle = \frac{1}{\sqrt{2J+1}}\delta_{J\lambda}\delta_{M\mu}\langle mJ||F_{\lambda}||mJ\rangle,$$

hence

$$\langle mJ||F_{\lambda}||0\rangle = \sqrt{2J+1}\delta_{J\lambda}\delta_{M\mu}\langle mJM|F_{\lambda\mu}|0\rangle \tag{A.1}$$

The Wigner-Eckart theorem should be used to evaluate the reduced matrix elements corresponding to $F^{\lambda\mu}_{ph}$ and $F^{\lambda\mu}_{hp}$ as well.

$$F^{\lambda\mu}_{ph} = \langle j_p m_p|F_{\lambda\mu}|j_h m_h\rangle = \frac{1}{\sqrt{2j_p+1}}\langle j_h m_h \lambda\mu|j_p m_p\rangle\langle j_p|F_{\lambda}||j_h\rangle$$

$$= \frac{(-)^{j_h-m_h}}{\sqrt{2\lambda+1}}\langle j_p m_p j_h - m_h|\lambda\mu\rangle\langle j_p||F_{\lambda}||j_h\rangle$$

$$F^{\lambda\mu}_{hp} = \langle j_h m_h|F_{\lambda\mu}|j_p m_p\rangle = \frac{1}{\sqrt{2j_h+1}}\langle j_p m_p \lambda\mu|j_h m_h\rangle\langle j_h||F_{\lambda}||j_p\rangle$$

$$= \frac{(-)^{j_p-m_p}}{\sqrt{2\lambda+1}}\langle j_p m_p j_h - m_h|\lambda - \mu\rangle\langle j_h||F_{\lambda}||j_p\rangle$$

The contribution to the reduced matrix element (A.1) which includes the forward RPA amplitudes is

$$a_1 = \sum_{\substack{ph \\ m_p m_h}} X^{mJ}_{ph}\langle j_p m_p j_h - m_h|JM\rangle\langle j_p m_p j_h - m_h|JM\rangle\langle j_p||F_J||j_h\rangle = \sum_{ph} X^{mJ}_{ph}\langle j_p||F_J||j_h\rangle \tag{A.2}$$

The contribution that includes the backward RPA amplitudes is

$$\begin{aligned} a_2 &= \sum_{\substack{ph \\ m_p m_h}} (-)^{J+j_h+j_p} Y^{mJ}_{ph}\langle j_p m_p j_h - m_h|J - M\rangle\langle j_p m_p j_h - m_h|J - M\rangle\langle j_h||F_J||j_p\rangle \\ &= \sum_{ph} Y^{mJ}_{ph}\langle j_p||F_J||j_h\rangle \end{aligned} \tag{A.3}$$

In Eq. (A.2) and Eq. (A.3) the orthogonality relation for the Clebsch-Gordan coefficients has been used.

By summing the two contribution, we get the expression for the reduced transition probability of Eq. (3.31).

Appendix B

# Particle-hole Conjugation

In the first approximation, the ground state of a nucleus is assumed to be a set of completely filled single-particle levels. For a spherically symmetric system, these single-particle states can always be characterized by the quantum numbers $|\alpha\rangle = |nljm_j\rangle$ with $j = |l \pm \frac{1}{2}|$. The parity of these states is $(-)^l$. The ground (vacuum) state of Hartree-Fock theory can be written, in second quantization, as

$$|0\rangle_{\mathrm{HF}} = \prod_{i=1}^{A} c_i^\dagger |0\rangle_{\mathrm{F}}$$

We can then define the operators corresponding to the particle-hole vacuum separately below and above the Fermi surface. The orbitals above the Fermi surface are particle orbitals and those below are hole orbitals. The particle's creation and annihilation operators $(a^\dagger, a)$ and the hole's creation and annihilation operators $(b^\dagger, b)$ are

$$\begin{aligned} c_k &= \theta\,(k - \mathrm{F}) a_k + \theta\,(\mathrm{F} - k) b_k^\dagger \\ c_k^\dagger &= \theta\,(k - \mathrm{F}) a_k^\dagger + \theta\,(\mathrm{F} - k) b_k \end{aligned} \tag{B.1}$$

The particle state associated with the operators $(a^\dagger, a)$ is

$$|k\rangle = a_k^\dagger |0\rangle_{\mathrm{HF}}$$

while the one corresponding to the operators $(b^\dagger, b)$ is indicated as

$$|k^{-1}\rangle = b_k^\dagger |0\rangle_{\mathrm{HF}}.$$

From the anti-commutation relations of the operators $(c^\dagger, c)$, i.e. $\{c_i, c_j^\dagger\} = \delta_{i,j}$ and zero otherwise, similar relations for $(a^\dagger, a)$ and $(b^\dagger, b)$ can be derived:

$$\begin{aligned} \{c_i, c_j^\dagger\} &= \delta_{i,j} \\ &= \{\theta\,(i - \mathrm{F}) a_i + \theta\,(\mathrm{F} - i) b_i^\dagger, \theta\,(j - \mathrm{F}) a_j^\dagger + \theta\,(\mathrm{F} - j) b_j\} \\ &= \theta\,(i - \mathrm{F})\,\theta\,(j - \mathrm{F}) \{a_i, a_j^\dagger\} + \theta\,(\mathrm{F} - i)\,\theta\,(\mathrm{F} - j) \{b_i^\dagger, b_j\} \\ &\quad + \theta\,(i - \mathrm{F})\,\theta\,(\mathrm{F} - j) \{a_i, b_j\} + \theta\,(j - \mathrm{F})\,\theta\,(\mathrm{F} - i) \{b_i^\dagger, a_j^\dagger\}. \end{aligned}$$

Therefore,

$$\begin{cases} \{a_i, a_j^\dagger\} = \delta_{i,j} \\ \{a_i, a_j\} = 0 = \{a_i^\dagger, a_j^\dagger\} \end{cases} \quad \text{if} \quad i, j > \mathrm{F} \tag{B.2}$$

$$\begin{cases} \{b_i, b_j^\dagger\} = \delta_{i,j} \\ \{b_i, b_j\} = 0 = \{b_i^\dagger, b_j^\dagger\} \end{cases} \quad \text{if} \quad i, j < \mathrm{F} \tag{B.3}$$

because $\theta\,(i - \mathrm{F})\,\theta\,(\mathrm{F} - j)$ and $\theta\,(j - \mathrm{F})\,\theta\,(\mathrm{F} - i)$ vanish when $i = j$ due to the properties of the step function, while do not vanish when $i \neq j$ and then $\{a_i, b_j\}$ and $\{b_i^\dagger, a_j^\dagger\}$ must be zero for $i \neq j$.

The quantum numbers of a hole state are related to those of the annihilated particle by time reversal conjugation. In fact, to produce a hole state with angular momentum $jm$ (or linear momentum $p$), we must remove a particle with quantum numbers $j - m$ (or $-p$). To be more explicit, the time reversal operator $\mathscr{T}$ acting on a single-particle state $|jm\rangle$ changes the sign of the projection of the angular momentum leading to

$$\mathscr{T}|jm\rangle = (-)^{p+m}|j - m\rangle. \tag{B.4}$$

The $m$-dependence of the phase is necessary to maintain the correct angular momentum transformation properties but the phase $p$ can be chosen in various ways. We choose it according to Ref. [BM69], i.e. $p = j$. Then, the operation

$$b_{jm}^\dagger|0\rangle_{\mathrm{HF}} = (-)^{j+m} a_{j-m}|0\rangle_{\mathrm{HF}}$$

creates a hole state with angular momentum quantum numbers $(j, m)$.

The $m$-dependent phase in Eq. (B.4) guarantees that the operator $b_i^\dagger$ creates a hole with angular momentum $(j, m)$, which may be proved by showing that $b_i^\dagger$ is an irreducible tensor operator of rank $j$ and component $m$. It is first necessary to construct the angular-momentum operator for the system

$$\begin{aligned} \hat{\boldsymbol{J}} &= \sum_{\alpha\beta} \langle\alpha|\boldsymbol{J}|\beta\rangle c_\alpha^\dagger c_\beta \\ &= \sum_{nljmm'>\mathrm{F}} \langle jm|\boldsymbol{J}|jm'\rangle a_{nljm}^\dagger a_{nljm'} + \sum_{nljmm'<\mathrm{F}} \langle jm|\boldsymbol{J}|jm'\rangle (-)^{j+m} b_{nlj-m} (-)^{j+m'} b_{nlj-m'}^\dagger \end{aligned} \tag{B.5}$$

where the second line follows because the single-particle matrix elements of $\boldsymbol{J}$ are diagonal in $\{nlj\}$ and independent of $n$ and $l$. The last two operators can be written

$$(-)^{j+m} b_{nlj-m} (-)^{j+m'} b_{nlj-m'}^\dagger = (-)^{m-m'} (\delta_{mm'} - b_{nlj-m}^\dagger b_{nlj-m'}),$$

and the first term in parentheses makes no contribution because

$$\sum_m \langle m|\boldsymbol{J}|m\rangle = \hat{z} \sum_m m = 0$$

Furthermore, the Wigner-Eckart theorem shows that the matrix element of $J_{1q}$ satisfies $\langle jm|J_{1q}|jm'\rangle = (-)^{m'-m+1}\langle j{-}m'|J_{1q}|j{-}m\rangle$. With the change of summation variables $(m, m') \rightarrow (-m', -m)$, the angular momentum operator finally becomes

$$\hat{\boldsymbol{J}} = \sum_{nljmm'>\mathrm{F}} \langle jm|\boldsymbol{J}|jm'\rangle a_{nljm}^\dagger a_{nljm'} + \sum_{nljmm'<\mathrm{F}} \langle jm|\boldsymbol{J}|jm'\rangle b_{nljm}^\dagger b_{nljm'} \tag{B.6}$$

The proof that $b_i^\dagger$ is a tensor operator now follows immediately from the relations

$$[\hat{\boldsymbol{J}}, b_i^\dagger] = \sum_{kl} \langle k|\boldsymbol{J}|l\rangle [b_k^\dagger b_l, b_i^\dagger] = \sum_{kl} \langle k|\boldsymbol{J}|l\rangle \delta_{li} b_k^\dagger = \sum_{m_k} \langle jm_k|\boldsymbol{J}|jm_i\rangle b_{jm_k}^\dagger .$$

In general, the basic relation between particles and holes can be written in the form

$$\begin{aligned} b_i^\dagger &= a_{\tilde{i}} \\ b_i &= a_{\tilde{i}}^\dagger \end{aligned} \tag{B.7}$$

where $|i^{-1}\rangle = \mathscr{T}|i\rangle$ is the time reverse of the state $|i\rangle$. We note that, since $\mathscr{T}^2 = -\mathbb{I}$ (and then $b_{\tilde{i}}^\dagger = a_{\tilde{\tilde{i}}} = -a_i$), the inverse relation of Eq. (B.7) is

$$\begin{aligned} a_i^\dagger &= -b_{\tilde{i}} \\ a_i &= -b_{\tilde{i}}^\dagger \end{aligned} \tag{B.8}$$

APPENDIX C

# Calculation of second-order self-energy diagrams

The prototype of beyond mean-field calculation at PVC level is the second order self-energy $\Sigma^{(2)}(a, b; \omega)$ of a single-particle , drawn in Fig. 7.4. It is important to note that, following Ref. [DV08], the matrix elements of the interaction are always antisymmetrized. For this reason, when a pair of sp lines appears in a diagram, a factor $\frac{1}{2}$ should be added. To understand this point, we refer to Fig. C.1 where the first diagram of Fig. 7.4 with the antisymmetrized interaction (l.h.s. – squared vertices) is expanded in the corresponding diagrams with non-antisymmetrized interaction (r.h.s. – rounded vertices).

Therefore, the second order self-energy reads (we refer to Fig. C.2 for the labels)

$$\Sigma^{(2)}(\lambda, \mu; \omega) = (-1)\frac{1}{2}\left(\frac{i}{2\pi}\right)^2 \sum_{\alpha\beta\gamma} \int d\omega_1 d\omega_2 \langle \lambda\gamma|\bar{V}|\alpha\beta\rangle\langle\alpha\beta|\bar{V}|\mu\gamma\rangle \times G^{\rm HF}(\alpha; \omega_1)$$
$$\times G^{\rm HF}(\gamma; \omega_1 + \omega_2 - \omega) G^{\rm HF}(\beta; \omega_2),$$

where the factor $(-1)$ is due to the presence of one fermion loop and $\frac{1}{2}$ comes because $\alpha$ and $\beta$ are equivalent sp lines. Introducing the HF expression for the sp propagator we

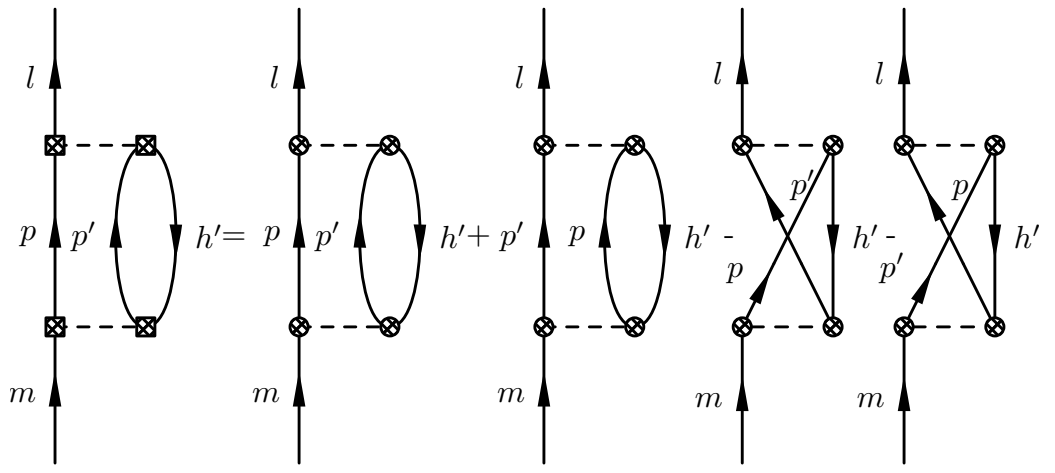

**FIGURE C.1**
Second order self-energy with matrix elements (shaded squares) of the antisymmetrized interaction (l.h.s.) expanded in the corresponding self-energy with matrix elements (shaded circles) of the non-antisymmetrized interaction.

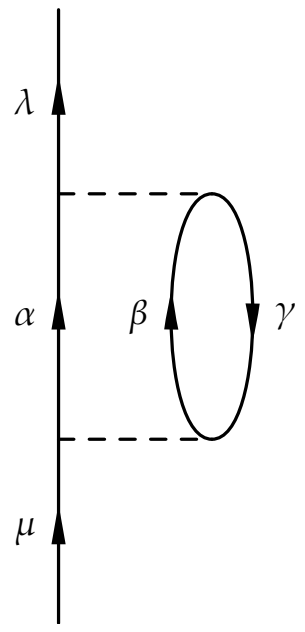

**Figure C.2**
Second order self-energy of the single-particle . Here $\alpha$, $\beta$ and $\gamma$ can be either particles or holes. See the main text for details.

get

$$\Sigma^{(2)}(\lambda,\mu;\omega) = \frac{1}{2(2\pi)^2}\sum_{\alpha\beta\gamma}\langle\lambda\gamma|\bar{V}|\alpha\beta\rangle\langle\alpha\beta|\bar{V}|\mu\gamma\rangle \times \int d\omega_1 d\omega_2 \left(\frac{1}{\omega_1-\varepsilon_\alpha+i\eta}+\frac{1}{\omega_1-\varepsilon_\alpha-i\eta}\right) \times\left(\frac{1}{\omega_2-\varepsilon_\beta+i\eta}+\frac{1}{\omega_2-\varepsilon_\beta-i\eta}\right) \times\left(\frac{1}{\omega_1+\omega_2-\omega-\varepsilon_\gamma+i\eta}+\frac{1}{\omega_1+\omega_2-\omega-\varepsilon_\gamma-i\eta}\right).$$

The energy integrals can be computed using the residue theorem:

$$\mathcal{I}_1(\omega_2;\omega) = \int d\omega_1 \left(\frac{1}{\omega_1-\varepsilon_\alpha+i\eta}+\frac{1}{\omega_1-\varepsilon_\alpha-i\eta}\right) \times\left(\frac{1}{\omega_1+\omega_2-\omega-\varepsilon_\gamma+i\eta}+\frac{1}{\omega_1+\omega_2-\omega-\varepsilon_\gamma-i\eta}\right) = \frac{-2\pi i}{\omega_2-\omega-\varepsilon_\gamma+\varepsilon_\alpha+i\eta}+\frac{2\pi i}{\omega_2-\omega-\varepsilon_\gamma+\varepsilon_\alpha-i\eta}$$

$$\mathcal{I}_2(\omega) = \int d\omega_2 \left(\frac{1}{\omega_2-\varepsilon_\beta+i\eta}+\frac{1}{\omega_2-\varepsilon_\beta-i\eta}\right)\mathcal{I}_1(\omega_2;\omega) = \frac{(2\pi)^2}{\omega-\left(\varepsilon_\alpha+\varepsilon_\beta-\varepsilon_\gamma\right)+i\eta}+\frac{(2\pi)^2}{\omega-\left(\varepsilon_\alpha+\varepsilon_\beta-\varepsilon_\gamma\right)-i\eta}$$

Inserting the expression for the energy integrals into the second order self-energy we have

$$\Sigma^{(2)}(\lambda,\mu;\omega) = \frac{1}{2}\sum_{\alpha\beta\gamma}\langle\lambda\gamma|\bar{V}|\alpha\beta\rangle\langle\alpha\beta|\bar{V}|\mu\gamma\rangle \times\left(\frac{1}{\omega-\left(\varepsilon_\alpha+\varepsilon_\beta-\varepsilon_\gamma\right)+i\eta}+\frac{1}{\omega-\left(\varepsilon_\alpha+\varepsilon_\beta-\varepsilon_\gamma\right)-i\eta}\right) \tag{C.1}$$

Since $\alpha$, $\beta$ and $\gamma$ can be either particles or holes, the only possible combination is $\alpha = p_1$, $\beta = p_2$ and $\gamma = h_3$ or $\alpha = h_1$, $\beta = h_2$ and $\gamma = p_3$ (see Fig. 7.4), we get Eq. (7.14), recalling that when we are dealing with a hole we have to change the sign of the energy of the particle. Actually, we can explicitly sum over the projection of the angular momenta of the intermediate states. Moreover, we have to sum on the angular momentum projections of the final state and make an average on the angular momentum projections of the initial one. In this way, we can express the matrix elements of the interaction as a function of the angular-momentum-coupled matrix elements

$$\langle ab|\bar{V}|cd\rangle = \sum_{JM} (-)^{j_c - m_c + j_b - m_b} \langle j_a m_a j_c - m_c|JM\rangle\langle j_d m_d j_b - m_b|JM\rangle V_J(abcd).$$

Eventually, we obtain Eq. (7.15).

Appendix D

# Schematic model for the NFT

The simple model considered consists of two single-particle levels, each with degeneracy $2\Omega$, and a schematic p-h interaction coupling the particles in the two levels. The total Hamiltonian $\mathcal{H}$ is equal to

$$\mathcal{H} = \mathcal{H}_{sp} + \mathcal{H}_{tb}$$

where

$$\begin{aligned}
\mathcal{H}_{sp} &= \frac{1}{2}\epsilon N_0 \\
N_0 &= \sum_{\substack{\sigma=\pm 1 \\ m=\pm 1,\pm 2\ldots}} \sigma a^\dagger_{m\sigma} a_{m\sigma} \\
\mathcal{H}_{tb} &= -V\Omega\left(A^\dagger_1 A_1 + A_1 A^\dagger_1\right) \\
A^\dagger_1 &= \frac{1}{\sqrt{2\Omega}} \sum_m a^\dagger_{m,1} a_{m,-1}
\end{aligned}$$

The index $\sigma$ labels the two levels, while $m$ labels the degenerate states within each level. The distance between the two levels is $\epsilon$ and the strength of the interaction is $V$.

The closed-shell system of this model corresponds to the lowest ($\sigma = -1$) level filled with $2\Omega$ particles while the upper ($\sigma = 1$) one remains empty. The particle and hole states are obtained by adding or removing a single-particle from this configuration. The corresponding wave function and energies are

$$|m,1\rangle = a^\dagger_{m,1}|0\rangle \qquad E(m,1) = \frac{1}{2}(\epsilon + V) \tag{D.1a}$$

$$|m,-1\rangle = a_{m,-1}|0\rangle \qquad E(m,-1) = \frac{1}{2}(\epsilon + V) \tag{D.1b}$$

The two-body interaction can be expressed in another way, by means of the canonical

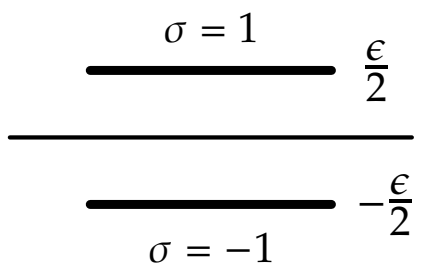

**Figure D.1**
Schematic representation of the two levels system anticommutation relations between the operators $a$ and $a^\dagger$ in the second part of the expression.

$$\mathcal{H}_{tb} = -\frac{V}{2}\left(\sum_{mm'} a^\dagger_{m,1}a_{m,-1}a^\dagger_{m',-1}a_{m',1} + \sum_{mm'} a^\dagger_{m',-1}a_{m',1}a^\dagger_{m,1}a_{m,-1}\right) =$$

$$= -\frac{V}{2}\left(2\sum_{mm'} a^\dagger_{m,1}a_{m,-1}a^\dagger_{m',-1}a_{m',1} - \sum_{m} a^\dagger_{m,1}a_{m,1}\sum_{m} a^\dagger_{m,-1}a_{m,-1}\right)$$

Then the Hamiltonian of the system can be rewritten as

$$\mathcal{H} = \frac{1}{2}(\epsilon + V)\, N_0 - 2V\Omega A_1^\dagger A_1$$

and Eq. (D.1a) and Eq. (D.1b) can be directly found:

$$\langle m,1|H|m,1\rangle = \frac{1}{2}(\epsilon + V)\sum_{\substack{\sigma=\pm1\\ m=\pm1,\pm2...}}\langle m,1|\sigma a^\dagger_{m\sigma}a_{m\sigma}|m,1\rangle$$

$$\langle m,-1|H|m,-1\rangle = \frac{1}{2}(\epsilon + V)\sum_{\substack{\sigma=\pm1\\ m=\pm1,\pm2...}}\langle m,-1|\sigma a^\dagger_{m\sigma}a_{m\sigma}|m,-1\rangle$$

Therefore, the unperturbed energy necessary to produce a p-h excitation with respect to the ground state is

$$\epsilon' = E\,(m,1) + E\,(m,-1) = \epsilon + V$$

where $V$ is the HF contribution to the p-h excitation.

For the collective states, having defined the normal mode creation operator

$$\beta^\dagger_\nu = \sum_m \lambda^\nu_m a^\dagger_{m,1}a_{m,-1},$$

the RPA yields

$$\omega_1 = \epsilon' - 2V\Omega \qquad\qquad \omega_\nu = \epsilon'\ (\nu = 2,3,\ldots,2\Omega)$$

corresponding to the normal modes

$$|n_\nu = 1\rangle = \beta^\dagger_1|0\rangle$$

and to the $2\Omega - 1$ other normal modes $i = 2,3,\ldots,2\Omega$ forming an orthogonal basis in the remaining space of p-h excitations.

Utilizing the normalization condition $\left[\beta_\nu, \beta^\dagger_\nu\right] = \delta\,(\nu,\nu')$, for the amplitudes $\lambda^1_m$ associated with the lowest mode

$$\lambda^1_m = \frac{1}{\sqrt{2\Omega}}$$

Being $\Lambda_1$ the matrix element of the coupling, this amplitude can be rewritten as $\lambda^1_m =$

$\frac{\Lambda_1}{\omega_1 - \epsilon'}$ and we obtain

$$\Lambda_1 = -V\sqrt{2\Omega}$$

This can also be seen by calculating the matrix element of the interaction Hamiltonian between the normal mode $|n_\nu = 1\rangle$ and the single p-h state.

$$\Lambda_\nu = \langle n_\nu = 1| \mathcal{H}_{tb} |m, 1; m', -1\rangle = -V\sqrt{2\Omega}\delta\left(m, m'\right)\delta\left(\nu, 1\right) \tag{D.2}$$

We postulate the independence of the collective mode $\beta_1^\dagger$ and of the p-h excitations; this implies the assumption of the orthogonality of both type of excitations.

$$\left[\beta_1, \sum_m a^\dagger_{m,1} a_{m,-1}\right] = 0$$

The interaction Hamiltonian can be expressed in terms of both $\beta_1^\dagger$ and $A_1^\dagger$ as

$$\mathcal{H}_{pv} = \mathcal{H}_{tb}\left(\beta_1^\dagger A_1, \beta_1 A_1^\dagger\right) = -V\Omega\left(\beta_1^\dagger A_1 + \beta_1 A_1^\dagger\right) + \text{h.c.} = -2V\Omega\left(\beta_1^\dagger A_1 + \text{h.c.}\right) =$$
$$= \Lambda_1\left(\beta_1^\dagger \sum_m a^\dagger_{m,1} a_{m,-1} + \text{h.c.}\right)$$

in which case we refer to it as the particle-vibration coupling Hamiltonian. Thus, the total Hamiltonian is the one in Eq. (7.21).

APPENDIX E

# Calculation of NFT diagrams

In this appendix we want to evaluate explicitly some of the relevant diagrams for this work (see chapter 7).

## E.1 The second order self-energy including vibrational states

### E.1.1 Calculation with the polarization propagator

In this section we want to show the complete calculation of the second-order self-energy of a nucleon including the particle-vibration coupling. The corresponding diagrams is drawn in Fig. E.1. We give here the expression only for the self-energy of a particle (Figs. E.1(A)-(B)), being the one for a hole equivalent. The intermediate sp state is called $\kappa$.

$$\Sigma^{\mathrm{PVC}}(\lambda,\mu;\omega) = \frac{i}{2\pi}\sum_{\substack{\alpha\beta\gamma\delta\kappa \\ \varepsilon_\kappa>\varepsilon_F}}\langle\lambda\beta|\bar{V}|\kappa\alpha\rangle\langle\kappa\gamma|\bar{V}|\mu\delta\rangle \times \int \mathrm{d}\omega_1\, G^{\mathrm{HF}}(\kappa;\omega_1)\Pi^{\mathrm{RPA}}(\alpha,\beta^{-1};\gamma,\delta^{-1};\omega-\omega_1)$$
$$\frac{i}{2\pi}\sum_{\substack{\alpha\beta\gamma\delta\kappa \\ \varepsilon_\kappa\le\varepsilon_F}}\langle\lambda\beta|\bar{V}|\kappa\alpha\rangle\langle\kappa\gamma|\bar{V}|\mu\delta\rangle$$

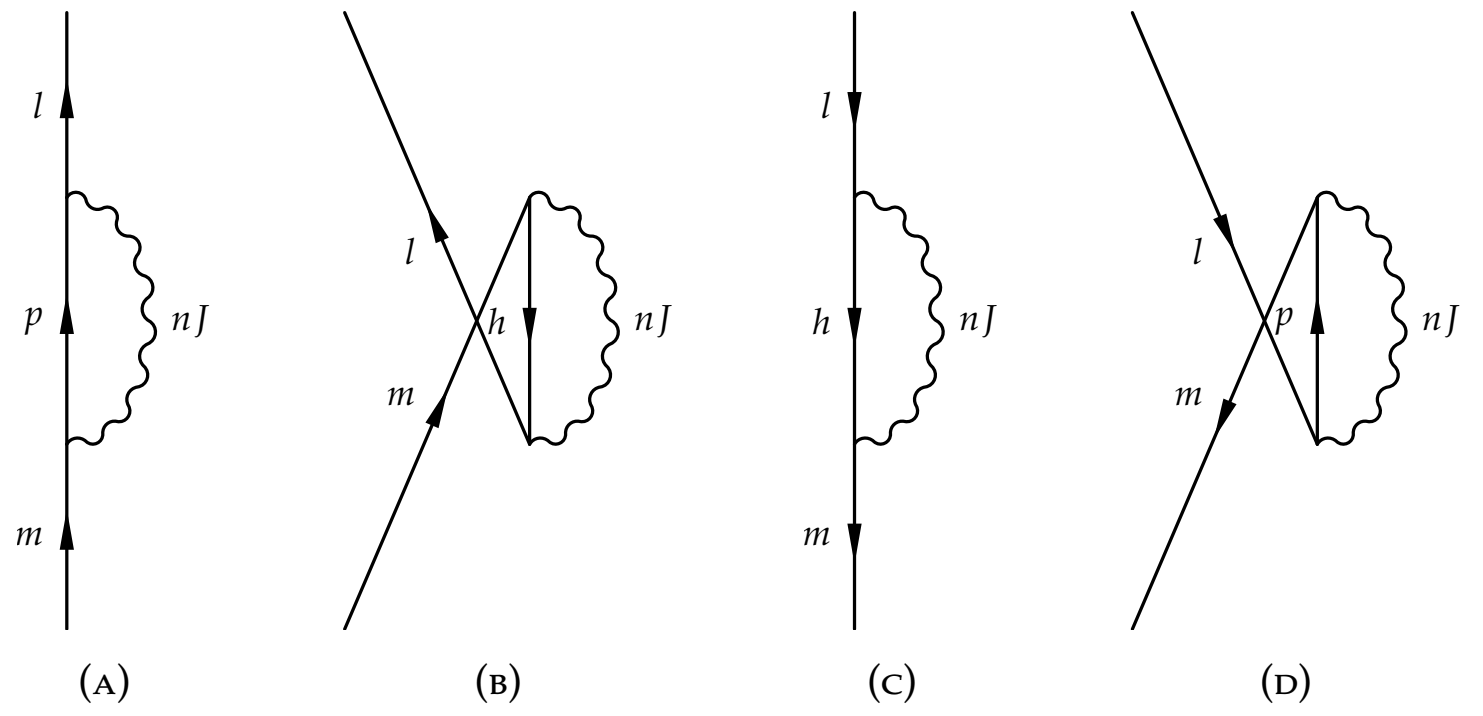

**FIGURE E.1**
Diagrams associated with the single-nucleon self-energy.

$$\times \int \mathrm{d}\omega_1\, G^{\mathrm{HF}}(\kappa;\omega_1)\Pi^{\mathrm{RPA}}(\delta,\gamma^{-1};\beta,\alpha^{-1};\omega-\omega_1)$$
$$= \frac{i}{2\pi}\sum_{\alpha\beta\gamma\delta\kappa}\sum_{n\neq 0}\langle\lambda\beta|\bar{V}|\kappa\alpha\rangle\langle\kappa\gamma|\bar{V}|\mu\delta\rangle$$
$$\times\left[\int \mathrm{d}\omega_1 \frac{\theta(\varepsilon_\kappa-\varepsilon_{\mathrm{F}})}{\omega_1-\varepsilon_\kappa+i\eta}\left(\frac{X_{\alpha\beta}X^*_{\gamma\delta}}{\omega-\omega_1-\omega_n+i\eta}-\frac{Y^*_{\alpha\beta}Y_{\gamma\delta}}{\omega-\omega_1+\omega_n-i\eta}\right)\right.$$
$$\left.+\int \mathrm{d}\omega_1 \frac{\theta(\varepsilon_{\mathrm{F}}-\varepsilon_\kappa)}{\omega_1-\varepsilon_\kappa-i\eta}\left(\frac{X_{\delta\gamma}X^*_{\beta\alpha}}{\omega-\omega_1-\omega_n+i\eta}-\frac{Y^*_{\delta\gamma}Y_{\beta\alpha}}{\omega-\omega_1+\omega_n-i\eta}\right)\right]$$

The energy integrals are again evaluated by means of the residue theorem

$$\mathcal{I}_1(\omega) = \int \mathrm{d}\omega_1 \frac{\theta(\varepsilon_\kappa-\varepsilon_{\mathrm{F}})}{\omega_1-\varepsilon_\kappa+i\eta}\left(\frac{X_{\alpha\beta}X^*_{\gamma\delta}}{\omega-\omega_1-\omega_n+i\eta}-\frac{Y^*_{\alpha\beta}Y_{\gamma\delta}}{\omega-\omega_1+\omega_n-i\eta}\right)$$
$$= \frac{-2\pi i X_{\alpha\beta}X^*_{\gamma\delta}\,\theta(\varepsilon_\kappa-\varepsilon_{\mathrm{F}})}{\omega-\varepsilon_\kappa-\omega_n+i\eta}$$
$$\mathcal{I}_2(\omega) = \int \mathrm{d}\omega_1 \frac{\theta(\varepsilon_{\mathrm{F}}-\varepsilon_\kappa)}{\omega_1-\varepsilon_\kappa-i\eta}\left(\frac{X_{\delta\gamma}X^*_{\beta\alpha}}{\omega-\omega_1-\omega_n+i\eta}-\frac{Y^*_{\delta\gamma}Y_{\beta\alpha}}{\omega-\omega_1+\omega_n-i\eta}\right)$$
$$= \frac{-2\pi i Y^*_{\delta\gamma}Y_{\beta\alpha}\,\theta(\varepsilon_{\mathrm{F}}-\varepsilon_\kappa)}{\omega-\varepsilon_\kappa+\omega_n-i\eta}.$$

As we have done in the previous section, we want to explicitly sum over the magnetic quantum number of intermediate and final state and to make the average of the initial states. It is then useful to introduce the coupled matrix elements and the coupled version of the polarization propagator. The angular momentum structure of the first part is (the one of the second part giving the same result)

$$\mathcal{A} = \sum_{\substack{J_1M_1\\J_2M_2\\J_3M_3}}\sum_{\substack{m_am_b\\m_cm_d\\m_km_mm_n}} \begin{array}{l}(-)^{j_b-m_b+j_k-m_k}\langle j_lm_l\,j_k-m_k|J_1M_1\rangle\langle j_am_a\,j_b-m_b|J_1M_1\rangle\\(-)^{j_c-m_c+j_m-m_m}\langle j_km_k\,j_m-m_m|J_2M_2\rangle\langle j_dm_d\,j_c-m_c|J_2M_2\rangle\\(-)^{j_b-m_b+j_d-m_d}\langle j_am_a\,j_b-m_b|J_3M_3\rangle\langle j_cm_c\,j_d-m_d|J_3M_3\rangle\end{array}$$
$$= \sum_J (2J+1)(-)^{j_k+j_c+j_m+j_d}\delta_{j_l,j_m}$$

The self-energy thus becomes

$$\Sigma^{\mathrm{PVC}}(l,m;\omega) = \frac{\delta_{j_l,j_m}}{(2j_m+1)}\sum_{abcdk}\sum_{Jn\neq 0}V_J(lbka)V_J(kcmd)$$
$$\times\left[\frac{X^J_{ab}\left(X^J_{cd}\right)^*\theta(\varepsilon_k-\varepsilon_{\mathrm{F}})}{\omega-\varepsilon_k-\omega_{nJ}+i\eta}+\frac{\left(Y^J_{dc}\right)^*Y^J_{ba}\,\theta(\varepsilon_{\mathrm{F}}-\varepsilon_k)}{\omega-\varepsilon_k+\omega_{nJ}-i\eta}\right]$$

As usual, $a, b, c, d$ can be either particles or holes and, in the particular case in which $l$ is equal to $m$ (diagonal approximation), the self-energy turns into (we rename $k$ as $p$ if it is above the Fermi level and as $h$ if it is below the Fermi level)

$$\Sigma^{\mathrm{PVC}}(m;\omega) = \frac{1}{2j_m+1}\sum_{\substack{p\\ Jn\neq 0}} \frac{|\langle m\|V\|p,nJ\rangle|^2}{\omega-\varepsilon_p-\omega_{nJ}+i\eta} + \frac{1}{2j_m+1}\sum_{\substack{h\\ Jn\neq 0}} \frac{|\langle m\|V\|h,nJ\rangle|^2}{\omega-\varepsilon_h+\omega_{nJ}-i\eta},$$

consistently with Ref. [CSB10].

### E.1.2 Calculation using rules (6)-(7) of section 7.3

The same calculation can be straightforwardly carried out using the expression for the vertices at the end of section 7.3. Following the same notation of Fig. E.1, in the diagonal approximation ($l = m$), the self-energy reads

$$\begin{aligned}
\Sigma^{\mathrm{PVC}}(m;\omega) =& \frac{1}{2j_m+1}\sum_{m_m}\sum_{nJM}\left[\sum_{pm_p}\frac{\Lambda^{nJ}_{mp}\Lambda'^{nJ}_{pm}}{\omega-\varepsilon_p-\omega_{nJ}+i\eta} + \sum_{hm_h}\frac{\Lambda^{nJ}_{hm}\Lambda'^{nJ}_{hm}}{\omega-\varepsilon_h+\omega_{nJ}-i\eta}\right]\\
=& \frac{1}{2j_m+1}\sum_{m_m}\sum_{pm_p}\sum_{nJM}\left[\frac{\langle m\|V\|p,nJ\rangle\langle p,nJ\|V\|m\rangle}{\omega-\varepsilon_p-\omega_{nJ}+i\eta}\right.\\
&\left.(-)^{j_m-m_m}\begin{pmatrix} j_m & j_p & J\\ m_m & -m_p & -M\end{pmatrix}(-)^{j_m-j_p+J+j_m-m_m}\begin{pmatrix} j_m & j_p & J\\ m_m & -m_p & -M\end{pmatrix}\right]\\
&+\frac{1}{2j_m+1}\sum_{m_m}\sum_{hm_h}\sum_{nJM}\left[\frac{\langle m\|V\|h,nJ\rangle\langle h,nJ\|V\|m\rangle}{\omega-\varepsilon_h+\omega_{nJ}-i\eta}\right.\\
&\left.(-)^{j_h-j_m+J}\begin{pmatrix} j_h & j_m & J\\ m_h & m_m & M\end{pmatrix}\begin{pmatrix} j_h & j_m & J\\ m_h & m_m & M\end{pmatrix}\right]\\
=& \frac{1}{2j_m+1}\sum_{\substack{p\\ Jn\neq 0}} \frac{|\langle m\|V\|p,nJ\rangle|^2}{\omega-\varepsilon_p-\omega_{nJ}+i\eta} + \frac{1}{2j_m+1}\sum_{\substack{h\\ Jn\neq 0}} \frac{|\langle m\|V\|h,nJ\rangle|^2}{\omega-\varepsilon_h+\omega_{nJ}-i\eta}.
\end{aligned}$$

The minus sign in the second contribution coming from the denominator is cancelled by the minus sign coming from the presence of a hole line.

## E.2 The lineshape of the GRs

In this section, we want to evaluate in detail the diagrams of Figs. 8.1(a) ($\Sigma^p_{s-e;X}(GR,E_{nJ})$) and 8.1(e) ($\Sigma^1_{v;X}(GR,E_{nJ})$), using the rules and the vertices of section 7.3. For the sake of simplicity, we report in Fig. E.2 the two diagrams. In the following $E_{nJ}$ and $E_{n'\lambda}$ are the energies of the phonons and $\varepsilon_{ph} = \varepsilon_p - \varepsilon_h$.

$$\begin{aligned}
\Sigma^p_{s-e;X}(GR,E_{nJ}) &= \sum_{\substack{pp'h\\ n'\lambda}}\sum_{\substack{m_pm_{p'}\\ m_h\mu}} \frac{\Lambda^{nJ}_{ph}\Lambda'^{nJ}_{ph}\Lambda^{n'\lambda}_{pp'}\Lambda'^{n'\lambda}_{p'p}}{(E_{nJ}-\varepsilon_{ph}+i\eta)^2(E_{nJ}-E_{n'\lambda}-\varepsilon_{p'h}+i\eta)}\\
&= \sum_{\substack{pp'h\\ n'\lambda}}\sum_{\substack{m_pm_{p'}\\ m_h\mu}} \frac{\langle p\|V\|h,nJ\rangle\langle h,nJ\|V\|p\rangle\langle p\|V\|p',n'\lambda\rangle\langle p',n'\lambda\|V\|p\rangle}{(E_{nJ}-\varepsilon_{ph}+i\eta)^2(E_{nJ}-E_{n'\lambda}-\varepsilon_{p'h}+i\eta)}
\end{aligned}$$

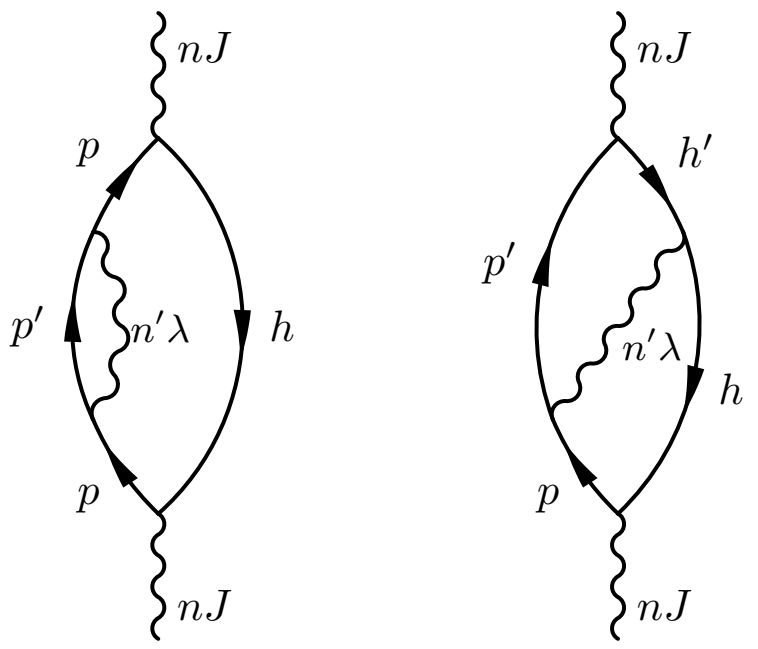

$\Sigma^{p}_{s-e;X}$ $\Sigma^{1}_{v;X}$

(A) (B)

**Figure E.2**
The two NFT diagrams contributing to the strength function of the giant resonance calculated in this appendix.

$$(-)^{J+M}\begin{pmatrix} j_h & j_p & J \\ m_h & m_p & -M \end{pmatrix}(-)^{j_p+m_p+j_h-m_h}\begin{pmatrix} j_h & j_p & J \\ m_h & m_p & -M \end{pmatrix}$$
$$(-)^{j_p-m_p}\begin{pmatrix} j_p & j_{p'} & \lambda \\ m_p & -m_{p'} & -\mu \end{pmatrix}(-)^{\lambda+\mu+j_{p'}-m_{p'}}\begin{pmatrix} j_p & j_{p'} & \lambda \\ m_p & -m_{p'} & -\mu \end{pmatrix}$$
$$= \sum_{\substack{pp'h \\ n'\lambda}} \frac{1}{(2J+1)(2j_p+1)} \frac{|\langle p\|V\|h, nJ\rangle|^2 |\langle p\|V\|p', n'\lambda\rangle|^2}{(E_{nJ}-\varepsilon_{ph}+i\eta)^2 (E_{nJ}-E_{n'\lambda}-\varepsilon_{p'h}+i\eta)}$$

$$\Sigma^{1}_{v;X}(GR, E_{nJ}) = \sum_{\substack{pp' \\ hh' \\ n'\lambda}} \sum_{\substack{m_p m_{p'} \\ m_h m_{h'} \\ \mu}} \frac{\Lambda^{nJ}_{ph}\Lambda'^{nJ}_{p'h'}\Lambda^{n'\lambda}_{pp'}\Lambda'^{n'\lambda}_{hh}}{(E_{nJ}-\varepsilon_{ph}+i\eta)^2 (E_{nJ}-E_{n'\lambda}-\varepsilon_{p'h}+i\eta)}$$
$$= \sum_{\substack{pp' \\ hh' \\ n'\lambda}} \sum_{\substack{m_p m_{p'} \\ m_h m_{h'} \\ \mu}} \frac{\langle p\|V\|h, nJ\rangle \langle h', nJ\|V\|p'\rangle \langle h\|V\|h', n'\lambda\rangle \langle p', n'\lambda\|V\|p\rangle}{(E_{nJ}-\varepsilon_{ph}+i\eta)(E_{nJ}-E_{n'\lambda}-\varepsilon_{p'h}+i\eta)(E_{nJ}-\varepsilon_{p'h'}+i\eta)}$$
$$(-)^{J+M}\begin{pmatrix} j_h & j_p & J \\ m_h & m_p & -M \end{pmatrix}(-)^{j_{p'}+m_{p'}+j_{h'}-m_{h'}}\begin{pmatrix} j_{h'} & j_{p'} & J \\ m_{h'} & m_{p'} & -M \end{pmatrix}$$
$$(-)^{\lambda+\mu+j_h+m_h}\begin{pmatrix} j_{h'} & j_h & \lambda \\ m_{h'} & -m_h & -\mu \end{pmatrix}(-)^{\lambda+\mu+j_{p'}-m_{p'}}\begin{pmatrix} j_p & j_{p'} & \lambda \\ m_p & -m_{p'} & -\mu \end{pmatrix}$$
$$= \sum_{\substack{pp' \\ hh' \\ n'\lambda}} \frac{1}{2J+1}(-)^{j_h+j_{h'}+j_p+j_{p'}} \begin{Bmatrix} j_h & j_p & J \\ j_{p'} & j_{h'} & \lambda \end{Bmatrix}$$
$$\frac{\langle p\|V\|h, nJ\rangle \langle h', nJ\|V\|p'\rangle \langle h\|V\|h', n'\lambda\rangle \langle p', n'\lambda\|V\|p\rangle}{(E_{nJ}-\varepsilon_{ph}+i\eta)(E_{nJ}-E_{n'\lambda}-\varepsilon_{p'h}+i\eta)(E_{nJ}-\varepsilon_{p'h'}+i\eta)}$$

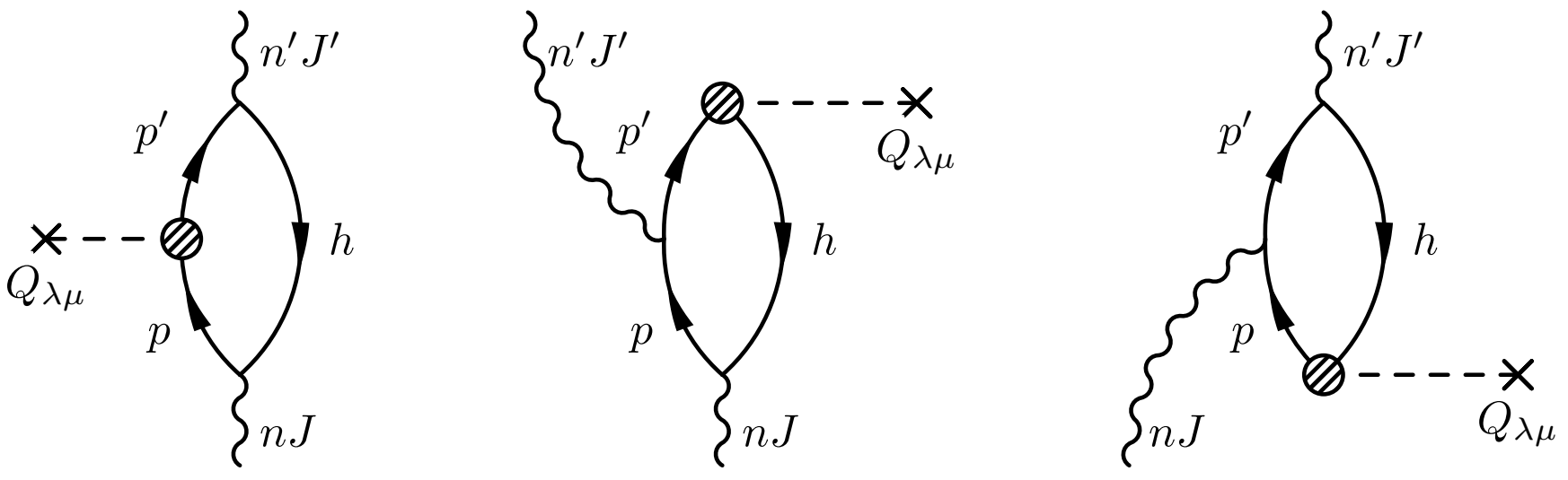

(A) (B) (C)

**Figure E.3**
The three NFT diagrams contributing to the $\gamma$ decay of GRs calculated in this appendix.

$$= \sum_{\substack{pp'\\hh'\\n'\lambda}} \frac{1}{2J+1}(-)^{J+\lambda-j'_h+j_p}\begin{Bmatrix} j_h & j_p & J\\ j_{p'} & j_{h'} & \lambda\end{Bmatrix}$$
$$\frac{\langle p\|V\|h,nJ\rangle\langle p'\|V\|h',nJ\rangle\langle h\|V\|h',n'\lambda\rangle\langle p\|V\|p',n'\lambda\rangle}{(E_{nJ}-\varepsilon_{ph}+i\eta)(E_{nJ}-E_{n'\lambda}-\varepsilon_{p'h}+i\eta)(E_{nJ}-\varepsilon_{p'h'}+i\eta)}$$

Exploiting the relations (7.26), it is possible to express $\Sigma^{p}_{s-e;X}(GR,E_{nJ})$ and $\Sigma^{1}_{v;X}(GR,E_{nJ})$ in terms of the $X$ and $Y$ RPA amplitudes:

$$\Sigma^{p}_{s-e;X}(GR,E_{nJ}) = \sum_{\substack{pp'h\\n'\lambda}} \frac{1}{(2j_p+1)} \frac{\left(X^{nJ}_{ph}\right)^2 |\langle p\|V\|p',n'\lambda\rangle|^2}{E_{nJ}-E_{n'\lambda}-\varepsilon_{p'h}+i\eta}$$

$$\Sigma^{1}_{v;X}(GR,E_{nJ}) = \sum_{\substack{pp'\\hh'\\n'\lambda}} (-)^{J+\lambda-j'_h+j_p}\begin{Bmatrix} j_h & j_p & J\\ j_{p'} & j_{h'} & \lambda\end{Bmatrix}$$
$$\frac{X^{nJ}_{ph}X^{nJ}_{p'h'}\langle h\|V\|h',n'\lambda\rangle\langle p\|V\|p',n'\lambda\rangle}{E_{nJ}-E_{n'\lambda}-\varepsilon_{p'h}+i\eta}$$

## E.3 The $\gamma$ decay to low-lying states

In this section, we want to evaluate in detail the diagrams of Figs. 8.2(A), 8.2(E) and 8.2(I), using the rules and the vertices of section 7.3. For the sake of simplicity, we report in Fig. E.3 the three diagrams. In the following $E_{nJ}$ and $E_{n'\lambda}$ are the energies of the phonons and $\varepsilon_{ph} = \varepsilon_p - \varepsilon_h$. We are interested in the reduced matrix element of the electromagnetic operator

$$\langle n'J'M'|Q_{\lambda\mu}|nJM\rangle = \frac{\langle JM\lambda\mu|J'M'\rangle}{2J'+1}\langle n'J'\|Q_{\lambda}\|nJ\rangle,$$

evaluated through the Wigner-Eckart theorem; hence, using the orthogonality relations for the Clebsch-Gordan coefficients,

$$\langle n'J'\|Q_\lambda\|nJ\rangle = (-)^{J-\lambda+M'}(2J'+1)\begin{pmatrix} J & \lambda & J' \\ M & \mu & -M' \end{pmatrix}\langle n'J'|Q_{\lambda\mu}|nJM\rangle. \tag{E.1}$$

The diagrams are evaluated as follows

$$\begin{aligned}
\langle n'J'\|Q_\lambda\|nJ\rangle_{(\mathrm{A})} =& \sum_{pp'h}\sum_{\substack{M\mu m_h\\ m_p m_{p'}}}(-)^{J-\lambda+M'}\begin{pmatrix} J & \lambda & J' \\ M & \mu & -M' \end{pmatrix}\frac{(2J'+1)\Lambda^{nJ}_{ph}\Lambda'^{n'J'}_{p'h}\langle p'|Q_{\lambda\mu}|p\rangle}{(E_{nJ}-\epsilon_{ph}+i\eta)(E_{n'J'}-\epsilon_{p'h}+i\eta')}\\
=& \sum_{pp'h}\sum_{\substack{M\mu m_h\\ m_p m_{p'}}}(-)^{J-\lambda+M'}(2J'+1)\begin{pmatrix} J & \lambda & J' \\ M & \mu & -M' \end{pmatrix}(-)^{J+M}\begin{pmatrix} j_h & j_p & J \\ m_h & m_p & -M \end{pmatrix}\\
&\times(-)^{J'+M'}\begin{pmatrix} j_h & j_{p'} & J' \\ m_h & m_{p'} & -M' \end{pmatrix}(-)^{j_{p'}-m_{p'}}\begin{pmatrix} j_{p'} & j_p & \lambda \\ m_{p'} & -m_p & -\mu \end{pmatrix}\\
&\times(-)^{J'+j_{p'}-j_h}X^{nJ}_{ph}X^{n'J'}_{p'h}Q^{\lambda pol}_{p'p}\\
=& \sum_{pp'h}(-)^{J+\lambda+j_{p'}+j_h}\begin{Bmatrix} J & \lambda & J' \\ j_{p'} & j_h & j_p \end{Bmatrix}X^{nJ}_{ph}X^{n'J'}_{p'h}Q^{\lambda pol}_{p'p}\\
\langle n'J'\|Q_\lambda\|nJ\rangle_{(\mathrm{E})} =& \sum_{pp'h}\sum_{\substack{M\mu m_h\\ m_p m_{p'}}}(-)^{J-\lambda+M'}\begin{pmatrix} J & \lambda & J' \\ M & \mu & -M' \end{pmatrix}\\
&\times\frac{(2J'+1)\Lambda^{nJ}_{ph}\Lambda'^{n'J'}_{p'p}\langle p(h)^{-1}|Q_{\lambda\mu}|0\rangle}{(E_{nJ}-\epsilon_{ph}+i\eta)(E_{nJ}-E_{n'J'}-\epsilon_{p'h}+i\eta')}\\
=& \sum_{ph}\sum_{\substack{M\mu m_h\\ m_p m_{p'}}}(-)^{J-\lambda+M'}(2J'+1)\begin{pmatrix} J & \lambda & J' \\ M & \mu & -M' \end{pmatrix}(-)^{J+M}\begin{pmatrix} j_h & j_p & J \\ m_h & m_p & -M \end{pmatrix}\\
&\times(-)^{J'+M'+j_{p'}-m_{p'}}\begin{pmatrix} j_{p'} & j_p & J' \\ m_{p'} & -m_p & M' \end{pmatrix}\begin{pmatrix} j_h & j_{p'} & \lambda \\ m_h & m_{p'} & \mu \end{pmatrix}\\
&\times\frac{-X^{nJ}_{ph}\langle p',n'J'\|V\|p\rangle Q^{\lambda pol}_{p'h}}{E_{nJ}-E_{n'J'}-\epsilon_{p'h}+i\eta}\\
=& \sum_{pp'h}(-)^{J'+j_p-j_{p'}}\begin{Bmatrix} J & \lambda & J' \\ j_{p'} & j_p & j_h \end{Bmatrix}\frac{X^{nJ}_{ph}\langle p\|V\|p',n'J'\rangle Q^{\lambda pol}_{hp'}}{E_{nJ}-E_{n'J'}-\epsilon_{p'h}+i\eta}\\
\langle n'J'\|Q_\lambda\|nJ\rangle_{(\mathrm{I})} =& \sum_{pp'h}\sum_{\substack{M\mu m_h\\ m_p m_{p'}}}(-)^{J-\lambda+M'}\begin{pmatrix} J & \lambda & J' \\ M & \mu & -M' \end{pmatrix}\\
&\times\frac{(2J'+1)\Lambda^{nJ}_{p'p}\Lambda'^{n'J'}_{p'h}\langle 0|Q_{\lambda\mu}|p(h)^{-1}\rangle}{(E_{n'J'}-\epsilon_{p'h}+i\eta)(E_{nJ}+\epsilon_{ph}-E_{n'J'}+i\eta')}
\end{aligned}$$

$$
\begin{aligned}
&= \sum_{pp'h} \sum_{\substack{M\mu m_h \\ m_p m_{p'}}} (-)^{J-\lambda+M'} (2J'+1) \begin{pmatrix} J & \lambda & J' \\ M & \mu & -M' \end{pmatrix} \\
&\times (-)^{j_{p'}-m_{p'}} \begin{pmatrix} j_{p'} & j_p & J \\ m_{p'} & -m_p & -M \end{pmatrix} \\
&\times (-)^{J'+M'} \begin{pmatrix} j_h & j_{p'} & J' \\ m_h & m_{p'} & -M' \end{pmatrix} (-)^{J'+j_{p'}-j_h} (-)^{\lambda+\mu} \begin{pmatrix} j_h & j_p & \lambda \\ m_h & m_p & -\mu \end{pmatrix} \\
&\times \frac{X^{n'J'}_{p'h} \langle p' \| V \| p, nJ \rangle Q^{\lambda pol}_{ph}}{(\hbar\omega_{J'} - \epsilon_{p'h})(E_J + \epsilon_{ph} - \hbar\omega_{J'} + i\eta)} \\
=&(-)^{J'+j_h+j_{p'}} \begin{Bmatrix} J & \lambda & J' \\ j_h & j_{p'} & j_p \end{Bmatrix} \frac{X^{n'J'}_{p'h} \langle p' \| V \| p, nJ \rangle Q^{\lambda pol}_{ph}}{E_{nJ} - E_{n'J'} + \epsilon_{ph} + i\eta}
\end{aligned}
$$

Appendix F

# Microscopic PVC vertex

In this appendix, we provide the derivation of the reduced matrix element associated with the PVC vertex, both in the case in which the phonon is in the initial state (see Fig. F.1(a)), following the appendix of Ref. [CSB10], and in the one in which the phonon is in the final state (see Fig. F.1(b)). We will use the RPA creation operator $\Gamma^\dagger_\nu(JM)$ (3.10) and its Hermitian conjugate $\Gamma_\nu(JM)$.

## F.1 Matrix element with the phonon in the initial state

We want to compute the matrix element

$$\Lambda^{nJ}_{ab} \equiv \langle j_a m_a | V | j_b m_b, nJM\rangle \langle 0| c_{j_a m_a} V c^\dagger_{j_b m_b} \Gamma^\dagger_n(JM)|0\rangle.$$

Since we are using RPA to treat phonons we have to make the following approximation

$$\begin{aligned}
\Lambda^J_{ab} &\approx \langle 0| c_{j_a m_b} [V, \Gamma^\dagger_n(JM)] c^\dagger_{j_b m_b}|0\rangle \\
&= \sum_{ph} X^{nJ}_{ph} \langle 0| c_{j_a m_b} [V, A^\dagger_{ph}(JM)] c^\dagger_{j_b m_b}|0\rangle - Y^{nJ}_{ph} \langle 0| c_{j_a m_b} [V, A_{ph}(\widetilde{JM})] c^\dagger_{j_b m_b}|0\rangle \\
&= v_1 + v_2.
\end{aligned}$$

Using the quasi-boson approximation (3.6), we get for the first commutator

$$\begin{aligned}
[V, A^\dagger_{ph}(JM)] &= \sum_{m_p m_h} (-)^{j_h - m_h} \langle j_p m_p j_h - m_h | JM\rangle [V, c^\dagger_p c_h] \\
&= \sum_{m_p m_h} (-)^{j_h - m_h} \langle j_p m_p j_h - m_h | JM\rangle \sum_{\alpha\beta} \bar{v}_{\alpha h \beta p} c^\dagger_\alpha c_\beta.
\end{aligned}$$

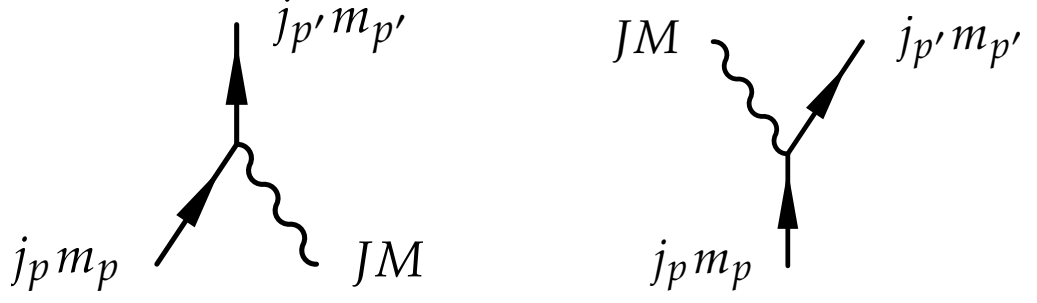

(a) (b)

**Figure F.1**
Particle-vibration coupling vertex in which the phonon is in the initial state (a) or in the final state (b).

Therefore,

$$v_1 = \sum_{ph} X_{ph}^{nJ} \sum_{m_p m_h} (-)^{j_h - m_h} \langle j_p m_p j_h - m_h | JM \rangle \sum_{\alpha\beta} \bar{v}_{ahbp}. \tag{F.1}$$

Similarly, the second commutator reads

$$\begin{aligned}
[V, A_{ph}(\widetilde{JM})] &= \sum_{m_p m_h} (-)^{J+M+j_h-m_h} \langle j_p m_p j_h - m_h | J - M \rangle [V, c_h^\dagger c_p] \\
&= \sum_{m_p m_h} (-)^{J+M+j_h-m_h} \langle j_p m_p j_h - m_h | J - M \rangle \sum_{\alpha\beta} \bar{v}_{\alpha p \beta h} c_\alpha^\dagger c_\beta,
\end{aligned}$$

which gives

$$v_2 = \sum_{ph} Y_{ph}^{nJ} \sum_{m_p m_h} (-)^{J+M+j_h-m_h} \langle j_p m_p j_h - m_h | J - M \rangle \sum_{\alpha\beta} \bar{v}_{apbh}. \tag{F.2}$$

We are interested in the reduced matrix element of the interaction. In order to compute this, we suppose that $V$ is a one-body operator carrying angular momentum $JM$ (because it effectively destroys a phonon):

$$\Lambda_{ab}^{nJ} = \frac{1}{\sqrt{2j_a+1}} \langle j_b m_b JM | j_a m_a \rangle \langle a \| V \| b, nJ \rangle = \frac{(-)^{j_b - m_b}}{\sqrt{2J+1}} \langle j_a m_a j_b - m_b | JM \rangle \langle a \| V \| b, nJ \rangle.$$

Eventually, using the definition of the angular-momentum-coupled matrix element (7.16), we obtain

$$\begin{aligned}
\langle a \| V \| b, nJ \rangle = \sqrt{2J+1} & \sum_{m_a m_b} (-)^{j_b - m_b} \langle j_a m_a j_b - m_b | JM \rangle \\
& \times \sum_{\substack{ph \\ m_p m_h}} \Big\{ X_{ph} (-)^{j_h - m_h} \langle j_p m_p j_h - m_h | JM \rangle \bar{v}_{ahbp} \\
& \qquad + Y_{ph} (-)^{J+M+j_h-m_h} \langle j_p m_p j_h - m_h | J - M \rangle \bar{v}_{apbh} \Big\} \\
= \sqrt{2J+1} & \sum_{ph} X_{ph} V_J(ahbp) + (-)^{J+j_h-j_p} Y_{ph} V_J(apbh).
\end{aligned}$$

## F.2 Matrix element with the phonon in the final state

The calculation of the PVC matrix element in which the phonon is in the final state is closely related to the previous one. Now we want to calculate the matrix element

$$\Lambda_{ba}'^{nJ} \equiv \langle j_b m_b, nJM | V | j_a m_a \rangle = \langle 0 | c_{j_b m_b} \Gamma_n(JM) V c_{j_a m_a}^\dagger | 0 \rangle.$$

The two contributions $v_1$ and $v_2$ can be obtained from Eqs. (F.2) and (F.1), respectively, changing the sign of $M$ and adding the phase $(-)^{J+M}$.

$$v_1 = \sum_{ph} X^{nJ}_{ph} \sum_{m_p m_h} (-)^{j_h - m_h} \langle j_p m_p j_h - m_h | JM \rangle \sum_{\alpha\beta} \bar{v}_{bpah}$$
$$v_2 = \sum_{ph} Y^{nJ}_{ph} \sum_{m_p m_h} (-)^{J+M+j_h - m_h} \langle j_p m_p j_h - m_h | J - M \rangle \sum_{\alpha\beta} \bar{v}_{bhap}.$$

In order to compute the reduced matrix element of the interaction, we treat $V$ as a one-body operator carrying angular momentum $J - M$ and a phase $(-)^{J+M}$ (because it effectively creates a phonon):

$$\Lambda'^{nJ}_{ba} = \frac{(-)^{J+M}}{\sqrt{2j_a+1}} \langle j_a m_a J - M | j_b m_b \rangle \langle b, nJ \| V \| a \rangle = \frac{(-)^{J+j_a-m_b}}{\sqrt{2J+1}} \langle j_a m_a j_b - m_b | JM \rangle \langle b, nJ \| V \| a \rangle.$$

Eventually, we get

$$\begin{aligned}
\langle b, nJ \| V \| a \rangle &= \sqrt{2J+1} \sum_{m_a m_b} (-)^{J+j_a-m_b} \langle j_a m_a j_b - m_b | J - M \rangle \\
&\quad \times \sum_{\substack{ph \\ m_p m_h}} \Big\{ X_{ph} (-)^{j_h - m_h} \langle j_p m_p j_h - m_h | JM \rangle \bar{v}_{ahbp} \\
&\qquad + Y_{ph} (-)^{J+M+j_h-m_h} \langle j_p m_p j_h - m_h | J - M \rangle \bar{v}_{apbh} \Big\} \\
&= \sqrt{2J+1} (-)^{j_a - j_b + J} \sum_{ph} X_{ph} V_J(ahbp) + (-)^{J+j_h-j_p} Y_{ph} V_J(apbh).
\end{aligned}$$

APPENDIX G

# The volume of the intersection between two spheres

Consider two spheres $\mathcal{S}$ and $\mathcal{S}'$, with radius $r$ and $r'$ respectively. Let $\boldsymbol{d}$ be the distance between the centers $C$ and $C'$. We can choose the frame of reference in such a way that $\mathcal{S}$ is centered in the origin and $\mathcal{S}'$ is centered in $C' = (d, 0, 0)$. The situation is depicted in Fig. G.1. We want to compute the volume of the intersection of the two spheres, as the sum of the volume $V_1$ of the spherical cap labeled with 1 in Fig. G.1, and the volume $V_2$ of the spherical cap 2. The equation of the two spheres are

$$\begin{aligned} x^2 + y^2 + z^2 &= r^2 \\ (x-d)^2 + y^2 + z^2 &= r'^2. \end{aligned} \tag{G.1}$$

Combining them,

$$\begin{aligned} (x-d)^2 + r^2 - x^2 &= r'^2 \\ -2dx + d^2 + r^2 - r'^2 &= 0 \\ x = x_0 &= \frac{d^2 + r^2 - r'^2}{2d} \end{aligned} \tag{G.2}$$

Therefore, the intersection is a curve in a plane $\pi$ parallel to the $yz$-plane with equation (G.2). At each distance x from the origin, the curve is a circle, having equation

$$y^2 + z^2 = r^2 - x^2 \tag{G.3}$$

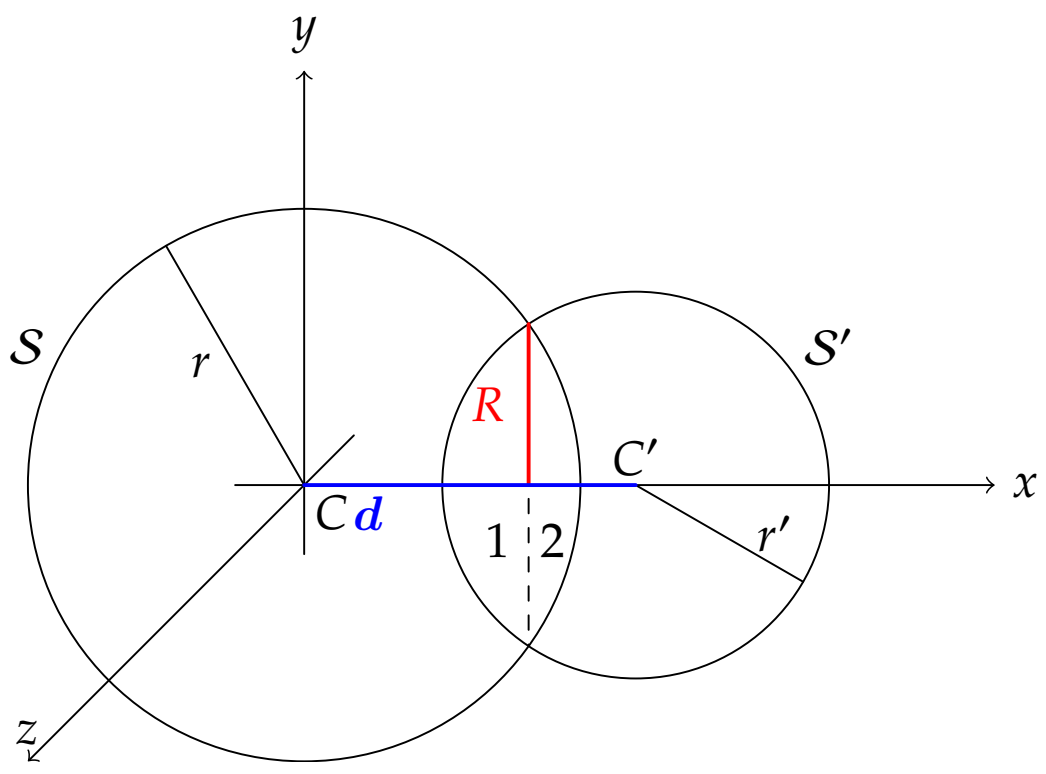

**FIGURE G.1**
The two intersecting spheres.

and surface

$$A = \pi \left( r^2 - x^2 \right). \tag{G.4}$$

To find the volume of $\mathcal{S} \cap \mathcal{S}'$, we have to sum the integrals of $A$ over 1 and 2.

$$\begin{aligned}
V_1 &= \pi \int_{x_0}^{r'} \mathrm{d}x \, \left( r^2 - x^2 \right) = \pi r^2 \left( r' - x_0 \right) - \frac{\pi}{3} \left( r'^3 - x_0^3 \right) \\
V_2 &= \pi \int_{x_0}^{r} \mathrm{d}x \, \left( r^2 - x^2 \right) = \pi r^2 \left( r - x_0 \right) - \frac{\pi}{3} \left( r^3 - x_0^3 \right) \\
V &= V_1 + V_2 = \pi r^2 \left( r + r' - 2x_0 \right) - \frac{\pi}{3} \left( r^3 + r'^3 - 2x_0^3 \right).
\end{aligned}$$

This expression can be simplified if $r = r'$. In this case,

$$V = \frac{1}{12} \pi (d - 2r)^2 (d + 4r) = \frac{4\pi}{3} r^3 \left[ 1 - \frac{3}{2} \frac{d}{2r} + \frac{1}{2} \left( \frac{d}{2r} \right)^3 \right]. \tag{G.5}$$

Appendix H

# Second order total energy in nuclear matter

In this appendix we compute in detail the integral of Eq. (10.11) of section 10.2 in the case in which the cutoff $\lambda$ on the relative momenta is larger than $2\sqrt{2}k_{\mathrm{F}}$. In Fig. H.1 is depicted the integration domain $\mathcal{D}(\boldsymbol{k},\boldsymbol{k}',\boldsymbol{k}'')$. The problem is symmetric under rotation around the $y$ axis and under reflection with respect to the $z$ axis. Thus, the domain of the integral in Eq. (10.11) reduces to the shaded regions in Fig. H.1 and the result of the integral has to be multiplied by four (two for the halving of the domain of $\boldsymbol{k}$, times two for the halving of the domain of $\boldsymbol{k}'$). We can use cylindrical coordinates with respect to the $y$ axis for the variable $\boldsymbol{k}'$. In this coordinate system, the integral in Eq. (10.11) reads (all the momenta are in units of $\sqrt{2}k_{\mathrm{F}}$):

$$
\begin{aligned}
&\frac{\sqrt{2}}{4\pi^3}\int_{\mathcal{D}(\boldsymbol{k},\boldsymbol{k}',\boldsymbol{k}'')} \mathrm{d}_3k\,\mathrm{d}_3k'\,\mathrm{d}_3k''\,\frac{1}{k'^2-k^2}\\
&\quad=\frac{\sqrt{2}}{4\pi^3}\int \mathrm{d}_3k''\int \mathrm{d}_3k\,\theta(1-k'')\,\theta(1-k)\,\theta(1-|\boldsymbol{k}''+\boldsymbol{k}|)\,\theta(1-|\boldsymbol{k}''-\boldsymbol{k}|)\\
&\qquad 2\times\left\{\int_0^{2\pi}\mathrm{d}\theta'\int_0^{1+k''}\mathrm{d}k'_y\int_{\sqrt{1-(k'_y-k'')^2}}^{+\infty}\mathrm{d}k'_\perp\,\frac{k'_\perp}{k'^2_y+k'^2_\perp-k^2}\right.\\
&\qquad\quad\left.+\int_0^{2\pi}\mathrm{d}\theta'\int_{1+k''}^{+\infty}\mathrm{d}k'_y\int_0^{+\infty}\mathrm{d}k'_\perp\,\frac{k'_\perp}{k'^2_y+k'^2_\perp-k^2}\right\}\\
&\quad=\frac{\sqrt{2}}{\pi^2}\int \mathrm{d}_3k''\int \mathrm{d}_3k\,\theta(1-k'')\,\theta(1-k)\,\theta(1-|\boldsymbol{k}''+\boldsymbol{k}|)\,\theta(1-|\boldsymbol{k}''-\boldsymbol{k}|)\\
&\qquad\times\left\{\int_0^{1+k''}\mathrm{d}k'_y\int_{\sqrt{1-(k'_y-k'')^2}}^{+\infty}\mathrm{d}k'_\perp\,\frac{k'_\perp}{k'^2_y+k'^2_\perp-k^2}\right.\\
&\qquad\quad\left.+\int_{1+k''}^{+\infty}\mathrm{d}k'_y\int_0^{+\infty}\mathrm{d}k'_\perp\,\frac{k'_\perp}{k'^2_y+k'^2_\perp-k^2}\right\}.
\end{aligned}
\tag{H.1}
$$

The integral on $k'$ has to be divided into two parts in order to properly exclude the internal spheres of Fig. H.1 from the domain.

If we want to introduce a cutoff $\lambda$ on the momenta $\boldsymbol{k}$ and $\boldsymbol{k}'$, we multiply the integrand function by $\theta(\lambda-k)\,\theta(\lambda-k')$, where $\lambda$ is in units of $\sqrt{2}k_{\mathrm{F}}$. Actually, since $k\leq 1$, the cutoff $\lambda$ on the momentum $\boldsymbol{k}$ has no effect. The cutoff $\lambda$ on $\boldsymbol{k}$ would have a physical meaning only if we consider also particle-particle contribution.

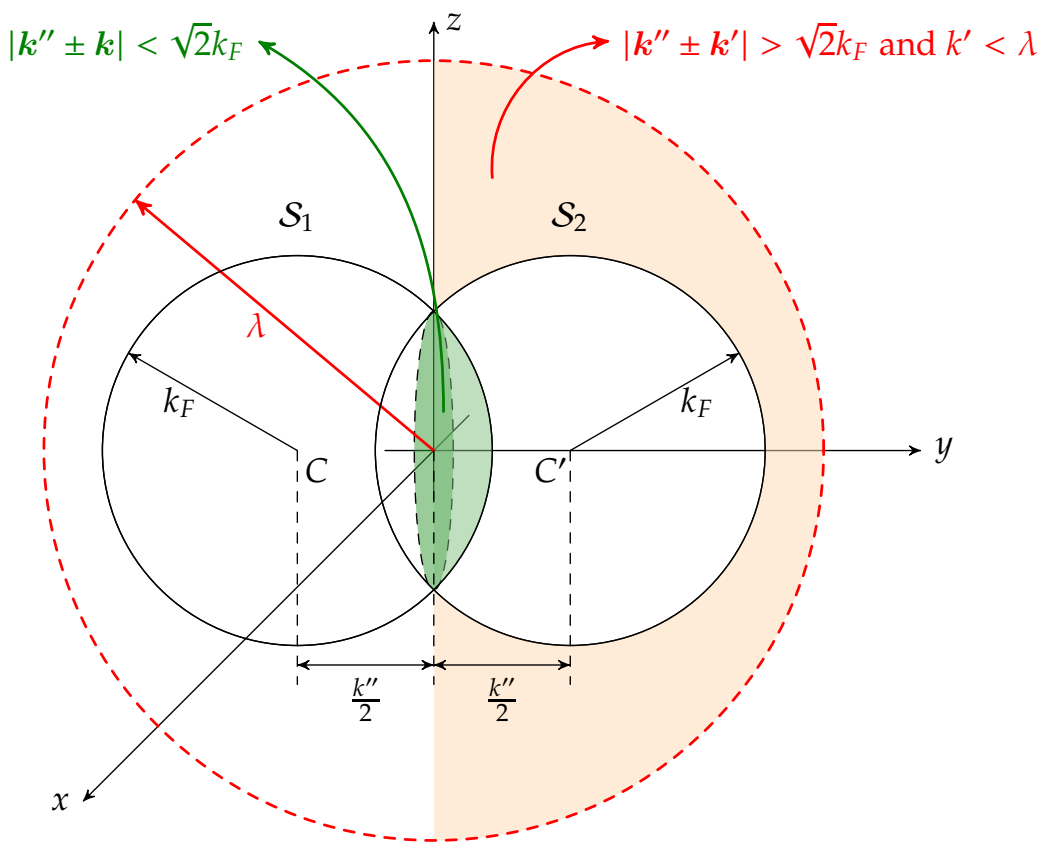

**Figure H.1**
The pictorial representation of the domain $\mathcal{D}(\boldsymbol{k},\boldsymbol{k}',\boldsymbol{k}'')$.

$$
\begin{aligned}
&\frac{\sqrt{2}}{4\pi^3}\int_{\mathcal{D}(\boldsymbol{k},\boldsymbol{k}',\boldsymbol{k}'')} \mathrm{d}_3k\,\mathrm{d}_3k'\,\mathrm{d}_3k''\,\frac{\theta(\lambda-k)\,\theta(\lambda-k')}{k'^2-k^2}\\
&\quad=\frac{\sqrt{2}}{\pi^2}\int \mathrm{d}_3k''\int \mathrm{d}_3k\,\theta(1-k'')\,\theta(1-k)\,\theta(1-|\boldsymbol{k}''+\boldsymbol{k}|)\,\theta(1-|\boldsymbol{k}''-\boldsymbol{k}|)\\
&\quad\times\left\{\int_0^{1+k''}\mathrm{d}k'_y\int_{\sqrt{1-\left(k'_y-k''\right)^2}}^{\sqrt{\lambda^2-k'^2_y}}\mathrm{d}k'_\perp\,\frac{k'_\perp}{k'^2_y+k'^2_\perp-k^2}\right.\\
&\quad\left.+\int_{1+k''}^{\lambda}\mathrm{d}k'_y\int_0^{\sqrt{\lambda^2-k'^2_y}}\mathrm{d}k'_\perp\,\frac{k'_\perp}{k'^2_y+k'^2_\perp-k^2}\right\}.
\end{aligned}
\tag{H.2}
$$

First consider the integral on $k'$. The two integrals are

$$
\begin{aligned}
I_1 &= \int_0^{1+k''}\mathrm{d}k'_y\int_{\sqrt{1-\left(k'_y-k''\right)^2}}^{\sqrt{\lambda^2-k'^2_y}}\mathrm{d}k'_\perp\,\frac{k'_\perp}{k'^2_y+k'^2_\perp-k^2} = \int_0^{1+k''}\mathrm{d}k'_y\left[\frac{1}{2}\ln\left(k'^2_y+k'^2_\perp-k^2\right)\right]_{\sqrt{1-\left(k'_y-k''\right)^2}}^{\sqrt{\lambda^2-k'^2_y}}\\
&=\int_0^{1+k''}\mathrm{d}k'_y\left[\frac{1}{2}\ln\left(\lambda^2-k^2\right)-\frac{1}{2}\ln\left(1-k''^2+2k''k'_y-k^2\right)\right]\\
&=\frac{1+k''}{2}\ln\left(\lambda^2-k^2\right)+\frac{1-k^2-k''^2+2k''k'_y}{4k''}\left[1-\ln\left(1-k^2-k''^2+2k''k'_y\right)\right]\Bigg|_0^{1+k''}\\
&=\frac{1+k''}{2}\ln\left(\lambda^2-k^2\right)-\frac{1-k^2-k''^2}{4k''}\left[1-\ln\left(1-k^2-k''^2\right)\right]\\
&\quad+\frac{1+2k''+k''^2-k^2}{4k''}\left[1-\ln\left(1-k^2+2k''+k''^2\right)\right].
\end{aligned}
$$

$$
\begin{aligned}
I_2 = & \int_{1+k''}^{\lambda} \mathrm{d}k'_y \int_0^{\sqrt{\lambda^2-k'^2_y}} \mathrm{d}k'_\perp \frac{k'_\perp}{k'^2_y + k'^2_\perp - k^2} = \int_{1+k''}^{\lambda} \mathrm{d}k'_y \left[\frac{1}{2}\ln\left(k'^2_y + k'^2_\perp - k^2\right)\right]_0^{\sqrt{\lambda^2-k'^2_y}} \\
= & \int_{1+k''}^{\lambda} \mathrm{d}k'_y \left[\frac{1}{2}\ln\left(\lambda^2 - k^2\right) - \frac{1}{2}\ln\left(k'_y - k^2\right)\right] \\
= & \frac{\lambda - 1 - k''}{2}\ln\left(\lambda^2 - k^2\right) + \frac{k'_y}{2}\left[2 - \ln\left(k'^2_y - k^2\right)\right]\Big|_{1+k''}^{\lambda} + \frac{k}{2}\ln\frac{k'_y - k}{k'_y + k}\Big|_{1+k''}^{\lambda} \\
= & \lambda - \frac{1+k''}{2}\left[2 + \ln\left(\lambda^2 - k^2\right)\right] + \frac{k}{2}\left(\ln\frac{1+k''+k}{1+k''-k} + \ln\frac{\lambda - k}{\lambda + k}\right) \\
& + \frac{(1+k'')}{2}\ln\left(1 + 2k'' + k''^2 - k^2\right). \\
I_1 + I_2 = & \frac{1+k''}{2}\ln\left(\lambda^2 - k^2\right) - \frac{1 - k^2 - k''^2}{4k''}\left[1 - \ln\left(1 - k^2 - k''^2\right)\right] \\
& + \frac{(1+k'')^2 - k^2}{4k''}\left\{1 - \ln\left[(1+k'')^2 - k^2\right]\right\} \\
& + \lambda - \frac{1+k''}{2}\left[2 + \ln\left(\lambda^2 - k^2\right)\right] + \frac{k}{2}\left(\ln\frac{1+k''+k}{1+k''-k} + \ln\frac{\lambda-k}{\lambda+k}\right) \\
& + \frac{(1+k'')}{2}\ln\left[(1+k'')^2 - k^2\right] \\
= & -\frac{1+k''}{2} + \frac{\left(1 - k''^2 - k^2\right)}{4k''}\ln\frac{1 - k''^2 - k^2}{(1+k'')^2 - k^2} + \frac{k}{2}\ln\frac{1+k''+k}{1+k''-k} \\
& + \lambda + \frac{k}{2}\ln\frac{\lambda - k}{\lambda + k} \\
= & a(k, k'') + b(k, \lambda).
\end{aligned}
\tag{H.3}
$$

Let us now consider the integral of $b(k,\lambda)$. In particular, this function does not depend on $k''$, then the integral on $k''$ can be done straightforwardly.

$$
\begin{aligned}
& \frac{\sqrt{2}}{\pi^2}\int \mathrm{d}_3k'' \int \mathrm{d}_3k\, \theta(1-k)\,\theta(1 - |\boldsymbol{k}'' + \boldsymbol{k}|)\,\theta(1 - |\boldsymbol{k}'' - \boldsymbol{k}|)\, b(k,\lambda) \\
= & \frac{\sqrt{2}}{\pi^2}\frac{4\pi}{3}\int \mathrm{d}_3k\, \theta(1-k)\left(\lambda + \frac{k}{2}\ln\frac{\lambda-k}{\lambda+k}\right)\left(1 - \frac{3}{2}k + \frac{1}{2}k^3\right) \\
= & \frac{16\sqrt{2}}{3}\int_0^1 \mathrm{d}k\, k^2\left(\lambda + \frac{k}{2}\ln\frac{\lambda-k}{\lambda+k}\right)\left(1 - \frac{3}{2}k + \frac{1}{2}k^3\right) \\
= & \frac{2}{35}\lambda - \frac{11}{35}\lambda^3 - \frac{2}{21}\lambda^5 - \left(\frac{4\lambda^5}{5} - \frac{4\lambda^7}{21}\right)\ln(\lambda) \\
& + \left(\frac{1}{35} - \frac{\lambda^4}{3} + \frac{2\lambda^5}{5} - \frac{2\lambda^7}{21}\right)\ln(\lambda - 1) \\
& - \left(\frac{1}{35} - \frac{\lambda^4}{3} - \frac{2\lambda^5}{5} + \frac{2\lambda^7}{21}\right)\ln(\lambda + 1)
\end{aligned}
$$

On the other hand, to integrate $a(k,k'')$ we need to use cylindrical coordinates for $k$.

The integral reads,

$$
\begin{aligned}
&\frac{1}{2\pi^2}\int \mathrm{d}_3k''\int \mathrm{d}_3k\,\theta(1-k'')\,\theta(1-|\boldsymbol{k}''+\boldsymbol{k}|)\,\theta(1-|\boldsymbol{k}''-\boldsymbol{k}|)\,a(k,\lambda)\\
=&\frac{2}{\pi}\int \mathrm{d}_3k''\int_0^{1-k2}\mathrm{d}k_y\int_0^{\sqrt{1-(ky+k2)^2}}\mathrm{d}k_\perp\,k_\perp\\
&\left(-\frac{1+k''}{2}+\frac{\left(1-k''^2-k_y^2-k_\perp^2\right)}{4k''}\ln\frac{1-k''^2-k_y^2-k_\perp^2}{(1+k'')^2-k_y^2-k_\perp^2}\right.\\
&\qquad\left.+\frac{\sqrt{k_y^2+k_\perp^2}}{2}\ln\frac{1+k''+\sqrt{k_y^2+k_\perp^2}}{1+k''-\sqrt{k_y^2+k_\perp^2}}\right)\\
=&\frac{2}{\pi}\int \mathrm{d}_3k''\int_0^{1-k2}\mathrm{d}k_y\\
&\left\{-\frac{5}{24}(1+k'')\left(1-k''^2-2k''k_y-k_y^2\right)+\frac{2+3k''-k''^3}{12}\ln(2k'')-\frac{1}{4}k''k_y^2\ln k_y\right.\\
&-\frac{1}{6}\left[(1+k'')^3-k_y^3\right]\left[\ln(1+k''-k_y)-\ln(1+k''+k_y)\right]\\
&-\frac{2+9k''+12k''^2+5k''^3-3k''ky^2}{12}\ln(1+k''+k_y)\\
&+\frac{\left(1-k''^2-k_y^2\right)^2}{16k''}\left\{\ln\left(1-k''^2-k_y^2\right)-\ln\left[(1+k'')^2-k_y^2\right]\right\}\\
&+\frac{1}{4}k''(1+k'')^2\ln\left[(1+k'')^2-k_y^2\right]\\
&\left.-\frac{\left(1-k''^2-2k''k_y\right)^{\frac{3}{2}}}{6}\ln\frac{1+k''-\sqrt{1-k''^2-2k''k_y}}{1+k''+\sqrt{1-k''^2-2k''k_y}}\right\}\\
=&\frac{2}{\pi}\int \mathrm{d}_3k''\\
&\frac{2k''}{15}-\frac{2k''^3}{15}-\left(\frac{1}{15k''}-\frac{k''}{3}\right)\ln 2+\left(\frac{1}{15k''}+\frac{k''}{3}\right)\ln k''\\
&+\left(\frac{1}{30k''}-\frac{k''}{6}+\frac{k''^2}{6}-\frac{k''^4}{30}\right)\ln[1-k'']\\
&+\left(\frac{1}{30k''}-\frac{k''}{6}-\frac{k''^2}{6}+\frac{k''^4}{30}\right)\ln[1+k'']\\
&+\frac{\left(1-k''^2\right)^{\frac{5}{2}}}{30k''}\ln\left[\frac{1-k''^2+\sqrt{1-k''^2}}{-1+k''^2+\sqrt{1-k''^2}}\right]\\
=&-\frac{11}{105}+\frac{2}{105}\ln 2.
\end{aligned}
$$

Summing up all the contributions, we get Eq. (10.12).

# Detailed computation of the matrix element of the interaction

In this appendix, we evaluate in detail the antisymmetryzed matrix element of the interaction (10.14) and (10.16) of section 10.5 and 10.6, respectively.

## I.1 Matrix element of the Skyrme interaction

$$\begin{aligned}
\langle ab|\bar{V}|cd\rangle &= \langle n_a l_a j_a m_a \tau_a, n_b l_b j_b m_b \tau_b|\bar{V}|n_c l_c j_c m_c \tau_c, n_d l_d j_d m_d \tau_d\rangle = \qquad \text{(I.1)}\\
&= \sum_{\substack{J_i\Lambda_i\Sigma_i\\ J_f\Lambda_f\Sigma_f}} \sum_{\substack{M_{J_i}M_{\Sigma_i}\\ M_{\Lambda_i}}} \sum_{\substack{M_{J_f}M_{\Sigma_f}\\ M_{\Lambda_f}}} \sum_{\substack{m_{l_a}m_{l_b}\\ m_{l_c}m_{l_d}}} \sum_{\substack{\sigma_a\sigma_b\\ \sigma_c\sigma_d}} i^{-l_a-l_b+l_c+l_d}\hat{J}_i^2\hat{\Lambda}_i^2\hat{\Sigma}_i^2\hat{J}_f^2\hat{\Lambda}_f^2\hat{\Sigma}_f^2\hat{j}_a\hat{j}_b\hat{j}_c\hat{j}_d\\
&(-)^{j_a-j_b+M_{J_f}+\Lambda_f-\Sigma_f+M_{J_f}+M_{\Sigma_f}+l_a-l_b+M_{\Lambda_f}+j_c-j_d+M_{J_i}+\Lambda_i-\Sigma_i+M_{J_i}+M_{\Sigma_i}+l_c-l_d+M_{\Lambda_i}}\\
&\begin{Bmatrix} l_a & l_b & \Lambda_f\\ \frac{1}{2} & \frac{1}{2} & \Sigma_f\\ j_a & j_b & J_f\end{Bmatrix}\begin{Bmatrix} l_c & l_d & \Lambda_i\\ \frac{1}{2} & \frac{1}{2} & \Sigma_i\\ j_c & j_d & J_i\end{Bmatrix}\\
&\begin{pmatrix} j_a & j_b & J_f\\ m_a & m_b & -M_{J_f}\end{pmatrix}\begin{pmatrix} \Lambda_f & \Sigma_f & J_f\\ M_{\Lambda_f} & M_{\Sigma_f} & -M_{J_f}\end{pmatrix}\begin{pmatrix} \frac{1}{2} & \frac{1}{2} & \Sigma_f\\ \sigma_a & \sigma_b & -M_{\Sigma_f}\end{pmatrix}\begin{pmatrix} l_a & l_b & \Lambda_f\\ m_{l_a} & m_{l_b} & -M_{\Lambda_f}\end{pmatrix}\\
&\begin{pmatrix} j_c & j_d & J_i\\ m_c & m_d & -M_{J_i}\end{pmatrix}\begin{pmatrix} \Lambda_i & \Sigma_i & J_i\\ M_{\Lambda_i} & M_{\Sigma_i} & -M_{J_i}\end{pmatrix}\begin{pmatrix} \frac{1}{2} & \frac{1}{2} & \Sigma_i\\ \sigma_c & \sigma_d & -M_{\Sigma_i}\end{pmatrix}\begin{pmatrix} l_c & l_d & \Lambda_i\\ m_{l_c} & m_{l_d} & -M_{\Lambda_i}\end{pmatrix}\\
&\int \mathrm{d}_3 r_1 \mathrm{d}_3 r_2 \mathrm{d}_3 r_1' \mathrm{d}_3 r_2'\, R_{n_a l_a}(r_1') Y^*_{l_a m_{l_a}}(\hat{r}_1') R_{n_b l_b}(r_2') Y^*_{l_b m_{l_b}}(\hat{r}_2') \qquad \text{(I.2)}\\
&\qquad g\left(\frac{\boldsymbol{r}_1+\boldsymbol{r}_2}{2}\right)\delta_3(\boldsymbol{r}_1-\boldsymbol{r}_2)\delta_3(\boldsymbol{r}_1-\boldsymbol{r}_1')\delta_3(\boldsymbol{r}_2-\boldsymbol{r}_2')\\
&\qquad R_{n_c l_c}(r_1) Y_{l_c m_{l_c}}(\hat{r}_1) R_{n_d l_d}(r_2) Y_{l_d m_{l_d}}(\hat{r}_2)\\
&\Big(\frac{3}{4}\langle\frac{1}{2}\sigma_a,\frac{1}{2}\sigma_b|\mathbb{1}_\sigma|\frac{1}{2}\sigma_c,\frac{1}{2}\sigma_d\rangle\langle\frac{1}{2}\tau_a,\frac{1}{2}\tau_b|\mathbb{1}_\tau|\frac{1}{2}\tau_c,\frac{1}{2}\tau_d\rangle\\
&-\frac{1}{4}\langle\frac{1}{2}\sigma_a,\frac{1}{2}\sigma_b|\boldsymbol{\sigma}(1)\boldsymbol{\sigma}(2)|\frac{1}{2}\sigma_c,\frac{1}{2}\sigma_d\rangle\langle\frac{1}{2}\tau_a,\frac{1}{2}\tau_b|\mathbb{1}_\tau|\frac{1}{2}\tau_c,\frac{1}{2}\tau_d\rangle\\
&-\frac{1}{4}\langle\frac{1}{2}\sigma_a,\frac{1}{2}\sigma_b|\mathbb{1}_\sigma|\frac{1}{2}\sigma_c,\frac{1}{2}\sigma_d\rangle\langle\frac{1}{2}\tau_a,\frac{1}{2}\tau_b|\boldsymbol{\tau}(1)\boldsymbol{\tau}(2)|\frac{1}{2}\tau_c,\frac{1}{2}\tau_d\rangle\\
&-\frac{1}{4}\langle\frac{1}{2}\sigma_a,\frac{1}{2}\sigma_b|\boldsymbol{\sigma}(1)\boldsymbol{\sigma}(2)|\frac{1}{2}\sigma_c,\frac{1}{2}\sigma_d\rangle\langle\frac{1}{2}\tau_a,\frac{1}{2}\tau_b|\boldsymbol{\tau}(1)\boldsymbol{\tau}(2)|\frac{1}{2}\tau_c,\frac{1}{2}\tau_d\rangle\Big).
\end{aligned}$$

The integral in Eq. (I.2) can be re-written as:

$$\begin{aligned}
\text{Eq. (I.2)} = &\int \mathrm{d}_3 r\ R_{n_a l_a}(r) Y^*_{l_a m_{l_a}}(\hat{r}) R_{n_b l_b}(r) Y^*_{l_b m_{l_b}}(\hat{r}) \\
&\quad g(\boldsymbol{r}) \\
&\quad R_{n_c l_c}(r) Y_{l_c m_{l_c}}(\hat{r}) R_{n_d l_d}(r) Y_{l_d m_{l_d}}(\hat{r}) \\
= &\int \mathrm{d}r\, r^2 R_{n_a l_a}(r) R_{n_b l_b}(r) g(r) R_{n_c l_c}(r) R_{n_d l_d}(r) \qquad \text{(I.3)} \\
&\int \mathrm{d}\hat{r}\ Y^*_{l_a m_{l_a}}(\hat{r}) Y^*_{l_b m_{l_b}}(\hat{r}) Y_{l_c m_{l_c}}(\hat{r}) Y_{l_d m_{l_d}}(\hat{r}) \\
= &\int \mathrm{d}r\, r^2 R_{n_a l_a}(r) R_{n_b l_b}(r) g(r) R_{n_c l_c}(r) R_{n_d l_d}(r) \\
&\sum_{\lambda\mu} \frac{\hat{l}_a \hat{l}_b \hat{\lambda}}{\sqrt{4\pi}} \begin{pmatrix} l_a & l_b & \lambda \\ 0 & 0 & 0 \end{pmatrix} \begin{pmatrix} l_a & l_b & \lambda \\ m_{l_a} & m_{l_b} & \mu \end{pmatrix} \\
&\int \mathrm{d}\hat{r}\ Y_{\lambda\mu}(\hat{r}) Y_{l_c m_{l_c}}(\hat{r}) Y_{l_d m_{l_d}}(\hat{r}) \\
= &\int \mathrm{d}r\, r^2 R_{n_a l_a}(r) R_{n_b l_b}(r) g(r) R_{n_c l_c}(r) R_{n_d l_d}(r) \\
&\sum_{\lambda\mu} \frac{\hat{l}_a \hat{l}_b \hat{l}_c \hat{l}_d \hat{\lambda}^2}{4\pi} \begin{pmatrix} l_a & l_b & \lambda \\ 0 & 0 & 0 \end{pmatrix} \begin{pmatrix} l_c & l_d & \lambda \\ 0 & 0 & 0 \end{pmatrix} \\
&\qquad \begin{pmatrix} l_a & l_b & \lambda \\ m_{l_a} & m_{l_b} & \mu \end{pmatrix} \begin{pmatrix} l_c & l_d & \lambda \\ m_{l_c} & m_{l_d} & \mu \end{pmatrix}
\end{aligned}$$

$$\begin{aligned}
\text{Eq. (I.1)} = &\sum_{\substack{J_i M_{J_i} \\ J_f M_{J_f}}} \sum_{\substack{\Sigma_i M_{\Sigma_i} \\ \Sigma_f M_{\Sigma_f}}} \sum_{\Lambda M_\Lambda} \sum_{\substack{\sigma_a \sigma_b \\ \sigma_c \sigma_d}} i^{-l_a - l_b + l_c + l_d} \hat{J}_i^2 \hat{\Sigma}_i^2 \hat{J}_f^2 \hat{\Sigma}_f^2 \hat{\Lambda}^2 \frac{\hat{j}_a \hat{j}_b \hat{j}_c \hat{j}_d \hat{l}_a \hat{l}_b \hat{l}_c \hat{l}_d}{4\pi} \\
&(-)^{j_a - j_b - \Sigma_f + M_{\Sigma_f} + l_a - l_b + j_c - j_d - \Sigma_i + M_{\Sigma_i} + l_c - l_d} \\
&\begin{Bmatrix} l_a & l_b & \Lambda \\ \frac{1}{2} & \frac{1}{2} & \Sigma_f \\ j_a & j_b & J_f \end{Bmatrix} \begin{Bmatrix} l_c & l_d & \Lambda \\ \frac{1}{2} & \frac{1}{2} & \Sigma_i \\ j_c & j_d & J_i \end{Bmatrix} \\
&\begin{pmatrix} j_a & j_b & J_f \\ m_a & m_b & -M_{J_f} \end{pmatrix} \begin{pmatrix} \Lambda & \Sigma_f & J_f \\ M_\Lambda & M_{\Sigma_f} & -M_{J_f} \end{pmatrix} \begin{pmatrix} \frac{1}{2} & \frac{1}{2} & \Sigma_f \\ \sigma_a & \sigma_b & -M_{\Sigma_f} \end{pmatrix} \begin{pmatrix} l_a & l_b & \Lambda \\ 0 & 0 & 0 \end{pmatrix} \\
&\begin{pmatrix} j_c & j_d & J_i \\ m_c & m_d & -M_{J_i} \end{pmatrix} \begin{pmatrix} \Lambda & \Sigma_i & J_i \\ M_\Lambda & M_{\Sigma_i} & -M_{J_i} \end{pmatrix} \begin{pmatrix} \frac{1}{2} & \frac{1}{2} & \Sigma_i \\ \sigma_c & \sigma_d & -M_{\Sigma_i} \end{pmatrix} \begin{pmatrix} l_c & l_d & \Lambda \\ 0 & 0 & 0 \end{pmatrix} \\
&\int \mathrm{d}r\, r^2 R_{n_a l_a}(r) R_{n_b l_b}(r) g(r) R_{n_c l_c}(r) R_{n_d l_d}(r)
\end{aligned}$$

$$\begin{aligned}&\Big(\frac{3}{4}\langle\frac{1}{2}\sigma_a,\frac{1}{2}\sigma_b|\mathbb{1}_\sigma|\frac{1}{2}\sigma_c,\frac{1}{2}\sigma_d\rangle\langle\frac{1}{2}\tau_a,\frac{1}{2}\tau_b|\mathbb{1}_\tau|\frac{1}{2}\tau_c,\frac{1}{2}\tau_d\rangle\\&-\frac{1}{4}\langle\frac{1}{2}\sigma_a,\frac{1}{2}\sigma_b|\boldsymbol{\sigma}(1)\boldsymbol{\sigma}(2)|\frac{1}{2}\sigma_c,\frac{1}{2}\sigma_d\rangle\langle\frac{1}{2}\tau_a,\frac{1}{2}\tau_b|\mathbb{1}_\tau|\frac{1}{2}\tau_c,\frac{1}{2}\tau_d\rangle\\&-\frac{1}{4}\langle\frac{1}{2}\sigma_a,\frac{1}{2}\sigma_b|\mathbb{1}_\sigma|\frac{1}{2}\sigma_c,\frac{1}{2}\sigma_d\rangle\langle\frac{1}{2}\tau_a,\frac{1}{2}\tau_b|\boldsymbol{\tau}(1)\boldsymbol{\tau}(2)|\frac{1}{2}\tau_c,\frac{1}{2}\tau_d\rangle\\&-\frac{1}{4}\langle\frac{1}{2}\sigma_a,\frac{1}{2}\sigma_b|\boldsymbol{\sigma}(1)\boldsymbol{\sigma}(2)|\frac{1}{2}\sigma_c,\frac{1}{2}\sigma_d\rangle\langle\frac{1}{2}\tau_a,\frac{1}{2}\tau_b|\boldsymbol{\tau}(1)\boldsymbol{\tau}(2)|\frac{1}{2}\tau_c,\frac{1}{2}\tau_d\rangle\Big).\end{aligned}\tag{I.4}$$

In Eq. (I.4) there are matrix elements like ($\varphi\equiv\sigma,\tau$ and $\Phi\equiv S,T$)

$$\begin{aligned}\langle\frac{1}{2}\varphi_a,\frac{1}{2}\varphi_b|\boldsymbol{\varphi}(1)\boldsymbol{\varphi}(2)|\frac{1}{2}\varphi_c,\frac{1}{2}\varphi_d\rangle&=4\sum_\mu(-)^\mu\langle\frac{1}{2}\varphi_a,\frac{1}{2}\varphi_b|\Phi_{1\mu}(1)\Phi_{1-\mu}(2)|\frac{1}{2}\varphi_c,\frac{1}{2}\varphi_d\rangle\\=&4\sum_\mu(-)^\mu\langle\frac{1}{2}\varphi_a|\Phi_{1\mu}|\frac{1}{2}\varphi_c\rangle\langle\frac{1}{2}\varphi_b|\Phi_{1-\mu}|\frac{1}{2}\varphi_d\rangle\\=&4\sum_\mu(-)^\mu\frac{1}{2}\langle\frac{1}{2}\varphi_c1\mu|\frac{1}{2}\varphi_a\rangle\langle\frac{1}{2}\varphi_b1-\mu|\frac{1}{2}\varphi_d\rangle\langle\frac{1}{2}\|\Phi_1\|\frac{1}{2}\rangle\langle\frac{1}{2}\|\Phi_1\|\frac{1}{2}\rangle\\=&6\sum_\mu(-)^{1+\varphi_a+\varphi_b+\mu}\begin{pmatrix}\frac{1}{2}&1&\frac{1}{2}\\\varphi_c&\mu&-\varphi_a\end{pmatrix}\begin{pmatrix}\frac{1}{2}&1&\frac{1}{2}\\\varphi_d&-\mu&-\varphi_b\end{pmatrix}\end{aligned}\tag{I.5}$$

We shall treat each term in Eq. (I.4) one by one.

$$\frac{3}{4}\langle\frac{1}{2}\sigma_a,\frac{1}{2}\sigma_b|\mathbb{1}_\sigma|\frac{1}{2}\sigma_c,\frac{1}{2}\sigma_d\rangle\langle\frac{1}{2}\tau_a,\frac{1}{2}\tau_b|\mathbb{1}_\tau|\frac{1}{2}\tau_c,\frac{1}{2}\tau_d\rangle=\frac{3}{4}\delta_{\sigma_a\sigma_c}\delta_{\sigma_b\sigma_d}\delta_{\tau_a\tau_c}\delta_{\tau_b\tau_d}\tag{I.6}$$

$$\begin{aligned}&-\frac{1}{4}\langle\frac{1}{2}\sigma_a,\frac{1}{2}\sigma_b|\boldsymbol{\sigma}(1)\boldsymbol{\sigma}(2)|\frac{1}{2}\sigma_c,\frac{1}{2}\sigma_d\rangle\langle\frac{1}{2}\tau_a,\frac{1}{2}\tau_b|\mathbb{1}_\tau|\frac{1}{2}\tau_c,\frac{1}{2}\tau_d\rangle=\\&=-\frac{6}{4}\delta_{\tau_a\tau_c}\delta_{\tau_b\tau_d}\sum_\mu(-)^{1+\sigma_a+\sigma_b+\mu}\begin{pmatrix}\frac{1}{2}&1&\frac{1}{2}\\\sigma_c&\mu&-\sigma_a\end{pmatrix}\begin{pmatrix}\frac{1}{2}&1&\frac{1}{2}\\\sigma_d&-\mu&-\sigma_b\end{pmatrix}\\&=-\frac{3}{2}\delta_{\tau_a\tau_c}\delta_{\tau_b\tau_d}\sum_\mu(-)^{1+\sigma_a+\sigma_b+\mu}\begin{pmatrix}\frac{1}{2}&1&\frac{1}{2}\\\sigma_c&\mu&-\sigma_a\end{pmatrix}\begin{pmatrix}\frac{1}{2}&1&\frac{1}{2}\\\sigma_d&-\mu&-\sigma_b\end{pmatrix}\end{aligned}\tag{I.7}$$

$$\begin{aligned}&-\frac{1}{4}\langle\frac{1}{2}\sigma_a,\frac{1}{2}\sigma_b|\mathbb{1}_\sigma|\frac{1}{2}\sigma_c,\frac{1}{2}\sigma_d\rangle\langle\frac{1}{2}\tau_a,\frac{1}{2}\tau_b|\boldsymbol{\tau}(1)\boldsymbol{\tau}(2)|\frac{1}{2}\tau_c,\frac{1}{2}\tau_d\rangle=\\&=-\frac{3}{2}\delta_{\sigma_a\sigma_c}\delta_{\sigma_a\sigma_c}\sum_\mu(-)^{1+\tau_a+\tau_b+\mu}\begin{pmatrix}\frac{1}{2}&1&\frac{1}{2}\\\tau_c&\mu&-\tau_a\end{pmatrix}\begin{pmatrix}\frac{1}{2}&1&\frac{1}{2}\\\tau_d&-\mu&-\tau_b\end{pmatrix}\end{aligned}\tag{I.8}$$

$$\begin{aligned}&-\frac{1}{4}\langle\frac{1}{2}\sigma_a,\frac{1}{2}\sigma_b|\boldsymbol{\sigma}(1)\boldsymbol{\sigma}(2)|\frac{1}{2}\sigma_c,\frac{1}{2}\sigma_d\rangle\langle\frac{1}{2}\tau_a,\frac{1}{2}\tau_b|\boldsymbol{\tau}(1)\boldsymbol{\tau}(2)|\frac{1}{2}\tau_c,\frac{1}{2}\tau_d\rangle=\\&=-9\sum_\mu(-)^{1+\sigma_a+\sigma_b+\mu}\begin{pmatrix}\frac{1}{2}&1&\frac{1}{2}\\\sigma_c&\mu&-\sigma_a\end{pmatrix}\begin{pmatrix}\frac{1}{2}&1&\frac{1}{2}\\\sigma_d&-\mu&-\sigma_b\end{pmatrix}\\&\quad\sum_{\mu'}(-)^{1+\tau_a+\tau_b+\mu'}\begin{pmatrix}\frac{1}{2}&1&\frac{1}{2}\\\tau_c&\mu'&-\tau_a\end{pmatrix}\begin{pmatrix}\frac{1}{2}&1&\frac{1}{2}\\\tau_d&-\mu'&-\tau_b\end{pmatrix}\end{aligned}\tag{I.9}$$

Therefore, the matrix element (I.1) is composed by four terms with different angular structure.

### $\langle ab|\bar{V}|cd\rangle_0$

$$
\begin{aligned}
\langle ab|\bar{V}|cd\rangle_0 = \frac{3}{4} \sum_{\substack{J_i M_{J_i} \\ J_f M_{J_f}}} \sum_{\substack{\Sigma_i M_{\Sigma_i} \\ \Sigma_f M_{\Sigma_f}}} \sum_{\Lambda M_\Lambda} \sum_{\substack{\sigma_a \sigma_b \\ \sigma_c \sigma_d}} i^{-l_a - l_b + l_c + l_d} \hat{J}_i^2 \hat{\Sigma}_i^2 \hat{J}_f^2 \hat{\Sigma}_f^2 \hat{\Lambda}^2 \frac{\hat{j}_a \hat{j}_b \hat{j}_c \hat{j}_d \hat{l}_a \hat{l}_b \hat{l}_c \hat{l}_d}{4\pi} \\
(-)^{j_a - j_b - \Sigma_f + M_{\Sigma_f} + l_a - l_b + j_c - j_d - \Sigma_i + M_{\Sigma_i} + l_c - l_d} \\
\delta_{\sigma_a \sigma_c} \delta_{\sigma_b \sigma_d} \delta_{\tau_a \tau_c} \delta_{\tau_b \tau_d} \begin{Bmatrix} l_a & l_b & \Lambda \\ \frac{1}{2} & \frac{1}{2} & \Sigma_f \\ j_a & j_b & J_f \end{Bmatrix} \begin{Bmatrix} l_c & l_d & \Lambda \\ \frac{1}{2} & \frac{1}{2} & \Sigma_i \\ j_c & j_d & J_i \end{Bmatrix} \\
\begin{pmatrix} j_a & j_b & J_f \\ m_a & m_b & -M_{J_f} \end{pmatrix} \begin{pmatrix} \Lambda & \Sigma_f & J_f \\ M_\Lambda & M_{\Sigma_f} & -M_{J_f} \end{pmatrix} \begin{pmatrix} \frac{1}{2} & \frac{1}{2} & \Sigma_f \\ \sigma_a & \sigma_b & -M_{\Sigma_f} \end{pmatrix} \begin{pmatrix} l_a & l_b & \Lambda \\ 0 & 0 & 0 \end{pmatrix} \\
\begin{pmatrix} j_c & j_d & J_i \\ m_c & m_d & -M_{J_i} \end{pmatrix} \begin{pmatrix} \Lambda & \Sigma_i & J_i \\ M_\Lambda & M_{\Sigma_i} & -M_{J_i} \end{pmatrix} \begin{pmatrix} \frac{1}{2} & \frac{1}{2} & \Sigma_i \\ \sigma_c & \sigma_d & -M_{\Sigma_i} \end{pmatrix} \begin{pmatrix} l_c & l_d & \Lambda \\ 0 & 0 & 0 \end{pmatrix} \\
\int \mathrm{d}r\, r^2 R_{n_a l_a}(r) R_{n_b l_b}(r) g(r) R_{n_c l_c}(r) R_{n_d l_d}(r) \\
= \frac{3}{4} \delta_{\tau_a \tau_c} \delta_{\tau_b \tau_d} \sum_{\substack{J_i M_{J_i} \\ J_f M_{J_f}}} \sum_{\Sigma M_\Sigma} \sum_{\Lambda M_\Lambda} i^{-l_a - l_b + l_c + l_d} \hat{J}_i^2 \hat{J}_f^2 \hat{\Lambda}^2 \hat{\Sigma}^2 \frac{\hat{j}_a \hat{j}_b \hat{j}_c \hat{j}_d \hat{l}_a \hat{l}_b \hat{l}_c \hat{l}_d}{4\pi} \\
(-)^{j_a - j_b + j_c - j_d + l_a - l_b + l_c - l_d} \begin{Bmatrix} l_a & l_b & \Lambda \\ \frac{1}{2} & \frac{1}{2} & \Sigma \\ j_a & j_b & J_f \end{Bmatrix} \begin{Bmatrix} l_c & l_d & \Lambda \\ \frac{1}{2} & \frac{1}{2} & \Sigma \\ j_c & j_d & J_i \end{Bmatrix} \begin{pmatrix} l_a & l_b & \Lambda \\ 0 & 0 & 0 \end{pmatrix} \begin{pmatrix} l_c & l_d & \Lambda \\ 0 & 0 & 0 \end{pmatrix} \\
\begin{pmatrix} j_a & j_b & J_f \\ m_a & m_b & -M_{J_f} \end{pmatrix} \begin{pmatrix} j_c & j_d & J_i \\ m_c & m_d & -M_{J_i} \end{pmatrix} \begin{pmatrix} \Lambda & \Sigma & J_f \\ M_\Lambda & M_\Sigma & -M_{J_f} \end{pmatrix} \begin{pmatrix} \Lambda & \Sigma & J_i \\ M_\Lambda & M_\Sigma & -M_{J_i} \end{pmatrix} \\
\int \mathrm{d}r\, r^2 R_{n_a l_a}(r) R_{n_b l_b}(r) g(r) R_{n_c l_c}(r) R_{n_d l_d}(r)
\end{aligned}
$$

### $\langle ab|\bar{V}|cd\rangle_\sigma$

$$
\begin{aligned}
\langle ab|\bar{V}|cd\rangle_\sigma = -\frac{3}{2} \sum_{\mu} \sum_{\substack{J_i M_{J_i} \\ J_f M_{J_f}}} \sum_{\substack{\Sigma_i M_{\Sigma_i} \\ \Sigma_f M_{\Sigma_f}}} \sum_{\Lambda M_\Lambda} \sum_{\substack{\sigma_a \sigma_b \\ \sigma_c \sigma_d}} i^{-l_a - l_b + l_c + l_d} \hat{J}_i^2 \hat{\Sigma}_i^2 \hat{J}_f^2 \hat{\Sigma}_f^2 \hat{\Lambda}^2 \frac{\hat{j}_a \hat{j}_b \hat{j}_c \hat{j}_d \hat{l}_a \hat{l}_b \hat{l}_c \hat{l}_d}{4\pi} \\
(-)^{j_a - j_b - \Sigma_f + M_{\Sigma_f} + l_a - l_b + j_c - j_d - \Sigma_i + M_{\Sigma_i} + l_c - l_d} \\
\delta_{\tau_a \tau_c} \delta_{\tau_b \tau_d} (-)^{1 + \sigma_a + \sigma_b + \mu} \begin{pmatrix} \frac{1}{2} & 1 & \frac{1}{2} \\ \sigma_c & \mu & -\sigma_a \end{pmatrix} \begin{pmatrix} \frac{1}{2} & 1 & \frac{1}{2} \\ \sigma_d & -\mu & -\sigma_b \end{pmatrix} \\
\begin{Bmatrix} l_a & l_b & \Lambda \\ \frac{1}{2} & \frac{1}{2} & \Sigma_f \\ j_a & j_b & J_f \end{Bmatrix} \begin{Bmatrix} l_c & l_d & \Lambda \\ \frac{1}{2} & \frac{1}{2} & \Sigma_i \\ j_c & j_d & J_i \end{Bmatrix}
\end{aligned}
$$

$$\begin{pmatrix} j_a & j_b & J_f \\ m_a & m_b & -M_{J_f} \end{pmatrix} \begin{pmatrix} \Lambda & \Sigma_f & J_f \\ M_\Lambda & M_{\Sigma_f} & -M_{J_f} \end{pmatrix} \begin{pmatrix} \frac{1}{2} & \frac{1}{2} & \Sigma_f \\ \sigma_a & \sigma_b & -M_{\Sigma_f} \end{pmatrix} \begin{pmatrix} l_a & l_b & \Lambda \\ 0 & 0 & 0 \end{pmatrix}$$
$$\begin{pmatrix} j_c & j_d & J_i \\ m_c & m_d & -M_{J_i} \end{pmatrix} \begin{pmatrix} \Lambda & \Sigma_i & J_i \\ M_\Lambda & M_{\Sigma_i} & -M_{J_i} \end{pmatrix} \begin{pmatrix} \frac{1}{2} & \frac{1}{2} & \Sigma_i \\ \sigma_c & \sigma_d & -M_{\Sigma_i} \end{pmatrix} \begin{pmatrix} l_c & l_d & \Lambda \\ 0 & 0 & 0 \end{pmatrix}$$
$$\int \mathrm{d}r\, r^2 R_{n_a l_a}(r) R_{n_b l_b}(r) g(r) R_{n_c l_c}(r) R_{n_d l_d}(r)$$

We can add the four $3-j$ coefficients which contain the information on the spin into a $6-j$.

$$\sum_{\substack{\sigma_a\sigma_b \\ \sigma_c\sigma_d \\ \mu}} (-)^{1+\sigma_a+\sigma_b+\mu} \begin{pmatrix} \frac{1}{2} & 1 & \frac{1}{2} \\ \sigma_c & \mu & -\sigma_a \end{pmatrix} \begin{pmatrix} \frac{1}{2} & 1 & \frac{1}{2} \\ \sigma_d & -\mu & -\sigma_b \end{pmatrix} \begin{pmatrix} \frac{1}{2} & \frac{1}{2} & \Sigma_f \\ \sigma_a & \sigma_b & -M_{\Sigma_f} \end{pmatrix} \begin{pmatrix} \frac{1}{2} & \frac{1}{2} & \Sigma_i \\ \sigma_c & \sigma_d & -M_{\Sigma_i} \end{pmatrix}$$
$$= \sum_{\substack{\sigma_a\sigma_b \\ \sigma_c\sigma_d \\ \mu}} (-)^{1+\sigma_a+\sigma_b+\mu} \begin{pmatrix} \frac{1}{2} & \frac{1}{2} & \Sigma_f \\ \sigma_a & \sigma_b & -M_{\Sigma_f} \end{pmatrix} \begin{pmatrix} \frac{1}{2} & 1 & \frac{1}{2} \\ -\sigma_b & -\mu & \sigma_d \end{pmatrix} \begin{pmatrix} 1 & \frac{1}{2} & \frac{1}{2} \\ \mu & -\sigma_a & \sigma_c \end{pmatrix} \begin{pmatrix} \frac{1}{2} & \frac{1}{2} & \Sigma_i \\ \sigma_d & \sigma_c & -M_{\Sigma_i} \end{pmatrix}$$
$$= \frac{(-)^{1+\Sigma_i}}{\hat{\Sigma}_i} \delta_{\Sigma_i \Sigma_f} \delta_{M_{\Sigma_i} M_{\Sigma_f}} \begin{Bmatrix} \frac{1}{2} & \frac{1}{2} & \Sigma_i \\ \frac{1}{2} & \frac{1}{2} & 1 \end{Bmatrix}$$

$$\langle ab|\bar{V}|cd\rangle_\sigma = -\frac{3}{2} \sum_{\substack{J_i M_{J_i} \\ J_f M_{J_f}}} \sum_{\Sigma M_\Sigma} \sum_{\Lambda M_\Lambda} i^{-l_a-l_b+l_c+l_d} \hat{J}_i^2 \hat{J}_f^2 \hat{\Lambda}^2 \hat{\Sigma}^2 \frac{\hat{j}_a \hat{j}_b \hat{j}_c \hat{j}_d \hat{l}_a \hat{l}_b \hat{l}_c \hat{l}_d}{4\pi}$$
$$(-)^{j_a-j_b+j_c-j_d+l_a-l_b+l_c-l_d} (-)^{1+\Sigma} \begin{Bmatrix} \frac{1}{2} & \frac{1}{2} & \Sigma \\ \frac{1}{2} & \frac{1}{2} & 1 \end{Bmatrix}$$
$$\begin{Bmatrix} l_a & l_b & \Lambda \\ \frac{1}{2} & \frac{1}{2} & \Sigma \\ j_a & j_b & J_f \end{Bmatrix} \begin{Bmatrix} l_c & l_d & \Lambda \\ \frac{1}{2} & \frac{1}{2} & \Sigma \\ j_c & j_d & J_i \end{Bmatrix} \begin{pmatrix} l_a & l_b & \Lambda \\ 0 & 0 & 0 \end{pmatrix} \begin{pmatrix} l_c & l_d & \Lambda \\ 0 & 0 & 0 \end{pmatrix}$$
$$\begin{pmatrix} j_a & j_b & J_f \\ m_a & m_b & -M_{J_f} \end{pmatrix} \begin{pmatrix} j_c & j_d & J_i \\ m_c & m_d & -M_{J_i} \end{pmatrix} \begin{pmatrix} \Lambda & \Sigma & J_f \\ M_\Lambda & M_\Sigma & -M_{J_f} \end{pmatrix} \begin{pmatrix} \Lambda & \Sigma & J_i \\ M_\Lambda & M_\Sigma & -M_{J_i} \end{pmatrix}$$
$$\int \mathrm{d}r\, r^2 R_{n_a l_a}(r) R_{n_b l_b}(r) g(r) R_{n_c l_c}(r) R_{n_d l_d}(r)$$

### $\langle ab|\bar{V}|cd\rangle_\tau$

The only difference between this term and $\langle ab|\bar{v}|cd\rangle_0$ is the isospin part, then

$$\langle ab|\bar{V}|cd\rangle_\tau = -\frac{3}{2} \sum_{\mu} (-)^{1+\tau_a+\tau_b+\mu} \begin{pmatrix} \frac{1}{2} & 1 & \frac{1}{2} \\ \tau_c & \mu & -\tau_a \end{pmatrix} \begin{pmatrix} \frac{1}{2} & 1 & \frac{1}{2} \\ \tau_d & -\mu & -\tau_b \end{pmatrix}$$

$$\sum_{\substack{J_iM_{J_i}\\J_fM_{J_f}}}\sum_{\Sigma M_\Sigma}\sum_{\Lambda M_\Lambda} i^{-l_a-l_b+l_c+l_d}\hat{J}_i^2\hat{J}_f^2\hat{\Lambda}^2\hat{\Sigma}^2\frac{\hat{j}_a\hat{j}_b\hat{j}_c\hat{j}_d\hat{l}_a\hat{l}_b\hat{l}_c\hat{l}_d}{4\pi}$$

$$(-)^{j_a-j_b+j_c-j_d+l_a-l_b+l_c-l_d}\begin{Bmatrix} l_a & l_b & \Lambda\\ \frac{1}{2} & \frac{1}{2} & \Sigma\\ j_a & j_b & J_f\end{Bmatrix}\begin{Bmatrix} l_c & l_d & \Lambda\\ \frac{1}{2} & \frac{1}{2} & \Sigma\\ j_c & j_d & J_i\end{Bmatrix}\begin{pmatrix} l_a & l_b & \Lambda\\ 0 & 0 & 0\end{pmatrix}\begin{pmatrix} l_c & l_d & \Lambda\\ 0 & 0 & 0\end{pmatrix}$$

$$\begin{pmatrix} j_a & j_b & J_f\\ m_a & m_b & -M_{J_f}\end{pmatrix}\begin{pmatrix} j_c & j_d & J_i\\ m_c & m_d & -M_{J_i}\end{pmatrix}\begin{pmatrix} \Lambda & \Sigma & J_f\\ M_\Lambda & M_\Sigma & -M_{J_f}\end{pmatrix}\begin{pmatrix} \Lambda & \Sigma & J_i\\ M_\Lambda & M_\Sigma & -M_{J_i}\end{pmatrix}$$

$$\int \mathrm{d}r\, r^2 R_{n_al_a}(r)R_{n_bl_b}(r)g(r)R_{n_cl_c}(r)R_{n_dl_d}(r)$$

### $\langle ab|\bar{V}|cd\rangle_{\sigma\tau}$

The only difference between this term and $\langle ab|\bar{v}|cd\rangle_\sigma$ is the isospin part, then

$$\langle ab|\bar{V}|cd\rangle_{\sigma\tau} = -9\sum_\mu(-)^{1+\tau_a+\tau_b+\mu}\begin{pmatrix} \frac{1}{2} & 1 & \frac{1}{2}\\ \tau_c & \mu & -\tau_a\end{pmatrix}\begin{pmatrix} \frac{1}{2} & 1 & \frac{1}{2}\\ \tau_d & -\mu & -\tau_b\end{pmatrix}$$

$$\sum_{\substack{J_iM_{J_i}\\J_fM_{J_f}}}\sum_{\Sigma M_\Sigma}\sum_{\Lambda M_\Lambda} i^{-l_a-l_b+l_c+l_d}\hat{J}_i^2\hat{J}_f^2\hat{\Lambda}^2\hat{\Sigma}^2\frac{\hat{j}_a\hat{j}_b\hat{j}_c\hat{j}_d\hat{l}_a\hat{l}_b\hat{l}_c\hat{l}_d}{4\pi}$$

$$(-)^{j_a-j_b+j_c-j_d+l_a-l_b+l_c-l_d}(-)^{1+\Sigma}\begin{Bmatrix} \frac{1}{2} & \frac{1}{2} & \Sigma\\ \frac{1}{2} & \frac{1}{2} & 1\end{Bmatrix}$$

$$\begin{Bmatrix} l_a & l_b & \Lambda\\ \frac{1}{2} & \frac{1}{2} & \Sigma\\ j_a & j_b & J_f\end{Bmatrix}\begin{Bmatrix} l_c & l_d & \Lambda\\ \frac{1}{2} & \frac{1}{2} & \Sigma\\ j_c & j_d & J_i\end{Bmatrix}\begin{pmatrix} l_a & l_b & \Lambda\\ 0 & 0 & 0\end{pmatrix}\begin{pmatrix} l_c & l_d & \Lambda\\ 0 & 0 & 0\end{pmatrix}$$

$$\begin{pmatrix} j_a & j_b & J_f\\ m_a & m_b & -M_{J_f}\end{pmatrix}\begin{pmatrix} j_c & j_d & J_i\\ m_c & m_d & -M_{J_i}\end{pmatrix}\begin{pmatrix} \Lambda & \Sigma & J_f\\ M_\Lambda & M_\Sigma & -M_{J_f}\end{pmatrix}\begin{pmatrix} \Lambda & \Sigma & J_i\\ M_\Lambda & M_\Sigma & -M_{J_i}\end{pmatrix}$$

$$\int \mathrm{d}r\, r^2 R_{n_al_a}(r)R_{n_bl_b}(r)g(r)R_{n_cl_c}(r)R_{n_dl_d}(r)$$

In the following the notation

$$\mathscr{F}(\tau) = \begin{cases} \delta_{\tau_a\tau_c}\delta_{\tau_b\tau_d} & \text{for } \langle ab|\bar{V}|cd\rangle_{0,\sigma}\\ \sum_\mu(-)^{1+\tau_a+\tau_b+\mu}\begin{pmatrix} \frac{1}{2} & 1 & \frac{1}{2}\\ \tau_c & \mu & -\tau_a\end{pmatrix}\begin{pmatrix} \frac{1}{2} & 1 & \frac{1}{2}\\ \tau_d & -\mu & -\tau_b\end{pmatrix} & \text{for } \langle ab|\bar{V}|cd\rangle_{\tau,\sigma\tau}\end{cases}$$

$$\mathscr{G}(\Sigma) = \begin{cases} 1 & \text{for } \langle ab|\bar{V}|cd\rangle_{0,\tau}\\ (-)^{1+\Sigma}\begin{Bmatrix} \frac{1}{2} & \frac{1}{2} & \Sigma\\ \frac{1}{2} & \frac{1}{2} & 1\end{Bmatrix} & \text{for } \langle ab|\bar{V}|cd\rangle_{\sigma,\sigma\tau}\end{cases}$$

$$\mathscr{N} = \begin{cases} \frac{3}{4} & \text{for } \langle ab|\bar{V}|cd\rangle_{0}\\ -\frac{3}{2} & \text{for } \langle ab|\bar{V}|cd\rangle_{\sigma,\tau}\\ -9 & \text{for } \langle ab|\bar{V}|cd\rangle_{\sigma\tau}\end{cases}$$

is used.

$$\langle ab|\bar{V}|cd\rangle_{0,\sigma,\tau,\sigma\tau} = \mathcal{N}\mathcal{F}(\tau)\sum_{\substack{J_iM_{J_i}\\J_fM_{J_f}}}\sum_{\Sigma M_\Sigma}\sum_{\Lambda M_\Lambda} i^{-l_a-l_b+l_c+l_d}\hat{J}_i^2\hat{J}_f^2\hat{\Lambda}^2\hat{\Sigma}^2\frac{\hat{j}_a\hat{j}_b\hat{j}_c\hat{j}_d\hat{l}_a\hat{l}_b\hat{l}_c\hat{l}_d}{4\pi}$$
$$(-)^{j_a-j_b+j_c-j_d+l_a-l_b+l_c-l_d}\mathcal{G}(\Sigma)$$
$$\begin{Bmatrix} l_a & l_b & \Lambda\\ \frac{1}{2} & \frac{1}{2} & \Sigma\\ j_a & j_b & J_f\end{Bmatrix}\begin{Bmatrix} l_c & l_d & \Lambda\\ \frac{1}{2} & \frac{1}{2} & \Sigma\\ j_c & j_d & J_i\end{Bmatrix}\begin{pmatrix} l_a & l_b & \Lambda\\ 0 & 0 & 0\end{pmatrix}\begin{pmatrix} l_c & l_d & \Lambda\\ 0 & 0 & 0\end{pmatrix}$$
$$\begin{pmatrix} j_a & j_b & J_f\\ m_a & m_b & -M_{J_f}\end{pmatrix}\begin{pmatrix} j_c & j_d & J_i\\ m_c & m_d & -M_{J_i}\end{pmatrix}\begin{pmatrix} \Lambda & \Sigma & J_f\\ M_\Lambda & M_\Sigma & -M_{J_f}\end{pmatrix}\begin{pmatrix} \Lambda & \Sigma & J_i\\ M_\Lambda & M_\Sigma & -M_{J_i}\end{pmatrix}$$
$$\int \mathrm{d}r\, r^2 R_{n_al_a}(r)R_{n_bl_b}(r)g(r)R_{n_cl_c}(r)R_{n_dl_d}(r)$$
$$=\mathcal{N}\mathcal{F}(\tau)\sum_{JM_J}\sum_{\Sigma\Lambda} i^{-l_a-l_b+l_c+l_d}\hat{J}^2\hat{\Lambda}^2\hat{\Sigma}^2\frac{\hat{j}_a\hat{j}_b\hat{j}_c\hat{j}_d\hat{l}_a\hat{l}_b\hat{l}_c\hat{l}_d}{4\pi}(-)^{l_a-l_b+l_c-l_d}\mathcal{G}(\Sigma)$$
$$\begin{Bmatrix} l_a & l_b & \Lambda\\ \frac{1}{2} & \frac{1}{2} & \Sigma\\ j_a & j_b & J\end{Bmatrix}\begin{Bmatrix} l_c & l_d & \Lambda\\ \frac{1}{2} & \frac{1}{2} & \Sigma\\ j_c & j_d & J\end{Bmatrix}\begin{pmatrix} l_a & l_b & \Lambda\\ 0 & 0 & 0\end{pmatrix}\begin{pmatrix} l_c & l_d & \Lambda\\ 0 & 0 & 0\end{pmatrix}$$
$$(-)^{j_a-j_b+j_c-j_d}\begin{pmatrix} j_a & j_b & J\\ m_a & m_b & -M_J\end{pmatrix}\begin{pmatrix} j_c & j_d & J\\ m_c & m_d & -M_J\end{pmatrix}$$
$$\int \mathrm{d}r\, r^2 R_{n_al_a}(r)R_{n_bl_b}(r)g(r)R_{n_cl_c}(r)R_{n_dl_d}(r)$$

which is Eq. (10.26).

## I.2 Matrix element of the interaction in center of mass and relative motion coordinates

$$\langle ab|\bar{V}|cd\rangle = \langle n_al_aj_am_a\tau_a, n_bl_bj_bm_b\tau_b|\bar{V}|n_cl_cj_cm_c\tau_c, n_dl_dj_dm_d\tau_d\rangle = \tag{I.10}$$
$$= \sum_{\substack{J_i\Lambda_i\Sigma_i\\J_f\Lambda_f\Sigma_f}}\sum_{\substack{M_{J_i}M_{\Sigma_i}\\M_{\Lambda_i}\\M_{l_i}M_{L_i}}}\sum_{\substack{M_{J_f}M_{\Sigma_f}\\M_{\Lambda_f}\\M_{l_f}M_{L_f}}}\sum_{\substack{l_il_f\\L_iL_f}}\sum_{\substack{\sigma_a\sigma_b\\\sigma_c\sigma_d}} i^{-l_a-l_b+l_c+l_d}\hat{J}_i^2\hat{\Lambda}_i^2\hat{\Sigma}_i^2\hat{J}_f^2\hat{\Lambda}_f^2\hat{\Sigma}_f^2\hat{j}_a\hat{j}_b\hat{j}_c\hat{j}_d$$
$$(-)^{j_a-j_b+\Lambda_f+M_{\Lambda_f}+\Sigma_f+M_{\Sigma_f}+l_f+L_f+j_c-j_d+\Lambda_i+M_{\Lambda_i}+\Sigma_i+M_{\Sigma_i}+l_i+L_i}$$
$$\begin{Bmatrix} l_a & l_b & \Lambda_f\\ \frac{1}{2} & \frac{1}{2} & \Sigma_f\\ j_a & j_b & J_f\end{Bmatrix}\begin{Bmatrix} l_c & l_d & \Lambda_i\\ \frac{1}{2} & \frac{1}{2} & \Sigma_i\\ j_c & j_d & J_i\end{Bmatrix}$$
$$\begin{pmatrix} j_a & j_b & J_f\\ m_a & m_b & -M_{J_f}\end{pmatrix}\begin{pmatrix} \Lambda_f & \Sigma_f & J_f\\ M_{\Lambda_f} & M_{\Sigma_f} & -M_{J_f}\end{pmatrix}\begin{pmatrix} \frac{1}{2} & \frac{1}{2} & \Sigma_f\\ \sigma_a & \sigma_b & -M_{\Sigma_f}\end{pmatrix}\begin{pmatrix} L_f & l_f & \Lambda_f\\ M_{L_f} & M_{l_f} & -M_{\Lambda_f}\end{pmatrix}$$

$$
\begin{pmatrix} j_c & j_d & J_i \\ m_c & m_d & -M_{J_i} \end{pmatrix}\begin{pmatrix} \Lambda_i & \Sigma_i & J_i \\ M_{\Lambda_i} & M_{\Sigma_i} & -M_{J_i} \end{pmatrix}\begin{pmatrix} \frac{1}{2} & \frac{1}{2} & \Sigma_i \\ \sigma_c & \sigma_d & -M_{\Sigma_i} \end{pmatrix}\begin{pmatrix} L_i & l_i & \Lambda_i \\ M_{L_i} & M_{l_i} & -M_{\Lambda_i} \end{pmatrix}
$$

$$
\sum_{\substack{n_i n_f \\ N_i N_f}} M_{\Lambda_f}(N_f L_f n_f l_f; n_a l_a n_b l_b) M_{\Lambda_i}(N_i L_i n_i l_i; n_c l_c n_d l_d)
$$

$$
\frac{\sqrt{2}}{4}\int d_3 R d_3 R' \, R_{N_f L_f}(\alpha R') Y^*_{L_f M_{L_f}}(\hat{R}') g\left(\frac{\boldsymbol{R}}{\sqrt{2}}\right)\delta_3(\boldsymbol{R}-\boldsymbol{R}') R_{N_i L_i}(\alpha R) Y_{L_i M_{L_i}}(\hat{R}) \tag{I.11}
$$

$$
\int d_3 r d_3 r' \, R_{n_f l_f}(\alpha r') Y^*_{l_f M_{l_f}}(\hat{r}') v(\boldsymbol{r}', \boldsymbol{r}) R_{n_i l_i}(\alpha r) Y_{l_i M_{l_i}}(\hat{r})
$$

$$
\begin{aligned}
&\Big[\langle \tfrac{1}{2}\sigma_a, \tfrac{1}{2}\sigma_b | \mathbb{1}_\sigma | \tfrac{1}{2}\sigma_c, \tfrac{1}{2}\sigma_d\rangle\langle \tfrac{1}{2}\tau_a, \tfrac{1}{2}\tau_b | \mathbb{1}_\tau | \tfrac{1}{2}\tau_c, \tfrac{1}{2}\tau_d\rangle \left(1-\frac{1}{4}(-1)^{l_c+l_d-L_i}\right) \\
&-\frac{1}{4}(-1)^{l_c+l_d-L_i}\langle \tfrac{1}{2}\sigma_a, \tfrac{1}{2}\sigma_b | \boldsymbol{\sigma}(1)\boldsymbol{\sigma}(2) | \tfrac{1}{2}\sigma_c, \tfrac{1}{2}\sigma_d\rangle\langle \tfrac{1}{2}\tau_a, \tfrac{1}{2}\tau_b | \mathbb{1}_\tau | \tfrac{1}{2}\tau_c, \tfrac{1}{2}\tau_d\rangle \\
&-\frac{1}{4}(-1)^{l_c+l_d-L_i}\langle \tfrac{1}{2}\sigma_a, \tfrac{1}{2}\sigma_b | \mathbb{1}_\sigma | \tfrac{1}{2}\sigma_c, \tfrac{1}{2}\sigma_d\rangle\langle \tfrac{1}{2}\tau_a, \tfrac{1}{2}\tau_b | \boldsymbol{\tau}(1)\boldsymbol{\tau}(2) | \tfrac{1}{2}\tau_c, \tfrac{1}{2}\tau_d\rangle \\
&-\frac{1}{4}(-1)^{l_c+l_d-L_i}\langle \tfrac{1}{2}\sigma_a, \tfrac{1}{2}\sigma_b | \boldsymbol{\sigma}(1)\boldsymbol{\sigma}(2) | \tfrac{1}{2}\sigma_c, \tfrac{1}{2}\sigma_d\rangle\langle \tfrac{1}{2}\tau_a, \tfrac{1}{2}\tau_b | \boldsymbol{\tau}(1)\boldsymbol{\tau}(2) | \tfrac{1}{2}\tau_c, \tfrac{1}{2}\tau_d\rangle\Big].
\end{aligned} \tag{I.12}
$$

The integral in Eq. (I.11) can be written scaling the $\boldsymbol{R}$ coordinate.

$$
\begin{aligned}
\text{Eq. (I.11)} =& \frac{\sqrt{2}}{4}\int d_3 R R_{N_f L_f}(\alpha R) Y^*_{L_f M_{L_f}}(\hat{R}) g\left(\frac{\boldsymbol{R}}{\sqrt{2}}\right) R_{N_i L_i}(\alpha R) Y_{L_i M_{L_i}}(\hat{R}) \\
=& \frac{2\sqrt{2}\sqrt{2}}{4}\int d_3 R R_{N_f L_f}(\sqrt{2}\alpha R) Y^*_{L_f M_{L_f}}(\hat{R}) g(\boldsymbol{R}) R_{N_i L_i}(\sqrt{2}\alpha R) Y_{L_i M_{L_i}}(\hat{R}) \\
=& \int d_3 R R_{N_f L_f}(\sqrt{2}\alpha R) Y^*_{L_f M_{L_f}}(\hat{R}) g(\boldsymbol{R}) R_{N_i L_i}(\sqrt{2}\alpha R) Y_{L_i M_{L_i}}(\hat{R})
\end{aligned} \tag{I.13}
$$

In Eq. (I.12) there are matrix elements like ($\varphi \equiv \sigma, \tau$ and $\Phi \equiv S, T$)

$$
\begin{aligned}
&\langle \tfrac{1}{2}\varphi_a, \tfrac{1}{2}\varphi_b | \boldsymbol{\varphi}(1)\boldsymbol{\varphi}(2) | \tfrac{1}{2}\varphi_c, \tfrac{1}{2}\varphi_d\rangle = 4\sum_\mu (-)^\mu \langle \tfrac{1}{2}\varphi_a, \tfrac{1}{2}\varphi_b | \Phi_{1\mu}(1)\Phi_{1-\mu}(2) | \tfrac{1}{2}\varphi_c, \tfrac{1}{2}\varphi_d\rangle \\
&=4\sum_\mu (-)^\mu \langle \tfrac{1}{2}\varphi_a | \Phi_{1\mu} | \tfrac{1}{2}\varphi_c\rangle\langle \tfrac{1}{2}\varphi_b | \Phi_{1-\mu} | \tfrac{1}{2}\varphi_d\rangle \\
&=4\sum_\mu (-)^\mu \frac{1}{2}\langle \tfrac{1}{2}\varphi_c 1\mu | \tfrac{1}{2}\varphi_a\rangle\langle \tfrac{1}{2}\varphi_b 1-\mu | \tfrac{1}{2}\varphi_d\rangle\langle \tfrac{1}{2} \| \Phi_1 \| \tfrac{1}{2}\rangle\langle \tfrac{1}{2} \| \Phi_1 \| \tfrac{1}{2}\rangle \\
&=6\sum_\mu (-)^{1+\varphi_a+\varphi_b+\mu}\begin{pmatrix} \frac{1}{2} & 1 & \frac{1}{2} \\ \varphi_c & \mu & -\varphi_a \end{pmatrix}\begin{pmatrix} \frac{1}{2} & 1 & \frac{1}{2} \\ \varphi_d & -\mu & -\varphi_b \end{pmatrix}
\end{aligned} \tag{I.14}
$$

We shall treat each term in Eq. (I.12) one by one.

$$
\langle \tfrac{1}{2}\sigma_a, \tfrac{1}{2}\sigma_b | \mathbb{1}_\sigma | \tfrac{1}{2}\sigma_c, \tfrac{1}{2}\sigma_d\rangle\langle \tfrac{1}{2}\tau_a, \tfrac{1}{2}\tau_b | \mathbb{1}_\tau | \tfrac{1}{2}\tau_c, \tfrac{1}{2}\tau_d\rangle = \delta_{\sigma_a\sigma_c}\delta_{\sigma_b\sigma_d}\delta_{\tau_a\tau_c}\delta_{\tau_b\tau_d} \tag{I.15}
$$

$$\langle \frac{1}{2}\sigma_a, \frac{1}{2}\sigma_b|\boldsymbol{\sigma}(1)\boldsymbol{\sigma}(2)|\frac{1}{2}\sigma_c, \frac{1}{2}\sigma_d\rangle\langle \frac{1}{2}\tau_a, \frac{1}{2}\tau_b|\mathbb{1}_\tau|\frac{1}{2}\tau_c, \frac{1}{2}\tau_d\rangle =$$

$$= 6\delta_{\tau_a\tau_c}\delta_{\tau_b\tau_d}\sum_{\mu}(-)^{1+\sigma_a+\sigma_b+\mu}\begin{pmatrix}\frac{1}{2} & 1 & \frac{1}{2}\\ \sigma_c & \mu & -\sigma_a\end{pmatrix}\begin{pmatrix}\frac{1}{2} & 1 & \frac{1}{2}\\ \sigma_d & -\mu & -\sigma_b\end{pmatrix} \tag{I.16}$$

$$\langle \frac{1}{2}\sigma_a, \frac{1}{2}\sigma_b|\mathbb{1}_\sigma|\frac{1}{2}\sigma_c, \frac{1}{2}\sigma_d\rangle\langle \frac{1}{2}\tau_a, \frac{1}{2}\tau_b|\boldsymbol{\tau}(1)\boldsymbol{\tau}(2)|\frac{1}{2}\tau_c, \frac{1}{2}\tau_d\rangle =$$

$$= 6\delta_{\sigma_a\sigma_c}\delta_{\sigma_a\sigma_c}\sum_{\mu}(-)^{1+\tau_a+\tau_b+\mu}\begin{pmatrix}\frac{1}{2} & 1 & \frac{1}{2}\\ \tau_c & \mu & -\tau_a\end{pmatrix}\begin{pmatrix}\frac{1}{2} & 1 & \frac{1}{2}\\ \tau_d & -\mu & -\tau_b\end{pmatrix} \tag{I.17}$$

$$\langle \frac{1}{2}\sigma_a, \frac{1}{2}\sigma_b|\boldsymbol{\sigma}(1)\boldsymbol{\sigma}(2)|\frac{1}{2}\sigma_c, \frac{1}{2}\sigma_d\rangle\langle \frac{1}{2}\tau_a, \frac{1}{2}\tau_b|\boldsymbol{\tau}(1)\boldsymbol{\tau}(2)|\frac{1}{2}\tau_c, \frac{1}{2}\tau_d\rangle =$$

$$= 36\sum_{\mu}(-)^{1+\sigma_a+\sigma_b+\mu}\begin{pmatrix}\frac{1}{2} & 1 & \frac{1}{2}\\ \sigma_c & \mu & -\sigma_a\end{pmatrix}\begin{pmatrix}\frac{1}{2} & 1 & \frac{1}{2}\\ \sigma_d & -\mu & -\sigma_b\end{pmatrix} \tag{I.18}$$

$$\sum_{\mu'}(-)^{1+\tau_a+\tau_b+\mu'}\begin{pmatrix}\frac{1}{2} & 1 & \frac{1}{2}\\ \tau_c & \mu' & -\tau_a\end{pmatrix}\begin{pmatrix}\frac{1}{2} & 1 & \frac{1}{2}\\ \tau_d & -\mu' & -\tau_b\end{pmatrix}$$

Therefore, the matrix element (I.10) is composed by four terms with different angular structure.

### $\langle ab|\bar{V}|cd\rangle_0$

$$\langle ab|\bar{V}|cd\rangle_0 = \sum_{\substack{J_i\Lambda_i\Sigma_i\\ J_f\Lambda_f\Sigma_f}}\sum_{\substack{M_{J_i}M_{\Sigma_i}\\ M_{\Lambda_i}\\ M_{l_i}M_{L_i}}}\sum_{\substack{M_{J_f}M_{\Sigma_f}\\ M_{\Lambda_f}\\ M_{l_f}M_{L_f}}}\sum_{\substack{l_il_f\\ L_iL_f}}\sum_{\substack{\sigma_a\sigma_b\\ \sigma_c\sigma_d}} i^{-l_a-l_b+l_c+l_d}\hat{J}_i^2\hat{\Lambda}_i^2\hat{\Sigma}_i^2\hat{J}_f^2\hat{\Lambda}_f^2\hat{\Sigma}_f^2\hat{j}_a\hat{j}_b\hat{j}_c\hat{j}_d$$

$$(-)^{j_a-j_b+\Lambda_f+M_{\Lambda_f}+\Sigma_f+M_{\Sigma_f}+l_f+L_f+j_c-j_d+\Lambda_i+M_{\Lambda_i}+\Sigma_i+M_{\Sigma_i}+l_i+L_i}$$

$$\left(1-\frac{1}{4}(-1)^{l_c+l_d-L_i}\right)\delta_{\sigma_a\sigma_c}\delta_{\sigma_b\sigma_d}\delta_{\tau_a\tau_c}\delta_{\tau_b\tau_d}\begin{Bmatrix}l_a & l_b & \Lambda_f\\ \frac{1}{2} & \frac{1}{2} & \Sigma_f\\ j_a & j_b & J_f\end{Bmatrix}\begin{Bmatrix}l_c & l_d & \Lambda_i\\ \frac{1}{2} & \frac{1}{2} & \Sigma_i\\ j_c & j_d & J_i\end{Bmatrix}$$

$$\begin{pmatrix}j_a & j_b & J_f\\ m_a & m_b & -M_{J_f}\end{pmatrix}\begin{pmatrix}\Lambda_f & \Sigma_f & J_f\\ M_{\Lambda_f} & M_{\Sigma_f} & -M_{J_f}\end{pmatrix}\begin{pmatrix}\frac{1}{2} & \frac{1}{2} & \Sigma_f\\ \sigma_a & \sigma_b & -M_{\Sigma_f}\end{pmatrix}\begin{pmatrix}L_f & l_f & \Lambda_f\\ M_{L_f} & M_{l_f} & -M_{\Lambda_f}\end{pmatrix}$$

$$\begin{pmatrix}j_c & j_d & J_i\\ m_c & m_d & -M_{J_i}\end{pmatrix}\begin{pmatrix}\Lambda_i & \Sigma_i & J_i\\ M_{\Lambda_i} & M_{\Sigma_i} & -M_{J_i}\end{pmatrix}\begin{pmatrix}\frac{1}{2} & \frac{1}{2} & \Sigma_i\\ \sigma_c & \sigma_d & -M_{\Sigma_i}\end{pmatrix}\begin{pmatrix}L_i & l_i & \Lambda_i\\ M_{L_i} & M_{l_i} & -M_{\Lambda_i}\end{pmatrix}$$

$$\sum_{\substack{n_in_f\\ N_iN_f}} M_{\Lambda_f}(N_fL_fn_fl_f;n_al_an_bl_b)M_{\Lambda_i}(N_iL_in_il_i;n_cl_cn_dl_d)$$

$$\int \mathrm{d}3RR_{N_fL_f}(\sqrt{2}\alpha R)Y^*_{L_fM_{L_f}}(\hat{R})g(\boldsymbol{R})R_{N_iL_i}(\sqrt{2}\alpha R)Y_{L_iM_{L_i}}(\hat{R})$$

$$\int \mathrm{d}3r\mathrm{d}3r'\, R_{n_fl_f}(\alpha r')Y^*_{l_fM_{l_f}}(\hat{r}')v(\boldsymbol{r}',\boldsymbol{r})R_{n_il_i}(\alpha r)Y_{l_iM_{l_i}}(\hat{r})$$

$$
\begin{aligned}
=&\delta_{\tau_a\tau_c}\delta_{\tau_b\tau_d}\sum_{\substack{J_i\Lambda_i\Sigma_i\\J_f\Lambda_f\Sigma_f}}\sum_{\substack{M_{J_i}M_{\Sigma_i}\\M_{\Lambda_i}\\M_{l_i}M_{L_i}}}\sum_{\substack{M_{J_f}M_{\Sigma_f}\\M_{\Lambda_f}\\M_{l_f}M_{L_f}}}\sum_{\substack{l_il_f\\L_iL_f}}\sum_{\sigma_a\sigma_b} i^{-l_a-l_b+l_c+l_d}\hat{J}_i^2\hat{\Lambda}_i^2\hat{\Sigma}_i^2\hat{J}_f^2\hat{\Lambda}_f^2\hat{\Sigma}_f^2\hat{j}_a\hat{j}_b\hat{j}_c\hat{j}_d\\
&(-)^{j_a-j_b+\Lambda_f+M_{\Lambda_f}+\Sigma_f+M_{\Sigma_f}+l_f+L_f+j_c-j_d+\Lambda_i+M_{\Lambda_i}+\Sigma_i+M_{\Sigma_i}+l_i+L_i}\\
&\left(1-\frac{1}{4}(-1)^{l_c+l_d-L_i}\right)\begin{Bmatrix}l_a & l_b & \Lambda_f\\ \frac{1}{2} & \frac{1}{2} & \Sigma_f\\ j_a & j_b & J_f\end{Bmatrix}\begin{Bmatrix}l_c & l_d & \Lambda_i\\ \frac{1}{2} & \frac{1}{2} & \Sigma_i\\ j_c & j_d & J_i\end{Bmatrix}\\
&\begin{pmatrix}j_a & j_b & J_f\\ m_a & m_b & -M_{J_f}\end{pmatrix}\begin{pmatrix}\Lambda_f & \Sigma_f & J_f\\ M_{\Lambda_f} & M_{\Sigma_f} & -M_{J_f}\end{pmatrix}\begin{pmatrix}\frac{1}{2} & \frac{1}{2} & \Sigma_f\\ \sigma_a & \sigma_b & -M_{\Sigma_f}\end{pmatrix}\begin{pmatrix}L_f & l_f & \Lambda_f\\ M_{L_f} & M_{l_f} & -M_{\Lambda_f}\end{pmatrix}\\
&\begin{pmatrix}j_c & j_d & J_i\\ m_c & m_d & -M_{J_i}\end{pmatrix}\begin{pmatrix}\Lambda_i & \Sigma_i & J_i\\ M_{\Lambda_i} & M_{\Sigma_i} & -M_{J_i}\end{pmatrix}\begin{pmatrix}\frac{1}{2} & \frac{1}{2} & \Sigma_i\\ \sigma_a & \sigma_b & -M_{\Sigma_i}\end{pmatrix}\begin{pmatrix}L_i & l_i & \Lambda_i\\ M_{L_i} & M_{l_i} & -M_{\Lambda_i}\end{pmatrix}\\
&\sum_{\substack{n_in_f\\N_iN_f}} M_{\Lambda_f}(N_fL_fn_fl_f;n_al_an_bl_b)M_{\Lambda_i}(N_iL_in_il_i;n_cl_cn_dl_d)\\
&\int \mathrm{d}3R R_{N_fL_f}(\sqrt{2}\alpha R)Y^*_{L_fM_{L_f}}(\hat{R})g(\boldsymbol{R})R_{N_iL_i}(\sqrt{2}\alpha R)Y_{L_iM_{L_i}}(\hat{R})\\
&\int \mathrm{d}3r\mathrm{d}3r'\, R_{n_fl_f}(\alpha r')Y^*_{l_fM_{l_f}}(\hat{r}')v(\boldsymbol{r}',\boldsymbol{r})R_{n_il_i}(\alpha r)Y_{l_iM_{l_i}}(\hat{r})\\
=&\delta_{\tau_a\tau_c}\delta_{\tau_d\tau_d}\sum_{\substack{J_i\Lambda_i\\J_f\Lambda_f}}\sum_{\Sigma M_\Sigma}\sum_{\substack{M_{J_i}M_{\Lambda_i}\\M_{l_i}M_{L_i}}}\sum_{\substack{M_{J_f}M_{\Lambda_f}\\M_{l_f}M_{L_f}}}\sum_{\substack{l_il_f\\L_iL_f}} i^{-l_a-l_b+l_c+l_d}\hat{J}_i^2\hat{\Lambda}_i^2\hat{\Sigma}^2\hat{J}_f^2\hat{\Lambda}_f^2\hat{j}_a\hat{j}_b\hat{j}_c\hat{j}_d\\
&(-)^{j_a-j_b+\Lambda_f+M_{\Lambda_f}+l_f+L_f+j_c-j_d+\Lambda_i+M_{\Lambda_i}+l_i+L_i}\\
&\left(1-\frac{1}{4}(-1)^{l_c+l_d-L_i}\right)\begin{Bmatrix}l_a & l_b & \Lambda_f\\ \frac{1}{2} & \frac{1}{2} & \Sigma_f\\ j_a & j_b & J_f\end{Bmatrix}\begin{Bmatrix}l_c & l_d & \Lambda_i\\ \frac{1}{2} & \frac{1}{2} & \Sigma_i\\ j_c & j_d & J_i\end{Bmatrix}\\
&\begin{pmatrix}j_a & j_b & J_f\\ m_a & m_b & -M_{J_f}\end{pmatrix}\begin{pmatrix}\Lambda_f & \Sigma & J_f\\ M_{\Lambda_f} & M_\Sigma & -M_{J_f}\end{pmatrix}\begin{pmatrix}L_f & l_f & \Lambda_f\\ M_{L_f} & M_{l_f} & -M_{\Lambda_f}\end{pmatrix}\\
&\begin{pmatrix}j_c & j_d & J_i\\ m_c & m_d & -M_{J_i}\end{pmatrix}\begin{pmatrix}\Lambda_i & \Sigma & J_i\\ M_{\Lambda_i} & M_\Sigma & -M_{J_i}\end{pmatrix}\begin{pmatrix}L_i & l_i & \Lambda_i\\ M_{L_i} & M_{l_i} & -M_{\Lambda_i}\end{pmatrix}\\
&\sum_{\substack{n_in_f\\N_iN_f}} M_{\Lambda_f}(N_fL_fn_fl_f;n_al_an_bl_b)M_{\Lambda_i}(N_iL_in_il_i;n_cl_cn_dl_d)\\
&\int \mathrm{d}3R R_{N_fL_f}(\sqrt{2}\alpha R)Y^*_{L_fM_{L_f}}(\hat{R})g(\boldsymbol{R})R_{N_iL_i}(\sqrt{2}\alpha R)Y_{L_iM_{L_i}}(\hat{R})\\
&\int \mathrm{d}3r\mathrm{d}3r'\, R_{n_fl_f}(\alpha r')Y^*_{l_fM_{l_f}}(\hat{r}')v(\boldsymbol{r}',\boldsymbol{r})R_{n_il_i}(\alpha r)Y_{l_iM_{l_i}}(\hat{r})
\end{aligned}
$$

**$\langle ab|\bar{V}|cd\rangle_\sigma$**

$$\langle ab|\bar{V}|cd\rangle_\sigma = -\frac{3}{2}\sum_{\mu}\sum_{\substack{J_i\Lambda_i\Sigma_i\\J_f\Lambda_f\Sigma_f}}\sum_{\substack{M_{J_i}M_{\Sigma_i}\\M_{\Lambda_i}\\M_{l_i}M_{L_i}}}\sum_{\substack{M_{J_f}M_{\Sigma_f}\\M_{\Lambda_f}\\M_{l_f}M_{L_f}}}\sum_{\substack{l_il_f\\L_iL_f}}\sum_{\substack{\sigma_a\sigma_b\\\sigma_c\sigma_d}} i^{-l_a-l_b+l_c+l_d}\hat{J}_i^2\hat{\Lambda}_i^2\hat{\Sigma}_i^2\hat{J}_f^2\hat{\Lambda}_f^2\hat{\Sigma}_f^2\hat{j}_a\hat{j}_b\hat{j}_c\hat{j}_d$$

$$(-)^{j_a-j_b+\Lambda_f+M_{\Lambda_f}+\Sigma_f+M_{\Sigma_f}+l_f+L_f+j_c-j_d+\Lambda_i+M_{\Lambda_i}+\Sigma_i+M_{\Sigma_i}+l_i+L_i}$$

$$(-1)^{l_c+l_d-L_i}\delta_{\tau_a\tau_c}\delta_{\tau_b\tau_d}(-)^{1+\sigma_a+\sigma_b+\mu}\begin{pmatrix}\frac{1}{2} & 1 & \frac{1}{2}\\ \sigma_c & \mu & -\sigma_a\end{pmatrix}\begin{pmatrix}\frac{1}{2} & 1 & \frac{1}{2}\\ \sigma_d & -\mu & -\sigma_b\end{pmatrix}$$

$$\begin{Bmatrix}l_a & l_b & \Lambda_f\\ \frac{1}{2} & \frac{1}{2} & \Sigma_f\\ j_a & j_b & J_f\end{Bmatrix}\begin{Bmatrix}l_c & l_d & \Lambda_i\\ \frac{1}{2} & \frac{1}{2} & \Sigma_i\\ j_c & j_d & J_i\end{Bmatrix}$$

$$\begin{pmatrix}j_a & j_b & J_f\\ m_a & m_b & -M_{J_f}\end{pmatrix}\begin{pmatrix}\Lambda_f & \Sigma_f & J_f\\ M_{\Lambda_f} & M_{\Sigma_f} & -M_{J_f}\end{pmatrix}\begin{pmatrix}\frac{1}{2} & \frac{1}{2} & \Sigma_f\\ \sigma_a & \sigma_b & -M_{\Sigma_f}\end{pmatrix}\begin{pmatrix}L_f & l_f & \Lambda_f\\ M_{L_f} & M_{l_f} & -M_{\Lambda_f}\end{pmatrix}$$

$$\begin{pmatrix}j_c & j_d & J_i\\ m_c & m_d & -M_{J_i}\end{pmatrix}\begin{pmatrix}\Lambda_i & \Sigma_i & J_i\\ M_{\Lambda_i} & M_{\Sigma_i} & -M_{J_i}\end{pmatrix}\begin{pmatrix}\frac{1}{2} & \frac{1}{2} & \Sigma_i\\ \sigma_c & \sigma_d & -M_{\Sigma_i}\end{pmatrix}\begin{pmatrix}L_i & l_i & \Lambda_i\\ M_{L_i} & M_{l_i} & -M_{\Lambda_i}\end{pmatrix}$$

$$\sum_{\substack{n_in_f\\N_iN_f}} M_{\Lambda_f}(N_fL_fn_fl_f;n_al_an_bl_b)M_{\Lambda_i}(N_iL_in_il_i;n_cl_cn_dl_d)$$

$$\int \mathrm{d}3RR_{N_fL_f}(\sqrt{2}\alpha R)Y^*_{L_fM_{L_f}}(\hat{R})g(\boldsymbol{R})R_{N_iL_i}(\sqrt{2}\alpha R)Y_{L_iM_{L_i}}(\hat{R})$$

$$\int \mathrm{d}3r\mathrm{d}3r'\,R_{n_fl_f}(\alpha r')Y^*_{l_fM_{l_f}}(\hat{r}')v(\boldsymbol{r}',\boldsymbol{r})R_{n_il_i}(\alpha r)Y_{l_iM_{l_i}}(\hat{r})$$

We can add the four $3-j$ coefficients which contain the information on the spin into a $6-j$.

$$\sum_{\substack{\sigma_a\sigma_b\\\sigma_c\sigma_d\\\mu}}(-)^{1+\sigma_a+\sigma_b+\mu}\begin{pmatrix}\frac{1}{2} & 1 & \frac{1}{2}\\ \sigma_c & \mu & -\sigma_a\end{pmatrix}\begin{pmatrix}\frac{1}{2} & 1 & \frac{1}{2}\\ \sigma_d & -\mu & -\sigma_b\end{pmatrix}\begin{pmatrix}\frac{1}{2} & \frac{1}{2} & \Sigma_f\\ \sigma_a & \sigma_b & -M_{\Sigma_f}\end{pmatrix}\begin{pmatrix}\frac{1}{2} & \frac{1}{2} & \Sigma_i\\ \sigma_c & \sigma_d & -M_{\Sigma_i}\end{pmatrix}$$

$$=\sum_{\substack{\sigma_a\sigma_b\\\sigma_c\sigma_d\\\mu}}(-)^{1+\sigma_a+\sigma_b+\mu}\begin{pmatrix}\frac{1}{2} & \frac{1}{2} & \Sigma_f\\ \sigma_a & \sigma_b & -M_{\Sigma_f}\end{pmatrix}\begin{pmatrix}\frac{1}{2} & 1 & \frac{1}{2}\\ -\sigma_b & -\mu & \sigma_d\end{pmatrix}\begin{pmatrix}1 & \frac{1}{2} & \frac{1}{2}\\ \mu & -\sigma_a & \sigma_c\end{pmatrix}\begin{pmatrix}\frac{1}{2} & \frac{1}{2} & \Sigma_i\\ \sigma_d & \sigma_c & -M_{\Sigma_i}\end{pmatrix}$$

$$=\frac{(-)^{1+\Sigma_i}}{\hat{\Sigma}_i}\delta_{\Sigma_i\Sigma_f}\delta_{M_{\Sigma_i}M_{\Sigma_f}}\begin{Bmatrix}\frac{1}{2} & \frac{1}{2} & \Sigma_i\\ \frac{1}{2} & \frac{1}{2} & 1\end{Bmatrix}$$

$$\langle ab|\bar{V}|cd\rangle_\sigma = -\frac{3}{2}\delta_{\tau_a\tau_c}\delta_{\tau_b\tau_d}\sum_{\substack{J_i\Lambda_i\\J_f\Lambda_f}}\sum_{\Sigma M_\Sigma}\sum_{\substack{M_{J_i}M_{\Lambda_i}\\M_{l_i}M_{L_i}}}\sum_{\substack{M_{J_f}M_{\Lambda_f}\\M_{l_f}M_{L_f}}}\sum_{\substack{l_il_f\\L_iL_f}} i^{-l_a-l_b+l_c+l_d}$$

$$(-)^{j_a-j_b+\Lambda_f+M_{\Lambda_f}+l_f+L_f+j_c-j_d+\Lambda_i+M_{\Lambda_i}+l_i+L_i}(-1)^{l_c+l_d-L_i}$$

$$\hat{J}_i^2\hat{\Lambda}_i^2\hat{\Sigma}^2\hat{J}_f^2\hat{\Lambda}_f^2\hat{j}_a\hat{j}_b\hat{j}_c\hat{j}_d \begin{Bmatrix} l_a & l_b & \Lambda_f \\ \frac{1}{2} & \frac{1}{2} & \Sigma \\ j_a & j_b & J_f \end{Bmatrix} \begin{Bmatrix} l_c & l_d & \Lambda_i \\ \frac{1}{2} & \frac{1}{2} & \Sigma \\ j_c & j_d & J_i \end{Bmatrix} (-)^{1+\Sigma} \begin{Bmatrix} \frac{1}{2} & \frac{1}{2} & \Sigma \\ \frac{1}{2} & \frac{1}{2} & 1 \end{Bmatrix}$$

$$\begin{pmatrix} j_a & j_b & J_f \\ m_a & m_b & -M_{J_f} \end{pmatrix} \begin{pmatrix} \Lambda_f & \Sigma & J_f \\ M_{\Lambda_f} & M_\Sigma & -M_{J_f} \end{pmatrix} \begin{pmatrix} L_f & l_f & \Lambda_f \\ M_{L_f} & M_{l_f} & -M_{\Lambda_f} \end{pmatrix}$$

$$\begin{pmatrix} j_c & j_d & J_i \\ m_c & m_d & -M_{J_i} \end{pmatrix} \begin{pmatrix} \Lambda_i & \Sigma & J_i \\ M_{\Lambda_i} & M_\Sigma & -M_{J_i} \end{pmatrix} \begin{pmatrix} L_i & l_i & \Lambda_i \\ M_{L_i} & M_{l_i} & -M_{\Lambda_i} \end{pmatrix}$$

$$\sum_{\substack{n_i n_f \\ N_i N_f}} M_{\Lambda_f}(N_f L_f n_f l_f; n_a l_a n_b l_b) M_{\Lambda_i}(N_i L_i n_i l_i; n_c l_c n_d l_d)$$

$$\int \mathrm{d}_3 R R_{N_f L_f}(\sqrt{2}\alpha R) Y^*_{L_f M_{L_f}}(\hat{R}) g(\boldsymbol{R}) R_{N_i L_i}(\sqrt{2}\alpha R) Y_{L_i M_{L_i}}(\hat{R})$$

$$\int \mathrm{d}_3 r \mathrm{d}_3 r' \, R_{n_f l_f}(\alpha r') Y^*_{l_f M_{l_f}}(\hat{r}') v(\boldsymbol{r}', \boldsymbol{r}) R_{n_i l_i}(\alpha r) Y_{l_i M_{l_i}}(\hat{r})$$

### $\langle ab|\bar{V}|cd\rangle_\tau$

The only difference between this term and $\langle ab|\bar{V}|cd\rangle_0$ is the isospin part, then

$$\langle ab|\bar{V}|cd\rangle_\tau = -\frac{3}{2}\sum_\mu (-)^{1+\tau_a+\tau_b+\mu} \begin{pmatrix} \frac{1}{2} & 1 & \frac{1}{2} \\ \tau_c & \mu & -\tau_a \end{pmatrix} \begin{pmatrix} \frac{1}{2} & 1 & \frac{1}{2} \\ \tau_d & -\mu & -\tau_b \end{pmatrix}$$

$$\sum_{\substack{J_i \Lambda_i \\ J_f \Lambda_f}} \sum_{\Sigma M_\Sigma} \sum_{\substack{M_{J_i} M_{\Lambda_i} \\ M_{l_i} M_{L_i}}} \sum_{\substack{M_{J_f} M_{\Lambda_f} \\ M_{l_f} M_{L_f}}} \sum_{\substack{l_i l_f \\ L_i L_f}} i^{-l_a-l_b+l_c+l_d} \hat{J}_i^2\hat{\Lambda}_i^2\hat{\Sigma}^2\hat{J}_f^2\hat{\Lambda}_f^2\hat{j}_a\hat{j}_b\hat{j}_c\hat{j}_d$$

$$(-)^{j_a-j_b+\Lambda_f+M_{\Lambda_f}+l_f+L_f+j_c-j_d+\Lambda_i+M_{\Lambda_i}+l_i+L_i}$$

$$(-1)^{l_c+l_d-L_i} \begin{Bmatrix} l_a & l_b & \Lambda_f \\ \frac{1}{2} & \frac{1}{2} & \Sigma_f \\ j_a & j_b & J_f \end{Bmatrix} \begin{Bmatrix} l_c & l_d & \Lambda_i \\ \frac{1}{2} & \frac{1}{2} & \Sigma_i \\ j_c & j_d & J_i \end{Bmatrix}$$

$$\begin{pmatrix} j_a & j_b & J_f \\ m_a & m_b & -M_{J_f} \end{pmatrix} \begin{pmatrix} \Lambda_f & \Sigma & J_f \\ M_{\Lambda_f} & M_\Sigma & -M_{J_f} \end{pmatrix} \begin{pmatrix} L_f & l_f & \Lambda_f \\ M_{L_f} & M_{l_f} & -M_{\Lambda_f} \end{pmatrix}$$

$$\begin{pmatrix} j_c & j_d & J_i \\ m_c & m_d & -M_{J_i} \end{pmatrix} \begin{pmatrix} \Lambda_i & \Sigma & J_i \\ M_{\Lambda_i} & M_\Sigma & -M_{J_i} \end{pmatrix} \begin{pmatrix} L_i & l_i & \Lambda_i \\ M_{L_i} & M_{l_i} & -M_{\Lambda_i} \end{pmatrix}$$

$$\sum_{\substack{n_i n_f \\ N_i N_f}} M_{\Lambda_f}(N_f L_f n_f l_f; n_a l_a n_b l_b) M_{\Lambda_i}(N_i L_i n_i l_i; n_c l_c n_d l_d)$$

$$\int \mathrm{d}_3 R R_{N_f L_f}(\sqrt{2}\alpha R) Y^*_{L_f M_{L_f}}(\hat{R}) g(\boldsymbol{R}) R_{N_i L_i}(\sqrt{2}\alpha R) Y_{L_i M_{L_i}}(\hat{R})$$

$$\int \mathrm{d}_3 r \mathrm{d}_3 r' \, R_{n_f l_f}(\alpha r') Y^*_{l_f M_{l_f}}(\hat{r}') v(\boldsymbol{r}', \boldsymbol{r}) R_{n_i l_i}(\alpha r) Y_{l_i M_{l_i}}(\hat{r})$$

### $\langle ab|\bar{V}|cd\rangle_{\sigma\tau}$

The only difference between this term and $\langle ab|\bar{V}|cd\rangle_{\sigma}$ is the isospin part, then

$$\langle ab|\bar{V}|cd\rangle_{\sigma\tau} = -9\sum_{\mu}(-)^{1+\tau_a+\tau_b+\mu}\begin{pmatrix}\frac{1}{2} & 1 & \frac{1}{2}\\ \tau_c & \mu & -\tau_a\end{pmatrix}\begin{pmatrix}\frac{1}{2} & 1 & \frac{1}{2}\\ \tau_d & -\mu & -\tau_b\end{pmatrix}$$
$$\sum_{\substack{J_i\Lambda_i\\ J_f\Lambda_f}}\sum_{\Sigma M_\Sigma}\sum_{\substack{M_{J_i}M_{\Lambda_i}\\ M_{l_i}M_{L_i}}}\sum_{\substack{M_{J_f}M_{\Lambda_f}\\ M_{l_f}M_{L_f}}}\sum_{\substack{l_il_f\\ L_iL_f}} i^{-l_a-l_b+l_c+l_d}$$
$$(-)^{j_a-j_b+\Lambda_f+M_{\Lambda_f}+l_f+L_f+j_c-j_d+\Lambda_i+M_{\Lambda_i}+l_i+L_i}(-1)^{l_c+l_d-L_i}$$
$$\hat{J}_i^2\hat{\Lambda}_i^2\hat{\Sigma}^2\hat{J}_f^2\hat{\Lambda}_f^2\hat{j}_a\hat{j}_b\hat{j}_c\hat{j}_d\begin{Bmatrix}l_a & l_b & \Lambda_f\\ \frac{1}{2} & \frac{1}{2} & \Sigma\\ j_a & j_b & J_f\end{Bmatrix}\begin{Bmatrix}l_c & l_d & \Lambda_i\\ \frac{1}{2} & \frac{1}{2} & \Sigma\\ j_c & j_d & J_i\end{Bmatrix}(-)^{1+\Sigma}\begin{Bmatrix}\frac{1}{2} & \frac{1}{2} & \Sigma\\ \frac{1}{2} & \frac{1}{2} & 1\end{Bmatrix}$$
$$\begin{pmatrix}j_a & j_b & J_f\\ m_a & m_b & -M_{J_f}\end{pmatrix}\begin{pmatrix}\Lambda_f & \Sigma & J_f\\ M_{\Lambda_f} & M_\Sigma & -M_{J_f}\end{pmatrix}\begin{pmatrix}L_f & l_f & \Lambda_f\\ M_{L_f} & M_{l_f} & -M_{\Lambda_f}\end{pmatrix}$$
$$\begin{pmatrix}j_c & j_d & J_i\\ m_c & m_d & -M_{J_i}\end{pmatrix}\begin{pmatrix}\Lambda_i & \Sigma & J_i\\ M_{\Lambda_i} & M_\Sigma & -M_{J_i}\end{pmatrix}\begin{pmatrix}L_i & l_i & \Lambda_i\\ M_{L_i} & M_{l_i} & -M_{\Lambda_i}\end{pmatrix}$$
$$\sum_{\substack{n_in_f\\ N_iN_f}} M_{\Lambda_f}(N_fL_fn_fl_f;n_al_an_bl_b)M_{\Lambda_i}(N_iL_in_il_i;n_cl_cn_dl_d)$$
$$\int \mathrm{d}_3RR R_{N_fL_f}(\sqrt{2}\alpha R)Y^*_{L_fM_{L_f}}(\hat{R})g(\boldsymbol{R})R_{N_iL_i}(\sqrt{2}\alpha R)Y_{L_iM_{L_i}}(\hat{R})$$
$$\int \mathrm{d}_3r\mathrm{d}_3r'\,R_{n_fl_f}(\alpha r')Y^*_{l_fM_{l_f}}(\hat{r}')v(\boldsymbol{r}',\boldsymbol{r})R_{n_il_i}(\alpha r)Y_{l_iM_{l_i}}(\hat{r})$$

In the following, we use also the notation

$$\mathscr{F}(\tau) = \begin{cases}\delta_{\tau_a\tau_c}\delta_{\tau_b\tau_d} & \text{for } \langle ab|\bar{V}|cd\rangle_{0,\sigma}\\ \sum_\mu(-)^{1+\tau_a+\tau_b+\mu}\begin{pmatrix}\frac{1}{2} & 1 & \frac{1}{2}\\ \tau_c & \mu & -\tau_a\end{pmatrix}\begin{pmatrix}\frac{1}{2} & 1 & \frac{1}{2}\\ \tau_d & -\mu & -\tau_b\end{pmatrix} & \text{for } \langle ab|\bar{V}|cd\rangle_{\tau,\sigma\tau}\end{cases}$$

$$\mathscr{G}(\Sigma) = \begin{cases}1 & \text{for } \langle ab|\bar{V}|cd\rangle_{0,\tau}\\ (-)^{1+\Sigma}\begin{Bmatrix}\frac{1}{2} & \frac{1}{2} & \Sigma\\ \frac{1}{2} & \frac{1}{2} & 1\end{Bmatrix} & \text{for } \langle ab|\bar{V}|cd\rangle_{\sigma,\sigma\tau}\end{cases}$$

$$\mathscr{N} = \begin{cases}1 & \text{for } \langle ab|\bar{V}|cd\rangle_{0}\\ -\frac{3}{2} & \text{for } \langle ab|\bar{V}|cd\rangle_{\sigma,\tau}\\ -9 & \text{for } \langle ab|\bar{V}|cd\rangle_{\sigma\tau}\end{cases}$$

$$\mathscr{M}(L_i) = \begin{cases}\left(1-\frac{1}{4}(-1)^{l_c+l_d-L_i}\right) & \text{for } \langle ab|\bar{V}|cd\rangle_{0}\\ (-1)^{l_c+l_d-L_i} & \text{fpr } \langle ab|\bar{V}|cd\rangle_{\sigma,\tau,\sigma\tau}\end{cases}$$

$$
\begin{aligned}
\langle ab|\bar{V}|cd\rangle_{0,\sigma,\tau,\sigma\tau} = {} & \mathcal{N}\mathcal{F}(\tau) \sum_{\substack{J_i\Lambda_i\\ J_f\Lambda_f}} \sum_{\Sigma M_\Sigma} \sum_{\substack{M_{J_i}M_{\Lambda_i}\\ M_{l_i}M_{L_i}}} \sum_{\substack{M_{J_f}M_{\Lambda_f}\\ M_{l_f}M_{L_f}}} \sum_{\substack{l_il_f\\ L_iL_f}} \\
& i^{-l_a-l_b+l_c+l_d} \hat{J}_i^2\hat{\Lambda}_i^2\hat{\Sigma}^2\hat{J}_f^2\hat{\Lambda}_f^2 \hat{j}_a\hat{j}_b\hat{j}_c\hat{j}_d \begin{Bmatrix} l_a & l_b & \Lambda_f \\ \frac{1}{2} & \frac{1}{2} & \Sigma \\ j_a & j_b & J_f \end{Bmatrix} \begin{Bmatrix} l_c & l_d & \Lambda_i \\ \frac{1}{2} & \frac{1}{2} & \Sigma \\ j_c & j_d & J_i \end{Bmatrix} \\
& \mathcal{M}(L_i)(-)^{j_a-j_b+\Lambda_f+M_{\Lambda_f}+l_f+L_f+j_c-j_d+\Lambda_i+M_{\Lambda_i}+l_i+L_i} \\
& \mathcal{G}(\Sigma) \begin{pmatrix} j_a & j_b & J_f \\ m_a & m_b & -M_{J_f} \end{pmatrix} \begin{pmatrix} \Lambda_f & \Sigma & J_f \\ M_{\Lambda_f} & M_\Sigma & -M_{J_f} \end{pmatrix} \begin{pmatrix} L_f & l_f & \Lambda_f \\ M_{L_f} & M_{l_f} & -M_{\Lambda_f} \end{pmatrix} \\
& \begin{pmatrix} j_c & j_d & J_i \\ m_c & m_d & -M_{J_i} \end{pmatrix} \begin{pmatrix} \Lambda_i & \Sigma & J_i \\ M_{\Lambda_i} & M_\Sigma & -M_{J_i} \end{pmatrix} \begin{pmatrix} L_i & l_i & \Lambda_i \\ M_{L_i} & M_{l_i} & -M_{\Lambda_i} \end{pmatrix} \\
& \sum_{\substack{n_in_f\\ N_iN_f}} M_{\Lambda_f}(N_fL_fn_fl_f;n_al_an_bl_b)M_{\Lambda_i}(N_iL_in_il_i;n_cl_cn_dl_d) \\
& \int \mathrm{d}RR^2R_{N_fL_f}(\sqrt{2}\alpha R)(\hat{R})g(R)R_{N_iL_i}(\sqrt{2}\alpha R) \int \mathrm{d}\hat{R}\, Y^*_{L_fM_{L_f}}(\hat{R})Y_{L_iM_{L_i}}(\hat{R}) \\
& \sum_{lm} \frac{(-)^l}{\hat{l}} \int \mathrm{d}r\mathrm{d}r'\, r^2r'^2R_{n_fl_f}(\alpha r')v_{lm}(r',r)R_{n_il_i}(\alpha r) \\
& \int \mathrm{d}\hat{r}\, Y_{l_iM_{l_i}}(\hat{r})Y^*_{lm}(\hat{r}) \int \mathrm{d}\hat{r}' Y^*_{l_fM_{l_f}}(\hat{r}')Y_{lm}(\hat{r}') \\
= {} & \mathcal{N}\mathcal{F}(\tau) \sum_{\substack{J_iJ_f\\ M_{J_i}M_{J_f}}} \sum_{\substack{\Lambda_i\Lambda_f\\ M_{\Lambda_i}M_{\Lambda_f}}} \sum_{\Sigma M_\Sigma} \sum_{LM_L} \sum_{lm} \\
& i^{-l_a-l_b+l_c+l_d} \frac{\hat{J}_i^2\hat{\Lambda}_i^2\hat{J}_f^2\hat{\Lambda}_f^2\hat{\Sigma}^2\hat{j}_a\hat{j}_b\hat{j}_c\hat{j}_d}{\hat{l}} \mathcal{G}(\Sigma) \begin{Bmatrix} l_a & l_b & \Lambda_f \\ \frac{1}{2} & \frac{1}{2} & \Sigma \\ j_a & j_b & J_f \end{Bmatrix} \begin{Bmatrix} l_c & l_d & \Lambda_i \\ \frac{1}{2} & \frac{1}{2} & \Sigma \\ j_c & j_d & J_i \end{Bmatrix} \\
& \mathcal{M}(L)(-)^{j_a-j_b+j_c-j_d+\Lambda_f+M_{\Lambda_f}+\Lambda_i+M_{\Lambda_i}}(-)^l \\
& \begin{pmatrix} j_a & j_b & J_f \\ m_a & m_b & -M_{J_f} \end{pmatrix} \begin{pmatrix} \Lambda_f & \Sigma & J_f \\ M_{\Lambda_f} & M_\Sigma & -M_{J_f} \end{pmatrix} \begin{pmatrix} L & l & \Lambda_f \\ M_L & m & -M_{\Lambda_f} \end{pmatrix} \\
& \begin{pmatrix} j_c & j_d & J_i \\ m_c & m_d & -M_{J_i} \end{pmatrix} \begin{pmatrix} \Lambda_i & \Sigma & J_i \\ M_{\Lambda_i} & M_\Sigma & -M_{J_i} \end{pmatrix} \begin{pmatrix} L & l & \Lambda_i \\ M_L & m & -M_{\Lambda_i} \end{pmatrix} \\
& \sum_{\substack{n_iN_i\\ n_fN_f}} M_{\Lambda_f}(N_fLn_fl;n_al_an_bl_b)M_{\Lambda_i}(N_iLn_il;n_cl_cn_dl_d) \\
& \int \mathrm{d}RR^2R_{N_fL}(\sqrt{2}\alpha R)g(R)R_{N_iL}(\sqrt{2}\alpha R) \int \mathrm{d}r\mathrm{d}r'\, r^2r'^2R_{n_fl}(\alpha r')v_{lm}(r',r)R_{n_il}(\alpha r) \\
= {} & \mathcal{N}\mathcal{F}(\tau) \sum_{\substack{J_iJ_f\\ M_{J_i}M_{J_f}}} \sum_{\Lambda M_\Lambda} \sum_{\Sigma M_\Sigma} \sum_{Ll} i^{-l_a-l_b+l_c+l_d} \mathcal{M}(L)(-)^{j_a-j_b+j_c-j_d}(-)^l
\end{aligned}
$$

$$\frac{\hat{J}_i^2\hat{J}_f^2\hat{\Lambda}^2\hat{\Sigma}^2\hat{j}_a\hat{j}_b\hat{j}_c\hat{j}_d}{\hat{l}}\mathscr{G}(\Sigma)\begin{Bmatrix} l_a & l_b & \Lambda \\ \frac{1}{2} & \frac{1}{2} & \Sigma \\ j_a & j_b & J_f \end{Bmatrix}\begin{Bmatrix} l_c & l_d & \Lambda \\ \frac{1}{2} & \frac{1}{2} & \Sigma \\ j_c & j_d & J_i \end{Bmatrix}$$

$$\begin{pmatrix} j_a & j_b & J_f \\ m_a & m_b & -M_{J_f} \end{pmatrix}\begin{pmatrix} j_c & j_d & J_i \\ m_c & m_d & -M_{J_i} \end{pmatrix}\begin{pmatrix} \Lambda & \Sigma & J_f \\ M_\Lambda & M_\Sigma & -M_{J_f} \end{pmatrix}\begin{pmatrix} \Lambda & \Sigma & J_i \\ M_\Lambda & M_\Sigma & -M_{J_i} \end{pmatrix}$$

$$\sum_{\substack{n_iN_i \\ n_fN_f}} M_\Lambda(N_fLn_fl;n_al_an_bl_b)M_\Lambda(N_iLn_il;n_cl_cn_dl_d)$$

$$\int \mathrm{d}RR^2R_{N_fL}(\sqrt{2}\alpha R)g(R)R_{N_iL}(\sqrt{2}\alpha R)\int \mathrm{d}r\mathrm{d}r'\, r^2r'^2R_{n_fl}(\alpha r')v_{lm}(r',r)R_{n_il}(\alpha r)$$

$$=\mathscr{N}\mathscr{F}(\tau)\sum_{JM_J}\sum_{\substack{\Lambda\Sigma \\ Ll}} i^{-l_a-l_b+l_c+l_d}(-)^{j_a-j_b+j_c-j_d}(-)^l\mathscr{G}(\Sigma)\begin{Bmatrix} l_a & l_b & \Lambda \\ \frac{1}{2} & \frac{1}{2} & \Sigma \\ j_a & j_b & J \end{Bmatrix}\begin{Bmatrix} l_c & l_d & \Lambda \\ \frac{1}{2} & \frac{1}{2} & \Sigma \\ j_c & j_d & J \end{Bmatrix}$$

$$\mathscr{M}(L)\frac{\hat{J}^2\hat{\Lambda}^2\hat{\Sigma}^2\hat{j}_a\hat{j}_b\hat{j}_c\hat{j}_d}{\hat{l}}\begin{pmatrix} j_a & j_b & J \\ m_a & m_b & -M_J \end{pmatrix}\begin{pmatrix} j_c & j_d & J \\ m_c & m_d & -M_J \end{pmatrix}$$

$$\sum_{\substack{n_iN_i \\ n_fN_f}} M_\Lambda(N_fLn_fl;n_al_an_bl_b)M_\Lambda(N_iLn_il;n_cl_cn_dl_d)$$

$$\int \mathrm{d}RR^2R_{N_fL}(\sqrt{2}\alpha R)g(R)R_{N_iL}(\sqrt{2}\alpha R)\int \mathrm{d}r\mathrm{d}r'\, r^2r'^2R_{n_fl}(\alpha r')v_{lm}(r',r)R_{n_il}(\alpha r),$$

which is Eq. (10.29).

Appendix J

# Multipole expansion of functions

## J.1 Functions which depend on one vector

A collection of spherical harmonics $Y_{lm}$, if $l$ is integer and non-negative and $|m| < l$, constitute a complete and orthonormal set of functions of the two variables $(\theta, \varphi)$, with $\theta \in [0, \pi]$ and $\varphi \in [0, 2\pi)$. Because of this, a generic function $f(r, \theta, \varphi)$ which is defined in the interval $0 \le \theta \le \pi$, $0 \le \varphi < 2\pi$ and satisfies the condition

$$\int_0^{2\pi} \mathrm{d}\varphi \int_0^{\pi} \mathrm{d}\theta \, \sin\theta |f(r, \theta, \varphi)|^2 < +\infty$$

can be expanded in series of spherical harmonics

$$f(r, \theta, \varphi) = \sum_{l=0}^{+\infty} \sum_{m=-l}^{l} f_{lm}(r) Y_{lm}(\theta, \varphi).$$

The expansion coefficients are given by the relation

$$f_{lm}(r) = \int_0^{2\pi} \mathrm{d}\varphi \int_0^{\pi} \mathrm{d}\theta \, \sin\theta f(r, \theta, \varphi) Y^*_{lm}(\theta, \varphi).$$

### J.1.1 The multipole expansion of the Dirac delta

An example is the function $\delta_3(\boldsymbol{r})$. In spherical coordinates it can be written as

$$\delta_3(\boldsymbol{r}) = \frac{1}{r^2 \sin\theta} \delta(r)\delta(\theta)\delta(\varphi)$$

In this case, the $(lm)$-coefficients are

$$\begin{aligned}\delta_3(\boldsymbol{r}) &= \frac{\delta(r)}{r^2} \int_0^{2\pi} \mathrm{d}\varphi \int_0^{\pi} \sin\theta \mathrm{d}\theta \, \frac{1}{\sin\theta} Y^*_{lm}(\theta, \varphi)\delta(\theta)\delta(\varphi) = \frac{\delta(r)}{r^2} Y_{lm}(0, 0) \\ &= \sqrt{\frac{2l+1}{4\pi}} \frac{\delta(r)}{r^2} \delta_{m,0}\end{aligned}$$

Thus,

$$\begin{aligned}\delta_3(\boldsymbol{r}) &= \frac{\delta(r)}{r^2} \sum_{lm} \sqrt{\frac{2l+1}{4\pi}} \delta_{m,0} Y^*_{lm}(\theta, \varphi) = \frac{\delta(r)}{r^2} \sum_{l} \sqrt{\frac{2l+1}{4\pi}} Y_{l0}(\theta, \varphi) \\ &= \frac{1}{2\pi} \frac{\delta(r)\delta(\cos\theta - 1)}{r^2},\end{aligned}$$

having used the relation

$$\sum_l \sqrt{2l+1} Y_{l0}(\theta,\varphi) = \frac{1}{\sqrt{\pi}}\delta(\cos\theta - 1)$$

## J.2 Functions which depend on two vectors

Similarly, any function $f(\boldsymbol{r_1}, \boldsymbol{r_2})$ which depends on two arbitrary vectors $\boldsymbol{r_1}(r_1, \theta_1, \varphi_1)$ and $\boldsymbol{r_2}(r_2, \theta_2, \varphi_2)$ can be expanded in series of bipolar harmonics

$$\begin{aligned} f(\boldsymbol{r_1}, \boldsymbol{r_2}) &= \sum_{\substack{l_1 l_2 \\ LM}} f^{LM}_{l_1,l_2}(r_1, r_2) \left\{ \boldsymbol{Y}_{l_1}(\theta_1,\varphi_1) \otimes \boldsymbol{Y}_{l_2}(\theta_2,\varphi_2) \right\}_{LM} \\ &= \sum_{\substack{l_1 l_2 \\ LM}} f^{LM}_{l_1,l_2}(r_1, r_2) \sum_{m_1,m_2} \langle l_1 m_1 l_2 m_2 | LM \rangle Y_{l_1 m_1}(\theta_1,\varphi_1) Y_{l_2 m_2}(\theta_2,\varphi_2). \end{aligned}$$

The coefficients $f^{LM}_{l_1,l_2}(r_1, r_2)$ are given accordingly by the relation

$$f^{LM}_{l_1,l_2}(r_1, r_2) = \int d\Omega_1 \int d\Omega_2\, f(\boldsymbol{r_1}, \boldsymbol{r_2}) \left\{ \boldsymbol{Y}_{l_1}(\Omega_1) \otimes \boldsymbol{Y}_{l_2}(\Omega_2) \right\}^*_{LM}$$

The most interesting case happen when $f(\boldsymbol{r_1}, \boldsymbol{r_2})$ are invariant under rotation of coordinate system. The expansions of these functions contain only the bipolar harmonics of zero rank ($L = 0$). In particular,

$$\begin{aligned} \left\{ \boldsymbol{Y}_{l_1}(\theta_1,\varphi_1) \otimes \boldsymbol{Y}_{l_2}(\theta_2,\varphi_2) \right\}_{00} &= \sum_{m_1,m_2} \langle l_1 m_1 l_2 m_2 | 00 \rangle Y_{l_1 m_1}(\theta_1,\varphi_1) Y_{l_2 m_2}(\theta_2,\varphi_2) \\ &= \sum_m \frac{(-)^{l_1}}{\sqrt{2l_1+1}} \delta_{l_1,l_2} Y_{l_1 m}(\theta_1,\varphi_1) Y^*_{l_2 m}(\theta_2,\varphi_2) \end{aligned}$$

Therefore,

$$\begin{aligned} f(\boldsymbol{r_1}, \boldsymbol{r_2}) &= \sum_{l_1 l_2} f^{00}_{l_1,l_2}(r_1, r_2) \sum_m \frac{(-)^{l_1}}{\sqrt{2l_1+1}} \delta_{l_1,l_2} Y_{l_1 m}(\theta_1,\varphi_1) Y^*_{l_2 m}(\theta_2,\varphi_2) \\ &= \sum_{lm} \frac{(-)^l f_{lm}(r_1, r_2)}{\sqrt{2l+1}} Y_{lm}(\theta_1,\varphi_1) Y^*_{lm}(\theta_2,\varphi_2), \end{aligned}$$

with

$$f_{lm}(r_1, r_2) = \sum_m \int d\Omega_1 \int d\Omega_2 \frac{(-)^l}{\sqrt{2l+1}} f(\boldsymbol{r_1}, \boldsymbol{r_2}) Y^*_{lm}(\Omega_1) Y_{lm}(\Omega_2).$$

### J.2.1 The multipole expansion of the Dirac delta

As an example, we consider the function $\delta_3(\boldsymbol{r} - \boldsymbol{r}')$, we have first to write it in spherical coordinates:

$$\delta_3(\boldsymbol{r} - \boldsymbol{r}') = \frac{1}{r^2 \sin\theta} \delta(r - r') \delta(\theta - \theta') \delta(\varphi - \varphi')$$

The $(lm)$-coefficients of the Dirac delta read

$$\begin{aligned}\delta_{lm}(r-r') =&\frac{\delta(r-r')}{r^2}\sum_m \int_0^{2\pi}\mathrm{d}\varphi \int_0^{\pi}\mathrm{d}\theta \sin\theta \int_0^{2\pi}\mathrm{d}\varphi' \int_0^{\pi}\mathrm{d}\theta' \sin\theta' \frac{(-)^l}{\sqrt{2l+1}} \\ &\frac{1}{\sin\theta}\delta(\theta-\theta')\delta(\varphi-\varphi')Y^*_{lm}(\theta,\varphi)Y_{lm}(\theta',\varphi') \\ =&\frac{\delta(r-r')}{r^2}\sum_m \int_0^{2\pi}\mathrm{d}\varphi \int_0^{\pi}\mathrm{d}\theta \sin\theta \frac{(-)^l}{\sqrt{2l+1}} Y^*_{lm}(\theta,\varphi)Y_{lm}(\theta,\varphi) \\ =&\frac{\delta(r-r')}{r^2}\sum_m \delta_{mm}\frac{(-)^l}{\sqrt{2l+1}} \\ =&\frac{\delta(r-r')}{r^2}(-)^l\sqrt{2l+1}\end{aligned}$$

Therefore,

$$\delta_3(\boldsymbol{r}-\boldsymbol{r}') = \frac{\delta(r-r')}{r^2}\sum_{lm} Y_{lm}(\theta,\varphi)Y^*_{lm}(\theta',\varphi')$$